\pdfoutput=1 
\documentclass{these-dbl}
\citeToolTip{false} 

\hypersetup{
   pdfauthor   = {Pierre Champion},
   pdftitle    = {PhD Pierre Champion},
   pdfsubject  = {Th\`{e}se de doctorat de Pierre Champion},
}

\makeglossaries
\newacronym{aan}{ANN}{Artificial Neural Networks}
\newacronym{dnn}{DNN}{Deep Neural Network}
\newacronym{stft}{STFT}{Short-time Fourier Transform}
\newacronym{mfcc}{MFCCs}{Mel Frequency Cepstral Coefficients}
\newacronym{aam}{AAM}{Additive Angular Margin}
\newacronym{f0}{$F_0$}{fundamental frequency}
\newacronym{vpc}{VPC}{VoicePrivacy Challenge}
\newacronym{dct}{DCT}{Discrete Cosine Transform}
\newacronym{mlp}{MLP}{Multilayer perceptron}
\newacronym{tdnn}{TDNN}{Time-Delay Neural Network}
\newacronym{tdnnf}{TDNN-F}{Factored Time-Delay Neural Network}
\newacronym{pca}{PCA}{Principle Component Analysis}
\newacronym{pii}{PII}{Personally Identifiable Information}
\newacronym{svd}{SVD}{Singular Value Decomposition}
\newacronym{gan}{GAN}{Generative Adversarial Network}
\newacronym{asr}{ASR}{Automatic Speech Recognition}
\newacronym{asv}{ASV}{Automatic Speaker Verification}
\newacronym{hmm}{HMM}{Hidden Markov Models}
\newacronym{gmm}{GMM}{Gaussian Mixture Models}
\newacronym{wer}{WER}{Word Error Rate}
\newacronym{mmi}{MMI}{Maximum Mutual Information}
\newacronym{lfmmi}{LF-MMI}{Lattice-Free Maximum Mutual Information}
\newacronym{eelfmmi}{E2E-LF-MMI}{End-to-End Lattice-Free Maximum Mutual Information}
\newacronym{plda}{PLDA}{Probabilistic Linear Discriminant Analysis}
\newacronym{eer}{EER}{Equal Error Rate}
\newacronym{far}{FAR}{False Acceptance Rate}
\newacronym{frr}{FRR}{False Rejection Rate}
\newacronym{li}{$D_{\leftrightarrow}^{\mathrm{sys}}$}{Linkability measure}
\newacronym{vc}{VC}{Voice Conversion}
\newacronym{tts}{TTS}{Text-To-Speech}
\newacronym{asrbn}{ASR-BN}{bottleneck from automatic speech recognition}
\newacronym{vae}{VAE}{Variational Autoencoders}
\newacronym{ppg}{PPG}{Phonetic Posteriorgrams}
\newacronym{mos}{MOS}{Mean Opinion Score}
\newacronym{pav}{PAV}{Pool Adjacent Violators}
\newacronym{cllr}{Cllr}{log-likelihood-ratio cost function}
\newacronym{gdpr}{GDPR}{General Data Protection Regulation}
\newacronym{awgn}{AWGN}{Additive White Gaussian noise}
\newacronym{dp}{DP}{Differential Privacy}
\newacronym{vq}{VQ}{Vector Quantization}

\addto\captionsenglish{}
\dominitoc

\begin{document}

\renewcommand{\thepage}{\roman{page}}

\selectlanguage{french}


\makeatletter
\newcommand\addcase[3]{\expandafter\def\csname\string#1@case@#2\endcsname{#3}}
\newcommand\makeswitch[2][]{%
  \newcommand#2[1]{%
    \ifcsname\string#2@case@##1\endcsname\csname\string#2@case@##1\endcsname\else#1\fi%
  }%
}
\makeatother

\newcommand\hauteurlogos[3]{
    \hauteurlogoecole{#1}
    \hauteurlogoetablissementA{#2}
    \hauteurlogoetablissementB{#3}
}


\newcommand\addecoledoctorale[5]{\direcole{#1}\numeroecole{#2}\definecolor{color-ecole}{RGB}{#3}\nomecoleA{#4}\nomecoleB{#5}}

\makeswitch[default]\ecoledoctorale{}

\addcase\ecoledoctorale{MathSTIC}{\addecoledoctorale
    {ALL}
    {77}
    {236,115,127}
    {}
    {}
        \hauteurlogos{1.5cm}{1.2cm}{1.5cm}
}


\newcommand\addetablissement[5]{\logoetablissementB{#1}\logoetablissementA{#2}\nometablissementC{#3}\nometablissementD{#4}\nometablissementE{#5}}

\makeswitch[default]\etablissement{}

\addcase\etablissement{tulul}{\addetablissement
    {inria}
    {LMU}
    {}
    {Mémoire présenté en vue de l'obtention du}
    {\textbf{grade de Docteur de l’Université de Lorraine}}
}

\newcommand\addpairetablissements[7]{
    \logoetablissementA{#1}
    \logoetablissementB{#2}
    \nometablissementA{#3}
    \nometablissementB{#4}
    \nometablissementC{#5}
    \nometablissementD{#6}
    \nometablissementE{#7}
}

\addcase\etablissement{UR2-ENSAB}{\addpairetablissements
    {ENSAB}
    {UR2}
    {}
    {L'\'{E}COLE NORMALE SUP\'{E}RIEURE}
    {D'ARCHITECTURE DE BRETAGNE}
    {D\'{E}LIVR\'{E}E CONJOINTEMENT AVEC}
    {L'UNIVERSIT\'{E} DE RENNES 2}
    \hauteurlogos{2cm}{1.2cm}{}
}

\ecoledoctorale{MathSTIC}

\etablissement{tulul}

\spec{Informatique}


\author{Pierre Champion}

\title{Anonymizing Speech: Evaluating and Designing\\ Speaker Anonymization Techniques}
\lesoustitre{Anonymisation de la parole~:~Évaluation et Conception\\ de Techniques d'Anonymisation du Locuteur}

\date{20/04/2023}
\lieu{Nancy}

\uniterecherche{\\Laboratoire Lorrain de Recherche en Informatique et ses Applications (LORIA) - UMR~7503\\Laboratoire d'Informatique de l'Université du Mans (LIUM) - EA~4023}

\numthese{} 

\jury{

\vspace{\baselineskip}
{\normalTwelve \textbf{Encadrants :}}\\ \newline
\footnotesizeTwelve
\begin{tabular}{@{}lll}
    Dir. de th\`{e}se :    & Slim Ouni & Ma\^{i}tre de conférences, Nancy Université de Lorraine/LORIA\\
    Co-encadrant :    & Denis Jouvet & Directeur de recherche, Nancy INRIA/LORIA\\
    Co-dir. de th\`{e}se : & Anthony Larcher & Professeur, Le Mans Université LIUM\\
\end{tabular}

\vspace{\baselineskip}
{\normalTwelve \textbf{Composition du Jury :}}\\ \newline
\footnotesizeTwelve
\begin{tabular}{@{}lll}

Pr\'{e}sident :        & Lori Lamel & Directrice de recherche, Université Paris-Scalay\\
Rapporteurs :          & Luciana Ferrer & Chargée de Recherche, University of Buenos Aires   \\
                       & Luk\'a\v{s} Burget & Associate professor, Brno University of Technology  \\
Examinateurs :         & Jean-Francois Bonastre & Professeur, Université d'Avignon\\

Encadrants : &      Slim Ouni & Ma\^{i}tre de conférences, Nancy Université de Lorraine/LORIA\\
     & Anthony Larcher & Professeur, Le Mans Université LIUM\\

\end{tabular}

\vspace{\baselineskip}
{\normalTwelve \textbf{Invit\'{e}s :}}\\ \newline
\footnotesizeTwelve
\begin{tabular}{@{}ll}
    Denis Jouvet & Directeur de recherche, Nancy INRIA/LORIA\\
    Nicholas Evans & Professeur, EURECOM \\
\end{tabular}

}

\maketitle

%

\ifSubfilesClassLoaded{}{
	\includegraphics[scale=0.75,page=1]{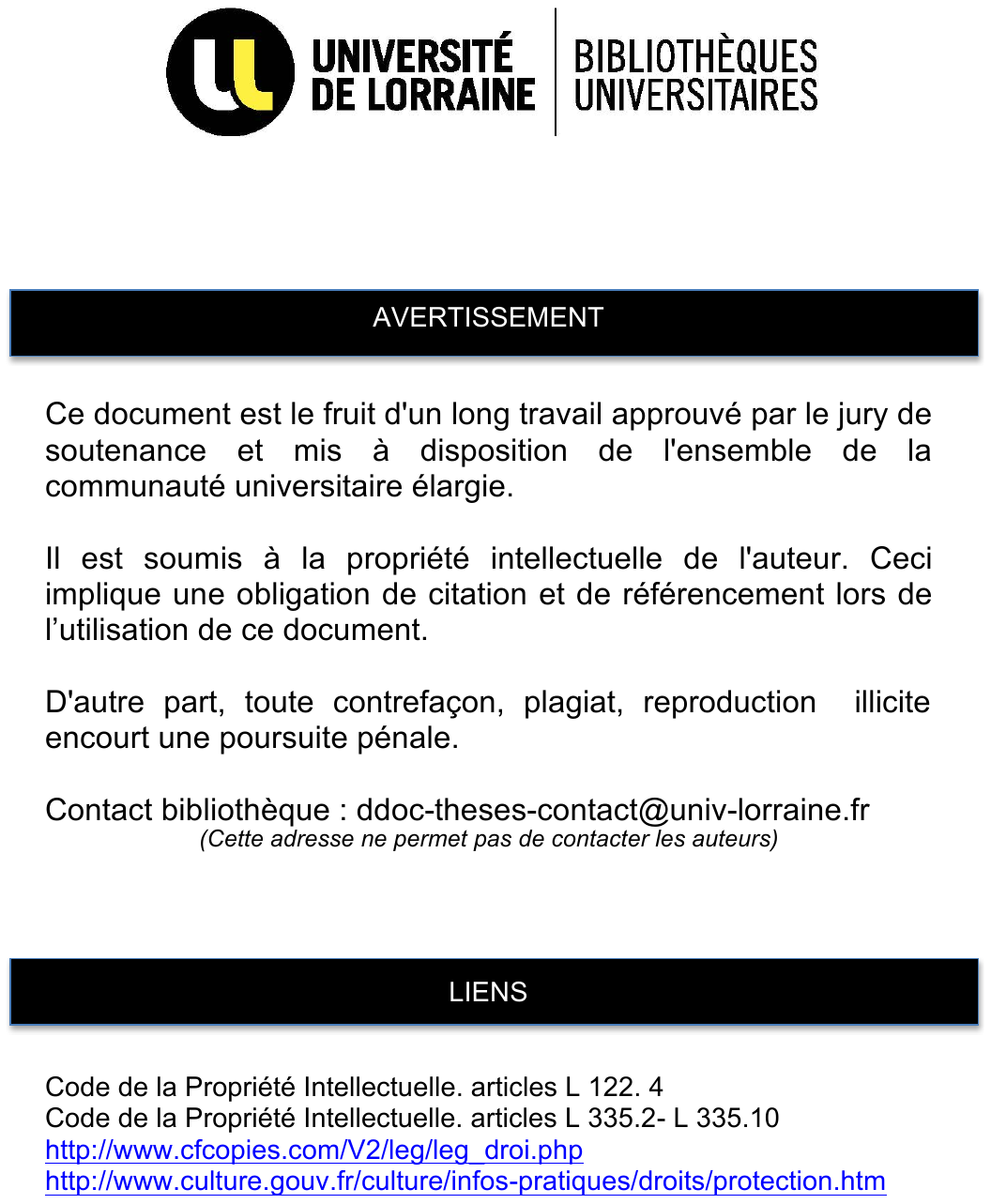}
}
\pagebreak
\vspace*{\fill} 
\begin{flushright}
	{Un hommage chaleureux, six ans plus tard, à la mémoire de Suzanne.}
\end{flushright}

\chapter*{Acknowledgement}

\selectlanguage{french}


\vspace{-0.8mm}
Ce n'est pas tous les jours que nous disposons d'un moment pour remercier les personnes de ce que nous sommes devenus. Je saisis cette occasion pour leur adresser ma gratitude car elles ont joué un rôle dans le fait que je sois arrivé à ce moment de ma vie, à la fois personnelle et scientifique, Merci, je n'en serais pas ici sans vous.

Pour commencer, je remercie Luciana Ferrer et Luk\'a\v{s} Burget, relecteurs de cette thèse, pour le temps qu'ils ont accepté d'y consacrer et pour
les remarques pertinentes et précises qui ont été formulées, qui m'ont permises d'affiner
davantage ce document et mes connaissances.
Je remercie aussi Lori Lamel, Jean-François Bonastre et Nicholas Evans pour avoir accepté de faire partie de mon jury et pour les
questions et commentaires au cours de la soutenance.

Je souhaiterais particulièrement remercier Denis Jouvet et Anthony Larcher, mes encadrants, de m'avoir donné cette opportunité et de m'avoir prodigué des conseils précieux tout au long de cette thèse.
Nos réunions ont toujours été à la fois excessivement bénéfiques pour mon travail et très agréables.
Je tiens aussi à remercier tous les enseignants, depuis le début avec Sylviane Viviant en maternelle, qui ont pris le temps de m'apprendre à lire/écrire (tâche qui s'est avérée bien difficile), et à compter, pour plus tard donné le goût à la science.

Faire partie du Loria/Inria-Nancy a été une formidable expérience. Pour les bons moments passés et les interactions enrichissantes, je tiens à remercier les membres de ce centre de recherche.
En particulier, mes remerciements les plus sincères vont à Noémie et Adrien, mes amis et collègues de bureau.
J'ai eu la chance de tomber sur vous, et je n'aurais pas pu rêver de meilleures personnes avec qui partager mon travail au quotidien, même si nous parlons plus que nous ne travaillons par moment.
Je tiens à remercier chaleureusement tout le personnel du Laboratoire d'Informatique de l'Université du Mans pour leur accueil bienveillant, alors même que ma présence était très sporadique.

Je tiens à exprimer ma gratitude à tous mes amis, du Mans de l'escalade et du lycée, de Nancy. Leur présence dans ma vie a été une source de joie et de soutien inestimable. Leurs rires, leurs encouragements et leur amitié sincère ont illuminé mes journées et m'ont aidé à surmonter les défis. Je suis reconnaissant d'avoir des personnes si merveilleuses à mes côtés.

Je tiens également à exprimer ma profonde gratitude envers ma famille, qui est un pilier solide dans ma vie. Leur amour inconditionnel, leur soutien constant et leurs valeurs transmises ont façonné la personne que je suis aujourd'hui. À travers les hauts et les bas, ils ont toujours été là pour moi, prêts à tendre la main et à m'offrir leur épaule réconfortante.
Je souhaite qu'ils sachent à quel point ils sont importants pour moi. Merci du fond du cœur.

\selectlanguage{english}

\clearemptydoublepage
\frontmatter

\cleartooddpage[\thispagestyle{empty}]
\begin{spacing}{1.3}
\tableofcontents 
\end{spacing}
\begin{mtchideinmaintoc}[-1]  
	\addstarredchapter{Table of Contents}
\end{mtchideinmaintoc} 

\clearemptydoublepage
\cleartooddpage[\thispagestyle{empty}]

\renewcommand{\thepage}{\arabic{page}}
\setcounter{page}{1}

\documentclass[../main.tex]{subfiles}

\ifSubfilesClassLoaded{
    \tableofcontentsfile
    \dominitoc
    \setcounter{chapter}{-1} 
    \externaldocument[]{../main}
    \def\locallabelprefix{intro}
}{}

\begin{document}

\selectlanguage{english}

\chapter*{Introduction}
\chaptermark{Introduction}
\addstarredchapter{Introduction}
\vspace{-1.5em}
\minitoc
\section*{Motivation}
\addcontentsline{toc}{section}{Motivation}
Speaking and listening, or verbal communication is a natural and convenient way for people to interact. During conversations, individuals exchange linguistic information through speech, but they also transmit additional information, such as identity, emotions, age, and gender, through paralinguistic cues. 
Human-machine interaction can benefit from the richness and convenience of speech allowing the machine to better understand humans and humans to easily share information with machines.
However, making the machine understand and reproduce speech remains a complex task that has been the subject of ongoing research for many years.
Still, over the last decade, great progress has been made in several speech tasks such as speech-to-text (automatic speech recognition), text-to-speech (speech synthesis), and more.
With these advancements came the introduction of a new product to the market: voice assistants.

The goal of many companies is to establish a seamless and natural communication experience between humans and machines powered by the latest developments in speech processing technology.
Currently, consumers are embracing various voice assistant devices.
The\citetitle{smart_home_adoption_rate} revealed that 50-60\% of the U.S. population has access to one or many voice assistant devices.

In order for companies to propose competitive services, advanced forms of deep learning techniques are used to power voice assistants.
Those algorithms are data-hungry meaning that the best performances are usually achieved when a large quantity of data is used to train the models.
This created a necessity for companies to collect, process and store the speech data of their user in centralized servers to continuously improve the proposed services and remain competitive.
As speech data contains a lot of personal information such as the identity of the speaker, the process of collecting speech data raises serious privacy concerns.

While the data collection practices of companies were initially unknown or not well understood by the public, this has recently changed.
Around 2017, more and more people became aware of the situation with press headlines\footnote{\citetitle{verge_data_listening}.} disclosing companies usage of the collected speech data.
Simultaneously, the \acrfull{gdpr} established the strictest privacy and security legislation in the world.
The \cite{gdpr} specifically stipulates that European citizens personal information must be handled with the utmost respect and privacy.

Privacy is an individual's right to keep confidential information or data private.
It is the ability to control who can access information about oneself, and for what purpose it is used.
In the context of the \acrshort{gdpr}, privacy is referred to as a fundamental human right, and it is essential for protecting citizens from having their personal information misused or abused.
In today's digital age, privacy is of utmost importance as the growing use of technology enables massive amounts of personal data to be collected and stored.
As technology continues to evolve, so must our understanding of the influence it has on privacy and our commitment to protecting it.

Speech is considered a highly sensitive type of personal data that must be protected.
Recent guidelines provided by the French \textit{Commission Nationale de l'Informatique et des Libertés} and European \textit{Data Protection Board} recalls that speech data is inherently biometric data given Article 4(14)\footnote{Article 4(14) \acrshort{gdpr} defines biometric data as \say{personal data resulting from specific technical processing relating to the physical, physiological or behavioral characteristics of a natural person, which allow or confirm the unique identification of that natural person, such as facial images or dactyloscopy data}.}
of the \acrshort{gdpr} \cite{white-paper-cnil,edpb-voice-assistant-guidelines}.
As the storage and processing of biometric data are even more regulated than that of personal data, it creates a necessity to develop more private data collection schemes.

With the above context about the state of speech and privacy, this thesis is an effort to propose more private speech collection solutions relying on data anonymization.
Data anonymization is the process of removing or altering personal identifying information from data to protect the privacy of individuals.
While the \acrshort{gdpr} does not impose systematic anonymization to personal data,
it is one solution, among others, that enables the processing of personal data in compliance with the rights and privacy of individuals.
In this thesis, we work on speaker anonymization methods that aim to remove speaker identity from speech signals while preserving linguistic content and speech quality.


\subsection*{Broader impact statement}
\addcontentsline{toc}{subsection}{Broader impact statement}
We believe that our work in the field of speaker anonymization is a new and important research domain in today's society.
With the increasing use of speech-enabled technology, there is a growing concern about the privacy and security of the speech of each individual.
Speaker anonymization has the potential to be used in a wide range of applications, such as in call centers, virtual assistants, or any smart devices that share speech information,
to better protect the privacy of individuals. 

\section*{Scope and objectives}
\addcontentsline{toc}{section}{Scope and objectives}

In order for the widespread adoption of privacy-preserving techniques to occur in current data processing pipelines, one of the main requirements deals with the ease of use of the technique.
Indeed, if the privacy-preserving technique to be implemented requires a complete modification of companies current data processing pipelines, it is likely that it will not be implemented.
A few privacy-preserving techniques have been proposed in the last decade.
They include encryption, distributed learning, and anonymization, in the following, we briefly present them to justify our choice to work with anonymization-based techniques.
Encryption solutions \cite{brasser18_interspeech} rely on cryptographic methods to transform clear (personal) data before sending it to the server.
The transformed data being sent is unreadable to any external observer, addressing security constraints.
For privacy constraints, the server should not be able to decrypt the data, which is possible with homomorphic encryption, where for example the only operation possible for the server is a neural network inference \cite{Pathak2012PrivacyPreservingML} on the encrypted data.
Distributed learning methods such as federated learning \cite{Leroy2018FederatedLF} enable multiple data sources to be used in the same learning process, this allows the clear data to remain on the devices instead of being centralized.
As a result, the clear data is not shared, allowing for more secure and private machine learning. 
Anonymization aims to remove personal information from the speech signal to improve privacy while reducing as much as possible the degradation of the other information necessary for other tasks (this is referred to as the utility).

The main advantage of encryption and distributed learning is that they are, today, the most effective techniques to ensure a sufficient level of privacy.
However, their main disadvantage is that the data shared between the client and server is no-longer speech.
In order for companies to implement such techniques, a complete re-work of the data processing pipeline is needed as there is a paradigm shift.
Additionally, it is not practical for companies or research labs to create large speech datasets with those techniques as speech signals are not shared.
For those reasons, anonymization techniques which take speech and produce privacy-preserving speech are quite interesting in the fact that they do not have an impact on current data processing pipelines.   

Over the last few years, speaker anonymization techniques have recently received interest with the release of the \acrlong{vpc}.
Most attempts rely on voice conversion systems to transform the identity of the speaker in the clear speech signal to another one in the anonymized signal.
The goal is that the anonymized signal can no longer be linked back to the real identity of the speaker.
To evaluate the privacy of an anonymization system, automatic speaker verification techniques are used on anonymized speech to assess its degree of linkability, the lower the better.
In contrast, the preservation of the linguistic content (utility) is currently evaluated with automatic speech recognition.

The first challenge of speaker anonymization relates to privacy evaluation.
Privacy evaluation depends on the target speaker identity parameter of voice conversion.
Indeed, voice conversion can increase linkability if requested by making a speaker's voice highly different from others speakers.
As such, there is a necessity to understand voice conversion such that the anonymized voices are not linkable and that the evaluation performed with the automatic speaker verification system corresponds to a privacy evaluation.

A second challenge of speaker anonymization relates to the voice conversion algorithm.
Initially, voice conversions were designed to transform the speaker of a signal, such that subjective (humans) listeners believe a signal was spoken by someone else.
While the quality of the transformation can be sufficient to fool humans, it is not the case for machines.
As such, more advanced forms of voice conversion systems need to be used for speaker anonymization.

A third challenge relates to the diversity and number of possible evaluations to assess privacy and utility.
Indeed there is not a single way to evaluate privacy and utility. 
For example, for privacy, a game involving an attacker and defender, where the attacker aims to break anonymization and the defender improves it has to take place to continuously adapt the technology to current threats.
Whereas for utility, the evaluation always depends on the intended purpose of sharing speech.
We focused on the preservation of the linguistic content, but there are many other valid application cases such as emotion recognition.

In summary, those three challenges give the objective of investigating the evaluation and design of speaker anonymization systems



\vspace{-1em}
\section*{Organization and main contributions of the thesis}
\addcontentsline{toc}{section}{Organization and main contributions of the thesis}

After this introduction, three state-of-the-art chapters present the current level of development in the field of machine learning applied to the speech, then follows three chapters detailing our contributions.
The remaining of the thesis is structured as follows:

\begin{itemize}
    \item In Chapter~\ref{main:chapt1} of the thesis, we introduce the fundamental of speech and how it can be categorized at different levels (e.g., semantic, prosodic, etc.).
    We also discuss various methods for numeric speech processing,
    and present artificial neural network architectures that recently revolutionized the field of speech processing.
        
    \item In Chapter~\ref{main:chapt2}, we focus on various speech-centric machine-learning disciplines that are necessary for building a speaker anonymization system.
    This includes automatic speech recognition, voice conversion, and objective evaluation methods using automatic speech recognition and automatic speaker verification.
        
    \item In Chapter~\ref{main:chapt3}, we delve into the specific topic of speaker anonymization including the current state-of-the-art techniques for removing personal information from speech signals, and the evaluation of these techniques.
    We also discuss the broader context of biometrics security and legislation.

    \item In Chapter~\ref{main:chapt4}, we focus on the role of the target speaker identity parameter in voice conversion-based speaker anonymization.
    Through our analysis, we found that the way current voice conversion-based anonymization systems are parameterized with target speakers creates a bias in privacy evaluation.
    We also analyze the effect of the target speaker on privacy and utility, to answer the question of whether there is a golden target speaker parameter that maximizes performance.
    Overall, our chapter highlights the importance of the target for proper privacy evaluation and utility preservation.
        
    \item In Chapter~\ref{main:chapt5}, we delve deeper into voice conversion techniques for speaker anonymization and carefully analyzed the input features.
    We found through in-depth analysis that traditional methods for disentangling linguistic and fundamental frequency features from speaker information were not fully effective.
    We experimented with adversarial learning to improve the level of disentanglement, however, this method showed limitations when it comes to removing speaker information.
   As a solution, we introduce a new approach using vector quantization to disentangle these features.
   Our results indicate that this method significantly improved privacy performance while producing anonymized speech that sounded more natural compared to noise-based alternatives.

    \item In Chapter~\ref{main:chapt6}, we present new measurement metrics and techniques for evaluating the privacy and utility of speaker anonymization systems.
    Specifically, we propose an invertibility attack and measurement that assesses the ability to reconstruct clear x-vectors from anonymized ones.
    We also propose a new utility measurement that aims to better evaluate the preservation of linguistic content after anonymization by taking into account only mispronunciation errors, rather than inherent recognition decoding errors.
\end{itemize}

This thesis was funded by the \href{https://anr.fr/Projet-ANR-18-CE23-0018}{ANR project DEEP-PRIVACY}, which promotes the development of distributed, personalized and privacy-preserving approaches for speech recognition and Région Grand Est.
We thank them for their financial support, which allowed us to conduct the research and experiments necessary to complete this work.

\ifSubfilesClassLoaded{
	\printglossary[title=Special Terms,type=\acronymtype]
	\printbibliography
}{}

\end{document}

\clearemptydoublepage
\mainmatter

\part{State of the art}

\clearemptydoublepage
\cleartooddpage[\thispagestyle{empty}]
\documentclass[../main.tex]{subfiles}

\ifSubfilesClassLoaded{
    \tableofcontentsfile
    \dominitoc
    \setcounter{chapter}{0} 
    \externaldocument[]{../main}
    \def\locallabelprefix{chapt_1}
}{}

\begin{document}

\selectlanguage{english}

\graphicspath{{./figures/dist}}

\chapter{Speech and neural networks} \locallabel{chapt1}
\minitoc
\section{Introduction}
This chapter introduces the very basic notions of speech, how it is produced and how it can be categorized at different levels ranging from low-level acoustics to high-level semantics.
As this thesis aims to remove the personal speaker characteristics from the lower levels, we will discuss methods for numeric speech processing.

In speech processing, acoustic features are numerical representations of speech signals that can be extracted using algorithms hand-crafted based on years of research, or automatically using machine learning techniques.
These features provide the foundation for further analysis and processing.

For this thesis, manual analysis of the acoustic features to determine and transform those that reveal personal speaker characteristics is impractical due to the convoluted nature of the representations.
Some works have experimentally identified acoustic features that are highly correlated with speaker identity \cite{,acoustic_spk_id} or emotions \cite{GEMAPS,accoustic_emotion_Review} but due to the high variability and convoluted properties of speech, making mechanistic conclusions is difficult.
As such, we rely on machine learning-based analysis using artificial neural networks to determine and transform the acoustic features that reveal personal speaker characteristics.
This chapter introduces the key concepts of artificial neural networks, deep learning and relevant network models for speech processing.


\section{Speech production}
Human communication is both a social and a cognitive process in which individuals exchange information through a standard system of codes and signs.
Humans communicate to ask for help, inform others, and share attitudes to integrate within a group of individuals.
The primary mode of communication among Humans is speech.
Humans have evolved a complex set of voice, hearing, and cognitive abilities allowing them to express a sophisticated natural language \cite{language_faculty}.
In this section, we introduce the different components and aspects of human speech and detail some key notions associated with speech production and their respective properties.

\subsection{Speech production mechanism} \locallabel{chapt_1:sec:speech_prod}

\begin{figure}[htbp]
	\begin{center}
		\includegraphics[width=1\linewidth]{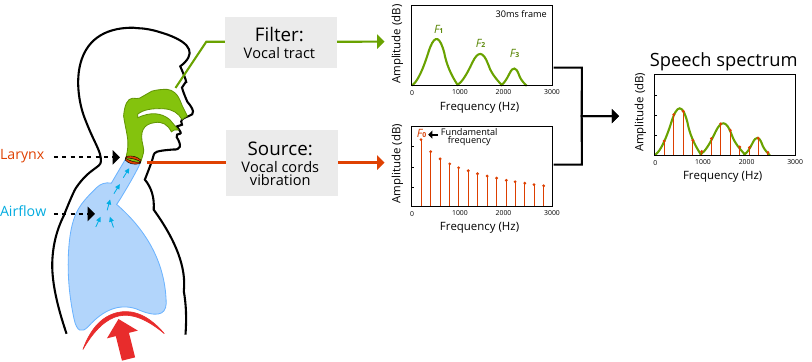}
	\end{center}
 \vspace{-8mm}
	\caption{
		The production mechanisms of voiced sounds as a source-filter model.
	}
  \vspace{-1mm}
	\locallabel{image_chapt1:speech_production}
\end{figure}

By definition, speech is an acoustic signal described by the modification of acoustic pressure over time.
Figure \localref{image_chapt1:speech_production} illustrates a schematic view of the primary organs involved in creating speech.
Speech generation starts with an airflow from the lungs, which goes through the larynx and the vocal tract to exit the lips.
The larynx contains two vocal cords, which vibrate rapidly with the
airflow to create the source of acoustic vibration.
Sounds created with the vibration of the vocal cords are called \textit{voiced} sounds, while sounds made without the vocal cords engaged are called \textit{unvoiced} sounds.
The vocal cord's vibrations are periodic and determine the sound's \acrfull{f0}.
\acrshort{f0} is expressed in Hertz (Hz), and the range of \acrshort{f0} correlates with the speaker's gender.
For the American population, the average \acrshort{f0} values are around 120 Hz for men and around 210 Hz for women. \cite{f0_men_women}.

Once the airflow exits the larynx, it is modulated by many organs in the vocal tract (for example, the tongue, the nasal cavity, etc.) to pronounce a multitude of speech sounds.
The shape of the vocal tract acts as a resonator and behaves as a frequency filter on the source signal.
This model of sound production is referred to as a source-filter model.
The frequency peaks from the vocal tract's shape are called formants and are denoted as $F_1$, $F_2$, $F_3$, etc.

\subsection{The different components and aspects of speech}
The capability of a Human to generate an acoustic signal is not the only requirement that defines speech.
Speech uses acoustic signals as a transmission medium.
Other aspects are essential to speech communication.

The comprehension process between a speaker and a listener requires both sides to have shared knowledge.
First, a common understanding of the semantics of the words in their context must be known to deduce the meaning of the communication.
Then, a mapping between words and their corresponding acoustic production and sounds has to be known.
Lastly, words should be arranged to match a defined syntax using the correct grammar.
In order to study speech, it is essential to break it down into several levels what constitutes speech.
The following list inspired by \cite{level_of_language,prosody_level} decomposes speech into different aspects:
\begin{itemize}
	\item \textbf{Semantics} concerns the content and meaning in language.
	      Semantics analysis can be applied to entire texts or single words.
	\item \textbf{Syntax} refers to the arrangement of words, phrases, clauses, and punctuation, in a specific order defined by a set of rules.
	\item \textbf{Phonetics and phonology} are the study of speech sounds and their function in a given language.
	      The production and classification of speech sounds according to their properties is the subject of phonetics, while their functions in a language are concerned with phonology.
	\item \textbf{Prosody} deals with how the semantic content is delivered.
	      It is important to structure the message (similar to the role of punctuation in writing), to emphasize certain terms or expressions, and to indicate the type of the sentences (statement, question, orders, exclamations).
	       \cite{Mary2018SignificanceOP}.
	      It helps to make speech more natural by carrying information about emotions, speaker intention, etc.
	\item \textbf{Acoustics} deals with the physical properties of the speech signal, including \acrshort{f0}, etc.
\end{itemize}

These levels allow humans to communicate with a very high degree of flexibility.
However, \acrfull{pii} can be extracted at different levels.
At the acoustic level, the voice characteristics of a speaker vary with the shape and size of the organs of the vocal tract allowing recognition of physical traits, and voice disorders (e.g., vocal cords paralysis, laryngeal cancer) \cite{cummins2018speech}.
At the prosody level, a speaker's manner of expression contains a rich array of \acrshort{pii}, including cues to the identity of the speaker, personality, emotions, age, gender, and many others \cite{biometic_ids,privacy_implication_voice}.
The syntax level contains information about potential language disorders (e.g., dysphasia, underdevelopment of vocabulary, or grammar).
At the semantic level, speakers may disclose private information, such as phone number, person's name, etc., in the linguistic content.

This thesis focuses on removing speakers \acrshort{pii} from the prosody and acoustic levels while keeping the linguistic content intact.
The other privacy aspects also need to be considered for complete privacy, such as removing phone numbers, person names, etc., from the linguistic content.
Still, they are out of the scope of this thesis.
The following section explains the basic processing blocks used to remove and evaluate the \acrshort{pii} contained in the prosody and acoustic levels.

\section{Speech processing}
\vspace{-2mm}
Speech or any other sound is composed of acoustic waves traveling through the air.
When recorded with a single microphone, acoustic waves are represented as the modification of pressure at a single point in space over time.
The microphone transforms acoustic waves into an electrical signal that is sampled and converted to a digital signal in order to be processed by computers.
This electrical signal is called a speech signal, and can also be referred to as a waveform or a time-domain signal.
A speech signal is, by nature, a convoluted, redundant, and highly variable signal, which makes it difficult to process directly.
Therefore, extracting a more compact representation that reduces redundancy is necessary.

The focus of this section is to introduce the process of creating traditional hand-crafted speech features.
Recent trends in research show the progressive abandoning of this step thanks to the increase in computational power and a new research field called \textit{Self-supervised Learning} \cite{ssl_survey}.
This aspect will later be explored in Chapter \ref{main:chapt:anon_me} of this thesis.

The ideal properties of the speech features should be:
\begin{itemize}
	\item relatively easy to extract
	\item resilient to background noise and transmission channel
	\item adapted to the downstream task, they should embed the information required while removing other information and noise
\end{itemize}
Satisfying all of these conditions simultaneously is difficult to accomplish in practice.
However, these criteria are considered idealistic design goals for speech processing.

\subsection{Time-frequency processing}

\begin{figure}[htbp]
	\vspace{-2mm}
	\begin{center}
		\includegraphics[width=0.8\linewidth]{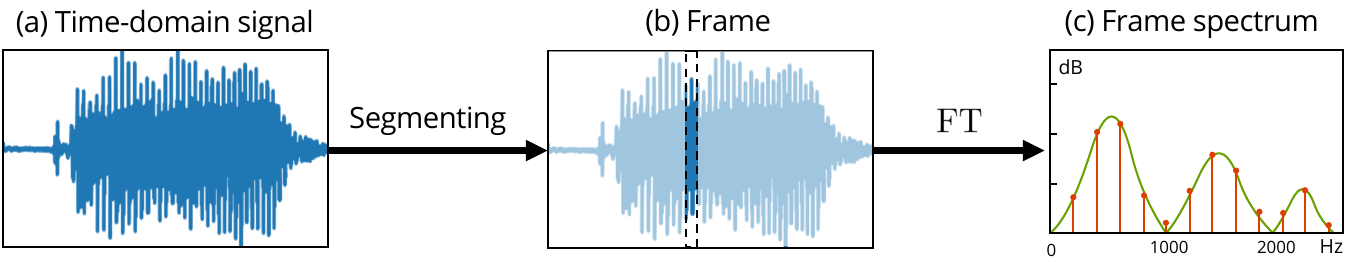}
	\end{center}
	\vspace{-7mm}
	\caption{
		The process of extracting a speech frame spectrum.
	}
	\locallabel{image_chapt1:speech_frame_spectrum}
	\vspace{-2mm}
\end{figure}

Traditional speech features are extracted by converting the time-domain signal
representation to its frequency-domain representation with the help of the Fourier transform.
The Fourier transform \cite{fourier1822theorie} decomposes a complex, non-sinusoidal speech signal into a series of sinusoidal sub-signals.
Figure \localref{image_chapt1:speech_frame_spectrum} shows an example of the application of the Fourier transform on a small speech segment (called a frame) illustrated by a spectrum.

A time-frequency representation of the whole speech signal is obtained by segmenting the source signal into overlapping frames of fixed length (usually 10 to 30 milliseconds).
Then, for each frame, a window function is applied that brings down, close to zero, the values near the edge of the window.
Lastly, a Fourier transform is applied on each frame independently, revealing their corresponding spectrum response.
This analysis is known as the \acrfull{stft}.
Plotting the changing spectrum response as a function of time generates a spectrogram plot.
An example of a power spectrogram is shown in Figure \localref{image_chapt1:signal_to_mfcc} (b).

\subsection{Short-term features}

Short-term features were initially developed in the 1980s for speech recognition.
They describe traits of human speech that are assumed to be stationary inside very short intervals.
At the time, the modeling approaches and available computational power were much more limited.
As such, low dimensional features were interesting as they require less computational effort for the subsequent processing, usually based on statistical models that could not handle very well high-dimensional data \cite{zeghidour_thesis}.

\begin{figure}[htbp]
	\begin{center}
		\includegraphics[width=0.95\linewidth]{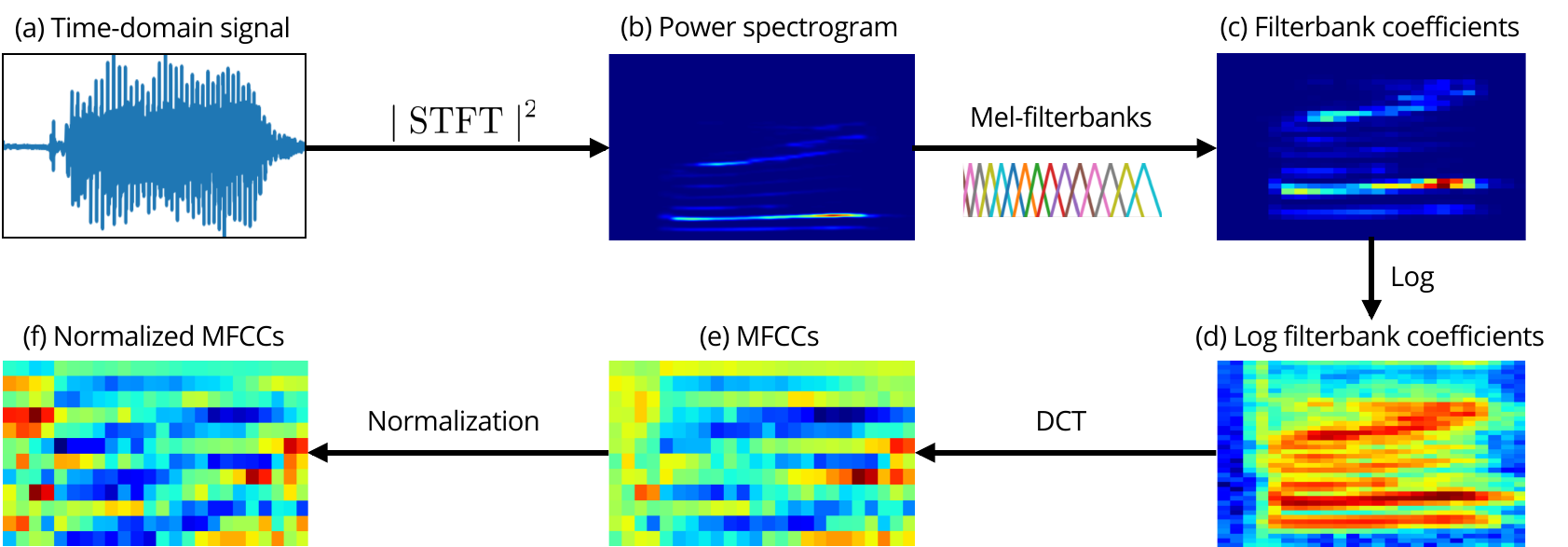}
	\end{center}
		\vspace{-4mm}
	\caption{
		The process of computing \acrfull{mfcc}.
	}
	\vspace{-2mm}
	\locallabel{image_chapt1:signal_to_mfcc}
\end{figure}

One of the most popular low-dimensional short-term features are the \acrfull{mfcc}.
To compute \acrshort{mfcc}, first, a power spectrogram is obtained by taking the squared magnitude of the \acrshort{stft} of the signal.
Having the \acrshort{stft} power spectrogram, the next step is to compute the energy in predefined frequency bands.
For \acrshort{mfcc}, the bands are arranged according to the Mel scale that mimics the frequency resolution of the human ear \cite{fbank_study}.
The filterbank coefficients can be visualized on a linear scale; however, this is not very informative for the human eye.
A logarithmic scale improves the quality, as observed in Figure (d).
The filterbank coefficients computed in the previous step are highly correlated, which could be problematic when modeling their behavior statistically.
Therefore, the {\acrfull{dct}} function is applied to decorrelate them and returns the \acrshort{mfcc} representation.
Finally, the \acrshort{mfcc} are normalized to lessen the influence of noise.
Cepstral mean normalization compensates for convolutional noise, while variance normalization (over time) compensates for additive noise.
With the advent of deep learning-based modeling systems, \acrshort{mfcc} are not the primary features anymore. Deep neural networks are less susceptible to highly correlated input, making the log filterbank coefficients or even the time-domain signal, which contains more information, a more relevant feature today.

\subsection{Long-term features}
Short-term features in speech or audio signals are gleaned from a brief interval of 10 to 30 milliseconds, while long-term features are drawn from a much longer period, ranging from 100 milliseconds to the full duration of an utterance - which can be a single word, a phrase, or an entire sentence.
They represent voice characteristics over longer intervals, for example, the mean and standard deviations of the \acrshort{f0}, or the \acrshort{f0} trajectory.
When the \acrshort{f0} is analyzed jointly with a spoken word, its variation models the prosody.
Long-term features encode speaker-related information as the speaker identity information in a signal is considered stationary for the utterance.

\vspace{-2mm}
\section{Artificial neural networks} \locallabel{chapt_1:section:nn}
\vspace{-2mm}

Previously, we introduced hand-crafted features as a means of representing speech through numerical values.
In this section, we will delve into the fundamental machine learning concept that utilizes these features as inputs to perform the analysis.
The field of artificial intelligence (AI), and more specifically machine learning, aims at teaching a computer to learn to perform a task for which it is not directly programmed.
It relies on mathematical approaches and data exploitation.
In this thesis, we utilize machine learning approaches to let the computer find adequate speech transformations that anonymize speech.
Interestingly, and more importantly, the evaluation procedure comprises a set of other machine learning methods that objectively assess the privacy and utility of the transformed (e.g., anonymized) speech signal.

Machine learning encompasses a multitude of methods, and among them are \acrfull{aan}, on which deep learning is based.
In recent years, \acrshort{aan} have become an indispensable tool for tackling a wide array of real-world problems.
At a fundamental level, \acrshort{aan} comprise many interconnected units called neurons.
The first formal neuron model was proposed in 1943 by Warren Sturgis McCulloch and Walter Pitts, inspired by the working of biological neurons.
It was then in \citeyear{rosenblatt1958perceptron} that Frank Rosenblatt set up a learning algorithm applicable to an artificial neuron, leading to the creation of the perceptron \cite{rosenblatt1958perceptron}.

The concept of the \acrfull{mlp} \cite{mlp} is to organize several simple perceptrons in such a way that the output of one neuron is the input of one or several other neurons.
Multiple neurons can be arranged side by side and connected to multiple perceptrons in a subsequent layer.
For example, the \acrshort{mlp} in Figure \localref{image_chapt1:mlp} has two hidden layers.
In this example, the outputs of the neurons of a given layer are connected to the inputs of the neurons of the next layer.
The information propagates from the input layer to the output layer; therefore, it is called a feedforward neural network.

\begin{figure}[htbp]
	\begin{center}
		\includegraphics[width=0.70\linewidth]{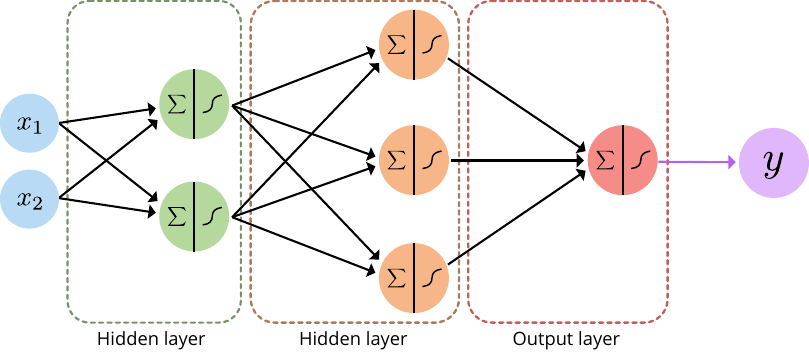}
	\end{center}
	\vspace{-5mm}
	\caption{
		Feedforward multilayer perceptron.
	}
	\vspace{-3mm}
	\locallabel{image_chapt1:mlp}
\end{figure}

The non-linear activation functions of the neurons are critical.
They allow the network to learn a complex mapping between the input and the output.
Without non-linearity, the outputs would be linear function of the input, resulting in an inability to learn patterns that are more complicated and potentially more meaningful.

A \acrfull{dnn} is a neural network architecture having many hidden layers.
The architecture in Figure \localref{image_chapt1:mlp} can be considered a simple \acrshort{dnn}, more complex architectures usually have more than a dozen layers.
The number of hidden layers defines the depth of the network.

In this thesis, we employ \acrshort{dnn} models to process acoustic features for various tasks.
For a long time, the training methods for this type of neural network did not allow to achieve good performance.
However, major advances in training methods and network structures have enabled the use of bigger and larger neural networks.
This section introduces fundamental neural network architectures and concepts that allow efficient modeling for many speech-related tasks.

\subsection{Bottleneck layer}
A bottleneck layer in a neural network is a layer with fewer neurons than the layers before or after.
Such layers encourage the network to compress its internal representations of the input \cite{he2016deep}.
A bottleneck representation can be extracted from the layer activations and used as a high-level representation of the input information.

\subsection{Residual network} \locallabel{chapt_1:residual_sub}
Upon training, the backpropagation calculates the partial derivatives of the error from the output to the first hidden layer.
Because of the chain rule, the first layers have much more non-linear activation functions on the way to compute their derivatives than the output layer.
If the derivatives are small the gradient will decrease exponentially as the error propagates backward until it eventually vanishes.
Vanishing gradients creates an issue as some weights of the network will not change throughout training.
As the number of layers in a deep neural network increases, the weights in the earlier layers are not effectively used as they are not effectively updated because of the vanishing gradient issue.
Reinforcing the error gradient of the first layers would solve this problem.
As such, a simple solution is to add skip connections between layers to allow the gradients to flow backward through the network easily.
This solution is referred to as residual or skip connections and was first proposed for image recognition \cite{resnet}, where the (residual) network had hundreds of layers for the first time.
The output of the residual connection can either be added or concatenated to the current layer.

\subsection{Time delayed neural and convolutional neural networks}
For many sequence classification tasks, it is interesting to include the context of the sequence to generate an output at a given time.
For example, to determine which phoneme is spoken at time $t$, it is helpful to know some parts of the sequences that precede and follow it.
In many speech-processing tasks, capturing long-term dependencies between acoustic events is helpful.
Many methods exist for capturing long-term sequences, and recurrent neural networks \cite{Graves2013SpeechRW} are one of them.
However, they are sensitive to the gradient vanishing problem and are hard to parallelize for efficient training.

In contrast, the \acrfull{tdnn} \cite{tdnn} is a type of architecture that can efficiently model temporal dependencies without suffering from the same problems.
It uses multiple stacked layers to encode an input sequence.
Each layer encodes a couple of spatially expanded elements from the previous layer.
The first layers learn a small context, while the higher layers learn a wider context.
The context resolution increases as the network gets deeper.
%
%
\begin{figure}[htbp]
	\begin{center}
		\includegraphics[width=0.9\linewidth]{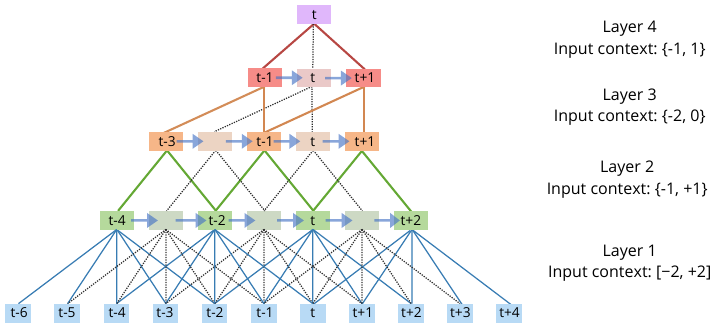}
	\end{center}
	\vspace{-2mm}
	\caption{
		Computations for a \acrshort{tdnn} in dotted and solid lines, and sub-sampled \acrshort{tdnn} in solid lines only.
		Frames $t - 2$ through $t + 2$ are spliced together at the input layer (which can be written as context \{-2, -1, 0, 1, 2\} or more compactly as [-2, 2]); and then at the upper three hidden layers with splice offsets of \{-1, 1\}, \{-2, 0\} and \{-1, 1\}.
	}
	\vspace{0mm}
	\locallabel{image_chapt1:tdnn}
\end{figure}

The hyperparameters of a \acrshort{tdnn} layer are defined by the length of the input context required to compute an output activation at a time step.
Figure \localref{image_chapt1:tdnn} shows the receptive field for each layer in dotted and solid lines.
The notation \{-2, 0\} for the third layer means that only the input at $t - 2$ and $t$ will be used to compute the activation.
In the example, the network does not take symmetrical past and future contexts to generate the final output at time $t$.
A smaller context on the right side of the network lowers the latency for online decoding use as fewer future elements are to be awaited before emitting an output.

\acrshort{tdnn} are, in fact, 1D Convolutional Neural Network (CNN) \cite{cnns} with specific settings for kernel size, dilatation, and strides.
Both network architectures process signals in a way that considers temporal dependencies between elements.
Similar to the convolution operation in CNN, \acrshort{tdnn} models nearby dependencies using a fixed-length context window to learn local correlations in the signal.
Additionally, both \acrshort{tdnn} and CNN can be stacked (like in Figure~\localref{image_chapt1:tdnn}) to form deep networks, allowing for complex feature extraction and modeling of hierarchical relationships in the data.

In \cite{PoveyTDNN2015}, it is proposed to use subsampling to reduce the number of hidden activations.
This method avoids unnecessary redundancy as large contexts overlap leading to highly similar subsequent activations.
A computation path of a subsampled \acrshort{tdnn} is shown by the solid line connections in Figure \localref{image_chapt1:tdnn}.
With subsampling, the overall computation is reduced leading to faster training and inference time.

\paragraph{Factorized version} \locallabel{chapt_1:tdnnf_svd}
In \cite{Povey2018SemiOrthogonal_TDNNF}, the author proposed a method to reduce the number of weights in a \acrshort{tdnn} model by using matrix decomposition.
The weight matrix $\boldsymbol{W}$ of each layer is approximated as a product of two low-rank matrices:
$\boldsymbol{W} = UV$, where $U \in \mathbb{R}^{u \times k}$ and $V \in \mathbb{R}^{k \times v}$ are obtained using \acrfull{svd} of $\boldsymbol{W} \in \mathbb{R}^{u \times v}$ \cite{svd} while discarding the basis corresponding to the smallest singular values.

Selecting a rank value $k$ smaller than $u$ and $v$ effectively inserts a bottleneck layer within a conventional \acrshort{tdnn} layer.
%
%
This bottleneck can be relatively strong, as it can reduce the dimension by a large factor. \locallabel{explanation:bottleneck}
For example, in real-world applications, the 1536x1536 weight matrix can be factorized as a product of 1536x160 and 160x1536 matrices.
Figure \localref{image_chapt1:tdnnf_layer} displays the differences between the architectures of the \acrshort{tdnn} and \acrfull{tdnnf} layers; in the \acrshort{tdnnf} layers, a skip connection is added between the input layer and the output projection.

\begin{figure}[htbp]
	\vspace{-3mm}
	\begin{center}
		\includegraphics[width=0.9\linewidth]{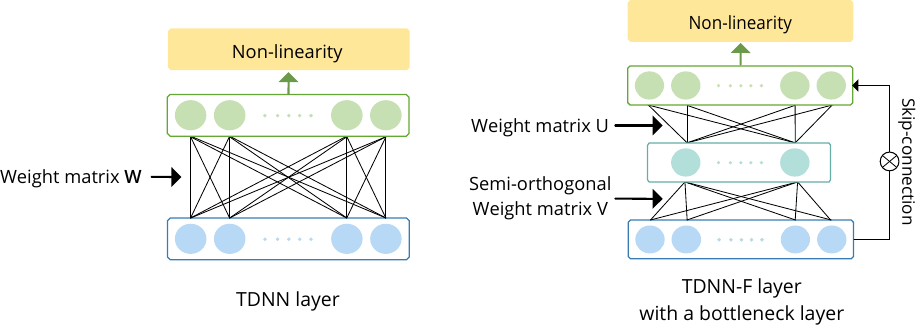}
	\end{center}
	\vspace{-8mm}
	\caption{\acrshort{tdnn} and \acrshort{tdnnf} layers.}
	\vspace{-2mm}
	\locallabel{image_chapt1:tdnnf_layer}
\end{figure}

In order to make the training converge, it is necessary to constrain the matrix $V$ to be semi-orthogonal, meaning that it must respect one of the following conditions $VV^T = I$ or $V^TV = I$, where $I$ is the identity matrix.
To ensure the $V$ matrix remains close to the semi-orthogonality, a constraint is applied every four backpropagation updates \cite{Povey2018SemiOrthogonal_TDNNF}.
%

In this thesis, we use both the Kaldi implementation \cite{KaldiPovey} and a PyTorch implementation \cite{pytorch,pkwrap} of the \acrshort{tdnnf} network.
The trick of factorizing the weight matrices has other valuable properties than just minimizing the number of weights.
As observed by \cite{RyffelPartiallyEncryptedDL2019} smaller bottleneck dimensions encode less personal information.

\subsection{Transformer} \locallabel{chapt_1:transformer}
The transformer model proposed by \cite{transformers} for sequence-to-sequence machine translation has received tremendous interest in recent years in many fields \cite{transformers_inter}, including speech processing.
This model consists of an encoder and a decoder, each with $L$ stacked processing blocks consisting of similar architectures.
In contrast with the recurrent neural network, the transformer architecture takes the whole input sequence at once rather than one symbol or frame at a time during training.
At inference, the transformer decoder is an autoregressive model\footnote[1]{
	Autoregressive models rely on previously generated output to predict the current ones.
}.
The decoder always generates the entire sequence from $y_1$ up to $y_{current}$, given the encoder state and the previous output sequence $\left[\textit{<sos>}, y_{1} \hdots y_{current - 1}\right]$ (starting with a start-of-sequence \textit{<sos>} symbol for the first prediction).
At each step, the current output sequence is fed to the bottom right decoder block (Figure~\localref{image_chapt1:transformer}) in the next time step until the \textit{<eos>} (end-of-sequence) symbol is generated by the network.
Interestingly, during training, the transformer model is not computationally autoregressive as it processes the input all at once instead of processing it one step at a time in a loop.
This is achieved by using masking techniques that prevent the network from depending on future information and allows it to be used during inference exactly like an autoregressive model.

The transformer model is agnostic to the order of the input (and output) sequence unless explicit positional embeddings are added to each input (and output).
Positional encoding allows to retain the position information necessary to keep the meaning of the sequence.
The encoder includes \textbf{multi-head self-attention} modules and a \acrshort{mlp} network.
Residual and normalization layers are employed around each module to facilitate backpropagation of gradients through the network during training.
Compared to the encoder blocks, the decoder blocks have a \textbf{multi-head attention} layer called \textbf{cross-attention module}.
The overall architecture of the transformer is well described in \cite{illustrated_trans} and displayed in Figure \localref{image_chapt1:transformer}.

\begin{figure}[ht]
	\vspace{-0mm}
	\begin{center}
		\includegraphics[width=0.63\linewidth]{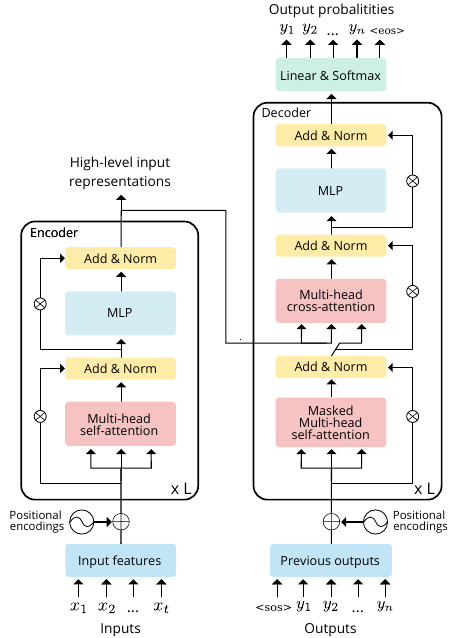}
	\end{center}
	\vspace{-4mm}
	\caption{\centering Overview of the transformer architecture, \protect\say{x~L} corresponds to the number of stacked encoders and decoders.}
	\vspace{-2mm}
	\locallabel{image_chapt1:transformer}
\end{figure}

Transformer architectures are very efficient at encoding the temporal dependency of a sequence with attention.
Additionally, this architecture is very versatile as it can be used in multiple manners:
\begin{itemize}
	\item Encoder-Decoder: The entire architecture is used.
	      This is typically used in sequence-to-sequence modeling (e.g., neural machine translation).
	\item Encoder only: Only the encoder is used, and the outputs of the encoder are utilized as a high-level representation of the input sequence.
    This encoder can be pretrained and fine-tuned to perform better on specific downstream tasks (for example, Wav2Vec-2.0, GPT-3 \cite{wav2vec2,gpt3}), or, jointly trained with different decoders (for example, \acrshort{asr} models with CTC decoder).
\end{itemize}

\subsection{Adversarial network} \locallabel{chapt_1:adv_net}

To enhance the capability of a network to encode interesting attributes, one can rely on adversarial machine learning.
In adversarial machine learning, a \textit{game theory} \cite{Osborne1994ACI} with two or more players is considered.
The loss of each player must be minimized, but the decision of other players impact their loss.
Applied to neural networks, this game enhances feature representation by promoting the emergence of representation relevant to a primary task while also being indiscriminate concerning one (or several) so-called adversarial task(s).

Such kind of models are interesting for training models capable of generating highly realistic data (e.g., image generation) \cite{goodfellow2014generative}, adapting models to out-of-domain data \cite{ganinDomainAdversarialTrainingNeural2017,domainAdaptationofEndtoendASR}, and extracting privacy-preserving features \cite{feutryLearningAnonymizedRepresentations2018,mohanPrivacyPreservingAdversarialRepresentation2019_reality_adversarial}.
The architecture comprises three modules: encoder, decoder the adversarial branch.
An encoder takes the input in the first module and generates a representation $f$.
Then the decoder uses this representation $f$ to predict the output  $y$ of the primary task.
This occurs while an additional adversarial branch, alongside the decoder, uses the representation $f$ to predict the output $d$ of the adversarial task.
This architecture is described in Figure \localref{image_chapt1:adversarial_training}.

\vspace{0mm}
\begin{figure}[ht]
	\begin{center}
		\includegraphics[width=0.93\linewidth]{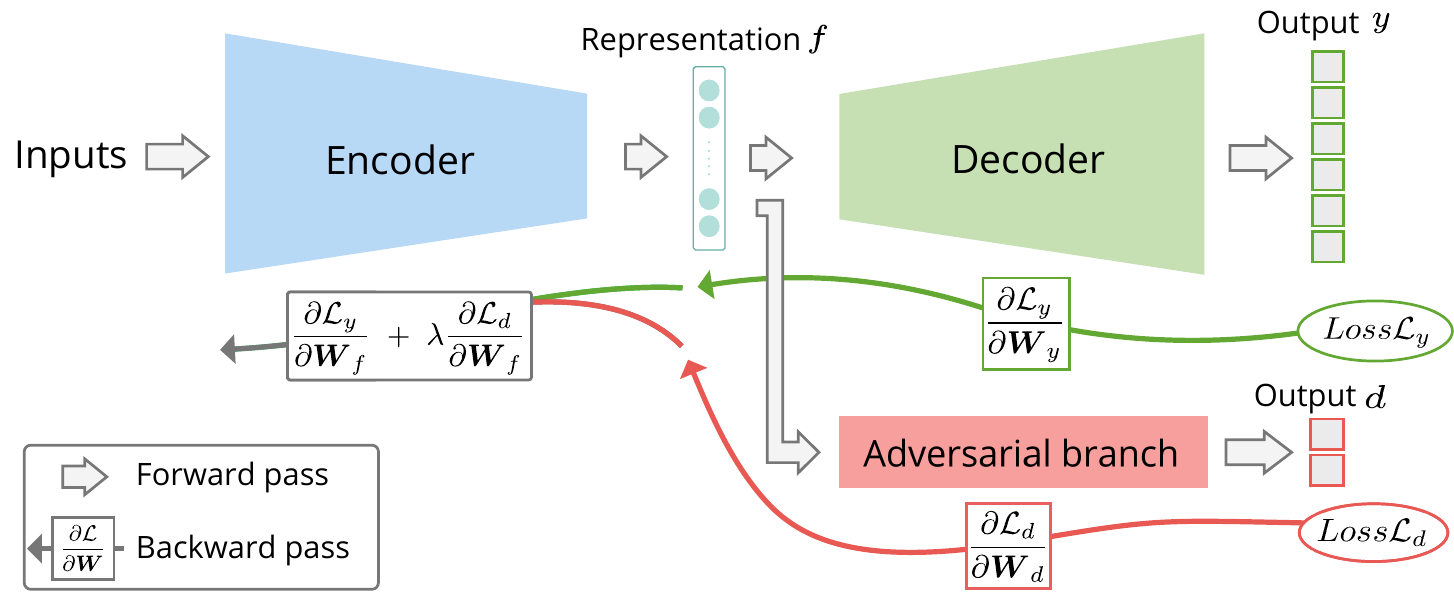}
	\end{center}
	\vspace{-1mm}
	\caption{
		Outline of an adversarial neural network.
	}
	\vspace{-1mm}
	\locallabel{image_chapt1:adversarial_training}
\end{figure}

During training, the forward pass computes the output of the primary and adversarial tasks.
The backward pass first optimizes the decoder and adversarial branch weights to minimize the two losses of primary and adversarial tasks.
Then, the encoder's weights are adjusted to maximize the adversarial task loss while minimizing the primary task loss.
A gradient reversal function achieves loss maximization by multiplying the gradient with a negative scalar during the backpropagation.
The following equation defines the gradient updates for each neural network module.
\vspace{-1mm}
\begin{equation}
	{\boldsymbol{w}_{f_{i}}}' = \boldsymbol{w}_{f_{i}} - \alpha\left(\frac{\partial\mathcal{L}_{y}}{\partial\boldsymbol{w}_{f_{i}}}\;+\;\lambda\frac{\partial\mathcal{L}_{d}}{\partial\boldsymbol{w}_{f_{i}}}\right)
\end{equation}
\vspace{-1mm}
\begin{equation}
	{\boldsymbol{w}_{y_{i}}}' = \boldsymbol{w}_{y_{i}} - \alpha\frac{\partial\mathcal{L}_{y}}{\partial\boldsymbol{w}_{y_{i}}}
\end{equation}
\vspace{-1mm}
\begin{equation}
	{\boldsymbol{w}_{d_{i}}}' = \boldsymbol{w}_{d_{i}} - \alpha\frac{\partial\mathcal{L}_{d}}{\partial\boldsymbol{w}_{d_{i}}}
\end{equation}
%
where $\lambda$ is the negative scalar coefficient, $\mathcal{L}_{d}$ the adversarial loss, $\mathcal{L}_{y}$ the primary loss, $\boldsymbol{w}_{f_{i}}$ $\boldsymbol{w}_{y_{i}}$, $\boldsymbol{w}_{d_{i}}$ are weights of the encoder, primary task decoder, and adversarial branch respectively.
This training scheme encourages the encoder to be invariant and, to some extent, disentangled from the adversarial task.
In this thesis, we will investigate if personal speaker information \acrshort{pii} can be removed using an adversarial network similarly as in \cite{mohanPrivacyPreservingAdversarialRepresentation2019_reality_adversarial}.
However, we addressed some limitations of their approach by, most notably, using a more advanced training scheme designed for privacy protection \cite{RyffelPartiallyEncryptedDL2019}.
Also, we will use adversarial networks with a generative model to synthesize speech.

\subsection{Generative adversarial network} \locallabel{chapt_1:gan}
\begin{figure}[ht]
	\begin{center}
		\includegraphics[width=0.85\linewidth]{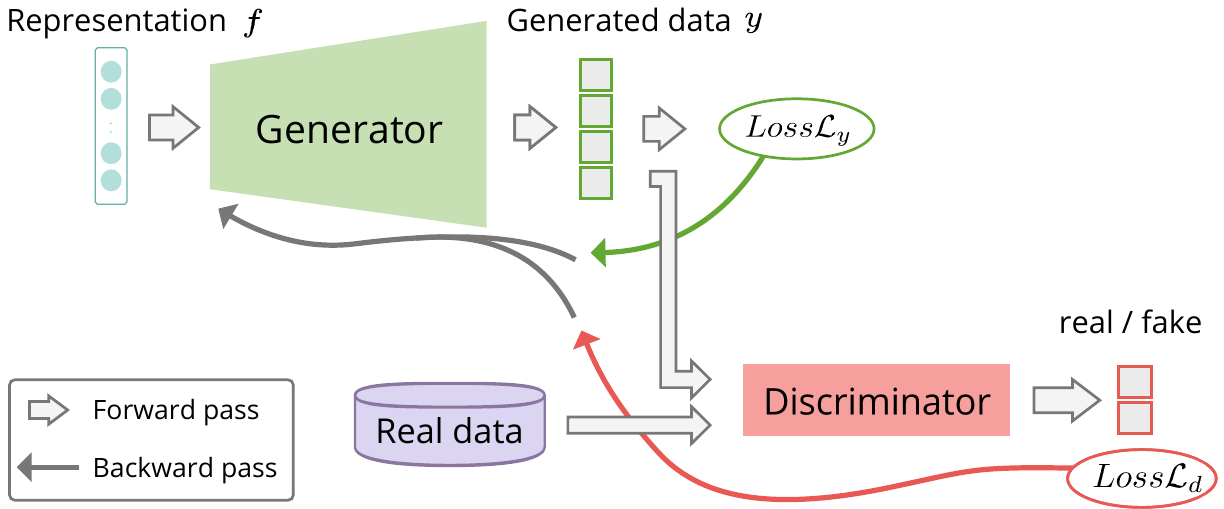}
	\end{center}
	\vspace{-2mm}
	\caption{
		Outline of a \acrfull{gan}.
	}
	\vspace{-0mm}
	\locallabel{image_chapt1:gan}
\end{figure}

The \acrfull{gan} is a network architecture and training scheme based on the adversarial network \cite{goodfellow2014generative}.
This type of network has seen broad interest as it is, as of today, the state-of-the-art method for generating realistic data.
The two players that compete and feed off each other involves:
\begin{enumerate}
	\vspace{-0mm}
	\item \textbf{A generator model} trained to generate new data aiming to reproduce the same distribution as real training data.
    \vspace{0mm}
	\item \textbf{A discriminator model} trained to classify inputs as real or fake, attempting to identify if an input originates from the training dataset or the generator model.
\end{enumerate}
During training, the discriminator is trained to distinguish real data from generated data (or fake in the \acrshort{gan} terminology).
On the other side, the generator is trained to generate output that can no longer be distinguished from real data by the discriminator.
This adversity between these two players drives both networks to improve until the fake data is indistinguishable from the real one.
Figure \localref{image_chapt1:gan} shows an example of a \acrshort{gan} taking as generator input a representation, which can be a latent noise variable or a representation of real data obtained by one or several encoders.
For voice conversion tasks, \acrshort{gan}s are particularly interesting because they can generate highly realistic speech data.

Each training step contained in the \acrshort{gan} framework
 consist of two mini-batches, the first coming from the real data $\boldsymbol{x}$ (labeled as 1) and the second coming from the generator $G(\boldsymbol{f}))$ (labeled as 0).
In the original \acrshort{gan} paper \cite{goodfellow2014generative}, the generator attempts to minimize the following function, whereas the discriminator attempts to maximize it:
%
\vspace{-1mm}
\begin{equation}
	\begin{aligned}
		\min _G \max _D\
		\mathcal{L}_D\left(D, G\right) & =
		\mathbb{E}_{\boldsymbol{x}} [\log D(\boldsymbol{x})]+\mathbb{E}_{\boldsymbol{f}}[\log (1-D(G(\boldsymbol{f})))]
	\end{aligned}
	\vspace{-2mm}
\end{equation}
%
where $\mathbb{E}_{\boldsymbol{x}}$ is the expected value according to the real distribution, $\mathbb{E}_{\boldsymbol{f}}$ the expected value with respect to the distribution of the generator, $D(\boldsymbol{x})$ the discriminator estimate of a probability that a real data instance $x$ comes from the real distribution rather than the fake one, and $D(G(\boldsymbol{f}))$ the discriminator estimate of the probability that a fake instance is real.
This formula is derived from the binary cross-entropy between two distributions.
Because the generator cannot directly influence the $\log D(\boldsymbol{x})$ term in the function, minimizing the loss for the generator is equal to minimizing $\mathbb{E}_{\boldsymbol{f}}[\log (1-D(G(\boldsymbol{f})))]$.
In practice, a reformulation of this formula is necessary to make the network converge better while avoiding vanishing gradient or other training failures.
Techniques such as the \textit{non-saturating} generator loss \cite{goodfellow2014generative} or the least-square loss \cite{Mao2017LeastSG} have proven to avoid vanishing gradients problems.
Other losses can be applied to the generator to increase the generator training efficiency and the quality of the produced data; however, this depends on the dataset and the availability of labels.

\acrshort{gan}s have been successfully employed by the speech processing community \cite{speech_gan_review}, in particular for speech synthesis \cite{hifigan}, voice conversion \cite{cycle_gan}, speech enhancement \cite{Speech_Enhancement_gan}.
Chapters~\ref{main:chapt4}~and~\ref{main:chapt5} of this thesis rely on \acrshort{gan} architectures to generate anonymized, natural, and realistic-sounding speech.

\vspace{-0mm}
\section{Conclusion}
\vspace{-0mm}

In this chapter, the process of speech production is introduced and important key concepts in numeric signal processing and machine learning are covered.
The focus then shifts to specific network architectures like \acrshort{tdnn}, transformer, and \acrshort{gan}, which will play a crucial role in later chapters.
All of this knowledge will not only enable us to build speaker anonymization systems but also evaluate their effectiveness in terms of privacy and utility using speech-centric machine-learning disciplines that will be covered in the next chapter.

\ifSubfilesClassLoaded{
	\printglossary[title=Special Terms,type=\acronymtype]
	\printbibliography
}{}

\end{document}

\clearemptydoublepage
\cleartooddpage[\thispagestyle{empty}]
\documentclass[../main.tex]{subfiles}

\ifSubfilesClassLoaded{
    \tableofcontentsfile
    \dominitoc
    \setcounter{chapter}{1} 
    \def\locallabelprefix{chapt_2}
    \externaldocument[]{../main}
}{}

\begin{document}

\selectlanguage{english}

\graphicspath{{./figures/dist}}

\chapter{Speech-centric machine-learning} \locallabel{chapt2}
\vspace{-3.7em}
\minitoc
\vspace{-1.8em}
\section{Introduction}
\vspace{-1em}
The task of speaker anonymization and its evaluation relies on many speech-centric machine-learning disciplines that will be presented in this chapter.
Knowledge of acoustic modeling is necessary to build a speaker anonymization system; as such, we will present how traditional \acrfull{asr} work.
\acrfull{vc} is another fundamental tool to generate an anonymized voice that will also be introduced.
The automatic evaluation of the performance of a speaker anonymization system relies on \acrshort{asr} and \acrfull{asv} systems to objectively assess the preservation of the linguistic content and the capability of the system to conceal the speaker identity.
That is why we will also present the fundamental of \acrshort{asv}.

\section{Automatic speech recognition}
\acrfull{asr} or speech-to-text aims at transforming an audio speech signal into the sequence of corresponding words.
In the scope of this thesis, \acrshort{asr} systems play a crucial role.
During the anonymization process, the acoustic model of an \acrshort{asr} system is used to extract a feature representing the linguistic content spoken by the speaker.
Precise modeling is necessary as the anonymization system will later use this representation to generate anonymized speech signals without distorting the content as best as possible.
Importantly as well, during the evaluation, an \acrshort{asr} system is used to assess unwanted degradation of the linguistic content intelligibility distortion.
First, this section defines the task of \acrshort{asr}, details the key models employed for speech recognition, and finally explains how to evaluate the performance of an \acrshort{asr} system.

As mentioned in Section \ref{main:image_chapt1:signal_to_mfcc}, speech can be considered as a finite sequence of acoustic features $X = [x_1 \hdots x_T]$ where $T$ is the number of audio frames.
Performing speech recognition consists in producing a sequence of $M$ words $W^{*} = [w_1 \hdots w_M]$ from the observation $X$.
This problem can be formulated as the maximization of the probability $P(W|X)$ \cite{statistical_asr}:
\vspace{-1mm}
\begin{equation} \locallabel{eq_argmax_asr}
    W^{*}=\operatorname*{argmax}_{W}{P}(W|X)
\end{equation}
%
The equation \localref{eq_argmax_asr} can be simplified since $P(X)$ is independent of $W$ (constant for all possible sequences of words).
Following Bayes' rule, the equation is decomposed into:
\begin{equation}
    \begin{aligned}
        W^{*} & =\operatorname*{argmax}_{W}\frac{{P}(X|W){P}(W)}{P(X)} \\
              & =\operatorname*{argmax}_{W}{{P}(X|W){P}(W)}
    \end{aligned}
\end{equation}
An \acrshort{asr} system usually relies on two models: an acoustic model, which models the probability $P(X|W)$ of a speech signal $X$, given a corresponding word sequence $W$; and a language model, which models the probability $P(W)$ of the sequence of the words $W$. 

\begin{figure}[htbp]
    \begin{center}
        \includegraphics[width=0.74\linewidth]{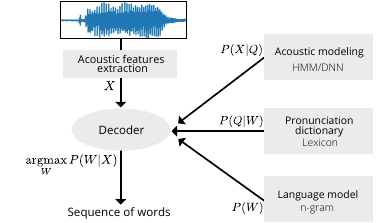}
    \end{center}
    \vspace{-4mm}
    \caption{
        Overview of an \acrshort{asr} system
    }
    \locallabel{image_chapt2:asr_pipeline}
\end{figure}

As shown in Figure \localref{image_chapt2:asr_pipeline}, a pronunciation lexicon may be employed to associate multiple discrete spoken sound units (for example phonemes) with a word.
In such a case, the acoustic model no longer models $P(X|W)$ but $P(X|Q)$, with $Q$ a sequence of phonemes.
The equation can be formulated as follows:
\vspace{-2mm}
\begin{equation}
    \begin{aligned}
        W^{*}=\operatorname*{argmax}_{W} P(X|Q) P(Q|W) P(W)
    \end{aligned}
    \vspace{-1mm}
\end{equation}
where $P(Q|W)$ is the probability of Q, the sequence of phonemes, knowing W, the sequence of words.
With a pronunciation lexicon, one or several pronunciations are mapped to each word.
This enables the acoustic model to identify spoken units smaller than words, for example, phonemes, which are the smallest unit of speech that differentiates between words in a language.
This mapping of phonemes to acoustic features is what the acoustic model learns.
Another benefit of the lexicon is that it allows for more efficient training of acoustic models when data is limited.
This is because the number of phonemes is typically much smaller than the number of words in the lexicon, resulting in a larger number of examples per phoneme in the training data.

\vspace{-2mm}
\subsection{Acoustic modeling} \locallabel{section:acoustic-modeling}
\vspace{-1mm}

The acoustic model is the most crucial element of \acrshort{asr} systems since it models the correspondence between the audio input and a sequence of spoken units.
This section details the methods based on, generative acoustic modeling combining \acrfull{gmm} and \acrfull{hmm}, hybrid \acrfull{dnn} \acrshort{hmm} modeling, and end-to-end and self-supervised learning.

\vspace{-1mm}
\subsubsection{Hidden Markov model}
\vspace{-1mm}
Hidden Markov Models were initially proposed in the 1970s for acoustic modeling \cite{hmm_70,hmm_tuto_1989}.
Today \acrshort{hmm}s are still a widespread approach but considered traditional, with the advent of end-to-end models.
\acrshort{hmm}s are statistical models that are used to model sequential or temporal data, by assuming that the observed data is generated by a hidden Markov process, where the current state only depends on the previous state, and the observation at the current time only depends on the current state.
\acrshort{hmm}s are composed of emitting states, transition probabilities between states, and probability densities for the emission of observations.
The latter can be obtained with \acrshort{gmm} or \acrshort{dnn} models.

In the context of acoustic modeling, the emitting states represent the probability of a particular sound being emitted from the model at a given time.
This sound can be a phoneme, possibly in a given surrounding context.
The three main units are the monophone, biphone, and triphone models, each referring to the different contexts that are taken into account when determining the probability of a sound being emitted.
A monophone model only reflects a single phoneme, the observation units do not consider the context in which it is being pronounced.
Monophones do not allow modeling co-articulation, which is how the configuration of the vocal tract gradually changes from one phoneme to another, causing a distortion of these phonemes in the process.
For example, the sound "a" ($/${æ}$/$) in the word "bat" ($/${bæt}$/$) will have a different probability of being emitted than the sound "a" in the word "bad" ($/${bæd}$/$), because the surrounding context is different.
A biphone model better captures co-articulation by taking into account the one phoneme that is before (or after) a given phoneme.
This allows the model to better capture the context in which a phoneme is being pronounced, and can improve the overall accuracy of the \acrshort{asr} system.
A triphone model goes even further, by considering the two phonemes that surround a given phoneme.
This provides even more contextual information, potentially improving the accuracy of the \acrshort{asr} system.

In practice, for each phonetic unit, the \acrshort{hmm} usually uses three-states corresponding to the beginning, middle, and end of a unit.
The following figure shows an example using monophones and triphones to represent the word "cup".
Both the monophone and triphone are modeled using three-states models.
\begin{figure}[htbp]
    \vspace{-0mm}
    \begin{center}
        \includegraphics[width=0.72\linewidth]{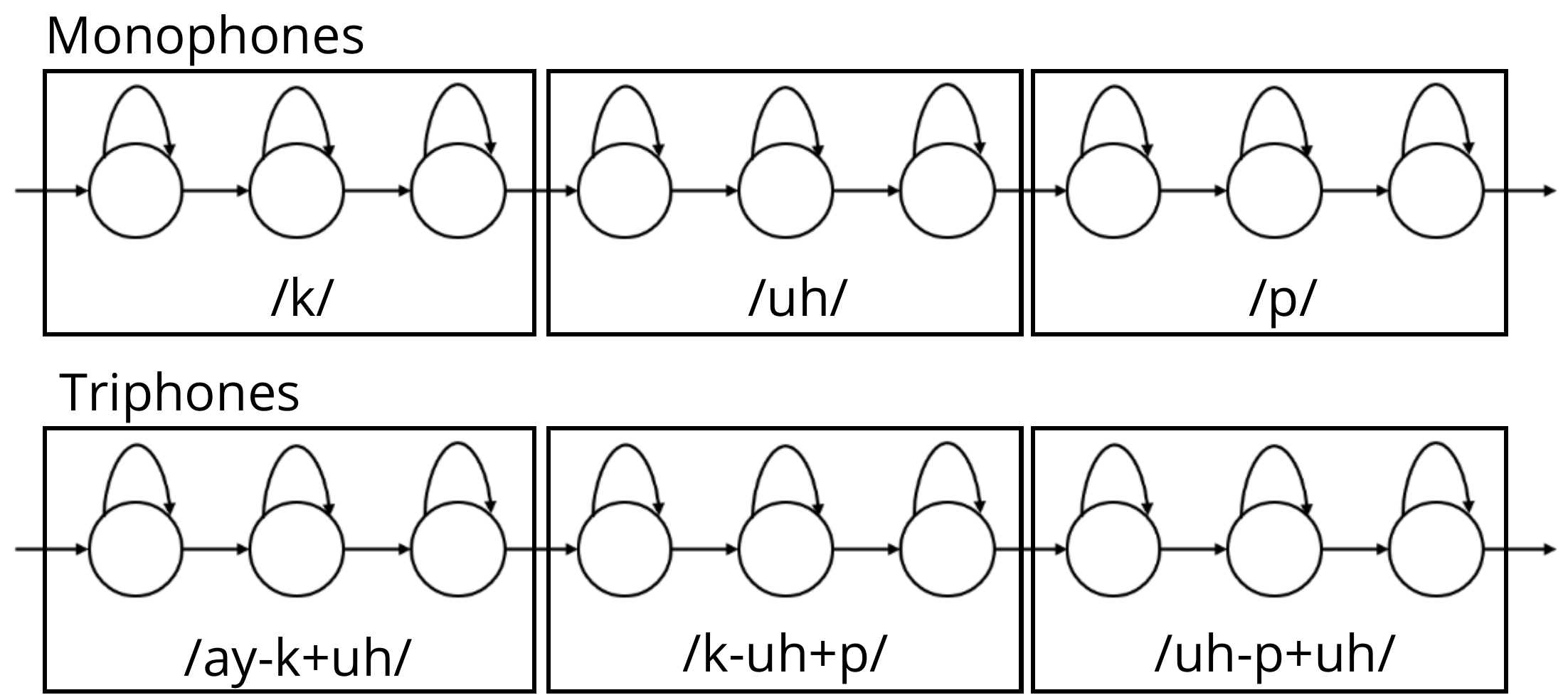}
    \end{center}
    \vspace{-2mm}
    \caption{
        Monophone and triphone three-state \acrshort{hmm} for the word "cup" in the expression of "a cup you".
    }
    \vspace{-0mm}
    \locallabel{image_chapt2:triphone}
\end{figure}
\noindent
If we consider 40 possible phonemes for monophones represented with three-state \acrshort{hmm}, there is a total of 120 possible emitting states.
When triphones are considered, the number of units increases to 40x40x40x3, resulting in 192~000 emitting states.
This explosion of number emitting states leads to two issues: insufficient data to train each triphone, and as such, some triphones will be appearing in testing but not in training.
One solution is clustering, where similar triphones are grouped into "senones" (reducing the emitting states to about 10~000).
This is done using a decision-tree clustering process where a decision tree is built for every state of every context-independent phone.
%
%
In general, the number of possible emitting states in an \acrshort{hmm} is a trade-off between the accuracy of the model and the computational resources/amount of data required to train it.

\vspace{-2mm}
\subsubsection{Generative acoustic models \texorpdfstring{\acrshort{gmm}-\acrshort{hmm}}{Generative acoustic models GMM-HMM}{}}
Traditionally, the probability densities of emitting observation of \acrshort{hmm} models were represented by generative \acrfull{gmm}.
The idea is to associate each \acrshort{hmm} state with a weighted sum of Gaussian probability densities.
However, \acrshort{gmm} has some limitations, such as its tendency to overfit training data and difficulty to deal with high-dimensional data. 

\vspace{-2mm}
\subsubsection{Hybrid neural network models \texorpdfstring{\acrshort{dnn}-\acrshort{hmm}}{DNN-HMM}} \locallabel{section:hmm-dnn}
The usage of neural networks as an alternative to \acrshort{gmm}s was first proposed with \acrshort{mlp}s \cite{mlp_speech_reco}.
However, it is not until \citeyear{dnn_speech_reco} that \acrfull{dnn} have over-performed \acrshort{gmm}s at estimating the emission probabilities of \acrshort{hmm} states \cite{dnn_speech_reco}.
The \acrshort{tdnn} \cite{tdnn} deep learning architecture presented in section \ref{main:chapt_1:section:nn}, which is particularly well-suited for speech processing, has been widely adopted in the community after the work of \cite{PoveyTDNN2015}, gradually replacing \acrshort{gmm}s.
\acrshort{dnn}s have the advantage over \acrshort{gmm}s of being able to model more complex and non-stationary representations thanks to how they are trained with their loss function.
As discriminative models, they are directly trained to model the posterior probabilities, allowing them to better distinguish classes.

The model \acrshort{dnn}-\acrshort{hmm} used in this thesis is a Kaldi \cite{KaldiPovey} \say{chain} \acrshort{tdnnf} model, trained with the \acrfull{lfmmi} cost function, the next paragraph explains into further details \acrfull{mmi} based cost functions.

\vspace{-2mm}
\paragraph{\texorpdfstring{\acrshort{mmi} training}{MMI training}}
The \acrshort{mmi} discriminative training method computes the cost function at the sequence level by maximizing the \acrshort{mmi} between the distribution of a predicted sequence and the distribution of the expected correct transcription. This is equivalent to maximizing the likelihood of the input sequence given (the state sequence corresponding to) the correct transcription, while simultaneously minimizing the marginal likelihood for any possible state sequence/transcription. This approach allows the model to better distinguish between correct and incorrect transcriptions, improving the overall performance of the \acrshort{asr} system \cite{seq_povey}.

\noindent
The loss function is defined as:
\begin{equation}
    \begin{aligned}
        \mathcal{L}_{\text{\scshape{mmi}}} & =\sum_{r=1}^{R} \log p\left(W_{r} \mid X_{r}\right)
        \\
        \mathcal{L}_{\text{\scshape{mmi}}} & =\sum_{r=1}^{R} \log \frac{p\left(X_{r} \mid W_{r}\right) P\left(W_{r}\right) }{\sum_{W}p\left(X_{r} \mid W\right) P\left(W\right)}
    \end{aligned}
    \locallabel{eq_chapt2:lfmmi}
\end{equation}
where $R$ is the total number of training segments,
$W_{r}$ is the correct transcription of the $r^{th}$ speech segment $X_{r}$,
$P(W_r)$ is the probability given by the language model for the sentence $W_r$.
For any sequence, $W$, $P\left(W\right)$ is estimated with a language model.
The numerator indicates the likelihood of the input sequence for the reference word sequence.
In contrast, the denominator indicates the total likelihood of the input sequence for all possible word sequences, which is equivalent to the sum of all possible word sequences estimated by the acoustic and a phoneme language model.
This cost function is optimized by maximizing the numerator (i.e., increasing the probability that the model predicts a sequence similar to the reference) and minimizing the denominator (decreasing the probability of other non-valid sequences).

\vspace{-2mm}
\paragraph{\texorpdfstring{\acrshort{lfmmi} training}{LF-MMI training}} \locallabel{sec:lfmmi}
The \acrfull{lfmmi} cost function is an extension of \acrshort{mmi} to make the computation of the numerator and denominator graphs faster and manageable in size \cite{purely_seq_lfmmi}.
Other implementation optimizations are also made, such as subsampling the output frame rate to one frame every 30ms instead of every 10ms which makes the denominator graph smaller and speeds up the forward/backward computation.
Consequently, the \acrshort{hmm} topology differs from other common three-state \acrshort{hmm}.
The \acrshort{hmm} has two states per phoneme unit and can be traversed in a single frame, instead of three, improving decoding speed.

In order to train the acoustic \acrshort{dnn} model with the \acrshort{lfmmi} loss, a phonetic alignment between the transcriptions and audio is required \cite{PoveyTDNN2015}.
This alignment is usually estimated with a \acrshort{gmm}-\acrshort{hmm} system.
Flat-start training of the \acrshort{dnn} in one stage (i.e., without using any previously trained model, alignment, or performing prior estimation) is possible, as shown in \cite{lfmmi_flat_start_interspeech} with the use of biphones.
This form of training can be assimilated into end-to-end training and is usually called \acrshort{eelfmmi}.

\subsection{N-gram language model} \locallabel{sec:ngram}
The language model estimates the probability $P(W)$ that a sequence of words $W$ corresponds to a sentence.
The goal is to estimate whether a sequence of words conforms to a particular grammar, specific to the language studied.
The probability of observing a sequence of words $W = [w_1 \hdots w_M]$ in a statistical language model is expressed in the equation below:
\begin{equation}
    \locallabel{chapt_2:eq_chapt2:ngram}
    \begin{aligned}
        P(W) & = P(w_1) P(w_2|w_1) P(w_3|w_1,w_2) \hdots P(w_i|w_1,w_2,\hdots,w_{m-1}) \\
        P(W) & = \prod_{m=1}^M P\left(w_m|w_1,\ldots,w_{m-1}\right)
    \end{aligned}
\end{equation}
where $M$ is the length of the sequence $W$ and $P(w)$ is the probability of observing the word $w$.
This equation indicates that the probability of observing a sequence of words is determined by the probability of observing each individual word given the past words in the sequence.

The most common n-gram models are 2-gram and 3-gram, which require a history of one or two words respectively \cite{ngram}.
For example, the probability of a word knowing the preceding word in a 2-gram model can be calculated using the following formula:
\begin{equation}
    P\left(w_j \mid w_i\right)=\frac{\operatorname{count}\left(w_i, w_j\right)}{\sum \operatorname{count}\left(w_i, w\right)}
\end{equation}
The probability $(w_m \mid w)$ is given by the number of occurrences of word $w_m$ followed by word $w_j$, divided by the number of occurrences of the same word $w_m$, followed by any other word.
The 2-gram model is then:
\begin{equation}
    P(W) \approx P\left(w_1\right) \times \prod_{m=2}^M P\left(w_m \mid w_{m-1}\right)
\end{equation}
In speech recognition, language models are generally n-gram models whose order is bounded between 2 and 4.


\subsection{End-to-end and self-supervised learning} \locallabel{chapt2:sec:wav2vec2}

With the increase of computational power, new methods to build \acrshort{asr} systems have emerged, called end-to-end automatic speech recognition \cite{e2e_asr_survey}.
End-to-end architectures eliminate the need to pre-align the data with an additional \acrshort{gmm} model for example.
They unify the training process with a single model.
Unlike models based on \acrshort{hmm}, language modeling is inherent to the architecture as end-to-end models can directly convert acoustic features into a sequence of words or subword units (e.g., byte-pair-encoding \cite{sennrich-etal-2016-neural} and unigram language model \cite{Kudo2018SubwordRI}), avoiding the need for a pronunciation lexicon.
In practice, an external language model is often used to improve recognition accuracy.

One paradigm from which end-to-end training benefits, is self-supervised learning for acoustic feature extraction.
Self-supervised learning enables to easily create systems that can, on one end, input raw speech signals and, on the other end, output words without necessitating large text-annotated corpora.
Here we will focus on the task of extracting speech features from raw audio data without using traditional hand-crafted features.
To accomplish this, large neural networks, usually based on the transformer architecture (see Section \ref{main:chapt_1:transformer}), learn from a very large quantity of unlabeled data the structure of the speech signal itself.
By doing so it replaces traditional hand-crafted features with models that can be fine-tuned later on for specific tasks.
One example of it is Wav2Vec-2.0 \cite{wav2vec2}, which is learned with a training objective similar to BERT \cite{bert} masked language modeling loss.
In the paper of Wav2Vec-2.0, the authors showed that fine-tuning the self-supervised pre-trained transformer with only one hour of labeled data outperforms existing state-of-the-art \acrshort{asr} systems trained on 100 times more labeled data.

The Wav2Vec-2.0 model is pre-trained to take raw audio data as input and generate latent speech representations using a multi-layer 1-d convolutional neural network.
Then, the model learns quantized vectors of the latent representations, where continuous latent representations are matched with discrete representations from a fixed-sized dictionary containing similar representations.
Then, the model selects the closest quantized representation from the dictionary for each latent representation.
During training, about half of the latent representations are masked before being fed to the transformer.
For each masked vector, 100 other quantized vectors from the same utterance are randomly selected as negative distractors (in red in Figure \localref{image_chapt2:wav2vec2}).
The model is trained to identify the real quantized vector (in green in Figure \localref{image_chapt2:wav2vec2}) among negative distractors from the output of the transformer at the masked positions, this kind of training is called contrastive learning \cite{contrastive_learning}.
Figure \localref{image_chapt2:wav2vec2} presents this pre-training process.

\begin{figure}[htbp]
    \vspace{-2.0mm}
    \begin{center}
        \includegraphics[width=0.89\linewidth]{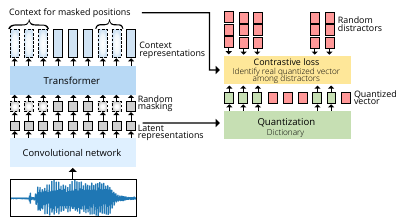}
    \end{center}
    \caption{
        \centering{
            Overview of the Wav2vec 2.0 architecture and its pre-training process.\linebreak
            (Modified from:\citetitle{illustrated_wav2vec2})
        }
    }
    \vspace{-4mm}
    \locallabel{image_chapt2:wav2vec2}
\end{figure}

\subsection{Evaluation}
Speech recognition is the process of transcribing an acoustic signal into words.
In order to evaluate the correctness of a predicted transcript, it is compared to a reference transcript.
The \acrfull{wer} metric can be employed to measure the number of errors between two transcripts.
The errors taken into account are substitution, deletion, and insertion errors.
Substitution errors are words that are incorrectly transcribed.
Insertions errors are words added during transcription, and deletion errors are words omitted.
The word error rate is computed as follows:

\begin{equation}
    \text{WER} \;=\; \frac{\operatorname{count}(\text{substitutions}) \;+\; \operatorname{count}(\text{deletions}) \;+\; \operatorname{count}(\text{insertions})}{\operatorname{count}(\text{words in the reference})}
\end{equation}
The \acrshort{wer} is a useful metric for comparing the performance of different speech recognition systems, as well as for tracking the improvements made within a single system over time.
Additionally, the \acrshort{wer} can be used to diagnose the specific types of errors that a speech recognition system is making, which can help identify areas for improvement.


\newpage
\section{Automatic speaker recognition} \locallabel{chapt_2:sec:asv}
Speaker recognition systems identify or verify the identity of a speaker based on a speaker's speech characteristics.
A distinction must be made between the automatic speaker identification task and the \acrfull{asv} task.
Speaker identification aims to find the most probable speaker among a set of known speakers (close-set), and speaker verification aims to assess whether a test sample matches a claimed speaker's identity.
This use case is more representative of real-world applications, where new identities are constantly being encountered.
As such, it is the one primarily used in this thesis.


In a \textit{text-dependent} scenario, the speaker must utter a predefined prompt. This helps to remove/control one of the major sources of acoustic variability, the linguistic content.
Having the user always utter similar passphrases allows better speaker representation, which is beneficial when high accuracy is required.
In contrast, in the \textit{text-independent} scenario, the linguistic content does not require to be imposed.
The \textit{text-independent} scenario is more generic and user-friendly.
In research, there is a larger amount of data available in the \textit{text-independent} scenario \cite{nagrani2017voxceleb}, and overall more research has been done in this scenario.
%
For this thesis, the application context chosen to evaluate how much the identity of a speaker is recognizable before and after anonymization is \textit{text-independent} speaker verification.
This context is also the de facto standard when it comes to evaluating biometric information protection \cite{jain2008biometric,iso24754,linkability}.

\begin{figure}[hbp]
    \begin{center}
        \includegraphics[width=0.99\linewidth]{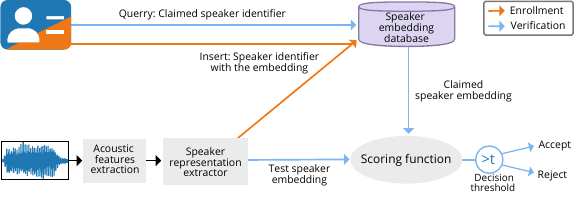}
    \end{center}
    \vspace{-4mm}
    \caption{
        Overview of the enrollment and verification steps in \acrshort{asv}.
    }
    \locallabel{image_chapt2:asv_pipeline}
\end{figure}


Speaker verification pipelines are composed of three main components, depicted in Figure \localref{image_chapt2:asv_pipeline}.
It starts with acoustic features extraction, as presented in chapter \ref{main:image_chapt1:signal_to_mfcc}.
The speech signal first needs to be converted to a time-frequency representation to obtain a sequence of acoustic features.
In contrast to traditional \acrshort{asr} systems where the 12 first \acrshort{mfcc} coefficient are used \cite{fbank_study}, traditional \acrshort{asv} requires more coefficients, above 24 \cite{mfcc_asv}, to better capture speaker characteristics.
Then, given a sequence of multiple acoustic feature frames, a speaker embedding extractor transforms this acoustic sequence into a fixed and compact vector called a speaker embedding.
An embedding is a high-level representation usually extracted from a bottleneck layer (see Section~\ref{main:explanation:bottleneck}).
However, in contrast to a bottleneck that can be extracted for each audio frame, speaker verification typically requires an embedding representation (speaker embedding) that describes the full audio segment.
%
%
%
Speaker embeddings should primarily encode speaker discriminative information while discarding incidental information not relevant to the speaker (e.g., microphone, recording session, etc.).
Finally, the speaker embeddings are compared to obtain a comparison score indicating how similar two speakers are for the system.
Given a comparison score, a binary decision can be made to confirm or deny the correspondence.

Figure \localref{image_chapt2:asv_pipeline} summarizes how a user interacts with the system during the enrollment and verification phases.
During enrollment, speaker embeddings are extracted from each user's speech sample and stored in a database indexed by a unique user identifier.
During verification, the user provides the system with a test speech sample and a unique user identifier which is used to get the associate speaker embeddings in the database.
If the database contains several speaker embeddings for the user (multiple samples), an average operation can be used to aggregate them (which increases robustness) to create a claimed speaker embedding.
A test speaker embedding is extracted from the test speech sample and used to compute a similarity score $s$ with the claimed speaker embedding.
The score is compared against a decision threshold $t$.
If $s>t$, then the verification trial is accepted; otherwise, rejected.

The following sections detail one method based on deep learning to train a speaker embedding extractor (x-vector), scoring function (\acrshort{plda} or cosine similarity), and metrics used to evaluate \acrshort{asv} systems performances (\acrlong{eer}, Linkability).

\vspace{-2mm}
\subsection{Speaker embedding extractor} \locallabel{section:x-vector}
The sequence of acoustic features contains a good deal of information about the signal.
They encode the spoken linguistic content, speaker voice characteristics (\acrshort{f0}, volume, timbre, speaking rate, etc.), speech characteristics (rhythm, accent), ambient noise, and background sounds.
Extracting a speaker embedding from the acoustic features invariant to the non-speaker-related characteristics is a complicated task \cite{speaker_reco_tutorial}.
In recent years, statistical and \acrshort{dnn}-based speaker models have primarily been used.
This section discusses the latter, as it has superior speaker modeling capabilities due to the advancements made in deep learning.

\begin{figure}[htbp]
    \begin{center}
        \includegraphics[width=0.50\linewidth]{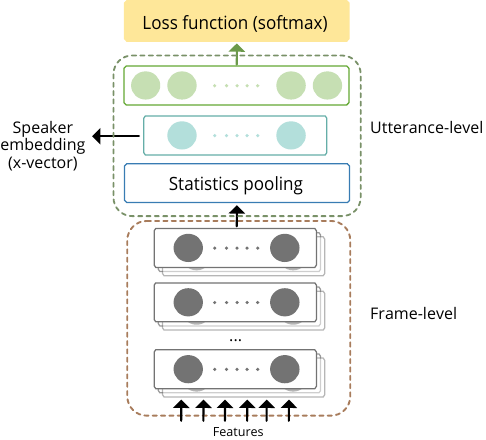}
    \end{center}
    \vspace{-6mm}
    \caption{
        The architecture of the model used to extract x-vector speaker embeddings.
    }
    \vspace{-1mm}
    \locallabel{image_chapt2:xvector}
\end{figure}
One of the first \acrshort{dnn}-based speaker embedding is called x-vector \cite{Snyder2017DeepNN,snyder2018xvector}.
The model used to extract x-vectors is trained in a supervised manner to classify the speakers of a training dataset given the sequences of acoustic features.
As shown in Figure~\localref{image_chapt2:xvector}, the model used to extract x-vectors can be separated into two main components: frame-level and utterance-level computations.
The network takes as input frame-by-frame acoustic features (e.g. \acrshort{mfcc}) which are transformed by multiple frame-level \acrshort{tdnn} layers.
The most common architecture is composed of 5 \acrshort{tdnn}  layers, having respectively the following input contexts: [t-2, t+2], \{t-2, t, t+2\}, \{t-3, t, t+3\}, \{t\}, \{t\} (refer to section \ref{main:image_chapt1:tdnn} for the notations).
After the frame-level layers, a statistic pooling layer computes the mean and standard deviation of intermediate features.
Then, five layers (including a bottleneck layer) are incorporated after the pooling layer to classify the speakers during training.
The cross-entropy loss function is employed to train the network to classify speakers.
During verification, the speaker embedding is typically extracted from the pre-final bottleneck layer of the network and usually has a dimension of 512.
After the extraction, the subsequent classification and softmax layers are not used.
Before scoring, the speaker embedding is length-normalized.

Although the cross-entropy loss aims to train the network to separate speakers, researchers have sought other loss functions that improve the speaker embedding representation.
Recent advancements in model training include the incorporation of angular/cosine-margin-based losses, which have the objective of minimizing intra-variance and maximizing inter-variance.
Among these losses, a particularly noteworthy one is the ArcFace loss, also known as the \acrfull{aam} loss \cite{aamlossArcMarginProduct}.
 This loss function has gained significant attention and has shown promising results in speaker verification. \locallabel{chapt_2:aam}


\vspace{-2mm}
\subsection{Scoring function}

The scoring function goal is to compare how close two speaker embeddings are.
Since it is almost impossible to have the exact same vector between the claimed (also called enrollment) and test embeddings (which would result in a binary yes/no same speaker), the scoring methods provide a score indicating how much the two vectors correspond to the same speaker.
If this score is higher (or lower) than a predefined threshold, the system accepts (or rejects) the trial.
The most straightforward comparison is to compute the score with the cosine similarity \cite{CosineASV} function.
Alternatively, the more sophisticated \acrfull{plda} \cite{plda_1,plda_2} can provide better scores on cross-entropy trained systems. 
%

\vspace{-2mm}
\subsubsection{Cosine Similarity}
Cosine similarity scoring is a computationally efficient method in many verification tasks.
The cosine similarity is a measure of the angle between the claimed ($\mathbf{x}^{C}$) and test ($\mathbf{x}^{T}$) embeddings.
This technique has the advantage of not requiring any training.
Scoring is performed directly in the speaker embedding space.

\subsubsection{\texorpdfstring{\acrlong{plda}}{PLDA}}
Unlike the cosine similarity, \acrshort{plda} is a supervised method where speaker labels are necessary to estimate the \acrshort{plda} parameters.
Several \acrshort{plda} variants exist \cite{pldasimple,pldatwocov,neuralplda}.
In this thesis, the same one as in \cite{pldatwocov} is employed.
It is known as the two-covariance \acrshort{plda} variant and is implemented in Kaldi.
The two-covariance \acrshort{plda} is used to model the distribution of speaker embeddings in a multidimensional space.
In the two-covariance \acrshort{plda}, the embedding $\mathbf{x}_{i}$ from the $i$-th speaker is assumed to be generated from a linear Gaussian model $p(\mathbf{x}_{i} \mid \mathbf{y}_i) = \mathcal{N}\left(\mathbf{x}_{i} \mid \mathbf{y}_i, \mathbf{W}^{-1}\right)$, where $\mathbf{y}_i$ represents the mean of the $i$-th speaker and is also assumed to follow a Gaussian distribution $p(\mathbf{y}_i) = \mathcal{N}\left(\mathbf{y}_i \mid \boldsymbol{\mu}, \mathbf{B}^{-1}\right)$.
$\mathbf{W}^{-1}$ and $\mathbf{B}^{-1}$ are the within-speaker and between-speaker covariance matrices that are used to capture the variation in speaker embeddings within and between different speakers, respectively.
$\boldsymbol{\mu}$ is the global mean in the speaker space.
The iterative E-M algorithm \cite{Dempster1977MaximumLF} is usually used to estimate the $\mathbf{W}$, $\mathbf{B}$ and $\boldsymbol{\mu}$ \acrshort{plda} parameters.
The score between two embeddings $x^C$ and $x^T$ is computed as a likelihood ratio based on two same/different class hypotheses as follows:
\begin{equation}
    s\left(\mathbf{x}^C, \mathbf{x}^T\right)= \frac{\operatorname{likelihood}\left(\text{same speaker}\right)}{\operatorname{likelihood}\left(\text{different speaker}\right)}= \frac{p\left(\mathbf{x}^C, \mathbf{x}^T\right)}{p\left(\mathbf{x}^C\right) p\left(\mathbf{x}^T\right)}
\end{equation}
where $p\left(\mathbf{x}^C, \mathbf{x}^T\right) = \int p\left(\mathbf{x}^C|\mathbf{y}\right)p\left(\mathbf{x}^T|\mathbf{y}\right)p\left(\mathbf{y}\right) \, dy$ and $p\left(\mathbf{x}^C\right) = \int p\left(\mathbf{x}^C|\mathbf{y}\right)p\left(\mathbf{y}\right) \, dy$.
One advantage of using \acrshort{plda} is that it explicitly takes into account speaker variability, which can improve the accuracy of speaker verification for the chosen system (\acrshort{tdnn}).
In contrast, cosine distance only measures the similarity between two vectors and does not consider the distribution of speaker embeddings.

In general, both \acrshort{plda} and cosine distance are useful for speaker verification, but the appropriate method depends on the speaker embedding extractor and the available data.


\subsection{Evaluation} \locallabel{chapt_2:asv_eval}
Evaluating the performance of \acrshort{asv} systems is a delicate process, as many parameters can influence the reliability of the system.
Those factors, listed in \cite{bimbot1997assessment}, can range from the quality of the speech signals to the quantity of speech for a given speaker or the size of the population of speakers.
Evaluation campaigns have been introduced to produce standardized datasets and establish assessment procedures that enable an objective comparison.

Speaker verification systems are evaluated using a test dataset containing multiple trials.
Each trial consists of a claimed identity and test audio segment.
If the claimed identity and the audio segment point to the same speaker, the trial is considered \textit{genuine}, else if they belong to different speakers, it is considered as \textit{impostor}.
The \acrshort{asv} systems generate a score for each trial; high scores values reflect a high similarity, and low values indicate a difference.
An example of the score distributions for genuine and impostor trials is illustrated in Figure \localref{image_chapt2:eer}.
In order to make a decision, a decision threshold $t$ must be set.
This threshold directly affects the two errors made by the \acrshort{asv} systems.
A low threshold value will accept too many impostor trials, while a high threshold increases the chance of rejecting genuine trials.
The two types of errors can be measured with two metrics:
\begin{itemize}
    \item The \acrfull{far} measures how many impostor trials are accepted.
    \item The \acrfull{frr} measures how many genuine trials are rejected.
\end{itemize}
\noindent
Increasing the detection threshold t decreases the \acrshort{far} while increasing the \acrshort{frr}, and vice versa.
\begin{figure}[hbtp]
    \vspace{-6mm}
    \begin{center}
        \includegraphics[width=0.75\linewidth]{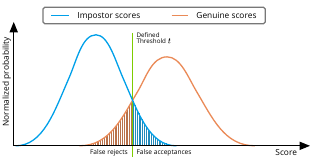}
    \end{center}
    \vspace{-6mm}
    \caption{
        Threshold-based decision-making on impostor/genuine scores
    }
    \locallabel{image_chapt2:eer}
    \vspace{-1mm}
\end{figure}


\subsubsection{\texorpdfstring{\acrlong{eer} metric}{EER metric}}
The \acrfull{eer} metric is one of the most popular metrics in speaker verification as it compares two systems based on a single threshold ($t_{\text{EER}}$).
The $t_{\text{EER}}$ is defined as the point where the \acrshort{far} and \acrshort{frr} values are equal.
This configuration is reached by varying the threshold until the two areas corresponding to the false acceptances and false rejections become equal (see example in Figure \localref{image_chapt2:eer}).
\begin{equation}
    \text{EER} \;=\; \text{FAR}(t_{\text{EER}}) = \text{FRR}(t_{\text{EER}})
\end{equation}
The \acrshort{eer} values range from 0\% to 50\%.
The best value of \acrshort{eer} is  0\% of error, meaning the impostor scores are entirely separated from the genuine one at threshold $t$.
Whereas 50\% of \acrshort{eer} means the system's performance equals random decision.


\vspace{-1em}
\paragraph{Limitation of the \texorpdfstring{\acrshort{eer} metric}{EER metric}  and uncalibrated scores}\leavevmode\newline \locallabel{chapt_2:lim_eer}
A limitation of threshold-based metrics is that they evaluate speaker verification given a decision threshold.
The \acrshort{eer} is considered to be an accurate indicator when the scores behave as they are expected, i.e. the higher score, the most likely the trial is genuine (like in Figure~\localref{image_chapt2:eer}).
However, when the scores are not as well distributed, evaluations based only on a single decision threshold will not necessarily be reflective of real-world application.
Other decision-based threshold metrics, in particular, the \acrfull{cllr} \cite{cllr}, evaluate every possible threshold value (considering all \acrshort{far} / \acrshort{frr} scores) and also considers calibration.

As shown in \cite{comparative_metric}, against uncalibrated scores, the \acrshort{eer} fails to evaluate the discriminative potential of \acrshort{asv} systems.
An extreme kind of uncalibrated score distributions is shown in Figure \localref{image_chapt2:bad_eer}, where the genuine score distribution is located between the two modes of the impostor scores distribution.
In this example, without calibration, the \acrshort{eer} equals 50\%, indicating the system performs random decisions.
A solution to better evaluate this example is to calibrate the scores to be proper log-likelihood ratio scores with a non-monotonic transformation \cite{non-linear_calibration}. 
After calibrating the scores, the \acrshort{eer} equals 15\%, a much lower value that better represents the discriminative potential of the system.
Such a situation is unlikely to occur in speaker verification assessment, where unmodified speech data is used.
However, in the field of speaker anonymization, where speech is anonymized, unexpected, more overlapping score distribution becomes frequent \cite{comparative_metric}, as the \acrshort{asv} system will have more difficulty recognizing the speaker's identity.

\begin{figure}[hbtp]
    \begin{center}
        \includegraphics[width=0.74\linewidth]{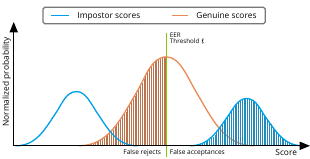}
    \end{center}
    \vspace{-6mm}
    \caption{
        \centering{
            Simulated uncalibrated impostor and genuine scores where the genuine scores are between the two modes of the impostor scores.
        }
    }
    \locallabel{image_chapt2:bad_eer}
    \vspace{-5mm}
\end{figure}

\subsubsection{Linkability metric}
The linkability metric introduced in \cite{linkability} for biometric template protection solves threshold-based metric limitations by analyzing the overlap between the impostor/genuine distributions (like the \acrshort{cllr}).
However, in contrast to the \acrshort{cllr} which usually performs calibration with a \acrfull{pav} \cite{pav}, the linkability metric uses histogram binning which works better with extremes uncalibrated score that requires a non-monotonic transformation.

The local linkability metric is based on a score-wise measure that depends on the likelihood ratio between scores distributions: $p(H|s) - p(\bar{H}|s)$, where $s$ is the score, $H$ is the binary variable for a genuine trial and $\bar{H}$ for an impostor trial.
When the local linkability is negative, it indicates with high confidence that the score corresponds to different speakers.
However, as this measure is targeted to describe the strength of linkability rather than the strength of unlinkability (different speakers), negative values are clipped:
\begin{equation}
    D_{\leftrightarrow}(s)=\operatorname*{max}(0,~ p(H|s)-p(\bar{{{H}}}|s))
\end{equation}
The global linkability measure $D_{\leftrightarrow}^{\mathrm{sys}}$ is the average value of $D_{\leftrightarrow}$ over all genuine scores:
\begin{equation}
    D_{\leftrightarrow}^{\mathrm{sys}}=\int p(s|H)\cdot D_{\leftrightarrow}(s)\;d s
\end{equation}
In practice, $D_{\leftrightarrow}(s)$ is rewritten as $(2 \cdot \omega \cdot \operatorname{lr}(s)) /(1+\omega \cdot \operatorname{lr}(s))-1$ where the likelihood ratio $\operatorname{lr}(s)$ is $p(s \mid H) / p(s \mid \bar{H})$ and the prior probability ratio $\omega$ is $p(H) / p(\bar{H})$, and $p(s \mid H)$ and $p(s \mid \bar{H})$ are computed via one-dimensional histograms.

The \acrshort{li} linkability values range from 0.00 to 1.00, and in opposition to the \acrshort{eer}, the higher the \acrshort{li}, the better the \acrshort{asv} system can accept or reject trials.

The advantage of the \acrshort{li} metric over the \acrshort{eer} is the non-single-threshold-based evaluation that it makes.
An advantage that is also shared with the \acrshort{cllr}.
The \acrshort{li} advantage over the \acrshort{cllr} is the default calibration function that it uses. As against scores non-monotonically related to the likelihood ratio, histogram binning is better than \acrshort{pav}.
It is for that reason that the \acrshort{li} is strongly advocated by the authors of \cite{linkability,comparative_metric} and used in this thesis for robust \acrshort{asv} evaluation under adversary and anonymization conditions.
It is worth noting that in practice, on real data, the \acrshort{cllr} and \acrshort{li} follow a clear relation \cite{comparative_metric}.

\newpage
\section{Voice conversion} \locallabel{chapt_2:vc}
\vspace{-1em}
\acrfull{vc} is a discipline where the goal is to modify the voice characteristics of a source speaker's speech to match those of a target speaker without changing the linguistic content.
The principle of \acrshort{vc} is to define a transposition function that converts the speech of one (or more) source speakers to the voice of one (or more) target speakers.

Traditional \acrshort{vc} systems require parallel speech corpus, in which speech recordings
come in pairs by the source speaker and the target speaker.
Direct relations between source and target pairs enable the creation of the transposition function.
Numerous approaches have been proposed, such as \acrshort{gmm} \acrshort{vc} \cite{parallel_vc_1,parallel_vc_2,parallel_vc_3}, frequency warping \acrshort{vc} \cite{vc_freq_wrap,vc_freq_wrap_2}, \acrshort{dnn} \acrshort{vc} \cite{vc_dnn_1,Nakashika2013VoiceCI}.
All the above approaches provide reasonably good results, the best being \acrshort{dnn}-based.
However, the requirement of a parallel corpus causes limitations as, in practical applications, parallel data is not easily available.
Hence, lately, the \acrshort{vc} research community has mainly focused on approaches where non-parallel data is used to build \acrshort{vc} systems.

Non-parallel trainable \acrshort{vc} systems are more valuable as training data acquisition is much easier, despite the fact that training is more complex.
Most non-parallel methods rely on separating the linguistic and speaker-related representations carried out by acoustic features.
During training, the model is asked to reproduce the speech of the source (same identity).
At the conversion stage, the linguistic content of the source speaker utterance is extracted by a linguistic encoder and kept unmodified.
In contrast, the speaker representation (usually a one-hot or x-vector embedding) is derived from the target speaker's speech.
A synthesis model is then used to generate speech with the target speaker's characteristics with the linguistic content of the source speaker (example in Figure \localref{image_chapt2:vc}).

\begin{figure}[htbp]
    \vspace{-2mm}
    \begin{center}
        \includegraphics[width=0.68\linewidth]{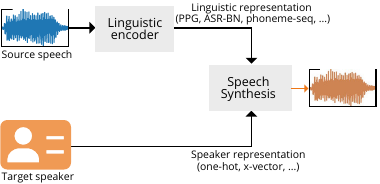}
    \end{center}
    \vspace{-6mm}
    \caption{
        A typical flow of a voice conversion system.
    }
    \vspace{-3mm}
    \locallabel{image_chapt2:vc}
\end{figure}

The following sections present some methods to separate linguistic and speaker information of speech signals.
Then we categorize voice conversion approaches based on the numbers and origins of the source and target speakers.
Last, we present the current state-of-the-art speech synthesis model.

\vspace{-2mm}
\subsection{Linguistic representation} \locallabel{chapt_2:vc_lin_rep}

Linguistic information extraction can be divided into two categories, supervised and self-supervised.
Supervised representation typically uses an \acrfull{asr} model to extract the linguistic representation.
The linguistic representation of speech signals can be expressed in a variety of forms, such as a sequence of discrete tokens (i.e., phonemes) as described in \cite{cascalde_vc_vcc}, a sequence of \acrfull{ppg} as described in \cite{ppgs,liu18d_interspeech}, or a sequence of bottleneck features from an acoustic model as described in \cite{vc_disentangled_2020}.
This category requires phoneme (or text) supervision during the training as an \acrshort{asr} model is necessary.
The second category does not require any label as it relies on self-supervised objectives.
Self-supervised objectives allow learning from unlabeled data by using the inherent structure of the data to generate labels for itself.
Models implicitly separate linguistic and speaker features in a single model using auto-encoder architectures \cite{vc_auto_enco,chou19_interspeech,qian2019autovc,one-shot-vc-vector-quant} or generative adversary networks \cite{gan_vc}.

For this thesis, the text supervision required to train the linguistic encoder is a realistic application setting.
Extracting linguist representation using \acrshort{asr} is an easier framework than letting an auto-encoder find the separation on its own while also having high-converted voice quality \cite{qian2019autovc}.
Text supervision helps when training the linguistic encoder as a clear indication of what constitutes an appropriate linguist representation is available.
Additionally, this method of converting voices has been proven relevant for speaker anonymization \cite{fangSpeakerAnonymizationUsing2019}.
The following section presents the main methods to obtain supervised linguistic representations for voice conversion.

\subsubsection{Discrete token sequence as linguistic features}

One of the most straightforward methods to implement \acrshort{vc} is to extract a sequence of discrete tokens (e.g., phonemes, text) from an \acrshort{asr} system and feed it to a \acrfull{tts} system, this method is referred to as cascade \acrshort{asr}+\acrshort{tts} \cite{vc_transformer_tts}.
The major advantage and drawback of cascade \acrshort{asr}+\acrshort{tts} is the discrete token used to transmit information from the \acrshort{asr} to the \acrshort{tts} system (example: \say{b ey s b {\usefont{T3}{cmr}{m}{n}\textopeno}h l} for the word \say{baseball}). 
The advantage is that no acoustic speaker information is contained in discrete token representation allowing the \acrshort{tts} more consistency in synthesizing specific target speaker characteristics across different source speakers.
The disadvantages are that early stages errors propagate to downstream models, meaning any recognition failure in the first \acrshort{asr} stage will harm linguistic consistency, creating mispronunciation errors in the \acrshort{vc} pipeline.
Significant degradation of emotions and speaker intention cues are also expected, as the discrete token representation does not encode the acoustic features which enable human understanding of the prosody.
Altering those aspects may affect the sentence's understanding as a statement may be synthesized similarly to a question. 
Even if \acrshort{tts} system can be conditioned on with on prosodic \cite{Raitio2020ControllableNT} features, prosody recognition stays a complicated and very subjective topic of research \cite{rosenberg18_speechprosody}.

\vspace{-2mm}
\subsubsection{Phonetic posteriorgrams sequence as linguistic features} \locallabel{chapt_2:ppg}
\vspace{-2mm}

To mitigate some mispronunciation and prosody transfer errors (like the speech rate, or the articulation) of the aforementioned discrete representation, \acrshort{ppg} \cite{ppgs} were introduced for voice conversion.
A \acrshort{ppg} is a time/class matrix representing the posterior probabilities of each phoneme class for each time frame of an utterance.
In other words, a soft label is generated for each speech frame, indicating the likelihood of each possible phoneme being uttered (see example in Figure \localref{image_chapt2:ppg_example}).
\begin{figure}[!htbp]
    \begin{center}
        \includegraphics[width=0.48\linewidth]{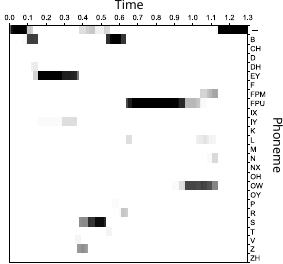}
    \end{center}
    \vspace{-2mm}
    \caption{
        An example \acrlong{ppg} for the spoken word ``baseball''.
        (reproduced from:\citetitle{ppg_image}, for display purposes, not all phonemes are listed on the y-axis).
    }
    \vspace{-2mm}
    \locallabel{image_chapt2:ppg_example}
\end{figure}

In contrast to discrete tokens, \acrshort{ppg} encodes pronunciation duration as the same consecutive labels are not merged.
This helps to keep the source speaker's emotions and intention information in the converted speech.
Furthermore, \acrshort{ppg} are less sensitive to modeling error than a sequence of discrete tokens because no decoding is made.
Operating at the frame level and extracting soft phoneme labels makes \acrshort{ppg} more resilient to the source speaker's pronunciation, as likelihood scores can encode non-common pronunciations.
However, \acrshort{ppg} are usually extracted after the softmax layer of a \acrshort{tdnnf}-based acoustic model (see Section \localref{section:acoustic-modeling}).
And, as a result of \acrshort{asr} training, the soft label values of \acrshort{ppg} are usually bimodal (highest values close to one and other values close to zero).
This creates a deficiency where mispronunciations can still occur in the pipeline if the acoustic model does not generalize (due to domain mismatch or noisy recording for example).


\subsubsection{\texorpdfstring{\acrshort{asr} bottleneck sequence as linguistic features}{ASR bottleneck sequence as linguistic features}} \locallabel{chapt_2:bn}
\vspace{-1mm}

To address the \acrshort{ppg} deficiencies mentioned above, linguistic features can be extracted from the lower intermediate bottleneck layer rather than at the softmax layer \cite{Liu2021AnytoManyVC}.
This simple modification makes bottleneck linguistic features similar to \acrshort{ppg} while greatly helping with noise, accent, and domain mismatch generalization.
However, \acrfull{asrbn} based linguistic features also encode more speaker-related information \cite{fangSpeakerAnonymizationUsing2019,adiReverseGradientNot2019} than \acrshort{ppg}, which can restrict downstream speaker modification in the \acrshort{vc} pipeline.
Bottleneck-based linguistic features for voice conversion are gaining popularity \cite{polyak20_interspeech,vc_disentangled_2020,liu21c_interspeech,Liu2021AnytoManyVC}, and in fact, we also use them for the approaches proposed in this thesis.



\subsection{Speaker representation}
\vspace{-1mm}

Similarly to the linguist representation, speaker representation can be extracted by supervised specialized models \cite{snyder2018xvector} or self-supervised in auto-encoder-based \acrshort{vc} models \cite{qian2019autovc}.
For this thesis, we focus on supervised methods where the two main speaker representations employed are x-vector (presented in Section \localref{section:x-vector}) and one-hot embedding.

One-hot embeddings represent categorical variables as binary vectors in a format compatible with machine-learning algorithms.
One-hot speaker embedding vectors have all values set at zero except for the speaker's index.
Thus, one dimension is allocated for each speaker in the dataset.

During training, \acrshort{vc} learns to synthesize speech from the linguistic and speaker representations.
Depending on the speaker representation, a \acrshort{vc} system might be constrained to a limited number of speakers \cite{Toda2007OnetoManyAM,Ohtani2009ManytomanyEC,Liu2021AnytoManyVC}.
For one-hot embeddings, the target speaker must be present in the training data, whereas for x-vectors, the target can be arbitrary.
One-hot speaker embedding performs the best in terms of synthesis quality because the target speaker is present in the training dataset.
However, when an application needs to consider new target speakers, x-vectors or a similar speaker representation is necessary as the target speaker might not be present in the training dataset.
In the following section, we describe the different variants that may lead a system to employ either one-hot or x-vector representations.

\subsubsection{Numbers and origins of the source and target speakers} \locallabel{any-many-one-vc}
\vspace{-1mm}

An essential characteristic of \acrshort{vc} methods is the number and origin of source speakers and target speakers that a single system can support.
\acrshort{vc} approaches can be categorized into \textit{one-to-one}, \textit{many-to-one}, \textit{many-to-many}, \textit{any-to-many} and \textit{any-to-any} \acrshort{vc}.
The following list briefly presents the differences between them:
\begin{itemize}
    \item
          \textit{One-to-one} \acrshort{vc} is the easiest model to build as the mapping function is limited to a specific pair of source and target speakers; thus, there is no need to generalize to multiple and unseen speakers \cite{cycle_gan}.

    \item
          \textit{Many-to-one} (or \textit{one-to-many}) \acrshort{vc} approaches extends the versatility by extending \acrshort{vc} to multiple source (or target) speakers \cite{Toda2007OnetoManyAM}.
          The source and target speakers must be present in the training dataset.

    \item
          \textit{Many-to-many} \acrshort{vc} approaches further improve the functionality as a single system can convert a known source speaker into a known target speaker voice \cite{Ohtani2009ManytomanyEC}.
          The source and target speakers must still be present in the training dataset.

    \item
          \textit{Any-to-many} \acrshort{vc} approaches allow converting an unseen source speaker into a known target speaker voice \cite{Liu2021AnytoManyVC}.

    \item
          \textit{Any-to-any} \acrshort{vc} approaches achieve conversion across unseen source-target speaker pair without prior knowledge of them \cite{liu18d_interspeech}.
          Generalizing to arbitrary sources and target speaker makes this approach more complicated than the others, and usually results in lower synthesized voice quality.

\end{itemize}
For this thesis, the most suited approaches rely on \textit{any-to-many} (using one-hot embedding) or \textit{any-to-any} (using x-vector embedding) voice conversion as the speech signals to anonymize comes from unknown speakers.

\subsection{Speech synthesizer} \locallabel{chapt2:speech_synt}
Given the linguistic and speaker representations, most pipeline separate speech synthesis into two steps: decoder and vocoder.
The decoder's purpose is to generate an output Mel spectrogram.
Then, the vocoder transforms the Mel spectrogram into the output waveform.
The linguistic and speaker representations need to be concatenated before being given to the decoder.
To do so, the utterance level speaker representation is incorporated into each linguistic frame.

The decoder can be built using recurrent neural networks \cite{qian2019autovc}, fully convolutional architectures \cite{one-shot-vc-vector-quant} or transformer \cite{vc_transformer_tts}.
On the other hand, popular vocoders include Griffin-Lim \cite{griffin} algorithm and neural network models such as WaveGlow \cite{waveglow} and HiFi-GAN \cite{hifigan}.
State-of-the-art approaches require fine-tuning the vocoder on the decoder's Mel spectrograms \cite{universal_vocoder}.
Training the vocoder this way helps to smooth out some decoder errors in contrast to training the vocoder only on real Mel spectrograms and using it with converted Mel spectrograms.

Recently, with the development of \acrshort{gan}-based techniques, this two steps synthesis pipeline has been compressed into one \cite{speech_Resynthesis,Kashkin2022HiFiVCHQ}.
The efficient HiFi-GAN vocoder framework can be adapted to perform both decoder and vocoder tasks removing the need for the intermediate Mel spectrogram in the synthesis phase.
Waveforms can be generated from the linguistic and speaker representation directly.
As a result, this technique does not necessitate training a decoder and then fine-tuning the vocoder.
The HiFi-GAN framework is presented in more detail in the subsequent section.

\vspace{-2mm}
\subsubsection{HiFi-GAN} \locallabel{sec:hifigan}
\vspace{-2mm}

Compared to other vocoders, the Hi-Fi-GAN (High Fidelity) \cite{hifigan} model delivers superior computational efficiency and sample quality.
Those results are achieved using a non-autoregressive network architecture and carefully crafted loss functions based on \acrshort{gan} (see \ref{main:chapt_1:gan}).
%
The HiFi-GAN framework consists of a generator, $G$, and a set of two types of discriminators, $D$.
The generator produces a speech signal from a sequence of features corresponding to linguistic and speaker representations (direct \acrshort{vc} with implicit decoder) or Mel spectrograms (vocoder only).
Then, various modules generate speech from the features using successive transposed convolutions with residual blocks having dilated layers.
The transposed convolutions upsample the features to match the temporal resolution of the corresponding waveform, while the dilated layers increase the receptive field.

Natural speech waveforms contain long-term dependencies.
For instance, if a phoneme lasts more than 100~ms, the raw waveform will exhibit a strong correlation between more than 1~600 adjacent samples\footnote{Depending on the sampling frequency.}.
Furthermore, waveforms consist of sinusoidal signals with various periods.
Modeling periodic patterns and long-term interdependence is essential to produce realistic speech audio.
Those problems have been addressed in the HiFi-GAN framework using two different types of discriminators.
During training, the multiscale and multi-period discriminators receive the generated audio sample $\hat{x}$ to determine if the produced audio is realistic.

\begin{figure}[!htbp]
    \begin{center}
        \includegraphics[width=.84\linewidth]{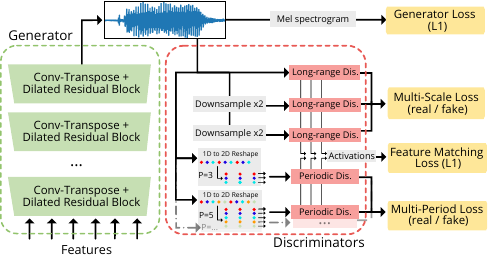}
    \end{center}
    \vspace{-6mm}
    \caption{
        Generator, discriminators, and training losses of a HiFi-GAN model.
    }
    \vspace{-2mm}
    \locallabel{image_chapt2:hifigan}
\end{figure}

The multi-scale discriminators are asked to explore long-range and consecutive audio interactions.
As presented in Figure \localref{image_chapt2:hifigan}, three long-range discriminators operating at different audio scales (raw audio, x2 downsampled audio, and x4 downsampled audio) are used to assess audio samples at various ranges.
On the other hand, five multi-period discriminators explore multiple periodic patterns (only two of them are displayed in the figure).
All period discriminators differ from each other based on the space between the samples.
As shown in Figure \localref{image_chapt2:hifigan}, the audio is first reshaped from a 1D structure to a 2D structure of length $T/p$ and height $p$, $T$ is the length of the raw audio file and $p$ the period to analyze.
The $p$ periods to analyze by each of the period discriminators are [2, 3, 5, 7, 11], they are chosen to avoid overlaps as much as possible.
The periodic samples are independently processed by the multi-period discriminators using convolution having a kernel size of one on the height axis (row by row in Figure \localref{image_chapt2:hifigan}).
%
%
The overall training losses of HiFi-GAN involve a set of adversarial and discriminator losses ($\mathcal{L}_{adv}$, $\mathcal{L}_D$), a generator loss $\mathcal{L}_G$, and a feature-matching loss.
For each $j$ discriminator (8 in total), $D_j$ is tasked with minimizing the following losses:
\vspace{-2mm}
\begin{align}
    \mathcal{L}_{adv}\left(D_j, G\right) & =\sum_{\boldsymbol{x}}\left\lVert 1-D_j(\hat{\boldsymbol{x}})\right\rVert _2^2 \locallabel{chapt2:eq_l_adv}                                                             \\
    \mathcal{L}_D\left(D_j, G\right)     & =\sum_{\boldsymbol{x}}\left[\left\lVert 1-D_j(\boldsymbol{x})\right\rVert _2^2+\left\lVert D_j(\hat{\boldsymbol{x}})\right\rVert _2^2\right]
     \locallabel{chapt2:eq_l_d}
\end{align}
\vspace{-8mm}

\noindent
where $\hat{x}$ is obtained from the generator $G$, $x$ is a real sample.
Equation \localref{chapt2:eq_l_adv} purpose is to encourage the generator to create (fake) audio indistinguishable from genuine (real) audio.
Equation \localref{chapt2:eq_l_d} goal is to train the discriminator to differentiate real and fake audio.
The least-square \acrshort{gan} loss variant is used \cite{Mao2017LeastSG}.

A reconstruction loss between the Mel-spectrogram of the real audio signal and the generated signal is used to train the generator efficiently.
The Mel-spectrogram loss is the L1 difference and is described as:
\vspace{-1mm}
\begin{equation}
    \vspace{-1mm}
    \mathcal{L}_{recon}(G)=\sum_{\boldsymbol{x}}\left\lVert\phi(\boldsymbol{x})-\phi(\hat{\boldsymbol{x}})\right\rVert _1
\end{equation}
where $\phi$ is a function that computes a Mel-spectrogram from a waveform.

Finally, a feature-matching loss \cite{feature_matching_loss} measures the distance between the discriminators internal activation for real audio and fake audio to help the generator better create fake signals.
    \vspace{-1mm}
\begin{equation}
    \vspace{0mm}
    \mathcal{L}_{fm}\left(D_j, G\right)=\sum_{\boldsymbol{x}} \sum_{i=1}^R \frac{1}{M_i}\left\lVert\psi_i(\boldsymbol{x})-\psi_i(\hat{\boldsymbol{x}})\right\rVert _1
\end{equation}
where $\psi$ is the operator extracting the discriminator activations at layer $i$, $M_i$ the number of weighs in layer $i$, and $R$ the total number of layers in $D_j$.

The final generator and discriminators losses to minimize can be written as:
\vspace{-3mm}
\begin{equation}
    \begin{aligned}
        \vspace{-2mm}
        \mathcal{L}_G^{\text {multi}}(G, D)= & \sum_{j=1}^J\left[L_{adv}\left(D_j, G\right)+\lambda_{fm} L_{fm}\left(D_j, G\right)\right] +\lambda_r L_{recon}(G), \\
        \mathcal{L}_D^{\text {multi}}(G, D)= & \sum_{j=1}^J L_D\left(D_j, G\right)
    \end{aligned}
\end{equation}
\vspace{-6mm}

\noindent
where $J$ is the number of discriminators (here 8), $\lambda_{fm}$ is set to $2$ and $\lambda_{recon}$ to $45$ to balance the adversarial losses, Mel-spectrogram loss, and the feature matching loss.

\subsection{\texorpdfstring{\acrshort{f0} conditioning}{F0 conditioning}} \locallabel{chapt_2:f0_cond}
\vspace{-2mm}
Speech converted using the above pipeline, where the two main features are the bottleneck linguistic and speaker representations, exhibit inconsistent \acrshort{f0} distribution compared to real speakers distribution \cite{F0_Huang2019InvestigationOF,F0_Qian2020F0ConsistentMN}.
These inconsistencies come from the bottleneck linguistic representation that encodes some prosody information (mainly \acrshort{f0}).
As a result, the synthesis model generates speech with unnatural \acrshort{f0} values based on the linguistic bottleneck and target speaker representation.
Inadequate encoding of the target speaker's prosodic style is the main limitation of speaker representations.
A straightforward method to address this problem is to have a normalized or transformed \acrshort{f0} sequence added to the linguist and speaker representations mix to help the synthesis model modify the \acrshort{f0} trajectory given the target speaker representation.

\vspace{-2mm}
\subsection{Evaluation}
\vspace{-2mm}
The quality of a voice conversion system (or any speech synthesis system) resides in its ability to convey a linguistic message by speaking the correct sequence of words while having suitable prosody and a similar speaking style as the target speaker.
Objective and subjective evaluations are necessary to assess the synthesized speech quality in all aspects.
For objective evaluations, a \acrshort{asr} system with the \acrshort{wer} metric can be a good proxy metric to asses if the linguistic message is still intelligible (understanding what is being said).
Concerning the target speaker identity, the use of a \acrshort{asv} can give some indication of whether or not the synthesized speech can be associated with the actual speech of the target speaker.
Those two evaluations are interesting because they provide a score without having human listeners.
However, human listeners must be present in the evaluation loop to measure the more subjective aspects of speech.
For instance, humans are needed to assess factors like naturalness (similarity to a natural voice), intonation, and other prosody-related features.
Even the intelligibility, which the \acrshort{asr} system can estimate, needs to be evaluated subjectively because a human's understanding of speech differs from that of an \acrshort{asr} model.

Many scoring exists for subjective evaluation, most notably the \acrfull{mos} score is the most commonly used in the literature.
The MOS score evaluates a system by candidates ranking from 1 (poor) to 5 (excellent) each sample in different dimensions (listening effort, intelligibility, naturalness, quality, rhythm, intonation, etc.).

\vspace{-3mm}
\section{Conclusion}
\vspace{-2mm}
This chapter presented the speech-centric machine learning disciplines crucial for the task of speaker anonymization.
It covers the basics of acoustic modeling, \acrfull{asr}, \acrfull{asv} and \acrfull{vc}.
In the next chapter, we delve into the topic of speaker anonymization and see how these disciplines combine together.


\ifSubfilesClassLoaded{
    \printglossary[title=Special Terms,type=\acronymtype]
    \printbibliography
}{}

\end{document}

\clearemptydoublepage
\cleartooddpage[\thispagestyle{empty}]
\documentclass[../main.tex]{subfiles}

\ifSubfilesClassLoaded{
    \tableofcontentsfile
    \dominitoc
    \setcounter{chapter}{2} 
    \def\locallabelprefix{chapt_3}
    \externaldocument[]{../main}
}{}

\begin{document}

\selectlanguage{english}

\graphicspath{{./figures/dist}}

\chapter{Speaker anonymization} \locallabel{chapt3}
\minitoc
\section{Introduction}

One modern technology that is used by many people is biometrics recognition \cite{Langenderfer2005TheEO}.
The most common examples are fingerprint or face recognition technologies used in smartphones.
However, online social platforms also used them to better track their users.
These technologies aim to extract individual personal characteristics from biometric data into a biometric template used to verify their identity.
The characteristics might be physical traits like fingerprints or behavioral traits like a particular method to solve a security-authentication puzzle \cite{Roth2004APM,Hirakawa2018ImprovementsIA}.
Although many end-users have become more familiar with this technology \cite{Habibu22}, recent federal regulations strictly restrict the use and storage of biometric and personal data.
The nature of data necessitates such laws to prevent biased decisions depending on the gender, origin, and other personal attributes of the user, which would raise privacy and ethical issues.
Additionally, users are more vulnerable in the event of a data breach since, unlike passwords and tokens, compromised unmodified biometric data cannot be revoked and reissued.

By nature, speech falls into the categories of physical and behavioral biometrics because its generation depends on physical traits like the shape of the vocal tract and personality traits like extroversion \cite{trilok2004establishing,5628942}.
Moreover, speech encapsulates a large amount of personal data, like age, gender, health and emotional state, racial or ethnic origin, etc. encouraging the need to develop privacy-enhancing solutions for speech technology \cite{privacy_implication_voice}.
One modern technology used to enhance the privacy of the user when sharing speech data is anonymization.
In this chapter, we present the legal aspects regarding data regulation, the application cases, the threat models and current systems and methods to anonymize speech signals.

\section{Legal perspectives} \locallabel{chapt3:sec:legal}

%
The \acrlong{gdpr} \cite{gdpr} obligates the entity storing personal data to implement all possible technical measures to enforce the protection of personal data by the principle of data protection by design and by default.
More specifically, Article 9 of the \acrshort{gdpr} prohibits the processing of biometric data that shows the gender, origin or health indications of a person.
This legislation is applicable for authentication platforms relying on fingerprint, face recognition, or other forms of template-based verification \cite{jain2008biometric} to avoid as much as possible unfair bias.
However, for certain modalities such as speech, this constraint is too restrictive as speech data inherently contains biometric information and additional useful information that does not fall under the same data regulation.
For that reason, Recital 26 of the \acrshort{gdpr} relaxes this constraint by allowing the processing of anonymous data that can not identify the user.
Article 5(c.1) of the \acrshort{gdpr} mentions the principle of data minimization which is to only collect data strictly necessary for the usage/improvement of the service.
In the case of a speech recognition service, being able to remove the other sensible attributes unnecessary to speech recognition will help to comply with data minimization.
To obtain anonymous and minimized data, one can rely on anonymization\footnote{From a legal standpoint, "anonymization" refers to methods that achieve complete concealment. In this thesis the term \say{privacy enhancement} is more appropriate as we do not fulfill full concealment, however, the term \say{anonymization} is more broadly used by the community.} methods to remove biometric clues from the data.
The strength of anonymization depends directly on the technical measures available, which motivates the pursuit of this thesis.

Related to biometric data protection, the European standard ISO/IEC 24745 \cite{iso24754} defines multiple criteria regarding the processing of biometric data.
The two criteria studied in the thesis that can be borrowed for data anonymization are \textit{invertibility} and \textit{unlinkability}.
First, the \textit{invertibility} criterion aims to prevent the use of biometric data for any purpose other than the ones originally intended, for that reason, biometric data must be processed by irreversible transformation before storage.
Secondly, the \textit{unlinkability} criterion aims to ensure that stored biometric templates can not be linkable across applications or databases.
Linkability (the opposite of \textit{unlinkability}) is considered the main threat in this thesis and will be explored in more detail in the following section.

It is worth mentioning two other federal publications, the white paper "Explore legal, technical and ethical issues associated with voice assistant" by \cite{white-paper-cnil}, which present multiple practical use cases and applications of the \acrshort{gdpr}, and the guidelines on virtual voice assistants by \cite{edpb-voice-assistant-guidelines}, which presents general recommendations regarding data retention, user profiling, data protection and more.

\section{Threat model} \locallabel{chapt_3:threat_model}
\begin{figure}[htbp]
    \vspace{-4mm}
    \begin{center}
        \includegraphics[width=0.79\linewidth]{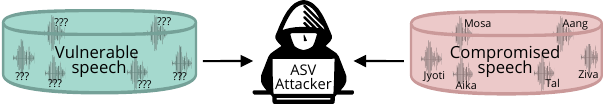}
    \end{center}
    \vspace{-4mm}
    \caption{
        The considered threat model, where the \textit{vulnerable} speech needs to be protected with privacy-enhancing solutions removing biometric \acrfull{pii} from the data.
        Whereas the \textit{compromised} speech is unprotected data full of biometric \acrshort{pii}, used by the attacker to find the original speakers of the \textit{vulnerable} speech.
    }
    \vspace{-2mm}
    \locallabel{image_chapt3:threat_model}
\end{figure}
The main threat model in this thesis is a linkability attack which corresponds to multiple tests where, given two speech samples, an attacker can distinguish if they come from the same speaker or different speakers,
this is similar to \acrfull{asv} (see Chapter \ref{main:chapt_2:sec:asv}).
An attacker refers to a formal or informal description of assumed capabilities that a security/privacy mechanism is designed to protect against.
In this thesis, we consider multiple users having their speech samples in a compromised/public dataset, where each sample is associated with a speaker identity.
The attacker has access to this data, and his goal is to link each known speaker's identity with an unknown speaker's speech coming from another dataset (Figure \localref{image_chapt3:threat_model}).


In this thesis, the goal is to remove biometric \acrfull{pii} from a speech signal such that it becomes unlinkable to the speaker's identity, this task is referred to as speaker anonymization (Figure \localref{image_chapt3:anon_plot_link}). 
This operation is judged to be necessary for publishing useful datasets for scientific and commercial use without compromising the users' privacy \cite{data_publish}.
\begin{figure}[htbp]
    \begin{center}
        \includegraphics[width=0.89\linewidth]{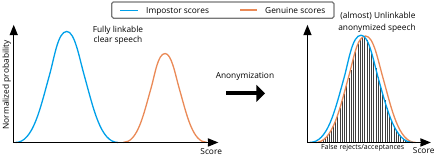}
    \end{center}
    \vspace{-4mm}
    \caption{
        Linkability assessment using an \acrshort{asv} system. In the example, before anonymization (left side), the speech is fully linkable, meaning the identity of the speakers could be verified all the time.
        To protect the identities, the speech should be as unlinkable as possible after anonymization (right side).
    }
    \locallabel{image_chapt3:anon_plot_link}
\end{figure}

The evaluation of this \acrshort{pii} removal (privacy performance) relies on the \acrshort{asv} model of the attacker.
Until the work of \cite{EvaluatingVoiceConversionbased2019}, most studies assumed the attacker had restricted capability i.e., the attacker was unaware of the anonymization.
This aspect leads to a high domain mismatch between the \textit{vulnerable} anonymized speech and \textit{compromized} clear speech datasets which the \acrshort{asv} model was not trained to overcome, additionally, no countermeasures were employed to defeat the anonymization system \cite{Hashimoto2016PrivacypreservingST,Qian2017VoiceMaskAA,magarinos2017reversible,F0_Bahmaninezhad2018ConvolutionalNN}.
In security and data protection, \say{clear data} refers to the original, unmodified and unprocessed data collected.
In the security/cryptography communities, evaluation based on \say{security by obscurity} \cite{security_by_obscurity}, where only weak attackers are considered is strongly refuted.
In the following section, we present and categorize more effective methods for evaluating privacy performance with various \acrshort{asv} attackers having varying levels of capability.

\subsection{Attacker capabilities} \locallabel{chapt_3:attacker_capa}

\begin{figure}[htbp]
    \begin{center}
        \vspace{-2mm}
        \includegraphics[width=0.63\linewidth]
        {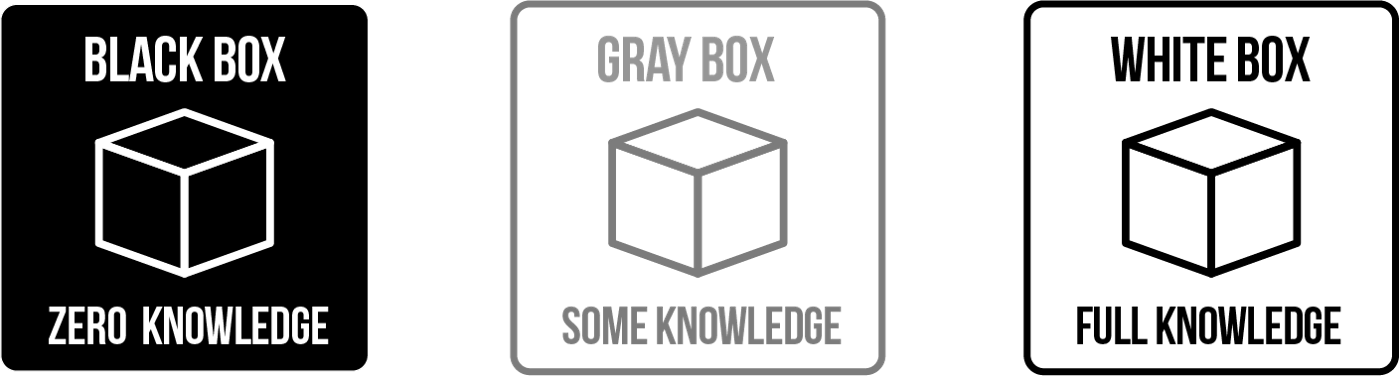}
    \end{center}
    \vspace{-6mm}
    \caption{Adversary's knowledge about the anonymization procedure.}
    \locallabel{chapt_3:image:black-gray-white_box}
\end{figure}

In software quality and security, testing techniques can be classified into three main categories, black-box, white-box and gray-box (see Figure \localref{chapt_3:image:black-gray-white_box}) \cite{white_box_black_box_gray_box,Guo2019SimpleBA}.
Black-box testing is a testing scenario where zero knowledge of the application is required.
Black-box testing provides a perspective where the attacker and the application are separated, however, it is an inefficient testing scenario, and in the case of speaker anonymization assessment can be assimilated to as an evaluation following the \say{security by obscurity} principle.
White-box testing is the case where full knowledge of the application and parameters is required.
White-box testing is the most efficient testing scenario to identify the largest amount of vulnerabilities, at the cost of being more expensive and requiring expertise.
In the middle, many gray-box testing scenarios can be crafted each having varying amounts of knowledge of the application.
Gray-box testing offers assessment from the point of view of the users rather than the application designer, for that reason it is a realistic scenario for speaker anonymization evaluation, however, it may not identify all vulnerabilities.
When this categorization is applied to test the privacy performance of speaker anonymization systems, \cite{EvaluatingVoiceConversionbased2019} multiple attacker schemes can be defined:
\begin{itemize}[leftmargin=*]
    \setlength\itemsep{0.2em}
    \item[\ding{109}] \textbf{Black-box}
        \begin{itemize}[leftmargin=4mm]
            \item The \textit{ignorant} attacker (II in Figure \localref{image_chapt3:attacker_type}) is unaware that speech was anonymized.
        \end{itemize}
    \item[\ding{109}] \textbf{Gray-box}
        \begin{itemize}[leftmargin=4mm]
            \item The \textit{lazy-informed} attacker (III in Figure \localref{image_chapt3:attacker_type}) is aware of the anonymization system but unaware of the parameter used to anonymize each utterance.
                  This attacker tries to reduce the domain mismatch between the clear (non-anonymized) \textit{compromized} speech and the anonymized \textit{vulnerable} speech datasets by anonymizing itself the \textit{compromized} speech dataset.


            \item The \textit{semi-informed} attacker (IV in Figure \localref{image_chapt3:attacker_type}) is aware of the anonymization system and some hyperparameters used but unaware of the exact parameters used to anonymize each utterance.
                This attacker reduces the domain mismatch between the \textit{compromized} speech and the anonymized datasets by anonymizing the \textit{compromized} speech (without using the most appropriate hyperparameters).
                  Additionally, this attacker adapts the \acrshort{asv} evaluation model to work with somewhat relevant anonymized data.
        \end{itemize}
    \item[\ding{109}] \textbf{White-box}
        \begin{itemize}[leftmargin=4mm]
            \item The \textit{informed} attacker (V in Figure \localref{image_chapt3:attacker_type}) is completely aware of the anonymization system, hyperparameters, and exact parameters used to anonymize each utterance.
                  This attacker minimizes the domain mismatch between the \textit{compromized} speech and the anonymized datasets by anonymizing the \textit{compromized} speech using the most appropriate hyperparameters.
                  Additionally, this attacker adapts the \acrshort{asv} evaluation model to specifically work against one anonymization system, hyperparameter and exact parameters used to anonymize each utterance.
        \end{itemize}
\end{itemize}

\begin{figure}[htbp]
    \begin{center}
        \vspace{-4mm}
        \includegraphics[width=0.99\linewidth]{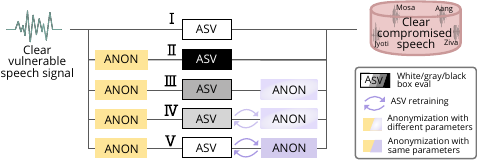}
    \end{center}
    \vspace{-6mm}
    \caption{
        Privacy evaluation in clear scenario I (baseline), \textit{ignorant} II, \textit{lazy-informed} III, \textit{semi-informed} IV, and \textit{informed} V attacker scenario.
    }
    \locallabel{image_chapt3:attacker_type}
\end{figure}

For an anonymization system implemented through the use of voice conversion (see Chapter~\ref{main:chapt2}), being aware of the anonymization system means having access to the \acrshort{vc} model and weights.
Being aware of the hyperparameters refers to knowing the target selection strategy used to select the target speaker to anonymize each utterance (but not the exact target speaker).
Being aware of the parameters means knowing the exact target speaker used to anonymize each utterance.
The creation of the target speaker is explained in more detail in Section \localref{chapt_3:anon_sys_x-vector} with the presentation of an anonymization system.


The {semi-informed} and {informed} attackers allow better privacy performance evaluation as adapting (through retraining) the \acrshort{asv} model on anonymized speech avoids a domain mismatch.
This process might also modifies the main goal of a classical \acrshort{asv} model.
Instead of directly recognizing each speaker from the others, the \acrshort{asv} model is trained to, to some extent, invert the anonymization to then identify the underlying speaker.

The {semi-informed} attacker is interesting in the fact that it is the most realistic attacker.
As anonymization should be performed by each user before sharing their data, the anonymization system is open to the public, which means the attacker can gain control over it.
With this anonymization system, the attacker can generate anonymized speech in a similar (but not exact) way as the users.
With anonymized speech, the attacker can adapt both the \textit{compromized} speech and \acrshort{asv} model to work with such kind of data.
However, depending on the hyperparameters used, the attacker-generated anonymized speech might not enable correct retraining of his/her adapted \acrshort{asv} model, this aspect is explored in Chapter~\ref{main:chapt4}.

In contrast, the \textit{informed} attacker might be unrealistic as having complete knowledge, even of a per utterance target speaker is impractical.
However, as this attacker falls into the category of white-box testing, it enables an evaluation from the point of view of the application designer (the specialist) who has more ideas about the limitation of his/her system and can identify more vulnerabilities.
With such capabilities, this attacker can precisely retrain his/her adapted \acrshort{asv} to ensure any improvement of privacy is indeed caused by a better anonymization system rather than a weak/unadapted \acrshort{asv} model.

\begin{figure}[htbp]
    \begin{center}
        \includegraphics[width=1\linewidth]{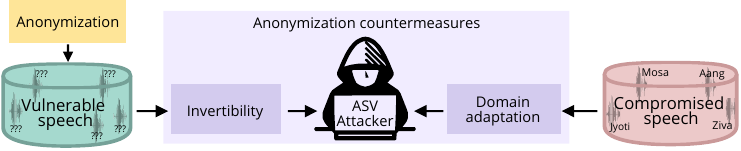}
    \end{center}
    \vspace{-8mm}
    \caption{
        Possible countermeasures an attacker may use to defeat anonymization.
    }
    \locallabel{image_chapt3:threat_model_2}
\end{figure}

The previously mentioned attackers rely on making sure the \acrshort{asv} system is properly adapted to recognize the original speaker from anonymized utterances, this process labeled \say{domain adaptation} in Figure \localref{image_chapt3:threat_model_2}  does not consider the anonymized \textit{vulnerable} speech as an attack vector.
New kinds of \say{invertibility} attacks recently developed \cite{invertibility_asru,invert_odyssey} consider using the anonymized \textit{vulnerable} speech to invert the anonymization procedure, such a form of attack is a contribution of this thesis and will be presented in more details in Chapter~\ref{main:chapt6}.

\section{Application cases}

While reducing linkability attacks, the anonymized speech also needs to be useful for other tasks, the terminology used to refer to this aspect is utility.
The utility of anonymized speech depends on the performance obtained when used for downstream tasks, i.e., recognizing speech content, emotions, and more.
Depending on the task, the anonymized speech utility requirements may differ from application to application.
For instance, in the case of a voice assistant, the service provider wants to receive speech where the linguistic content is preserved such that an \acrshort{asr} can decode and comprehend. 
Additionally, the voice assistant service provider might also want to preserve various usage conditions, i.e, noisy environment, to improve their service under adverse conditions.
Other requirements such as preservation of the intonation, naturalness, and voice distinctiveness \cite{tomashenko2020voiceprivacy_eval2022} might be necessary for some applications.
Some requirements like gender or emotion recognition are attributes that can fall into both a private aspect to suppress or a utility aspect to keep \cite{spsc,aloufi_privacy_prev}.
Figure \localref{image_chapt3:application_spectrum} presents a spectrum of a few privacy/utility requirements.
Speaker anonymization has well-defined privacy and utility requirements for some of them, while for others it depends on the targeted application.
Given this indecisive aspect, evaluations in this field should be done considering a specific use case.

\begin{figure}[htbp]
    \begin{center}
        \includegraphics[width=0.8\linewidth]{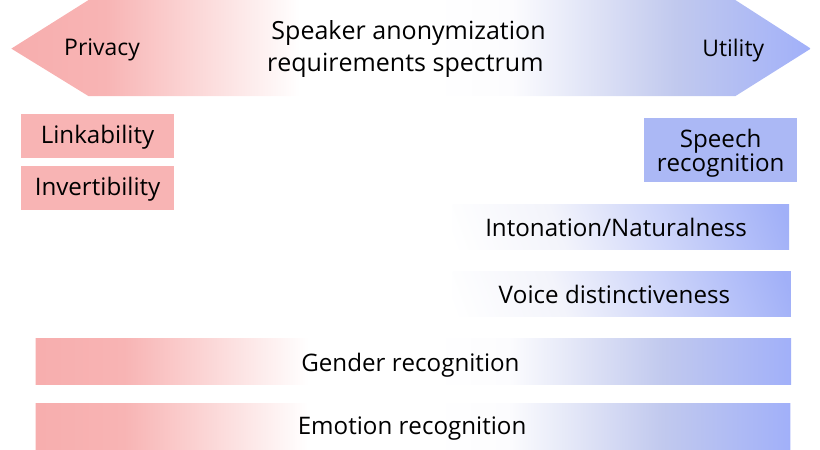}
    \end{center}
    \vspace{-5mm}
    \caption{
        Example of the range of requirements of a speaker anonymization system.
    }
    \locallabel{image_chapt3:application_spectrum}
\end{figure}

The use case considered in this thesis corresponds to the one where a service provider wants to collect a large dataset to, improve their service.
This thesis focuses on the data collection part, where the aim is to find a privacy-preserving transformation of the speech data that removes the biometric \acrshort{pii} speech information that an \acrshort{asv} system may capture while keeping high utility for the linguistic content such that an \acrshort{asr} system can still decode the content.
Additional requirements may be added while maintaining this core objective, which can be thought of as the fundamental objective that all speaker anonymization should share.

\section{Evaluation methods} \locallabel{sec:vpcdes}
To assess the privacy and utility against \acrshort{asv} linkability attack and \acrshort{asr} decoding, the \acrfull{vpc} provides a very relevant evaluation framework.
As such, in this thesis, the experimental setup used in all experiments comes from the \acrshort{vpc}  evaluation framework.
Established in 2020, the VoicePrivacy initiative~\cite{tomashenko2020voiceprivacy} is spearheading the effort to develop privacy preservation solutions for speech technology.
Up until now, two challenges have taken place one in 2020 and the other in 2022.
The most important difference between the two challenges is the use of a stronger semi-informed attacker in \acrshort{vpc} 2022 instead of a lazy-informed attacker in 2020.

\subsection{The VoicePrivacy challenge requirements}
The challenge aims to advance progress in the development of anonymization and pseudonymization\footnote{Pseudonymization is the process of replacing \acrshort{pii} with a pseudo or an alias, which can be used to link back to the original information if the original/pseudo mapping is available. This differs from anonymization, which aims to completely remove all \acrshort{pii}.} solutions that suppress personally identifiable information contained within recordings of speech while preserving linguistic content, paralinguistic attributes, subjective voice distinctiveness (make each individual's voice easily recognizable and distinguishable from others by human listeners), intelligibility and naturalness properties.

The anonymization goals of the \acrshort{vpc} are the following:
\begin{enumerate}[label=(\alph*)]
    \item output a speech waveform
    \item conceal the speaker's identity against the thread model presented in Section \localref{chapt_3:threat_model}
    \item leave the linguistic content and paralinguistic attributes unchanged
    \item ensure that all trial utterances from a given speaker are uttered by the same pseudo-speaker, while trial utterances from different speakers are uttered by different pseudo-speakers
\end{enumerate}

The requirement (d) is motivated by multi-party conversation application cases, where the anonymized voices of all speakers must be distinguishable from each other and should not change over time.
The terminology used to refer to this type of anonymization is \textit{speaker-level} target selection.
This is an alternative approach to \textit{utterance-level} target selection where different utterances of the same source speaker are anonymized using different parameters of the anonymization system, so that they may sound as if they were spoken by different speakers.
Requirement (d) is assessed subjectively for voice distinctiveness and objectively for pseudonymization.

\subsubsection{Data}
The training, development and evaluation datasets of the \acrshort{vpc} consist of several publicly available corpora:
\begin{itemize}
    \setlength\itemsep{0.2em}
    \item \textit{\textbf{LibriSpeech}} \cite{Librispeech} is a corpus of read English speech derived from audiobooks and designed for ASR research. It contains approximately {$1000$}~hours of speech sampled at 16~kHz.

    \item \textit{\textbf{LibriTTS}} \cite{libritts} is a corpus of English speech derived from LibriSpeech and designed for research in text-to-speech (TTS). It contains approximately 585~hours of read English speech sampled at 24~kHz.

    \item {\textbf{\textit{VCTK}}} \cite{vctk} is a corpus of read speech collected from 109 native speakers of English with various accents. It was originally aimed for research in TTS and contains approximately 44 hours of speech sampled at 48~kHz.

    \item \textit{\textbf{VoxCeleb-1,2}} \cite{nagrani2017voxceleb,Chung18b} is an audiovisual corpus extracted from videos uploaded to YouTube and designed for speaker verification research. It contains approximately {$2770$} hours of speech sampled at 16~kHz collected from {$7363$} speakers, covering a wide range of accents and languages.
\end{itemize}

\noindent
Below, we provide a detailed description of the datasets used for training the anonymization system, training the evaluation models (privacy/utility), and testing datasets.
Table \localref{chapt_2:dataset} presents statistics about the number of speakers and utterances per dataset, the table also indicates the evaluation usage and potential training usage for each of them.
\vspace{-1em}
\paragraph{Training data} The data allowed to train an anonymization system consists of \textit{VoxCeleb-1,2, LibriSpeech train-clean-100, LibriSpeech train-other-500, LibriTTS train-clean-100} and \textit{LibriTTS train-other-500}.
The \textit{LibriTTS train-other-500} dataset can be used to select the target speaker to anonymize each utterance.
The motivation for using this dataset is to have a wide variety of target speakers, which, at the time, was seen as an asset in order to generate many kinds of voices for the voice distinctiveness requirement.
In this thesis, we will challenge whether or not having the voice distinctiveness requirement (as well as the use of the \textit{LibriTTS train-other-500} dataset) is beneficial for speech privacy.
\vspace{-1em}
\paragraph{Testing data} The evaluation dataset consists of \textit{LibriSpeech test-clean} and \textit{VCTK test}.
For each of those two datasets, the challenge organizers divided them into \textit{vulnerable} and \textit{compromised} speech datasets.
\vspace{-1em}
\paragraph{Attacker data} To match a realistic scenario, the attacker has access to another dataset, here \textit{LibriSpeech train-clean-360}, which he uses to train the \acrshort{asv} linkability attack model.
Depending on the attacker's capability, he might train the \acrshort{asv} (privacy) evaluation model with anonymized data to reduce the domain mismatch as seen in Section \localref{chapt_3:attacker_capa}.
If anonymized data is used, the \acrshort{vpc} request that the \textit{LibriSpeech train-clean-360} is anonymized at the \textit{utterance-level} rather than at the \textit{speaker-level}, because it has been observed that training the \acrshort{asv} model with \textit{LibriSpeech train-clean-360} anonymized at \textit{speaker-level} leads to overfitting.
\vspace{-1em}
\paragraph{Application data} In the \acrshort{vpc} this data is the same as the attacker data, it is used to train the \acrshort{asr} model that assesses the preservation of the linguistic content (utility).
In the 2022 challenge, the \acrshort{asr} model is trained with anonymized data to maximize the utility as demonstrated in \cite{post_eval_vpc_2020}.

\vspace{-0.2cm}
\begin{table}[!ht]
    \caption{Statistics of the datasets.}
    \vspace{0.4cm}
    \hspace{-1.2cm}
    \resizebox{1.15\textwidth}{!}{
    \centering{
        \begin{tabular}{cc
                c@{\extracolsep{-4mm}}
                r@{\extracolsep{2mm}}
                r
r
            }
            \toprule
             & Dataset                                               & Usage                    & \# Speakers & \# Utterances & Avg duration \\
            \midrule
            \multirow{5}{*}{\rotatebox{90}{\textbf{Train}}}
             & \textit{LibriSpeech train-clean-100}                  & Linguistic extractor     & 251         & 28 539 & 12.70 sec.        \\
             & \textit{LibriSpeech train-other-500 }                 & Linguistic extractor     & 1 166        & 148 688 & 12.03 sec.       \\
             & \textit{LibriTTS train-clean-100}                     & Speech synthesizer       & 247         & 33 236  & 5.82 sec.       \\
             & \textit{LibriTTS train-other-500}                     & Pool of x-vectors        & 1 160        & 205 044  & 5.45 sec.      \\
             & \textit{VoxCeleb1,2}                                  & Speaker extractor        & 7 363       & 1 281 762& 8.10 sec.      \\
            \midrule
            \multirow{4}{*}{\rotatebox{90}{\textbf{Test}}}
             & \multirow{2}{*}{\textit{ LibriSpeech test-clean }}    & Compromised speech       & 29          & 438        & 6.18 sec.   \\
             &                                                       & Vulnerable  speech       & 40          & 1 496 & 8.67 sec.         \\
             & \multirow{2}{*}{ \textit{VCTK test} }                 & Compromised speech       & 30          & 600 & 3.05 sec.           \\
             &                                                       & Vulnerable  speech       & 30          & 11 448 & 3.22 sec.         \\
            \midrule
            \multirow{2}{*}{\begin{minipage}{2em}{\begin{center}
                            \rotatebox{90}{\textbf{Eval}}
                        \end{center}
                                    }\end{minipage} }
             & \multirow{2}{*}{\textit{LibriSpeech train-clean-360}} &
            \multirow{2}{*}{\begin{minipage}{120pt}{\begin{center}
                            Train privacy/utility eval models
                        \end{center}
                                    }\end{minipage} }
             & \multirow{2}{*}{921}                                  & \multirow{2}{*}{104 014} & \multirow{2}{*}{12.58 sec.}                               \\
            \\
            \bottomrule
        \end{tabular}
    }
    }
    \locallabel{chapt_2:dataset}
\end{table}


\vspace{\fill}
In this thesis, we only focus our work on the core objective of anonymization, so, we do not present any paralinguistic, or subjective analysis in contrast to the \acrshort{vpc} evaluations.
For similar reasons, we believe requirement (d) about \textit{speaker-level} target selection falls out of the scope of the core objective of speaker anonymization evaluation.
We believe the way it is objectively evaluated in the \acrshort{vpc} (described in \cite{similarity_matrices}) misses the opportunity to use watermarking/steganography methods \cite{watermarking_1,faundez2009digital,Djebbar2012ComparativeSO,Nematollahi2013AnOO,Nematollahi2017} that could embed a pseudo identifier into anonymized speech to comply with pseudonymization.
As for the subjective voice distinctiveness evaluation, it is missing real-world application cases defining the size of the multi-party conversation.
We believe having \say{all} voices distinguishable from each other is impractical knowing it can become difficult for a person to keep track of who is speaking and what was being said by whom, especially when hearing many unfamiliar voices.

\subsection{Privacy model and metrics} \locallabel{chap2:privacy_model_metric}
Like the \acrshort{vpc}, the primary objective privacy metric of this thesis is obtained through the \acrshort{asv} linkability attack between vulnerable and compromised speech for both \textit{LibriSpeech test-clean} and \textit{VCTK test} datasets.
The model used for all experiments in this thesis (involving target \acrshort{vc} x-vectors), is based on the Kaldi x-vector model with five \acrshort{tdnn} layers as presented in Section \ref{main:section:x-vector}.
The metrics presented in Section \ref{main:chapt_2:asv_eval} and used to assess the strength of linkability are the \acrshort{eer}, and the linkability \acrshort{li} metrics.
For comparison without anonymization (clear speech), a linkability test is performed without anonymizing the vulnerable speech, (i.e., scenario I (baseline) in Figure \localref{image_chapt3:attacker_type}).
After anonymization, the higher the \acrshort{eer}, or the lower the \acrshort{li}, the greater the privacy.

\subsection{Utility model and metric} \locallabel{chap2:utility_model_metric}
The capability of the anonymization method to maintain the linguistic content is objectively evaluated using an \acrshort{asr} model based on Kaldi.
This evaluation model is used for all experiments in this thesis and is a hybrid \acrshort{dnn}-\acrshort{hmm} triphone acoustic model with a 17 \acrshort{tdnnf} layers architecture presented in Section~\ref{main:section:hmm-dnn} and a large trigram language model presented in Section~\ref{main:sec:ngram}.
The \acrshort{asr} decodes the word sequence for both \textit{LibriSpeech test-clean} and \textit{VCTK test} anonymized speech datasets.
The \acrshort{wer} is then calculated.
For comparison, an \acrshort{asr} decoding is also performed on clear (non-anonymized) speech.
The lower the \acrshort{wer}, the greater the utility.

As improvement in privacy performance is usually correlated with utility degradation, good speaker anonymization must achieve a suitable privacy/utility trade-off \cite{privacy-utility-tradeoff,understanding_tradeoff_privacy}.
In the next section we present some methods to anonymize speech data, and when relevant, present their privacy/utility trade-off.

\section{Current methods for speaker anonymization}
This section presents several methods to perform speaker anonymization.
They can be classified into three main categories: signal processing, voice conversion, and adversarial \acrshort{asv} attack anonymization.

\vspace{-2mm}
\subsection{Signal processing anonymization}
\vspace{-1mm}
Signal-processing speaker anonymization aims to shift the perceived original speaker by manipulating speech features such as pitch, rhythm, tempo, pause, etc.
In contrast to other methods, this anonymization method does not require any training as only simple signal processing techniques are used.

In \cite{Qian2017VoiceMaskAA}, the authors present VoiceMask, a VTLN-based frequency warping technique to transform the spectral envelope before resynthesizing a waveform.
Evaluated with the informed attacker, the relative utility degradation is by 10\%, while the privacy does not change, \cite{EvaluatingVoiceConversionbased2019} argues this result comes from the parameters of the transformation always wrapping the spectra at each frame of the utterance in a single direction.

In \cite{8844600}, the authors present an \say{audio sanitizer} that modifies the pitch, tempo and pause features, then, they add white noise after resynthesizing the waveform from \acrshort{mfcc}.
In their paper, the utility decreased by 27\% (relative), while the privacy gain is not measured with a strong enough attacker to draw any conclusion.

In \cite{patino21_interspeech}, the authors present a speaker anonymization method based on McAdams transformation \cite{mcadams}.
The McAdams transformation relies on timbre modification using the McAdams coefficient which expands/contracts the frequency of each harmonic.
Under semi-informed evaluation, the utility decreased by 8.5\% (relative), while the privacy increased by 100\% (from 4\% to 8\% for EER).

In \cite{cascade_sp_anon}, the authors propose a cascade of previously mentioned techniques.
Multiple combinations are explored, the most notables are a combination of Resampling and Modulation Spectrum smoothing, and a combination of Resampling, Modulation Spectrum Smoothing, McAdams Transformation clipping, and Chorus.
The evaluation shows a utility degradation of 36\% and 600\% for each combination respectively, while privacy only slightly increases (from 4\% to 7 and 10\% of EER) for both methods when using a semi-informed attacker.

In \cite{mawalim22_spsc}, the authors proposed a phase vocoder-based time-scale modification anonymization that compresses or stretches audio signals.
Very interestingly, they perform multiple evaluations using Ignorant, Lazy-informed and semi-informed attackers.
With the Ignorant and Lazy informed attackers, the privacy metric reaches around 44\% of EER, while with the semi-informed attacker, the EER obtained is 16\%.
The utility remains very close to clear speech (10\% of degradation).
Note that these results are the ones highlighting the biggest overestimation of privacy protection when using under-capable attackers.

\vspace{-2mm}
\subsection{Voice conversion anonymization} \locallabel{chapt_3:vc_anon}
As presented in Section \ref{main:chapt_2:vc}, \acrfull{vc} is the process of adapting the characteristics of a source speaker's speech to match those of a target speaker without changing the linguistic content.
In contrast to signal-processing speaker techniques, \acrshort{vc} needs a linguistic representation and paralinguistic information to work.
Section \ref{main:chapt_2:vc_lin_rep} explains the two main approaches to obtaining the linguistic representation.
The {supervised} approaches typically use specialized models trained with text supervision like an \acrshort{asr} acoustic model and the {self-supervised} approaches typically have an implicit linguistic representation like auto-encoders for example.

\subsubsection{\texorpdfstring{Self-supervised linguistic representation}{Self-supervised linguistic representation}} \locallabel{chapt3:anon_self_sup}


Starting with the latest self-supervised linguistic representation extraction \acrshort{vc} approach, works of \cite{9247219} have presented a Cycle\acrshort{vae}-\acrshort{gan} which uses \acrlong{vae} to extract linguistic representation and a one-hot vector for the target speaker.
Evaluation with a lazy-informed attacker shows almost perfect privacy under the experiment where the one-hot represented a single speaker which is the farthest from the source in cosine similarity distance.
However, the nature of this type of attacker may not accurately reflect the true privacy performance of the system.
The relative utility degradation recorded is 22\%.

Similarly, the authors of \cite{aloufi_privacy_prev} propose a \acrshort{vae} to extract a representation accordingly to the user's preference about the characteristics to remove.
However, their evaluation lacks a linkability attack.
Their evaluation instead focuses on preserving the utility of the linguistic content and emotional state.

Following on a similar \acrshort{vc} approach, \cite{PRAJAPATI2022101353} proposed a Cycle-\acrshort{gan} and speed perturbation pipeline, where the target speaker is the opposite gender of the source.
Results under the semi-informed attacker show a privacy improvement of 360\% from (4\% of \acrshort{eer} to 17.5\% of \acrshort{eer}) while affecting the utility by less than 10\%.

\subsubsection{\texorpdfstring{Supervised linguistic representation}{Supervised linguistic representation}} \locallabel{chapt_3:anon_sys_x-vector}

Using a supervised linguistic representation extractor based on \acrshort{asr} acoustic model for \acrshort{vc}, \cite{fangSpeakerAnonymizationUsing2019} proposed the first x-vector-based speaker anonymization system used as a baseline for many following works including the \acrlong{vpc}.
This baseline and components of this baseline are used in our experiments and proposed systems.

\begin{figure}[htbp]
    \begin{center}
        \includegraphics[width=0.82\linewidth]{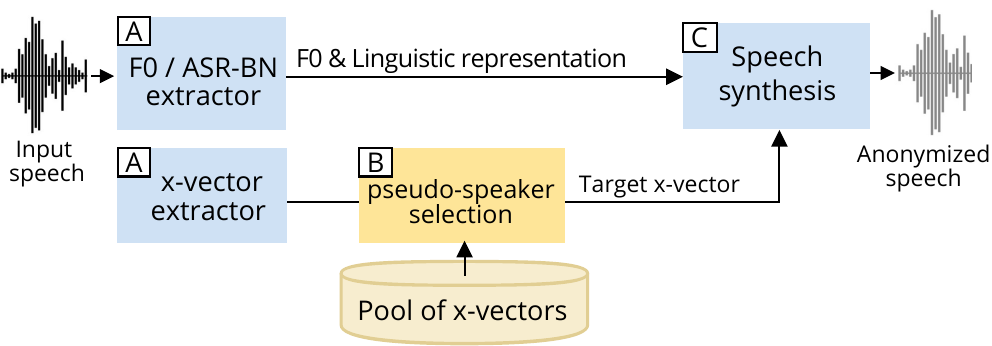}
    \end{center}
    \vspace{-5mm}
    \caption{
        X-vector-based speaker anonymization.
    }
    \locallabel{image_chapt3:vpc_baseline}
\end{figure}
In their model, speaker identity and linguistic content are first extracted from an input speech utterance.
Assuming that those features are disentangled, an anonymized speech waveform is generated by altering only the features that encode the speaker's identity.
The anonymization system depicted in Figure \localref{image_chapt3:vpc_baseline} can be decomposed into three groups of modules.
Modules from group A extract different features from the source signal: the fundamental frequency, the linguistic representation and the speaker's x-vector.
Module B derives a new target identity using a target selection strategy.
The target selection strategy is the following: the x-vector from each source input speaker is compared to a pool of external x-vectors in order to select the 200 furthest vectors; 100 of them are randomly selected and averaged to create a target x-vector identity.
Finally, module C synthesizes a speech waveform from the target x-vector together with the original linguistic representation and \acrshort{f0}.
Speaker anonymization is achieved by selecting a private target x-vector.
%
The fundamental frequency is extracted using the YAAPT algorithm \cite{yaapt}.
The linguistic representation is obtained using a 17 \acrshort{tdnnf} layers Kaldi triphone \acrshort{asrbn} extractor (dimension 256 extracted from the final factorized hidden layer) and trained with \textit{LibriSpeech train-clean-100} and \textit{LibriSpeech train-other-500}.
The x-vectors are also extracted using Kaldi, which is a five-layers \acrshort{tdnn} trained on \textit{VoxCeleb1,2}.
The speech synthesis model is HiFi-GAN (see Section~\localref{sec:hifigan}).
Table~\localref{chapt_2:dataset} describes the datasets used to train each model and the pool of x-vectors.
Evaluated with the informed attacker, the privacy increases by 175\% (from 4\% to 11\% of \acrshort{eer}), while the utility is not significantly decreased.
In the following sections, we describe some extensions of this system, however, they are not included in our baseline.

\paragraph{\texorpdfstring{X-vector modification}{X-vector modification}} \locallabel{chapt3_xvector_modef}

For this system, \cite{brij_vpc_design} experienced many design target selection strategies to generate the target x-vector based on the pool of x-vectors.
They observed that randomly picking half of the x-vectors from a random dense cluster and averaging them created the best privacy/utility trade-off under the lazy-informed attacker.
Other studies explored this topic of target vector generation for speaker anonymization: \cite{9247219} randomly modifies each component of a one-hot speaker embedding to generate target vectors,
\cite{Turner_xvector_vpc2020} uses a \acrshort{gmm} coupled with \acrshort{pca} to generate target x-vectors mirroring the distribution of those found in the real world (increasing naturalness), \cite{candy_xvector_vpc2020} uses singular value decomposition and statistical regression to generate target x-vectors, \cite{adv_xvector_vpc2020} uses adversarial learning to enhance the disentanglement of the x-vectors from gender and accent, \cite{meyer_gan_xvector} uses Wasserstein \acrshort{gan} to generate target x-vectors with the same naturalness property as Turner.

\paragraph{\texorpdfstring{\acrshort{asrbn} modification}{ASR-BN modification}}

Research in \acrshort{asr} technology by \cite{adiReverseGradientNot2019,mohanPrivacyPreservingAdversarialRepresentation2019_reality_adversarial} showed that the speaker/linguistic disentanglement property of the \acrshort{asrbn} representation is limited.
This directly impacts the performance of anonymization systems that rely on the \acrshort{asrbn} representation.
Compared to other features (namely the \acrshort{f0} and target x-vector), the \acrshort{asrbn} representation is the feature with the highest dimension and is sequential, as such, not disentangled \acrshort{asrbn} can substantially restrict performance in terms of speaker concealment.
In the following, we present valuable research that aims to improve the speaker/linguistic content disentanglement (or privacy) of \acrshort{asrbn} representation.

\vspace{0mm}
\subparagraph{{Dimensionality Reduction}{}}

One approach to enhance the privacy of bottleneck representation discussed in \cite{hybrid_privacy_framework,RyffelPartiallyEncryptedDL2019}, is to reduce their dimensionality.
By reducing bottleneck dimensionality, the encoding capacity is reduced, as such, only the most important information should be kept.
Training a network with a bottleneck of very low dimensionality can be challenging, so using \acrfull{pca} or \acrfull{svd} methods after training is usually easier.
\acrshort{pca} or \acrshort{svd} methods attempt to preserve the primary structure of bottlenecks and remove as many unnecessary components as possible.
When the bottleneck is extracted with a \acrfull{tdnnf} model at the factorized layer, which is the case for the model presented in Figure \localref{image_chapt3:vpc_baseline}, the bottleneck has already been reduced with a technique similar to an \acrshort{svd} as presented in Section \ref{main:chapt_1:tdnnf_svd}.

\vspace{-2mm}
\subparagraph{{Adversarial network}{}}
Another approach to enhance the representation of \acrshort{asrbn} is through the use of an adversarial network (see Section \ref{main:chapt_1:adv_net}).
Using the \acrshort{asrbn} representation, the main network is trained for the main task (e.g., triphone classification) while an additional adversarial model is trained alongside the same \acrshort{asrbn} representation to identify the speaker.
To improve the disentanglement of the \acrshort{asrbn} using this training scheme, the negative gradient of the adversarial model is used to encourage the \acrshort{asrbn} to encode less speaker information.
Work done by \cite{mohanPrivacyPreservingAdversarialRepresentation2019_reality_adversarial}, found at the feature level (no anonymized speech produced) that the speaker classification accuracy was reduced, but did translate to reduced linkability.
The hypothesis regarding this disparity may be due to the use of an atypical adversarial speaker verification model architecture, as well as the formulation of the adversarial loss.
Further research is needed to assess the benefits and limitations of this approach to speaker information removal in an \acrshort{asr} model using adversarial learning.
Work of \cite{Ericsson2020AdversarialRL}, has generated speech hiding the gender information using adversarial learning while maintaining a high quality of the anonymized audio.

\vspace{-2mm}
\subparagraph{\texorpdfstring{Noise perturbation}{Noise perturbation}}
Motivated by the field of Differential Privacy \cite{Dwork2006DifferentialP} and the \textit{Laplace mechanism} \cite{Dwork2014TheAF}, noise can be used to perturb the \acrshort{asrbn} representation such that the representation maintain a satisfactory privacy/utility trade-off.
Work of \cite{dp_vpc} has modified the \acrshort{f0} and \acrshort{asrbn} representation of the x-vector-based speaker anonymization system presented above with \textit{Laplace} noise.
Directly applying noise to the \acrshort{asrbn} representation (or raw speech) destroys linguistic and prosodic information.
Hence, the authors have adapted their model to work with noise, by training them to retain relevant information while adding the required level of noise to get the Differential Privacy guarantees.
For their model which adds the highest amount of noise, under semi-informed evaluation, the utility decreased by 48\%, while the privacy increased by 650\% (from 4\% to 30\% for EER).

In \cite{noise_vc}, the authors present a method that adds noise to specific regions of the \acrshort{asrbn} representation (from Wav2Vec-2.0) by using a transformer-based privacy-risk saliency estimator to estimate the regions.
Their evaluation is performed at the feature level (no anonymized speech produced), with an ignorant attacker which does not reflect the true privacy performance.
For the best noise parameters, the utility degradation recorded is 58\% with a privacy improvement of 366\%.

\vspace{-2mm}
\subparagraph{\texorpdfstring{Discrete token vs \acrshort{ppg} vs \acrshort{asrbn}}{Discrete token vs PPG vs ASR-BN}}
Up until now, the considered supervised linguistic representation has been \acrshort{asrbn}.
However, as presented in Section \ref{main:chapt_2:vc_lin_rep}, two other supervised representations, namely \acrlong{ppg} and discrete token, can also be used for voice conversion.
In the original paper of the x-vector-based speaker anonymization system \cite{fangSpeakerAnonymizationUsing2019}, the authors experimented with extracting both \acrshort{asrbn} and \acrshort{ppg} representations.
Accordingly, to the voice conversion literature, the use of \acrshort{ppg} results in better speaker transformation, so better privacy results, however, this comes at the cost of more mispronunciations errors.
In contrast, \acrshort{asrbn} preserves more of the speaker's original pronunciation resulting in significantly better utility, at the cost of lower privacy due to the encoding of more speaker attributes.
The \acrshort{asrbn} version of this system was considered to have a better privacy/utility trade-off and was used in the VoicePrivacy 2020 challenge.
This system has received an update in 2022 for the VoicePrivacy 2022 challenge using a HiFi-GAN synthesizer (directly taking linguistic and speaker representations rather than Mel spectrograms) \cite{tomashenko2020voiceprivacy_eval2022} instead of a two-stage spectrogram+vocoder pipeline \cite{nsf}.
Both versions have been broadly used as baselines for many other works.

The Preech pipeline \cite{preech}, performs many steps to ensure anonymization not only at the prosody and acoustic level but also at the semantics, and syntax level.
Among segmentation and shuffling, sensitive word scrubbing (requiring \acrshort{asr}), and dummy segment injection (requiring \acrshort{tts} to generate speech from dummy words), they convert voices (dummy and real) to a single target voice using \acrshort{ppg} representations.
The evaluation provided goes into deep detail about textual privacy, however, proper evaluation of the anonymized speech unlinkability is missing.
There is no guarantee that an adversary could not be able to distinguish dummy segments from real ones (cf the research in \cite{partialspoof}) and perform a linkability attack on the real segment.

In \cite{Turner2022ImH} the authors present AltVoice which uses discrete tokens as linguistic representation.
They manage to anonymize speech to the fullest extent possible (fully unlinkable), by producing just text as their linguistic representation, leaving no room for any incidental transfer of the source identity.
They are properly outlining the limit of their system, with this solution the privacy/utility trade-off archived is unfavorable for utility.
The cascade \acrshort{asr}+\acrshort{tts} leads to the anonymized speech having high \acrshort{wer}, limited perceived audio quality, and no intonation and emotion preservation.

Similarly, in \cite{meyer22b_interspeech} the authors have presented a cascade \acrshort{asr}+\acrshort{tts} using phonemes as supervised linguistic representation.
They achieve perfect unlinkability but degrade the utility by 82\%.

\vspace{-2mm}
\subsection{\texorpdfstring{Adversarial \acrshort{asv} attack anonymization}{Adversarial ASV attack anonymization}}
In contrast to traditional signal processing and voice conversion techniques presented above, very recent kinds of anonymization have been proposed.
They rely on adversarial attacks and are inspired by the use of adversarial examples to put into failure the deep neural network \acrshort{asv} models while remaining imperceptible to the human ear.

The authors of \cite{v-cloak} introduced {V-Cloak} which modulates and adjusts the bottleneck features of a Wave-U-Net audio source separation model at each frequency level according to the source speaker characteristics and requested constraints on the anonymization perturbations.
Their model not only preserves intelligibility, but also naturalness, intonation, and more.
They show excellent privacy scores, with \acrshort{eer} above 40\% on most experiments, while not having the utility significantly decreased.
Additionally, they evaluated subjective speaker verification with a limited number of test speakers and asses that the speaker timber is persevered in comparison to McAdams, VoiceMask, and x-vector-based speaker anonymization.
The authors also consider three kinds of attacks against their system, ignorant, gray-box with a denoising technique to invert the anonymization, and gray-box by adapting the compromised speech.
However, informed (white-box) evaluation where the \acrshort{asv} attack model is trained to accommodate specific modulation is missing.

The authors of \cite{oreilly2022voiceblock} presented VoiceBlock, which applies a time-varying finite impulse response filter to the speech signal, allowing inconspicuous perturbations.
The way they obtain the filter is similar to audio-to-audio tasks such as denoising and speech enhancement.
Objective and subjective evaluation of the quality of the signal is very high as well as intelligibility, however, under lazy-informed attack, VoiceBlock \say{de-identification\footnote{de-identification is the process of removing identifying information, same as speaker anonymization.} performance suffers significantly}.
Similarly to {V-Cloak}, VoiceBlock does not conceal speaker identity from human listeners.

\vspace{-3mm}
\section{Conclusion}
\vspace{-2mm}
In this chapter, we presented numerous techniques to protect the speech of a speaker against linkability attacks.
In addition to the technique, we also glanced at the attackers' capabilities the authors used to evaluate the privacy of their model.
The main observation regarding this aspect is that many kinds of attackers are used, all ranging from black-box to white-box, this makes decisive comparison difficult with the scores presented.
Besides, electing a unique superior anonymization technique is almost impossible as for each of the ones presented, a hidden privacy/utility trade-off is present, not recorded by the metric.
For instance, the cascade of \acrshort{asr}+\acrshort{tts} methods are the ones offering the best trade-off in today's measurement method, however, we can question if their utility performance generalizes to noisy environments, reverberated speech, strong accents, the latter one being a real issue as strong racial disparities in \acrshort{asr} exist \cite{fair_speech}.
Additionally, when the application case is to collect large speech corpora that are representative of the various usage conditions to improve a downstream \acrshort{asr} system for all users, the use of discrete tokens for anonymization makes it impossible to annotate downstream \acrshort{asr} error, as the speech will have mispronunciations from the anonymization procedure.
Adversarial attacks on \acrshort{asv} models are methods for intentionally altering input data in a way that causes the model to make incorrect predictions.
This can be used to reduce linkability attacks, however, as it does not remove \acrshort{pii} (human listeners can still link speakers) we do not classify them as anonymization solutions.
Finally, depending on the training data available one may be forced to use signal processing methods, or voice conversion with self-supervised linguistic representation if no label is available.
In summary, to select the best anonymization technique one has to first identify his application case, and given it, make an informed decision.

For this thesis, we use the evaluation framework of the VoicePrivacy challenge, hence we chose as our baseline system the same as the VoicePrivacy, e.g., the x-vector-based speaker anonymization system.
The x-vector-based speaker anonymization technique requires text supervision during training, which is more restrictive than self-supervised approaches but, as discussed in Section \ref{main:chapt_2:vc_lin_rep}, might be an easier framework to separate and modify speaker and linguistic information, providing better anonymization.
Additionally, we believe its privacy/utility trade-off, where the utility is very well preserved for objective and subjective linguistic recognition, is a good starting point to enhance the privacy of this system.
Overall, we also think the x-vector-based system has the most possible application cases as it uses the less restricted \acrshort{asrbn} representation and a defined \acrshort{f0} for intonation.

\ifSubfilesClassLoaded{
    \printglossary[title=Special Terms,type=\acronymtype]
    \printbibliography
}{}

\end{document}

\clearemptydoublepage
\mainmatter

\cleartooddpage[\thispagestyle{empty}]
\addtocontents{toc}{\protect\pagebreak} 
\part{Contributions}

\clearemptydoublepage
\cleartooddpage[\thispagestyle{empty}]
\documentclass[../main.tex]{subfiles}

\ifSubfilesClassLoaded{
    \tableofcontentsfile
    \dominitoc
    \setcounter{chapter}{3} 
    \def\locallabelprefix{chapt_4}
    \externaldocument[]{../main}
}{}

\begin{document}

\selectlanguage{english}

\graphicspath{{./figures/dist}}

\chapter{The role of the target speaker} \locallabel{chapt4}
\minitoc

\section{Introduction}

In this chapter, we are studying the impact of the target speaker in the x-vector-based speaker anonymization system \cite{fangSpeakerAnonymizationUsing2019} presented in Chapter \ref{main:chapt3}.
This system works by swapping the source speaker identity (here extracted as the x-vector) to a target identity (selected given a specific strategy) to later generate anonymized speech.
As such, the conclusions made in this chapter are potentially applicable to any voice-conversion-based anonymization system which relies on replacing the source speaker identity with a target speaker (see Section \ref{main:chapt_3:vc_anon}).
The goal of this chapter is to analyze how this speaker modification influences the performance of the anonymization system in both privacy and utility metrics.
In the first section, we will study the role of the target selection algorithm used to select the target speaker for voice conversion.
In particular, we study how different target selection algorithm affects the privacy evaluation.
In the second section, we will study how the selected target speaker influences the privacy and utility of each source speaker to anonymize.
\section{Impact of the target selection algorithm in privacy evaluation}

In speaker anonymization, the privacy of a user is evaluated using linkability attacks.
Before diving into the target selection algorithm used for anonymization, it is important to understand the relation between the privacy evaluation procedure and the test datasets.
Linkability attacks involve determining the extent to which an individual's speech can be linked to their identity.
This evaluation relies on \acrshort{asv} where the test dataset (compromised and vulnerable speech) strongly affects the privacy measurement.
For a speech signal to be linked to an identity, the \acrshort{asv} must provide high similarity scores for
the considered genuine trials and lower scores for all other impostor trials.

In this section, we aim to study target selection algorithms that were designed to increase privacy/unlinkability, some of them have been presented in Section~\ref{main:chapt3_xvector_modef}.
We propose to analyze target selection algorithms to query how they influence the creation of vulnerable and compromised anonymized datasets specifically for linkability attack evaluation (see Section~ \ref{main:chapt_3:threat_model}).
The goal is to better understand the impact of the speaker selection step necessary for \acrshort{vc}-based anonymization and its influence against linkability evaluation.

\subsection{Experimental setup}
This experiment mainly follows the VoicePrivacy challenge requirements and uses the x-vector-based anonymization system (Section \ref{main:chapt_3:anon_sys_x-vector}) trained with the dataset of the \acrshort{vpc} (Table \ref{main:chapt_2:dataset}).
Privacy is evaluated using the gray-box {semi-informed} attacker, trained on the anonymized version of \textit{LibriSpeech train-clean-360} to reduce as much as possible domain mismatch with the testing data.
In contrast to the \acrshort{vpc} \acrshort{asv}, where the x-vectors of an enrollment speaker are averaged over his/her utterances before \acrshort{plda} scoring, we obtain \acrshort{plda} scores for each possible pair of one vulnerable utterance and one compromised utterance.
Additionally, in contrast to the \acrshort{vpc} anonymization, anonymization is performed at the \textit{utterance-level} rather than at the \textit{speaker-level}. The difference between the two will be studied in this section.
The research will concentrate exclusively on the \textit{LibriSpeech test-clean} dataset and will feature only male subjects.
The choice of using only one gender is justified by the fact that some target selection algorithms are gender dependent.
It is important to note that the key findings will be relevant to any dataset and any gender, selecting male was arbitrary.

\vspace{-4mm}
\subsection{Clear speech linkability} \locallabel{sec:clean_speech_link}
\vspace{-1mm}

In this section, we present how linkable the clear (non-anonymized) speech is using the baseline linkability evaluation scenario presented in Section \ref{main:chapt_3:attacker_capa}, where the vulnerable and compromised speech datasets are clear and evaluated with an \acrshort{asv} system also trained on clear speech.
Figure \localref{fig:asv_clean} displays the \acrshort{plda} score distributions for genuine and impostor trials, for each vulnerable speaker of our test dataset (right part of the figure), and all speakers together (left part of the figure).
The \acrshort{eer} is computed over all speakers and the dotted red line represents the corresponding \acrshort{plda} \acrshort{eer} threshold.
From the distinct distributions shown and the low 3.1\% \acrshort{eer} score, we can conclude that clear speech is highly representative of the speaker identity as the linkability attack allow to highly separate genuine and impostor trials.
We observe that speaker ID 260 seems to be more difficult than the others to recognize whereas speaker ID 8224 had a distinct voice compared to the others.
%
\begin{figure}[!htbp]
  \centering
  \centerline{\includegraphics[width=1.03\linewidth]{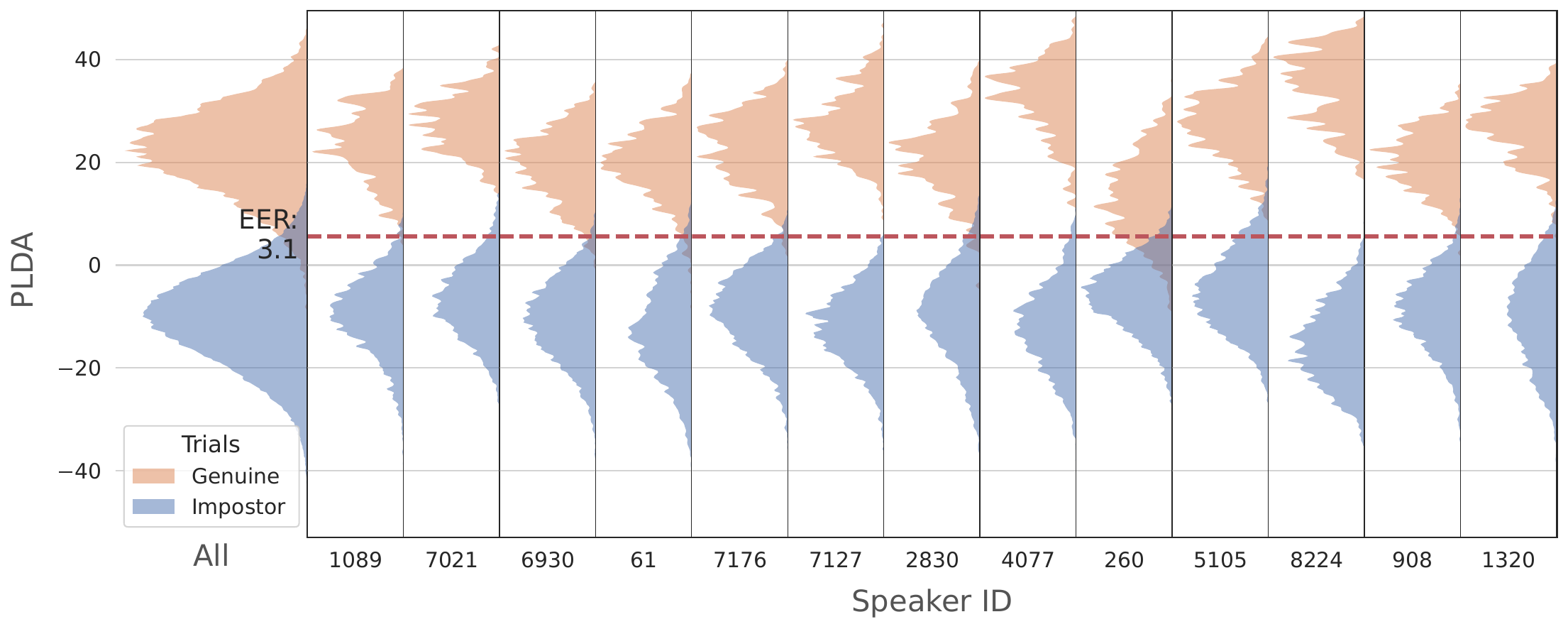}}
  \vspace{-4mm}
  \caption{Clear speech linkability.}
  \vspace{-5mm}
  \locallabel{fig:asv_clean}
\end{figure}

\vspace{-2mm}
\subsection{\texorpdfstring{\protect\say{Farther 200 random 100} target selection strategy}{Farther 200 random 100 target selection strategy}}
\vspace{-1mm}

In this section, we study the target selection strategy introduced by the authors of the x-vector-based system \cite{fangSpeakerAnonymizationUsing2019} and used as the selection strategy for the \acrshort{vpc} 2020 and 2022 baseline systems.
We will describe how this target selection strategy works, and how it is conceptually and experimentally flawed.
The following steps are followed to generate the target x-vector:
1) the process begins by extracting an x-vector from a clear signal;
2) the clear x-vector is then compared to each of the speakers in a pool of external x-vectors with a \acrshort{plda} scoring function;
3) the 200 x-vectors that are the least similar (farthest) to the clear x-vector are selected from the pool, and 4) 100 of these x-vectors are randomly chosen and averaged to create the anonymized target x-vector.
As presented in Table~\ref{main:chapt_2:dataset} of the VoicePrivacy requirements section, the pool of external x-vectors is \textit{LibriTTS train-other-500} and the x-vector extractor is trained on VoxCeleb1,2.



\begin{figure}[!htbp]
  \centering
  \centerline{\includegraphics[width=0.65\linewidth]{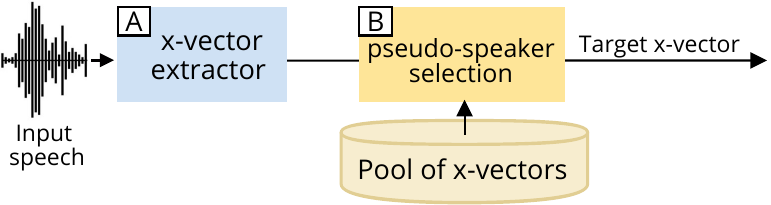}}
  \vspace{-3mm}
  \caption{X-vector anonymization defined in \cite{fangSpeakerAnonymizationUsing2019}.}
  \locallabel{fig:target_select_vpc}
  \vspace{-1mm}
\end{figure}

To evaluate how robust against linkability attacks this target selection algorithm is, we will start by analyzing how linkable the selected target x-vectors are by themselves, without using speech synthesis.
\acrshort{asv} is performed with the target x-vector obtained from the selection algorithm, see Figure~\localref{fig:target_select_vpc}.
Figure~\localref{fig:spk_select_100_200_feat} shows the \acrshort{plda} scores distributions for the selected target x-vectors for each speaker, the first thing that stood out is the low 4.8\% \acrshort{eer} value which indicates a strong link between the clear x-vector and anonymized target x-vector.
If we were to suppose the anonymization system can completely remove the source speaker identity and replace it with one of the targets, effectively performing the complete \acrshort{pii} removal needed for anonymization, this target selection strategy would still allows linkability attacks as the anonymized speech will have the same \acrshort{eer} as the target x-vector (4.8\%).
Hence, even the most perfect anonymization system fails as privacy is evaluated through linkability.
We can consider the \acrshort{asv} linkability results (\acrshort{eer} here) of target selection algorithms as the expected linkability results that anonymization systems should produce.

\begin{figure}[!htbp]
  \centering
  \vspace{1em}
  \centerline{\includegraphics[width=1.03\linewidth]{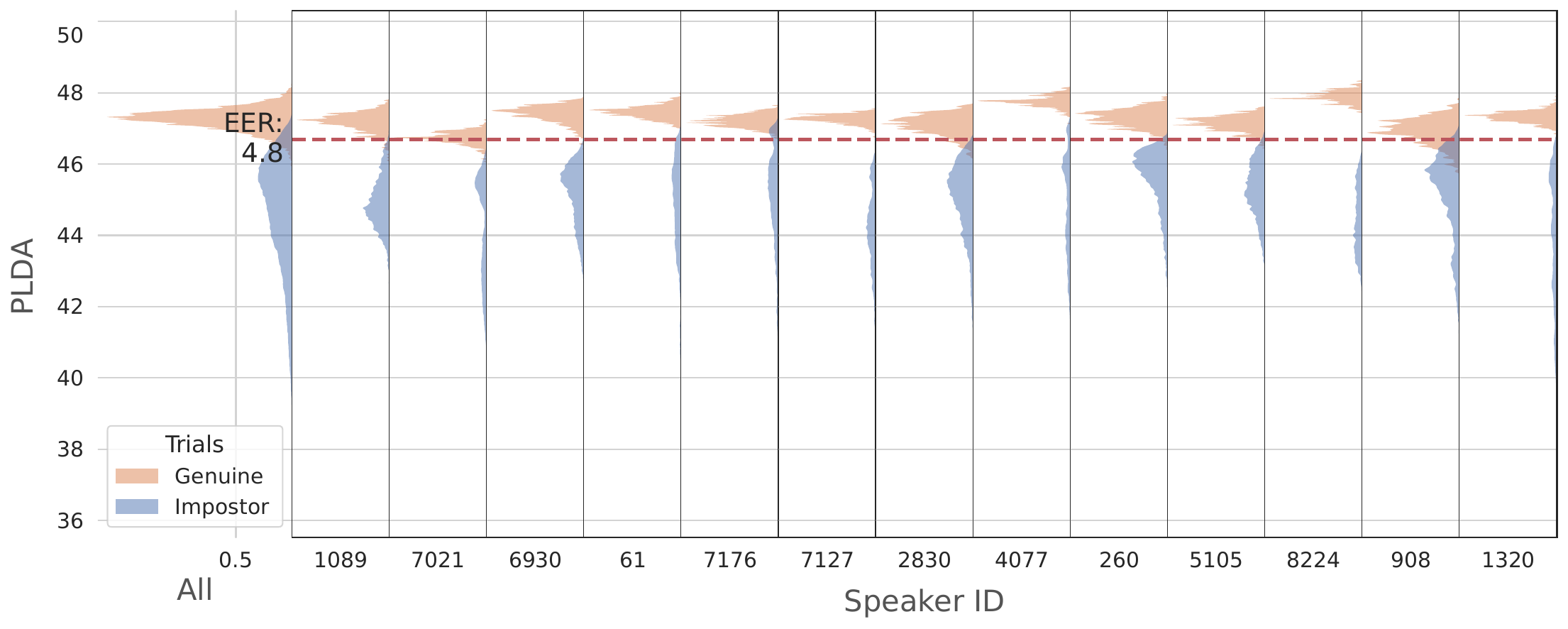}}
  \vspace{-3mm}
  \caption{\centering{\acrshort{plda} genuine and impostor scores distributions for the \protect\say{farther 200 random 100} target selection strategy. Scores are computed directly from target x-vectors, no speech synthesis is done. The \acrshort{plda} model is trained on clear data (\textit{VoxCeleb1,2}, see Table~\ref{main:chapt_2:dataset}).}}
  \locallabel{fig:spk_select_100_200_feat}
  \vspace{-3mm}
\end{figure}
Limitations of this approach can be explained by the target x-vector selection process.
Indeed, the attacker in the {semi-informed} scenario has access to the anonymization pipeline, hence a similar x-vector will be extracted for the vulnerable and compromised speech of a speaker as the x-vector extractor (of the anonymization pipeline) is the same.
Then given similar clear x-vectors from the vulnerable and compromised speech, the 200 farther x-vectors selected as candidates will also be similar or at least distributed in the same x-vector space as the pool of speakers (of the anonymization pipeline) is the same.
In the fourth stage, differences between vulnerable speech anonymization and compromised speech anonymization occur, the randomly 100 x-vectors selected from the 200 candidates differ and are very unlikely to be the same.
However, the last stage of this target selection strategy is to average the 100 x-vector together to generate the anonymized target x-vector.
This average operation cancels the fourth stage, as it is very likely that the average of the 100 x-vector is very close to the average of the 200 candidate x-vectors, explaining the very high \acrshort{plda} scores in Figure~\localref{fig:spk_select_100_200_feat}.
This means that the anonymized vulnerable speech and anonymized compromised speech for a speaker will have similar target x-vectors, while also being different from other speakers due to the farther 200 selection.
Still, the \acrshort{plda} \acrshort{eer} threshold score is very high at around 46, with most scores in the interval of 42 to 48, this indicates that all selected target x-vectors lay in a confined region space.

\begin{figure}[!htbp]
  \centering
  \centerline{\includegraphics[width=1.03\linewidth]{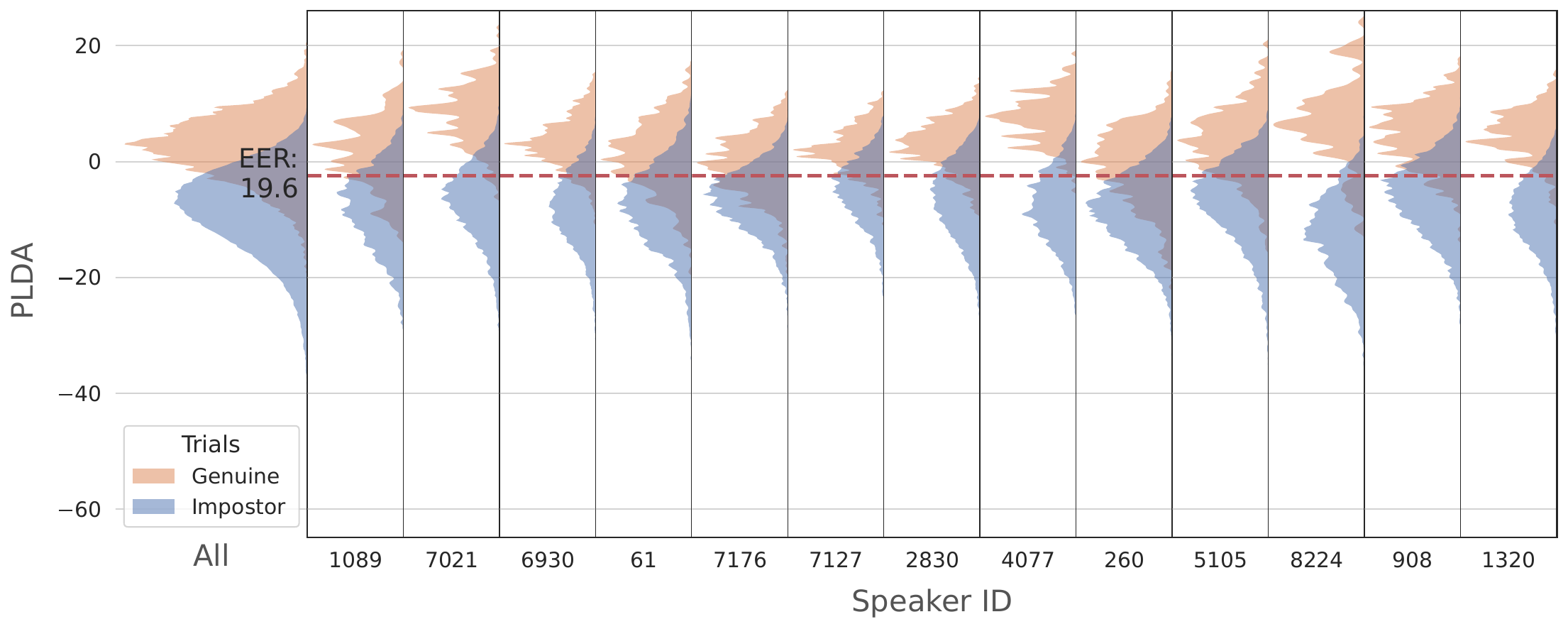}}
  \vspace{-3mm}
  \caption{\centering{\acrshort{plda} genuine and impostor scores distributions of the {semi-informed} attacker \acrshort{asv} model on anonymized speech for the \protect\say{farther 200 random 100} target selection strategy.}}
  \locallabel{fig:spk_select_100_200_speech}
  \vspace{-3mm}
\end{figure}
Figure \localref{fig:spk_select_100_200_speech} displays the attacker \acrshort{plda} scores distribution for anonymized speech (anonymized with x-vector-based speaker anonymization system).
The attacker \acrshort{asv} model is trained with anonymized speech as required by the {semi-informed} attacker definition.
The x-vector-based anonymization system was not able to effectively replace the source speaker identities with those of the target, as indicated by the \acrshort{eer} value of 19.6\%, higher than the expected \acrshort{eer} value of 4.8\% of the target x-vector.
Multiple factors can explain this disparity.
The first one is the ineffectiveness of the synthesis system to perform speaker replacement, resulting in a speech signal having an imprecise speaker identity (maybe it has a mixture of clear and target identities).
The second one is related to the target x-vectors.
The target x-vectors only occupy a confined region space indicated by the high \acrshort{plda} scores in Figure~\localref{fig:spk_select_100_200_feat}, and the cause of this is the average operation. 
This means that the target x-vectors are all close to each other, still, they are preserving inter-speaker and intra-speaker separation for \acrshort{asv}.
We hypothesize that for the synthesis system, the inter-speaker and intra-speaker x-vectors are so close to each other that after synthesis, the resulting speaker's identities are considered similar.
This aspect would lead to anonymized voices being less linkable.

Overall, linkability attack evaluations with the \say{farther 200 random 100} is confusing, we would like the \acrshort{plda} scores distributions to be as overlapping as possible to comply with unlinkability.
However, at the target x-vector level, there are no guarantees of unlinkability, and understanding the intended output of the synthesis system is intricate.
Hence, are we evaluating how unlinkable anonymized speech is? Or are we evaluating how distinguishable converted speakers are?

We observe that speaker ID 260 seems to be better anonymized than the others whereas speaker ID 8224 is harder to anonymize, this observation is correlated with the one made in Section \localref{sec:clean_speech_link} on clear speech, where speaker ID 260 was hard to recognize and speaker ID 8224 had a distinct voice.

\vspace{-3mm}
\subsection{\texorpdfstring{\protect\say{Dense} target selection strategy}{Dense target selection strategy}}
\vspace{-2mm}
In this section, we evaluate the \say{dense} target selection strategy introduced in \cite{brij_vpc_design} and slightly modified in \cite{dp_vpc}, the first version is used here.
To select the target x-vector for each utterance, the following steps are followed: 1) all x-vectors of the pool are grouped into clusters using the Affinity Propagation algorithm and \acrshort{plda}, resulting in a total of 80 clusters; 2) one cluster is randomly chosen from the 10 largest clusters (the cluster which is closest to the source speaker is filtered out); 3) half of the members of the chosen cluster are randomly selected to introduce more randomness in the selection of the x-vector, and 4) the selected candidate x-vectors are averaged to obtain the target x-vector.
This method is similar to the "farthest 200 random 100" strategy described above, but the x-vector chosen is less related to the input utterance so less linked to the original speaker.
As far as our understanding, in the modified version, the cluster which is closest to the source speaker is not filtered out, making it completely independent of the source utterance.

\begin{figure}[!htbp]
  \centering
  \vspace{-3mm}
  \centerline{\includegraphics[width=1.03\linewidth]{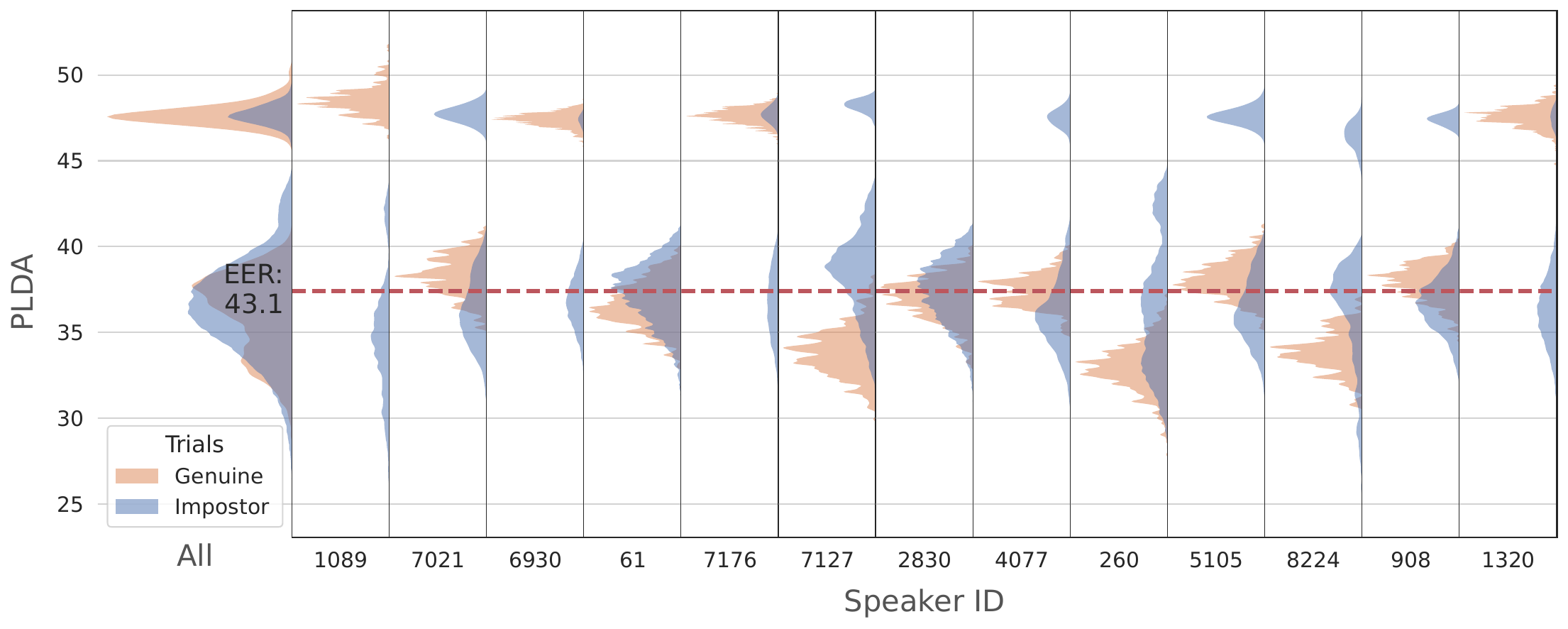}}
  \vspace{-3mm}
  \caption{\centering{\acrshort{plda} genuine and impostor scores distributions for the \protect\say{dense} target selection strategy. Scores are computed directly from target x-vectors, no speech synthesis is done. The \acrshort{plda} model is trained on clear data (\textit{VoxCeleb1,2}, see Table~\ref{main:chapt_2:dataset}).}}
  \locallabel{fig:spk_select_dense_feat}
  \vspace{-2mm}
\end{figure}

Figure \localref{fig:spk_select_dense_feat} shows the target x-vector \acrshort{plda} distributions obtained for this selection algorithm.
The first thing that stood out is the seemingly two Gaussians per distribution which we suppose is the result of the cluster selection in the algorithm.
The \acrshort{eer} value is 43.1\% indicating that at the target x-vector level, better unlinkability is guaranteed.
However, for specific speakers such as speaker ID 1089, we can observe that his target x-vectors genuine and impostor scores are not overlapping, which does not guarantee unlinkability for this speaker's speech.
This effect might be caused by the nearest cluster being filtered out.

Figure \localref{fig:spk_select_dense_speech} displays the attacker \acrshort{plda} scores distributions for anonymized speech.
The attacker \acrshort{asv} model is trained with anonymized speech data that was anonymized using the same \say{dense} target selection algorithm as required by the {semi-informed} attacker.
Overall we can observe a high \acrshort{eer} value of 39.2\%, indicating better performance against linkability attacks.

\begin{figure}[!htbp]
  \centering
  \centerline{\includegraphics[width=1.03\linewidth]{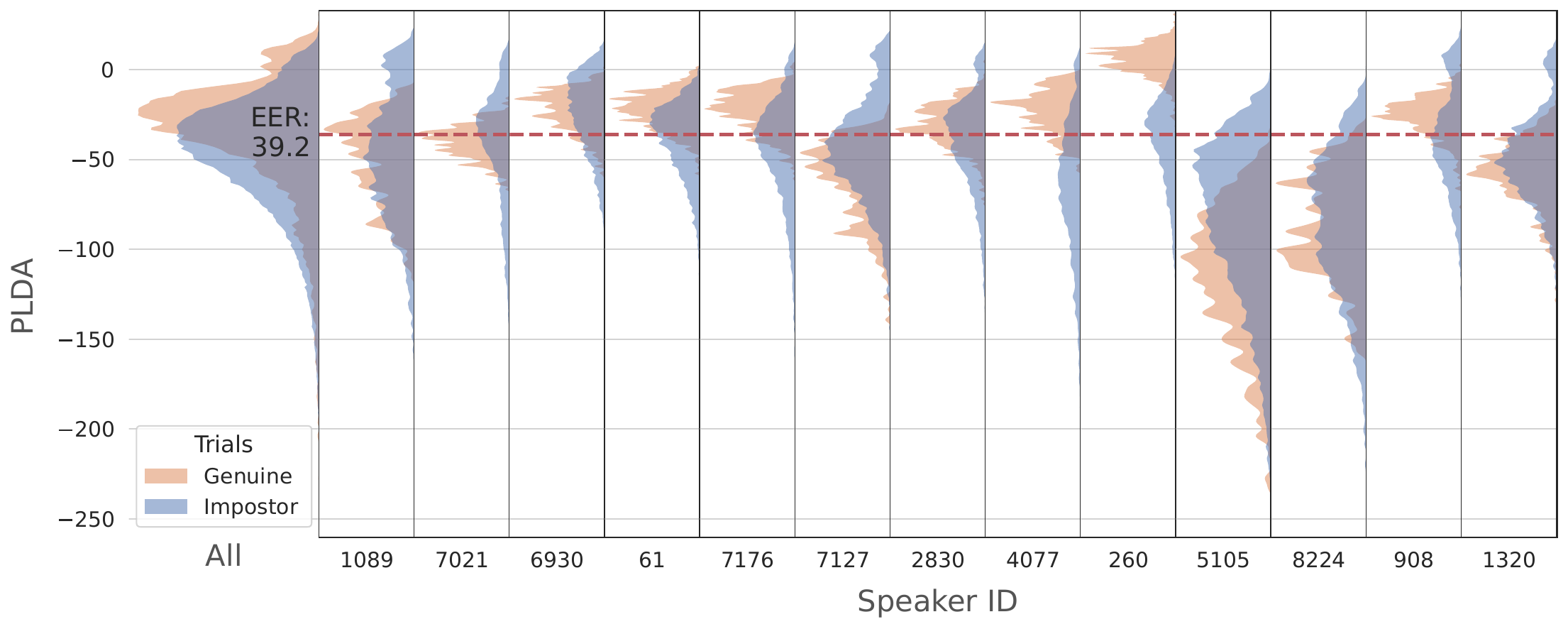}}
  \vspace{-1mm}
  \caption{\centering{\acrshort{plda} genuine and impostor scores distributions of the {semi-informed} attacker \acrshort{asv} model on anonymized speech for the \protect\say{dense} target selection strategy.}}
  \locallabel{fig:spk_select_dense_speech}
  \vspace{-3mm}
\end{figure}

However, the \acrshort{plda} \acrshort{eer} threshold equals $-$35, which might indicate that the \acrshort{plda} produces very miscalibrated scores due to a potential attacker \acrshort{asv} model improperly trained.
There is a possibility that this target selection algorithm generates data in a way that makes training the {semi-informed} attacker fail.
To evaluate this hypothesis, we retrained the \acrshort{asv} {semi-informed} attacker model with data anonymized with the \say{random~speaker} (presented below) target selection algorithm and obtained an \acrshort{eer} of~20.9\%.

A similar conclusion has been drawn from \cite{dp_vpc,tomashenko2020voiceprivacy_eval2022} where training the attacker with \textit{speaker-level} rather than \textit{utterance-level} target selection yielded a training overfit of the \acrshort{asv} model.
In this study, we have identified that the design of the target selection algorithm can also contribute to this issue.

\vspace{-1mm}
\subsection{\texorpdfstring{\protect\say{Random speaker} target selection strategy}{Random speaker target selection strategy}}
\vspace{-1mm}

In this target selection strategy, for each utterance, a random x-vector is sampled from the pool and used as the target identity \cite{EvaluatingVoiceConversionbased2019}.
Figure \localref{fig:spk_select_rand_feat} shows the target x-vector \acrshort{plda} distributions, as expected, they follow Gaussian distributions and are completely overlapping, leading to a target x-vector level \acrshort{eer} value of 50\%.
This guarantee that if the anonymization system can completely remove the source identity and replace it with the target, linkability attacks will fail against properly trained \acrshort{asv} models.

\begin{figure}[!htbp]
  \vspace{-1mm}
  \centering
  \centerline{\includegraphics[width=1.00\linewidth]{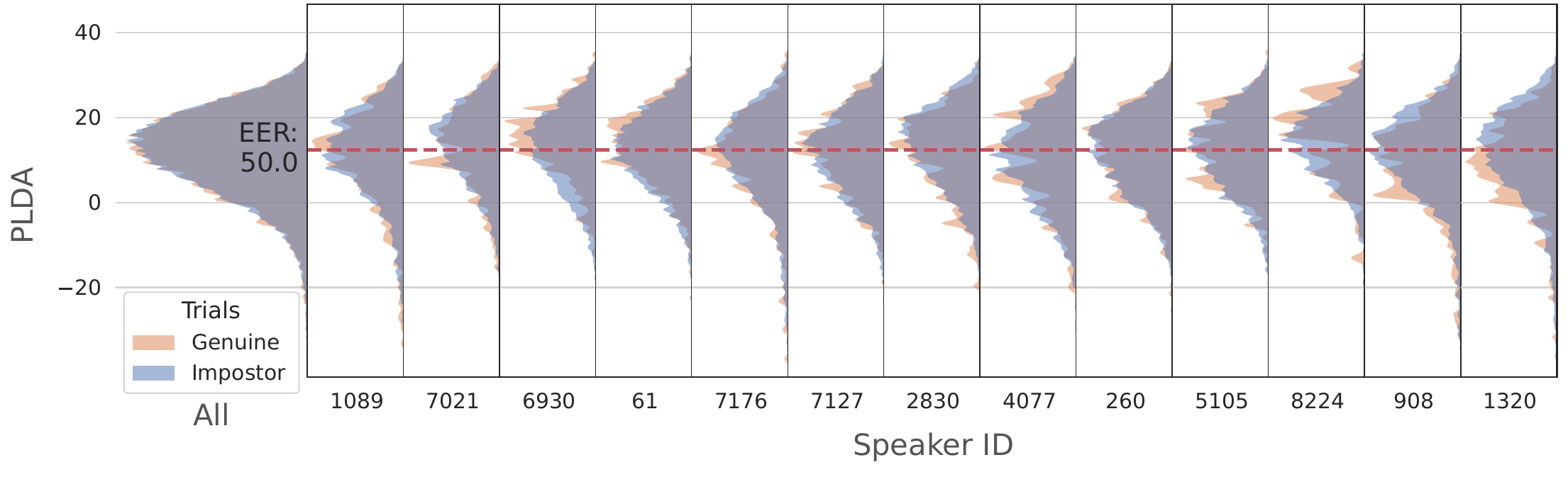}}
  \vspace{-5mm}
  \caption{\centering{\acrshort{plda} genuine and impostor scores distributions for the \protect\say{random speaker} target selection strategy. Scores are computed directly from target x-vectors, no speech synthesis is done. The \acrshort{plda} model is trained on clear data  (\textit{VoxCeleb1,2}, see Table~\ref{main:chapt_2:dataset}).}}
  \locallabel{fig:spk_select_rand_feat}
  \vspace{-2mm}
\end{figure}

When anonymizing speech and training the \acrshort{asv} attacker model with this target selection algorithm, the \acrshort{eer} value obtained is 21.6\%.
Figure \localref{fig:spk_select_rand_speech} displays the corresponding attacker \acrshort{plda} scores distributions, the \acrshort{plda} \acrshort{eer} threshold is close to~0 indicating the \acrshort{plda} is correctly calibrated.
This time, the \acrshort{eer} disparity is primarily caused by the ineffectiveness of the synthesis systems to perform speaker replacement.
Overall, in this study, we showed that this target selection strategy has a much better guarantee than the previous ones in terms of expected unlinkability, and, the creation of anonymized data for training the attacker.
We observe the same conclusion about speaker ID 260 being easier to anonymize than speaker ID 8224.
This disparity will be explored in Section~\localref{sec:radar}.

\begin{figure}[!htbp]
  \centering
  \vspace{-3mm}
  \centerline{\includegraphics[width=1.00\linewidth]{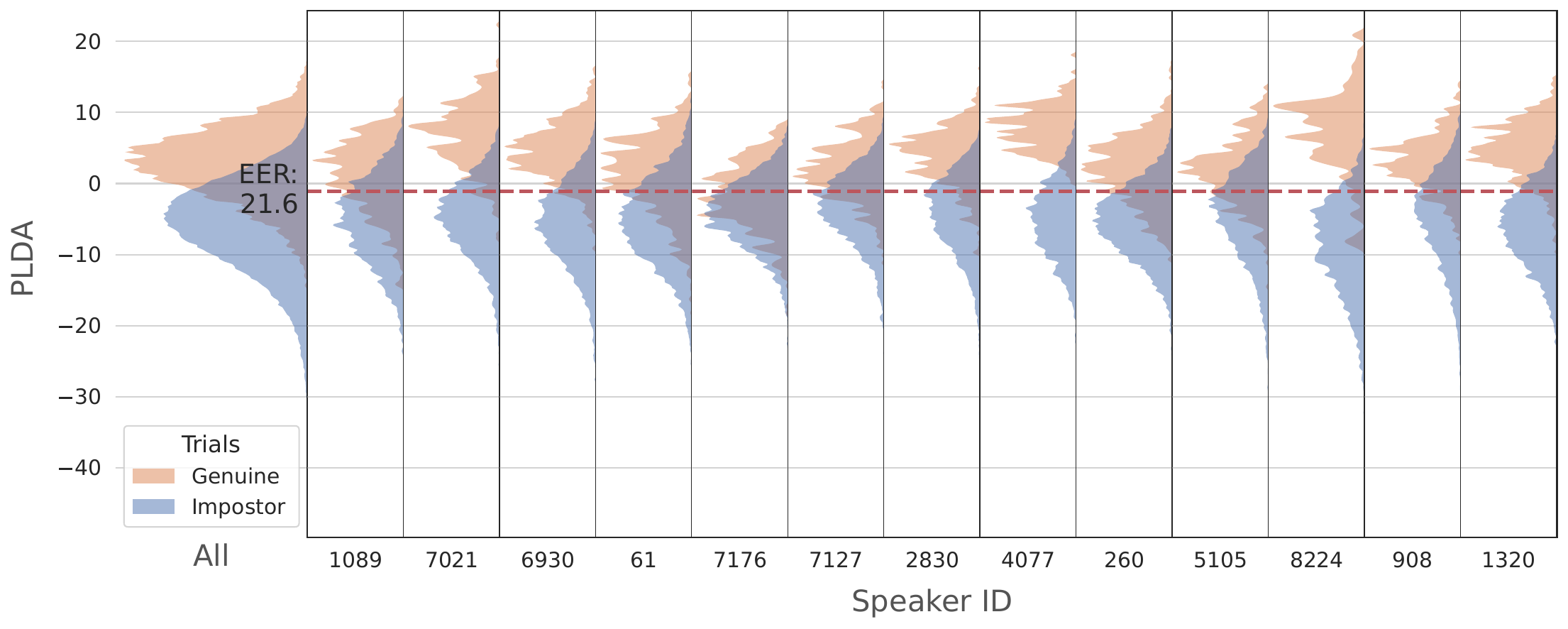}}
  \vspace{-4mm}
  \caption{\centering{\acrshort{plda} genuine and impostor scores distributions of the {semi-informed} attacker \acrshort{asv} model on anonymized speech for the \protect\say{random speaker} target selection strategy.}}
  \locallabel{fig:spk_select_rand_speech}
  \vspace{-4mm}
\end{figure}

\vspace{-3mm}
\subsection{\texorpdfstring{\protect\say{Random vector} target selection strategy}{Random vector target selection strategy}}
\vspace{-2mm}
This target selection strategy is similar to the previous one, but instead of sampling for each utterance to anonymize an x-vector from a pool, it samples it from a random Gaussian distribution.
At the target x-vector level, the \acrshort{eer} value is also 50\%.
And analogous to the previous strategy, the anonymized speech has an \acrshort{eer} value of 23.3\%, with a similar distribution to one of the \say{random speaker} in Figure~\localref{fig:spk_select_rand_speech}.
Again, this disparity can be explained by the ineffectiveness of the synthesis systems to perform speaker replacement.
The slightly higher \acrshort{eer} compared to the \say{random speaker} target strategy could be explained by a worse speech's naturalness.

\vspace{-3mm}
\subsection{\texorpdfstring{\protect\say{Constant speaker} target selection strategy}{Constant speaker target selection strategy}}
\vspace{-1mm}
The \say{constant speaker} target selection strategy might be the most simplistic one.
A single target identity is used to anonymize every utterance of every speaker.
At the target x-vector level, the \acrshort{eer} value is 50\% as everyone should sound like a single speaker.
Hence, the target level genuine/impostor scores follow a single Dirac-delta distribution (see~Figure~\localref{fig:spk_select_const_feat}).
Comparable to the previous strategies, the anonymized speech has an \acrshort{eer} value of 22.4\%.
The attacker \acrshort{plda} scores distributions are similar to the one of the \say{random speaker} in Figure~\localref{fig:spk_select_rand_speech}.

\begin{figure}[H]
  \centering
  \centerline{\includegraphics[width=1.00\linewidth]{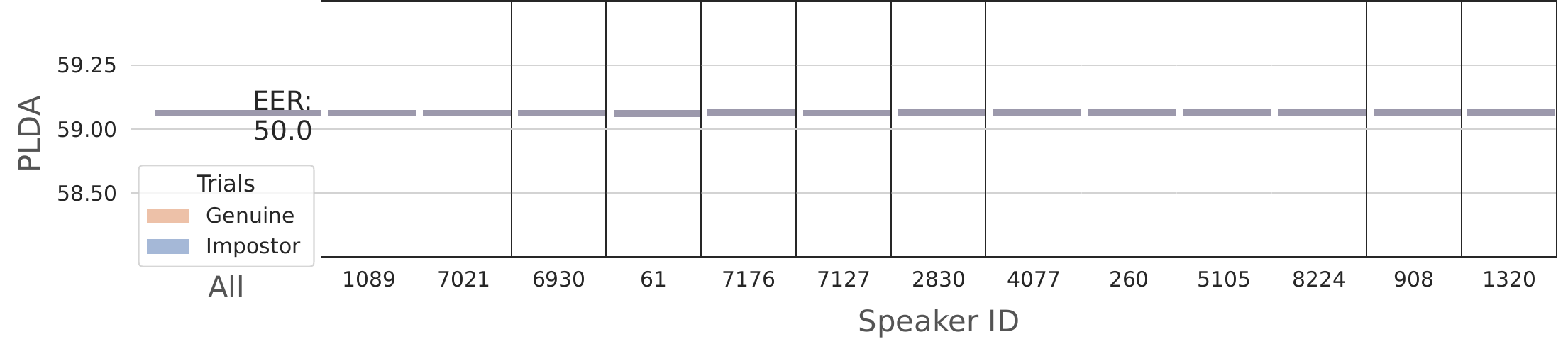}}
  \vspace{-6mm}
  \caption{\centering{\acrshort{plda} genuine and impostor scores distributions for the \protect\say{constant speaker} target selection strategy. Scores are computed directly from target x-vectors, no speech synthesis is done. The \acrshort{plda} model is trained on clear data  (\textit{VoxCeleb1,2}, see Table~\ref{main:chapt_2:dataset}).}}
  \locallabel{fig:spk_select_const_feat}
  \vspace{-2em}
\end{figure}


\subsection{\texorpdfstring{\textit{Speaker-level} target selection}{Speaker-level target selection}}
\vspace{-2mm}

\begin{figure}[!hbtp]
\vspace{-4mm}
  \centering
  \centerline{\includegraphics[width=1.0\linewidth]{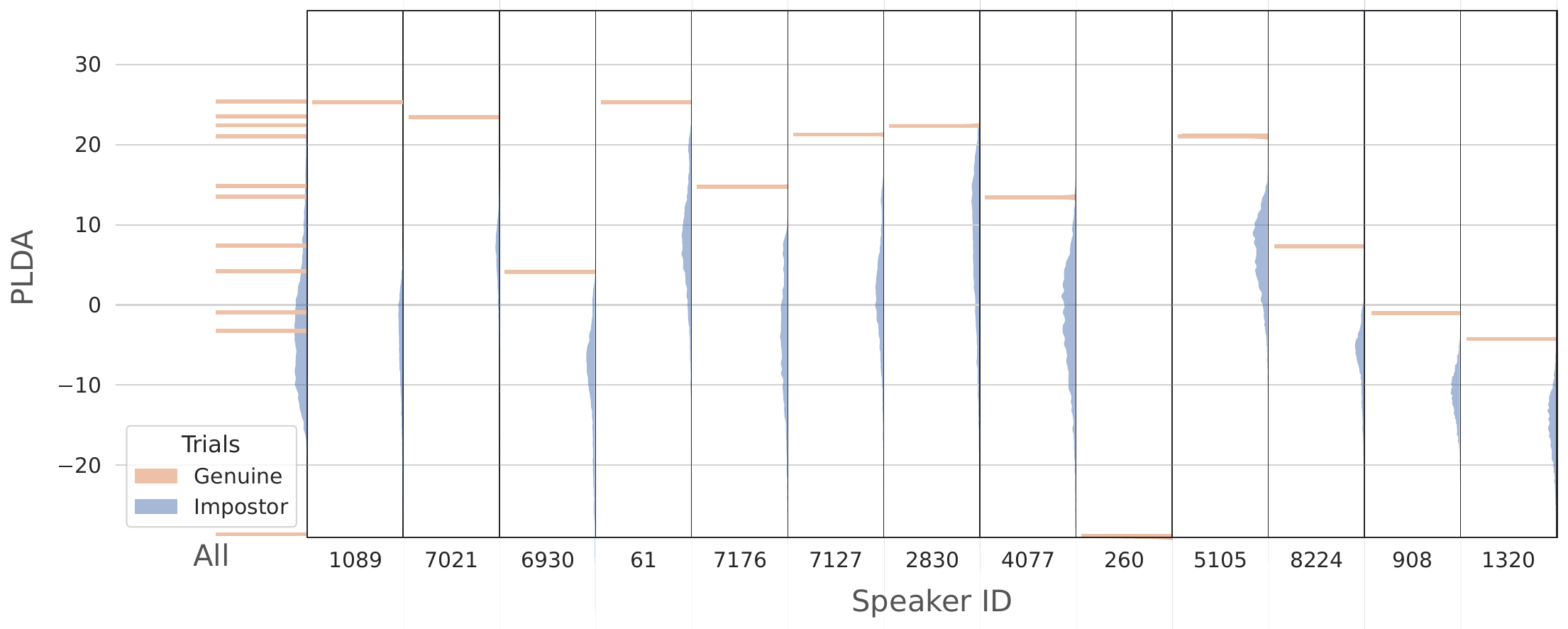}}
  \vspace{-4mm}
  \caption{\centering{\acrshort{plda} genuine and impostor scores distributions for the \protect\say{random speaker} x-vector target selection strategy using \textit{speaker-level} anonymization. Scores are computed directly from target x-vectors, no speech synthesis is done.}}
  \locallabel{fig:spk_select_rand_spk_feat}
  \vspace{-4mm}
\end{figure}

The level of anonymization can either be \textit{utterance-level} or \textit{speaker-level} target selection.
For the \textit{utterance-level} anonymization, the target x-vector is generated for each utterance that needs to be anonymized.
While for \textit{speaker-level} target selection, the target x-vector is generated once for each speaker, and then applied to every utterance made by that speaker.
In the above experiments, we only used \textit{utterance-level} as it is the one supposed to have better unlinkability capability \cite{tomashenko2020voiceprivacy_eval2022}.
However, \textit{speaker-level} target selection is required in the \acrshort{vpc} challenge as there is a requirement that the anonymized voices of all speakers must be distinguishable from each other and should not change over time.
Something interesting in multi-party conversations application cases.

In this experiment, we study how \textit{speaker-level} affects the creation of anonymized datasets used for privacy evaluation.
We chose to use the \say{random speaker} target selection strategy, as the \say{Farther 200 random 100} target selection strategy has already been proven to not be a suited one, and the \say{constant speaker} cannot fulfill the speaker distinctiveness requirement.
Figure \localref{fig:spk_select_rand_spk_feat} displays the target x-vector \acrshort{plda} distributions.
We can see that each utterance of a speaker has the same target x-vector as the genuine trials have all the same \acrshort{plda} score indicated by the per speaker Dirac-delta distributions.
We observe that the impostor trials are not overlapping the per-speaker Dirac-delta distributions indicating that a given speaker will not get confused by another, this follows the speaker distinctiveness requirement.
However, there is an inherent issue, by definition, \textit{speaker-level} target selections do not create datasets where speakers are unlinkable.
And this is troublesome as privacy evaluation is based on \acrshort{asv} linkability evaluation.
In Figure~\localref{fig:spk_select_rand_spk_feat}, the linkability metric equals 1.0.
For reference, the same target selection strategy applied at the \textit{utterance-level} has a linkability metric score of 0.0 (full overlap genuine/impostor distributions).

\subsection{\texorpdfstring{Discussion}{Discussion}} \locallabel{chapt4:vctk_libi_eer_diff}

\begin{table}[!htbp]
  \vspace{-0.4cm}
  \caption{Privacy results for the four main x-vector-based target selection strategies at the target x-vector level and anonymized speech x-vector level for the \textit{LibriSpeech test-clean} male set.
    The confidence interval stays within $\pm~0.40$~\acrshort{eer} for all experiments$^{1}$.
Evaluation models are retrained on anonymized speech when necessary, and the model architecture is described in \ref{main:chap2:privacy_model_metric}
    }
  \vspace{0.4cm}
  \centering{
    \begin{tabular}{
      l@{}@{\extracolsep{5.0mm}}
      S[table-format=1.2]@{\extracolsep{2.0mm}}
      S[table-format=2.1]@{\extracolsep{7.0mm}}
      S[table-format=1.2]@{\extracolsep{2.0mm}}
      S[table-format=2.1]@{\extracolsep{0.0mm}}
      }
      \toprule
      \multirow{1}{0pt}{\begin{minipage}{200pt}{Target selection algorithm}\end{minipage}}
                                                                                           & \multicolumn{2}{c}{Target x-vector}            & \multicolumn{2}{c}{\hspace{-8mm} Anonymized speech x-vector}                                                                                                  \\
      \midrule
      \multirow{2}{0pt}{\begin{minipage}{0pt}{\vspace{1mm}Metric}\end{minipage}} & \multicolumn{2}{c}{ \hspace{-2mm} Privacy}     & \multicolumn{2}{c}{ \hspace{-10mm} Privacy}                                                                                                                   \\
                                                                                           & \multicolumn{1}{c}{\acrshort{li}~$\downarrow$} & \multicolumn{1}{c}{\acrshort{eer}~$ \uparrow$}               & \multicolumn{1}{c}{\acrshort{li}$\downarrow$} & \multicolumn{1}{c}{\acrshort{eer}~$ \uparrow$} \\
      \midrule
      Clear speech                                                                         &                                                &                                                              & 0.92                                          & 3.1                                            \\
      \midrule
      VPC farther 200 random 100                                                           & 0.87                                           & 4.8                                                          & 0.54                                          & 19.5                                           \\
      \midrule
      Dense, \footnotesize{Dense \acrshort{asv} model}                                     & 0.19                                           & 43.1                                                         & 0.14                                          & 39.2                                           \\
      \multirow{1}{170pt}{Dense, \footnotesize{Random speaker \acrshort{asv} model}}       & \texttt{"}                                     & \texttt{"}                                                   & 0.50                                          & 20.9                                           \\
      \midrule
      Random speaker                                                                       & 0.00                                           & 50.0                                                         & 0.49                                          & 21.6                                           \\
      Random vector                                                                        & 0.00                                           & 50.0                                                         & 0.45                                          & 23.3                                           \\
      Constant speaker                                                                     & 0.00                                           & 50.0                                                         & 0.47                                          & 22.4                                           \\
      \bottomrule
    \end{tabular}
  }
  \locallabel{tab:target-select}
\end{table}

\footnotetext[1]{The FEERCI toolkit \protect\cite{Haasnoot2018FEERCIAP} is used to estimate the \protect\acrshort{eer} confidence interval with the bootstrapping method.}

Table \localref{tab:target-select} summarizes the previous \acrshort{eer} results accompanied by their corresponding \acrshort{li} results.
Out of the five target selections presented, only three of them have an \acrshort{eer} at the target x-vector level of 50\%, and one has a high target level \acrshort{eer} value but did not enable proper training of the \acrshort{asv} attacker model.
In our experiment, the best target selection algorithms were those that had overlapping impostor and genuine distributions.

The main takeaway of this section is that when building target selection strategies, the main goal should be to ensure unlinkability at the target x-vector level.
To do so, the designer has to make sure the target x-vector of an utterance/speaker will be confused by other target x-vector of other utterances/speakers.
By doing so, if the anonymization system is capable of completely removing the source identity and replacing it with the target one, the anonymized speech will have the same \acrshort{eer} as the target level \acrshort{eer}, which is a value of 50\% in the best case.

In contrast, if the target x-vector has a link with the source speaker, then the \acrshort{asv} attacker model will evaluate the strength of the link between the target and anonymized speech, rather than evaluating the strength of anonymized speech unlinkability.
For example, if the target level \acrshort{eer} value is 0\%, and the anonymized speech \acrshort{eer} value is 0\%, this means that we aim to achieve full linkability and that we achieve full linkability.
In this case, what is evaluated is the successfulness of speech synthesis to generate distinct voices.
Something that can be useful for assessing voice distinctiveness or the preservation of a pseudo-identifier in pseudonymization.
However, those voices could very well leak the source speakers' identities.
That is why trying to produce unlinkable speech at all stages and trying to build the strongest attacker to defeat linkability, is the only way to properly record the strength of the anonymization.
We conclude that \textit{speaker-level} target selection can be used for voice distinctiveness evaluation.
However, it should not be used for privacy linkability evaluation.

Similarly, in the case of the \say{farther 200 random 100} target selection strategy, we argue that what is evaluated is the strength of the voice distinctiveness rather than the strength of the anonymization.
The \acrshort{eer} of the anonymized speech of the \say{farther~200 random~100} strategy being close to the one obtained with the \say{constant speaker} algorithm in the \textit{LibriSpeech test-clean} male set does not help to assess the difference.
It seems that the strength of voice distinctiveness is equal to the strength of anonymization for this dataset.
However, in the \textit{VCTK test} dataset, moving from the \say{farther 200 random 100} target selection strategy to the \say{constant speaker} strategy shows an absolute increase by more than 10\% of \acrshort{eer}, highlighting the evaluation disparity, see privacy results Table~\localref{table:vctk_f_c_diff}.
Another issue of the \say{Farther 200 random 100} target selection strategy affects fairness.
As described in \cite{Turner_xvector_vpc2020}, male and female are not equally transformed when using the \say{farther 200 random 100} target selection strategy as a large \acrshort{eer} gap exists.
Using the \say{constant speaker} target selection strategy (with the same anonymization pipeline) fixes this issue.

In Table~\localref{table:vctk_f_c_diff}, the \say{constant speaker} algorithm has lower utility performance compared to the \say{farther 200 random 100} one.
This is likely to be caused by the selected x-vector having component characteristics different from the ones of the training dataset used to train the synthesis model.
The \say{farther 200 random 100} does not have this issue, as the average of multiple x-vector mitigates this effect.
Later in this chapter, we propose and evaluate other solutions to select the constant target speaker.

\newcolumntype{H}{>{\setbox0=\hbox\bgroup}c<{\egroup}@{}} 
\begin{table}[!ht]
  \vspace{-0.3cm}
  \caption{{Privacy and utility results disparity for the \protect\say{farther 200 random 100} and \protect\say{constant~speaker} target selection strategies. Evaluation models are retrained on anonymized speech, and the model architectures are described in \ref{main:chap2:privacy_model_metric} and \ref{main:chap2:utility_model_metric}}}
  \vspace{0.4cm}
  \centering
  \begin{tabular}{
    l@{}@{\extracolsep{0.15mm}}
    H
    H
    H
    S[table-format=1.2]@{\extracolsep{0.01mm}}
    S[table-format=2.1]@{\extracolsep{0.0in}}
    S[table-format=2.1]@{\extracolsep{0.00in}}
    }
    \toprule
    \multirow{1}{0pt}{\begin{minipage}{0pt}{Dataset}\end{minipage}}
                                                                                         &      & \multicolumn{5}{c}{\textit{ VCTK test}}                                                                                                                                                                                                 \\
    \midrule
    \multirow{2}{0pt}{\begin{minipage}{0pt}{\vspace{1mm}Method}\end{minipage}} &      &                                         & \multicolumn{3}{c}{ \hspace{-2mm} Privacy} & \multicolumn{1}{c}{ \hspace{-2.5mm} Utility}                                                                                                     \\
                                                                                         &      &                                         &                                            & \multicolumn{1}{c}{\acrshort{li}~$\downarrow$} & \multicolumn{1}{c}{\acrshort{eer}~$\uparrow$} & \multicolumn{1}{c}{\acrshort{wer}~$\downarrow$} \\
    \midrule
    Clear speech                                                                         & 0.93 & 4.1                                     & 4.1                                        & 0.93                                           & 2.7                                           & 12.8                                            \\
    \midrule

    Anonymized \say{Constant speaker}                                                    & 0.67 & 13.5                                    & 5.1                                        & 0.49                                           & 20.6                                          & 13.0                                            \\
    Anonymized \say{Farther 200 random 100}                                              & 0.78 & 8.3                                     & 4.4                                        & 0.73                                           & 10.2                                          & 10.7                                            \\
    \bottomrule
  \end{tabular}
  \locallabel{table:vctk_f_c_diff}
\end{table}

Finally, in this section, we confirmed that the anonymized data used to train the \acrshort{asv} linkability attack matters.
More specifically, we showed that not all target selection algorithms generate training anonymized data in a way that makes the {semi-informed} \acrshort{asv} system learn a linkability attack, in our experiment, the first version of the \say{dense} algorithm was one of them.

All of those pitfalls are common in today's speaker anonymization evaluation framework.
The study of x-vector generation for anonymization draws many interests and is constantly evolving with new methods to generate target x-vectors, see Section \ref{main:chapt3_xvector_modef}.
However, to help in making mechanistic conclusions about privacy performance, we recommend using the target selection strategies that have well-defined behavior such as \say{random speaker} or \say{constant speaker} at the \textit{utterance-level}.
We encourage target speaker strategy developers to make sure their strategy has the correct unlinkability criterion and ability to train \acrshort{asv} attacker.
We have found that plotting the target-level genuine/impostor distributions greatly helps.
The most important takeaway of this section is that the privacy evaluation of speaker anonymization requires assessing its unlinkability which is an indirect evaluation, a system could very well completely anonymize speech (that is remove \acrshort{pii}) but still be linkable to a pseudo-identity.

In this thesis, we focused the rest of our work on the \say{constant speaker} target selection strategy, as it is the most simplistic one, and allows us to isolate this variable to study it.
Interestingly enough, when using the \say{constant speaker} strategy, the attacker capability is upgraded from a gray-box {semi-informed} attacker to a white-box {informed} attacker, as the attacker has complete knowledge of the voice conversion parameters used to anonymize each utterance (see definition in Section \ref{main:chapt_3:image:black-gray-white_box}).
In practice, this section showed that a well-trained {semi-informed} attacker measures with the same capability the strength of anonymization as the {informed} attacker (when the target selection algorithm is appropriate).
However, a point can be made about the fact that it is possible to poorly train the {semi-informed} attacker, whereas, in our experiment, the {informed} attacker is always well-trained.
%


\vspace{\fill}
\pagebreak 

\section{A quest for the golden target speaker} \locallabel{sec:radar}
\vspace{-2mm}
In contrast to the previous section where analysis was mainly conducted at the dataset level.
In this section, we are much more interested to analyze the effect of the target speaker on a per-speaker basis.
The question that we are aiming to answer is the following:
\say{Is there a source \& target speaker combination that maximizes privacy and utility performance?}.
If this is to be the case, speaker anonymization could be personalized for each user.
(Note that the attacker could use this as an attack vector).
By trying to answer this question we will also answer the following question:
\say{Is there a target speaker that maximizes privacy and utility performance for everyone?}.

In order to study the influence of the target speaker, we need to isolate this variable in order to study it.
As such, the only target selection strategy that allows specifying the target speaker parameter is the \say{constant speaker} target selection algorithm.
Hence, anonymization will be performed by modifying all utterances of all speakers to sound like a single target speaker.

Experiments are performed multiple times with different target speaker identities to provide averaged global results, and detailed analysis of the target effect on multiple source speakers.

\vspace{-3mm}
\subsection{Experimental setup}
\vspace{-2mm}

For this experiment, we run the anonymization and evaluation 40 times, each having a specific target speaker.
To cover as best as possible the target speaker space, we identify 20 female and 20 male clusters in the \acrshort{vpc} speaker x-vector pool (\textit{LibriTTS~train-other-500}) using K-Means.
We then pick the speaker x-vector that is the closest to the centroid of each cluster.
The performance assessment is carried on for each of those~40 target speakers.
The following procedure methodology is used:
first, considering each of the 40 target speakers, we anonymize all utterances of \textit{Librispeech test-clean} (compromised and vulnerable speech) and \textit{Librispeech train-clean-360} using the \say{constant speaker} target selection strategy
using the x-vector-based anonymization system (Section \ref{main:chapt_3:anon_sys_x-vector}) trained with the dataset of the \acrshort{vpc} (Table \ref{main:chapt_2:dataset}).
In contrast to the previous experiment where we obtained \acrshort{plda} scores for each possible pair of one vulnerable utterance and one compromised utterance, here we follow the \acrshort{vpc} evaluation method
where the x-vectors of an enrollment speaker are averaged over his/her utterances before \acrshort{plda} scoring.
As such the results presented here slightly differ from the previous sections.
Then, we train the \acrshort{asv} model on the anonymized \textit{Librispeech train-clean-360} dataset to comply with the white-box {informed} attacker.
Lastly, we evaluate the privacy performance for each of the speakers of \textit{Librispeech test-clean} using the specially trained \acrshort{asv} attacker.
As a result, we obtain \acrshort{eer} and \acrshort{li} results for each compromised speaker and target x-vector, that is a total of $29\times40$ results.
In this section, the primary metric used to draw conclusions is the \acrshort{li} metric.
To evaluate the quality of the conversion process in terms of utility, we use the pre-trained (non-adapted) \acrshort{asr} model released by the \acrshort{vpc} 2020 organizers.

In the following sections, we start by presenting the privacy metric at the dataset level, much like previously.
This provides the average privacy performances for all the speakers of the \textit{Librispeech test-clean} dataset for all the target x-vectors.
Then we present privacy performances for a chosen clear speaker, which gives a more fine-grained insight into the role of the target x-vector.
We finish, by classifying the population of the compromised speakers into two groups, depending on their privacy performance.

\vspace{-3mm}
\subsection{Global privacy results}
\vspace{-2mm}

Table \localref{table:all_eer_dsys} compares the anonymization performance on a global scale.
The first line presents the linkability when no anonymization is performed (i.e., on clear speech data).
Clear speech encapsulates the speaker's information to a high degree as the \acrshort{li} scores are very high~$>$~0.90.
For clear speech, a disparity between male and female linkability performance is observed.
Females seem to be harder to recognize.
This observation may be a result of the training dataset having more male variety or the inherent differences in pitch and spectral content of female and male speech, which may make female speech more difficult to differentiate from one another.

\begin{table}[!ht]
\vspace{-0.4cm}
  \caption{Linkability (\acrshort{li}) and \acrshort{eer} scores
    for clear and anonymized speech. For~the anonymized data, the mean and standard deviation values are calculated over the 40 experiments (i.e., one for each target speaker).
  }
  \vspace{0.1cm}
  \begin{tabular}{
    l@{\hskip 0.2in}@{\extracolsep{0.1in}}
    S[table-format=1.2]@{\,\( \pm \)\,}@{\extracolsep{0.02in}}
    S[table-format=1.2]@{\extracolsep{0.1in}}
    S[table-format=2.1]@{\,\( \pm \)\,}@{\extracolsep{0.02in}}
    S[table-format=1.1]@{\extracolsep{0.25in}}
    S[table-format=1.2]@{\,\( \pm \)\,}@{\extracolsep{0.02in}}
    S[table-format=1.2]@{\extracolsep{0.1in}}
    S[table-format=2.1]@{\,\( \pm \)\,}@{\extracolsep{0.02in}}
    S[table-format=1.1]@{\extracolsep{0.25in}}
    S[table-format=1.2]@{\,\( \pm \)\,}@{\extracolsep{0.02in}}
    S[table-format=1.2]@{\extracolsep{0.1in}}
    S[table-format=2.1]@{\,\( \pm \)\,}@{\extracolsep{0.02in}}
    S[table-format=1.1]@{\extracolsep{0.3in}}
    }
    \toprule
               & \multicolumn{4}{c}{Female speakers}            & \multicolumn{4}{c}{Male speakers}             & \multicolumn{4}{c}{Average}                                                                                                                                                                                                                                                                                        \\[0.05in]
               & \multicolumn{2}{c}{\acrshort{li}~$\downarrow$} & \multicolumn{2}{c}{\acrshort{eer}~$\uparrow$} & \multicolumn{2}{c}{\acrshort{li}~$\downarrow$} & \multicolumn{2}{c}{\acrshort{eer}~$\uparrow$} & \multicolumn{2}{c}{\acrshort{li}~$\downarrow$} & \multicolumn{2}{c}{\acrshort{eer}~$\downarrow$}                                                                                                                  \\
    \midrule
    Clear speech      & \multicolumn{1}{c}{\hspace{-4mm}0.90}          &                                               & \multicolumn{1}{c}{\hspace{-3mm}7.7}           &                                               & \multicolumn{1}{c}{\hspace{-4mm}0.96}          &                                                 & \multicolumn{1}{c}{\hspace{-3mm}1.1} &     & \multicolumn{1}{c}{0.93} &  & \multicolumn{1}{c}{\hspace{2mm}4.4} \\
    Anonymized & 0.74                                           & 0.01                                          & 11.6                                           & 0.6                                           & 0.75                                           & 0.01                                            & 10.6                                 & 0.8 & \multicolumn{1}{c}{0.74} &  & \multicolumn{1}{c}{11.1}            \\
    \bottomrule
  \end{tabular}
  \centering
  \locallabel{table:all_eer_dsys}
  \vspace{-0.2cm}
\end{table}

For anonymized speech, the privacy metrics come from 40 \acrshort{asv} evaluations, each using a different target speaker identity and \acrshort{asv} model.
The mean and standard deviation values are calculated from the 40 evaluations.
From the linkability score difference between clear and anonymized lines of Table \localref{table:all_eer_dsys}, we can conclude that speakers are less linkable to their true identity after applying the x-vector-base anonymization system.
Interestingly, the female and male linkability performance disparity is not present anymore after anonymization, privacy performance is the same regardless of the speaker's gender.
Linkability scores drop by 0.19, meaning the anonymization system has some effectiveness.
The rather low standard deviation values across all scores show that there is no large variation when changing the target speaker.
This suggests that a given target is not more suited than another to anonymize the whole dataset.

\vspace{-3mm}
\subsection{Detailed privacy results}
\vspace{-2mm}
We conducted a detailed analysis to check whether a specific target identity is more suited to anonymize one or more speakers of our test dataset.
Figure \localref{fig:ASV40-res}~illustrates the visualization used for this study in the case of a single source speaker.
The linkability \acrshort{li} scores are computed for a speaker whose speech was anonymized 40 times, with different target x-vectors.

\begin{figure*}[!htbp]
  \centering
  \begin{minipage}[b]{0.8\linewidth}
    \centerline{\includegraphics[width=0.6\linewidth]{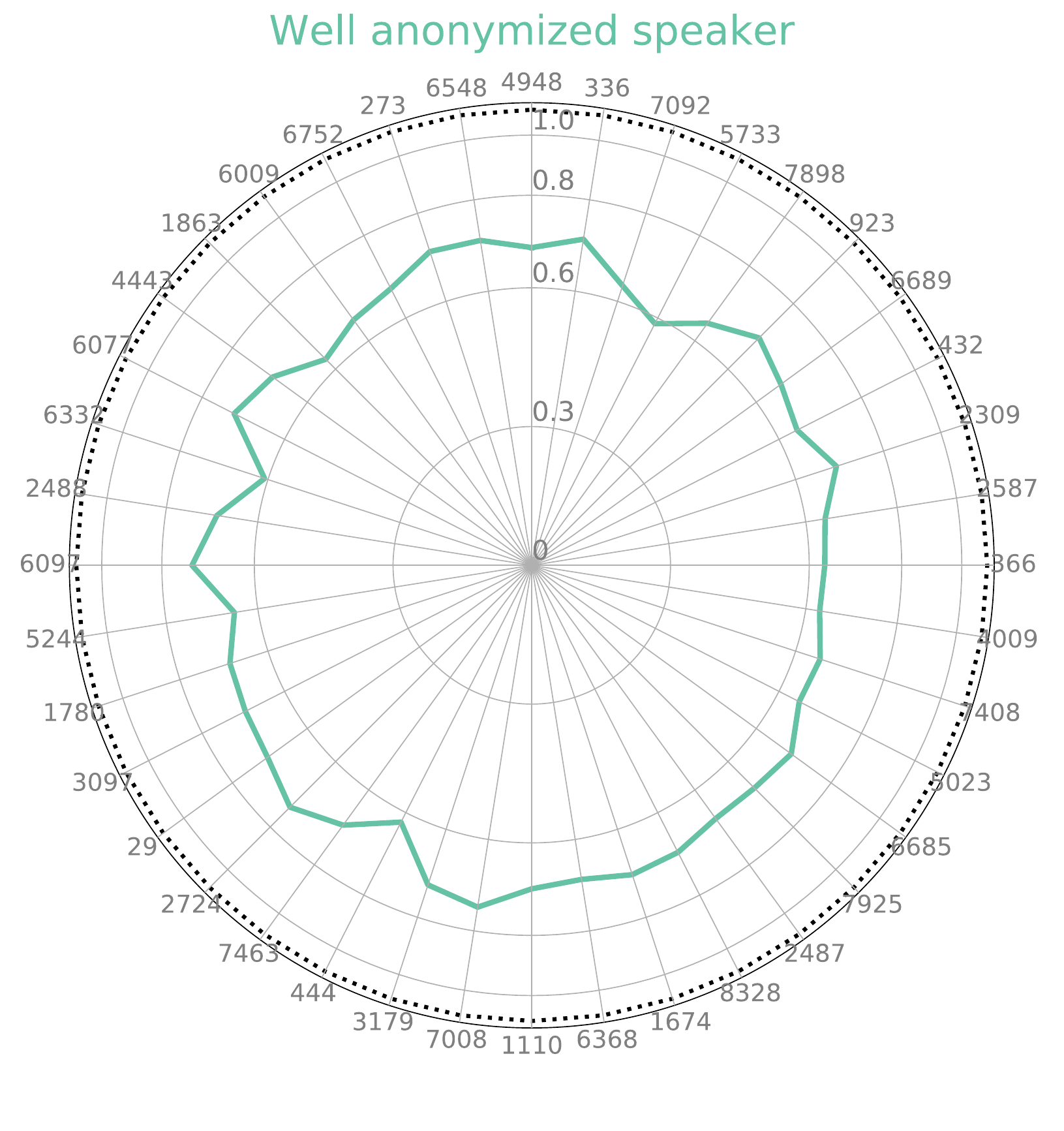}}
    \vspace{-4mm}
    \centerline{Results on a single test speaker (speaker ID 5105)}
  \end{minipage}
  \caption{Linkability scores (\acrshort{li}) obtained using 40 {informed} \acrshort{asv} attackers and baseline \acrshort{asv}, for each of the 40 target speakers (colored solid line) and on clear speech (dotted black circle).
    Each spoke corresponds to a target speaker (and \acrshort{asv} attacker model).
  }
  \locallabel{fig:ASV40-res}
  \vspace{-2mm}
\end{figure*}

The dotted black circle indicates the linkability of this speaker on clear speech (note that the clear speech evaluation does not depend on the target speakers, hence the circle). And, for each of the 40 target speakers, the linkability is presented by the colored solid line.
After transforming the speech with 40 target speakers, we can observe that none of the 40 targets are significantly better to anonymize this speaker's voice.
The variation between the anonymized linkability scores is more likely due to the difference between the \acrshort{asv} attacker model rather than a better target choice.
This observation also applies to the 29 other speakers of our test dataset.
It is also noteworthy that, out of the 40~target identities, 20 of them induce cross-gender voice conversion, as we have for the targets, 20 male centroids and 20 female centroids.
Results for same-gender and cross-gender voice anonymization were found similar.

Out of this experiment, we observed that some speakers had their privacy increased (e.g., speaker ID 5105), while others had not.
In the latter case, the colored solid line in Figure \localref{fig:ASV40-res} would be close to the dotted one.
The following figure presents this disparity.

Figure \localref{fig:ASV40-res-more} shows the linkability \acrshort{li} scores of two groups to illustrate an anonymization disparity behavior on two groups of speakers: one for which the anonymization system did not conceal \acrshort{pii}, and the other for which the anonymization did conceal some \acrshort{pii}.
Interestingly, both groups are of similar size.
For the poorly anonymized speakers, we observe that the distribution of linkability scores on anonymized speech (colored area) completely overlaps the distribution of linkability scores on clear speech (gray area).
The anonymization system did not conceal any speaker information for half of our test speakers regardless of the target speaker.
On the other hand, for the well anonymized speakers, the anonymized speech and clear speech score distributions diverge.
The difference is distinct, speaker information was removed by the anonymization system for the other half of our test speakers.
Similarly, as above, it does not appear that a specific target speaker is capable of better anonymizing clear speech from the poorly and well anonymized population.
However, we note that target speakers IDs 2487, 7463 and 6332 seem to do a slightly better job as the ranges of one standard deviation around the mean, represented by the gray and colored areas of the well anonymized speakers, do not overlap in Figure \localref{fig:ASV40-res-more}.

\begin{figure*}[!htbp]
  \centering
  \begin{minipage}[b]{0.45\linewidth}
    \centering
    \centerline{\includegraphics[width=1.0\linewidth]{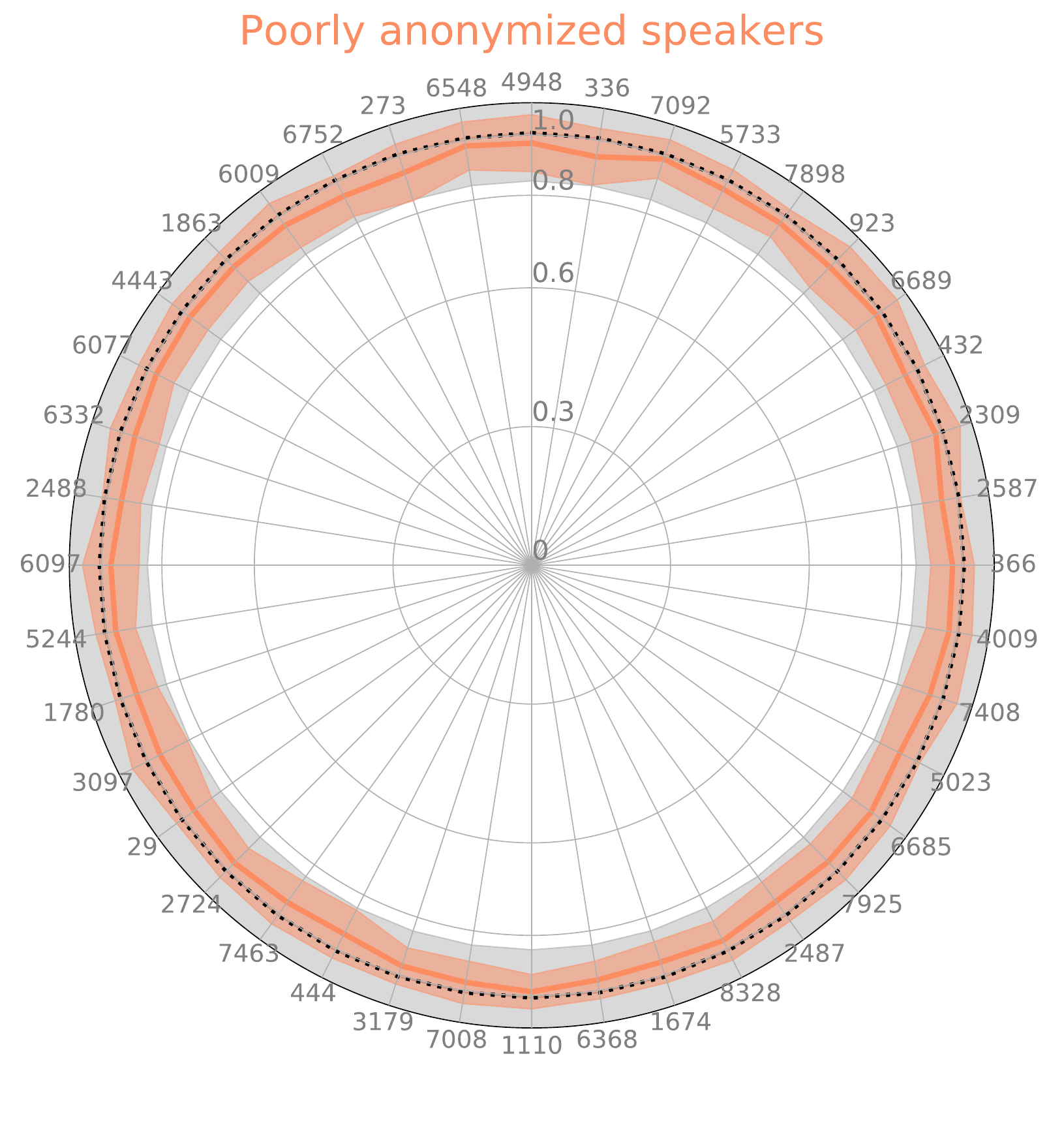}}
     \vspace{-5mm}
    \centerline{Statistics from $N=15$ speakers}
  \end{minipage}
  \hspace{0.02\linewidth}
  \begin{minipage}[b]{0.45\linewidth}
    \centering
    \centerline{\includegraphics[width=1.0\linewidth]{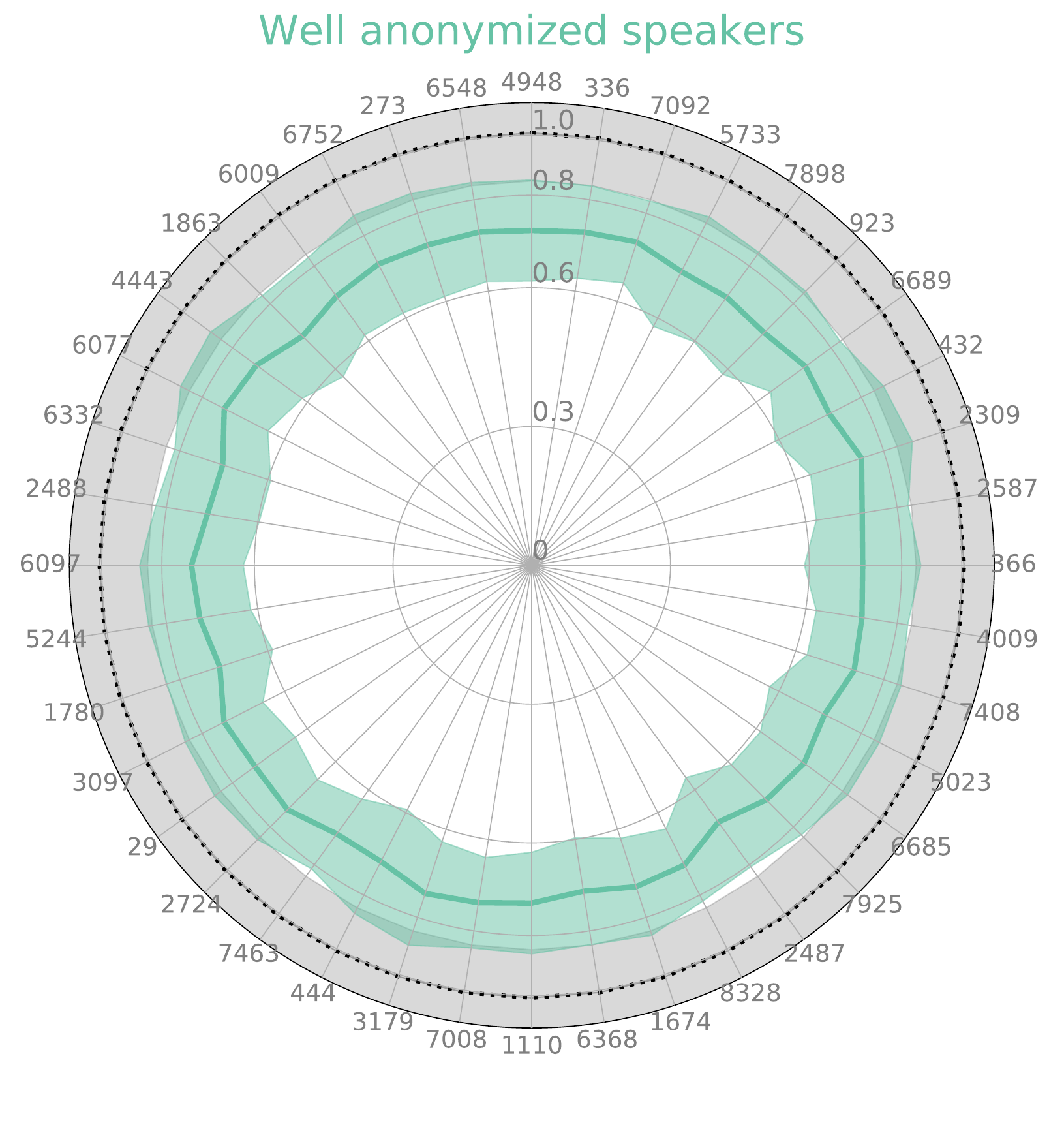}}
     \vspace{-4mm}
    \centerline{Statistics from $N=14$ speakers}
  \end{minipage}
  \caption{Mean and standard deviation \acrshort{li} scores obtained for $N$ speakers on anonymized speech for each of the 40 target speakers, and the baseline \acrshort{asv} model for clear speech.
    On clear speech, the mean value for $N$ speakers is displayed as the dotted black circle and the standard deviation as the corresponding gray area.
    Whereas for anonymized speech, the mean value for $N$ speakers, for each attacker/target of the circular axis, is displayed as the colored solid line with the standard deviation as the corresponding colored area.
  }
  \locallabel{fig:ASV40-res-more}
   \vspace{-4mm}
\end{figure*}

\vspace{-4mm}
\subsection{Utility results}
\vspace{-2mm}

\begin{figure*}[!htbp]
  \centering
  \centering
  \vspace{-1em}
  \centerline{\includegraphics[width=1.1\linewidth,trim=0 50 0 120,clip]{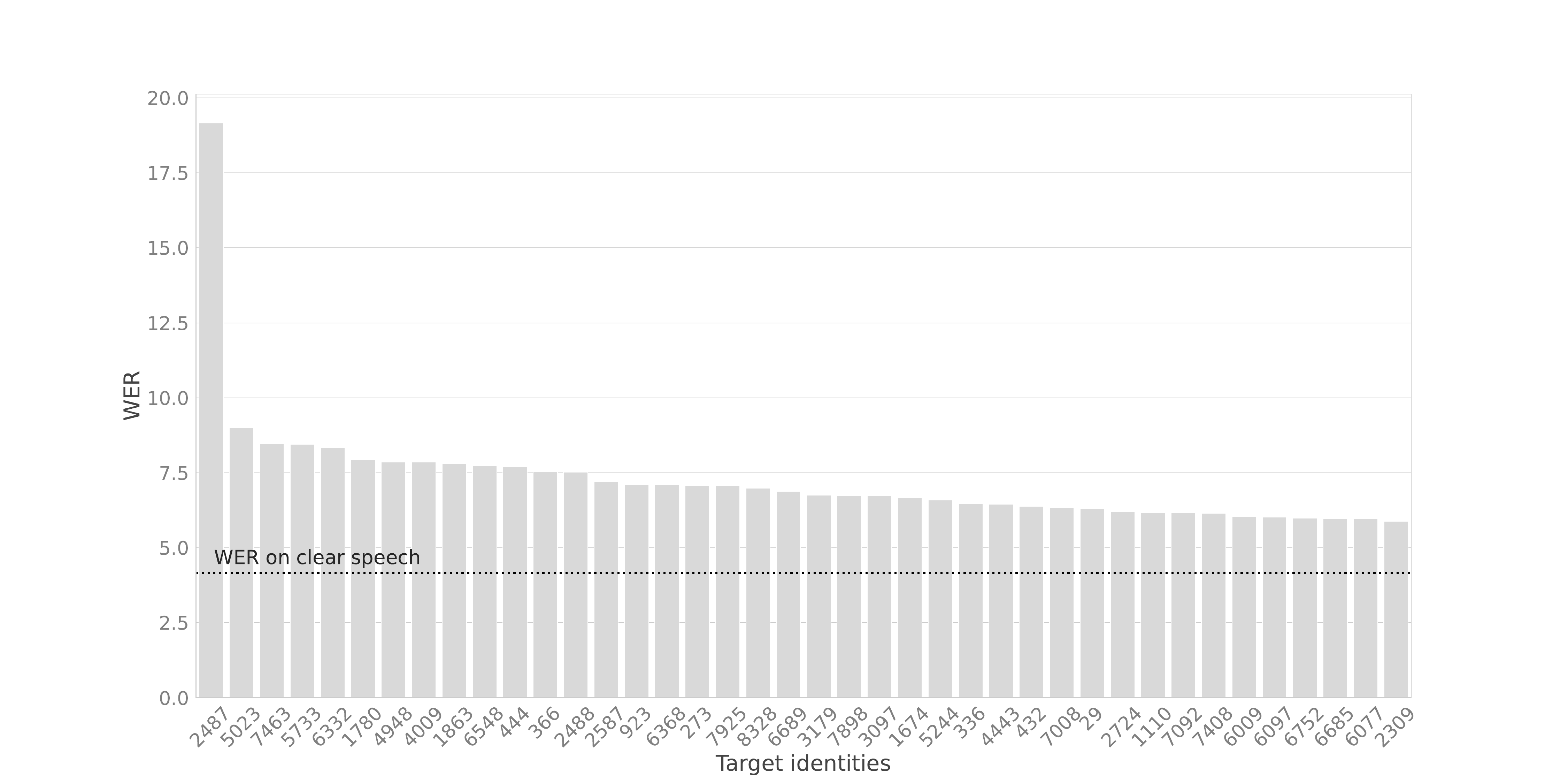}}
  \caption{\acrshort{wer} scores obtained by the VoicePrivacy \acrshort{asr} evaluation system, trained once on clear speech, for each of the 40 target speakers and on clear speech (dotted black line).}
  \locallabel{fig:ASR40-res}
  \vspace{-4mm}
\end{figure*}

Across all experiments, we evaluated the utility for each of the 40 target speakers.
We performed the \acrshort{vpc} objective intelligibility test with the pre-trained \acrshort{asr} system (see Section~\ref{main:chap2:utility_model_metric}).
Figure~\localref{fig:ASR40-res} shows the utility scores after anonymization with each of the 40 target identities.
On clear speech signal, the \acrshort{asr} systems score 4.15\% of \acrshort{wer}.
When using the same model (trained on clean data) across all 40 experiments the overall average \acrshort{wer} on the anonymized data reaches 7.30\% of \acrshort{wer}\footnote{Retraining \acrshort{asr} systems on anonymized speech improves the \acrshort{wer} significantly \cite{tomashenko2020voiceprivacy_eval2022}, however, to show an overall trend, retrain \acrshort{asr} systems is unnecessary and very expensive.}.

The very high utility loss yielded when using target speaker ID 2487 is due to a generalization issue of the anonymization pipeline, not the \acrshort{asr} model, in this case, non-intelligible speech was generated at the beginning of some segments.
We conducted an additional test using 100 randomly selected target speakers from the pool, and were able to find 3 target speaker x-vectors that have similar behavior, in the worse case the \acrshort{wer} reached a value of 59.37.
Informal listening tests reported that the target speaker's speech contained singing segments, producing faulty x-vectors. Further analysis needs to be conducted.
Interestingly, the target speakers IDs 2487, 7463 and 6332, which were slightly better in terms of privacy protection are all worse in terms of utility preservation as they are in the top 5 worse target speakers in Figure~\localref{fig:ASR40-res}.
Making mechanistic conclusions about privacy/utility trade-offs relative to the target speaker parameter is challenging due to the dependency on the speech synthesis system.
However, we observed that some target x-vectors are better suited to generate intelligible speech than others.

In the previous section, we outline that using the \say{constant speaker} target selection strategy was less effective in terms of utility compared to the \say{Farther 200 random 100} target selection strategy (see Tables~\localref{table:vctk_f_c_diff} and \localref{tab:pool-training}).
The hypothesis explaining this disparity might be a generalization issue of the speech synthesis against unseen (during training) x-vector, the average operation performed in the \say{Farther 200 random 100} seems to mitigate this issue.
Backing up this argumentation, in \cite{dp_vpc}, the authors identified that using too few speakers before the averaging in the \say{dense} strategy negatively affects utility.
The average of many target speakers appears to decrease the specificity of the x-vector, making it a better-suited target that preserves utility.
From Table~\localref{tab:pool-training}, the same conclusion is made.
The \acrshort{wer} is lower for the \textit{\say{Farther 200 random 100} from external pool} compared to the \textit{\say{Constant speaker} from external pool} for both \textit{LibriSpeech} and \textit{VCTK} datasets.
To obtain the same \acrshort{wer} for the \say{Constant speaker}, we suggest sampling the target speaker from the dataset used to train the speech synthesis model rather than an external one.
Table~\localref{tab:pool-training} shows that performing this simple modification, yields the same \acrshort{wer} for the \say{Constant speaker} and \say{Farther 200 random 100}.
One may argue that the \say{Constant speaker} strategy, using a real speaker's voice as the target, raises ethical concerns, compared to more random target selection strategies, however, there is no guarantee that the \say{Farther 200 random 100} does not also generate speech similar to the one of a real identity.
Overall, for research anonymization evaluation, the ethical concerns are nonexistent, whereas, in real applications the use of a pitch morphing algorithm (that most TVs/media already use) can be applied on top of the anonymized speech generated by the pipeline, to address any eventual ethical concern.

\begin{table}[!ht]
  \vspace{-0.3cm}
  \caption{Privacy and utility performance metrics of target speaker selection strategies. \acrshort{asv} and \acrshort{asr} models are trained on anonymized speech. Refer to Table~\ref{main:chapt_2:dataset} for the dataset/pool. }
  \vspace{0.1cm}
  \resizebox{1.00\textwidth}{!}{
    \begin{tabular}{
      l@{}@{\extracolsep{0.15mm}}
      S[table-format=1.2]@{\extracolsep{0.01mm}}
      S[table-format=2.1]@{\extracolsep{0.0in}}
      S[table-format=2.1]@{\extracolsep{3.0mm}}
      S[table-format=1.2]@{\extracolsep{0.01mm}}
      S[table-format=2.1]@{\extracolsep{0.0in}}
      S[table-format=2.1]@{\extracolsep{0.00in}}
      }
      \toprule
      \multirow{1}{0pt}{\begin{minipage}{0pt}{Dataset}\end{minipage}}
                                                                                            & \multicolumn{3}{c}{  LibriSpeech test-clean}
                                                                                            & \multicolumn{3}{c}{ \textit{VCTK test}}                                                                                                                                \\
      \midrule
      \multirow{2}{0pt}{\begin{minipage}{0pt}{\vspace{1mm}Method}\end{minipage}} & \multicolumn{2}{c}{ \hspace{-2mm} Privacy}     & \multicolumn{1}{c}{ \hspace{-2.5mm} Utility}
                                                                                            & \multicolumn{2}{c}{ \hspace{-2mm} Privacy}     & \multicolumn{1}{c}{ \hspace{-2.5mm} Utility}                                                                          \\
                                                                                            & \multicolumn{1}{c}{\acrshort{li}~$\downarrow$} & \multicolumn{1}{c}{\acrshort{eer}~$ \uparrow$} & \multicolumn{1}{c}{\acrshort{wer}~$\downarrow$}
                                                                                            & \multicolumn{1}{c}{\acrshort{li}~$\downarrow$} & \multicolumn{1}{c}{\acrshort{eer}~$ \uparrow$} & \multicolumn{1}{c}{\acrshort{wer}~$\downarrow$}                      \\
      \midrule
      Clean speech                                                                          & 0.93                                           & 4.1                                            & 4.1                                             & 0.93 & 2.7  & 12.8 \\
      \midrule
      \say{Farther 200 random 100} from external pool                                       & 0.78                                           & 8.3                                            & 4.4                                             & 0.73 & 10.2 & 10.7 \\
      \say{Constant speaker} from external pool                                             & 0.67                                           & 13.5                                           & 5.1                                             & 0.49 & 20.6 & 13.0 \\
      \say{Constant speaker} from training dataset                                          & 0.73                                           & 11.2                                           & 4.4                                             & 0.52 & 19.8 & 10.2 \\
      \bottomrule
    \end{tabular}
  }
  \locallabel{tab:pool-training}
  \vspace{-4mm}
\end{table}

\vspace{-3mm}
\subsection{Discussion}
\vspace{-1mm}

The conclusions presented in this section are multifaceted.
To answer the questions,
\say{Is there a source \& target speaker combination that maximizes privacy and utility performance?},
and \say{Is there a target speaker that maximizes privacy and utility performance for everyone?}.
%
In our experiment, privacy performance was not significantly increased by a particular choice of a target speaker.
There is no source \& target speaker combination that maximizes the privacy, as well as a target speaker that better conceals \acrshort{pii} for everyone.
We conclude that the linkability \acrshort{asv} {informed} attacker was robust to different target speakers.
As for utility, our experiment shows that there are target speakers' identities that better preserve objective intelligibility measured by \acrshort{asr} systems.

Hence, we further analyzed the \say{constant speaker} target selection strategy to optimize its utility performance.
We observed that using target speakers that were seen during the training of the synthesis system, generated anonymized speech better preserving the utility.
As such, the main conclusion about the influence of the target speaker relates to utility, not privacy.
We believe that using the \say{constant speaker} and cherry-picking a target speaker from the synthesis training dataset in such a way that the anonymized voice produced has good intelligibility is the most effective and secure method to maximize the utility.
This form of \acrshort{vc} is referred as to \textit{any-to-many} \acrshort{vc}.
As presented in Section \ref{main:any-many-one-vc}, \textit{any-to-many} \acrshort{vc} approaches cannot target \textit{any} target speaker, only the \textit{many} ones seen during training can be used as target speaker, this is not a problem for the \textit{\say{constant speaker} from training dataset} target selection strategy but rather an advantage.
As in the voice conversion field, it is well known that the \textit{any-to-many} \acrshort{vc} approaches are simpler than the \textit{any-to-any} one, in terms of designing and training, additionally, they usually produce better speech quality.
By using \textit{any-to-many} \acrshort{vc}, the use of x-vector as the encoding medium of the target speaker is no longer justified.
A one-hot encoding can very well be used to target \textit{many} speakers, this simplifies the target selection procedure as the x-vector extractor and pool of x-vectors are no longer necessary.

In this section, large privacy performance disparities between different speakers of the test dataset are exhibited, indicating an issue about fairness!
Tracing the cause of these disparities is a topic of research that has never been explored, however, we believe it is an essential topic.
One first step in this research could be to identify how individuals behave differently regarding linkability attacks on their anonymized speech.
The concept of the biometric menagerie in the literature formally recognizes the idea of categorizing and labeling user groups with animal names based on their characteristics when interacting with biometric systems \cite{menagerie_Doddington1998SHEEPGL,menagerie++}.
Extending the menagerie to the field of speaker anonymization could be interesting.
The privacy performance gaps between speakers also raise the question of the use of speaker average-based metrics such as the \acrshort{li} and \acrshort{eer} metrics.
While not used in this thesis, the ZEBRA \cite{zebra} metric seems to be a compelling solution to better evaluate privacy in worst-case privacy disclosure.

\vspace{-1.2em}
\section{Conclusion}
\vspace{-3mm}

In this chapter, we have challenged the target speaker attribute of voice conversion-based speaker anonymization.
We found that existing target selection strategies have two major issues: 1) no target-level unlinkability guarantees that compromise privacy evaluation and 2) potential difficulties to generate appropriate training data for the attacker.
Additionally, in our experiment, the target selection algorithm and target speaker do not affect the strength of \acrshort{pii} concealment, there is no golden target speaker for privacy.
However, in our experiment, we showed that there are multiple golden and bad target speakers when the utility is considered.
For those reasons, we promote the use of the \textit{\say{constant speaker} from training dataset} target selection strategy which provides many characteristics: 1) target-level unlinkability guarantees, 2) appropriate data generation for attacker training because of a 3) white-box {informed} attacker scenario, 4) cherry-picking targets that preserve utility, 5) simplicity and straightforwardness.
Ensuring proper privacy evaluation in \acrshort{vc}-based anonymization is a challenge.
And we conclude that the main role of the target speaker in \acrshort{vc}-based anonymization is firstly to ensure proper evaluation.

\ifSubfilesClassLoaded{
  \printglossary[title=Special Terms,type=\acronymtype]
  \printbibliography
}{}

\end{document}

\clearemptydoublepage
\cleartooddpage[\thispagestyle{empty}]
\documentclass[../main.tex]{subfiles}

\ifSubfilesClassLoaded{
    \tableofcontentsfile
    \dominitoc
    \setcounter{chapter}{4} 
    \def\locallabelprefix{chapt_5}
    \externaldocument[]{../main}
}{}

\begin{document}

\selectlanguage{english}

\graphicspath{{./figures/dist}}

\chapter{Voice conversion anonymization with feature-level disentanglement} \locallabel{chapt:anon_me} \locallabel{chapt5}
\vspace{-4.5em}
\minitoc
\vspace{-1.2em}
\section{Introduction}
\vspace{-0.8em}
Speaker anonymization aims to transform a speech signal to remove the source speaker's identity while leaving the spoken content unchanged and potentially other information.
Our baseline, the x-vector-based method performs the transformation by relying on the disentanglement of linguistic and F0 information from speaker information and voice conversion.
Usually, an acoustic model from an automatic speech recognition system extracts the linguistic representation while an x-vector system extracts the speaker representation.
Additionally, the fundamental frequency is also extracted for intonation preservation.
In this chapter, we identify the degree to which the linguistic and \acrshort{f0} features are disentangled from the speaker, and propose modification methods to improve the disentanglement, and thus the privacy of the converted anonymized speech.

\section{\texorpdfstring{Fundamental frequency}{Fundamental frequency} feature transformation} \locallabel{f0_modif}

Previously, under the x-vector-based methods, we have assumed that only the original x-vector extracted by the anonymization framework would be transformed into the target speaker x-vector while the other two sets of features (\acrshort{asrbn} linguistic representation and \acrshort{f0} trajectory) would remain unchanged.
However, intonation features contribute to speaker identity and the \acrshort{f0} trajectory contains some \acrfull{pii}, in \cite{f0_men_women} the authors were capable to distinguish female and male only with the \acrshort{f0}.

Additionally, maintaining the original \acrshort{f0} while potentially changing the gender of the x-vector can result in inconsistent features and affect the naturalness of synthesized speech.
Multiple works have investigated \acrshort{f0} conditioned voice conversion \cite{F0_Huang2019InvestigationOF,F0_Qian2020F0ConsistentMN} (see Section~\ref{main:chapt_2:f0_cond}) and concluded that \acrshort{f0} conversion effectively improves the naturalness of the output speech.
Motivated by those results, we propose to modify the \acrshort{f0} values of a source utterance (cf. module D in Figure \localref{image_chapt5:vpc_model_f0}) in the x-vector-based voice conversion system to see if it also improve privacy.

\begin{figure}[htbp]
  \begin{center}
    \includegraphics[width=0.84\linewidth]{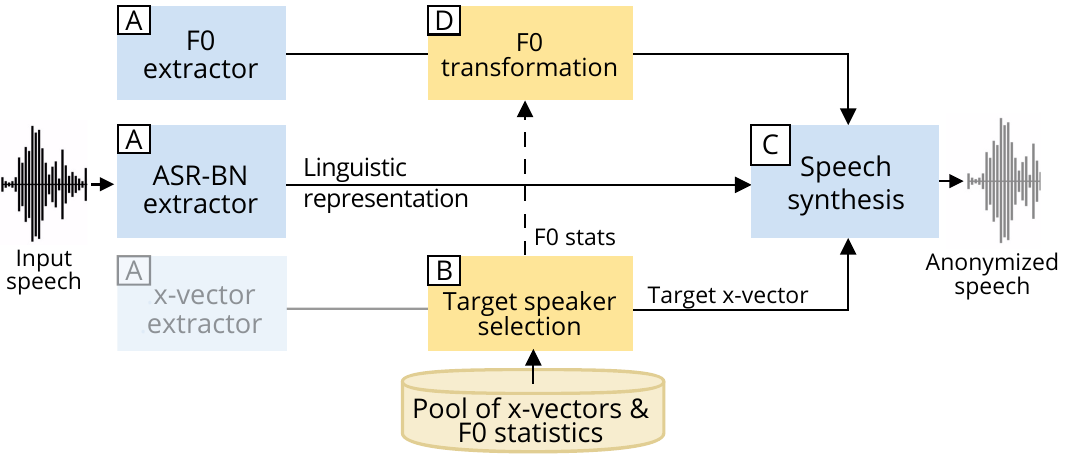}
  \end{center}
  \vspace{-5mm}
  \caption{
    X-vector-based speaker anonymization system with module $D$ added.
    The original x-vector of the input utterance might, or might not be used by the target selection strategy.
  }
  \locallabel{image_chapt5:vpc_model_f0}
\end{figure}

\subsection{Linear shift transformation} \locallabel{f0_shift_tran}
To convert the source \acrshort{f0} trajectory to be similar to the target speaker, we propose to use a simple linear transformation:
\begin{equation}
  \hat{f}_{t}=\mu_{y}+\frac{\sigma_{y}}{\sigma_{x}}\left(f_{t}-\mu_{x}\right)
\end{equation}
where $f_{t}$ represents the log-scaled \acrshort{f0} of the source speaker at frame $t$, $\mu_{x}$ and $\sigma_{x}$ represent the mean and standard deviation of the log-scaled \acrshort{f0} for the source speaker.
$\mu_{y}$ and $\sigma_{y}$ represent the mean and standard deviation of the log-scaled \acrshort{f0} for the target-speaker.
The linear transformation and statistical calculation are only performed on voiced frames.
When the target speaker x-vector is a combination of multiple speakers like in the \say{Farther 200 random 100} target selection strategy, the mean and standard deviation for the target speaker is calculated by averaging information from the same 100 speakers selected to derive the target x-vector.
For the target selection strategies, where a unique known speaker is selected as the target x-vector like in the \say{Constant speaker} strategy, the \acrshort{f0} statistics come from the same selected speaker.
These statistics are first extracted and stored for the pool of speakers when developing the pipeline and passed to the \acrshort{f0} conversion module (module $D$ in Figure~\localref{image_chapt5:vpc_model_f0}) before speech synthesis.

\subsection{Additive white Gaussian noise transformation} \locallabel{chapt5:awgn}
Inspired by the work of \cite{F0_trajectories_unal,dp_vpc}, we also experiment with noise based \acrshort{f0} modification.
Here the goal is not to improve naturalness as the noise will disturb the \acrshort{f0}, however, it will increase privacy.
Noise addition is one of the most used data privacy protection techniques.
When noise is added to data, it introduces random variations that can make it more difficult for an attacker to extract sensitive information from the data.
This is because the added noise can obscure or mask the underlying patterns in the data that might reveal sensitive information.
Work on additive noise was first published in \cite{kim1986method} with a method named \acrfull{awgn}.
Noise is additive because it is added to any noise that might be intrinsic to the information system, white because it has uniform power across the frequency band for the information system (an analogy to the color white which has uniform emissions at all frequencies in the visible spectrum), and Gaussian because it has a normal distribution in the time domain with an average time domain value of zero.
For each frame ${f}_{t}$, and a target noise power $D$ in dB, the transformation can be written as:
\vspace{-.6em}
\begin{equation}
  \hat{f}_{t}=f_{t} + \mathcal{N}(0, \sqrt{10^{(\text{D}/10)}})
\vspace{-0.8em}
\end{equation}

\subsection{Quantization transformation} \locallabel{chapt5:f0_vq}
Quantization is a technique for representing a continuous signal as a discrete set of values.
It works by mapping a set of input values to a set of output values, with each input value being replaced by the nearest output value from the set.
The quantization method that we propose to use to modify the \acrshort{f0} is the simple linear quantization method that reduces the values used to represent a signal using the equation described in \cite{Oppenheim1999DiscretetimeSP}:
\begin{equation}
  \hat{f}_{t} =  \operatorname{round} \left( 2^{B-1} \frac{f_{t} - f_{min}}{f_{max} - f_{min}} \right)
\end{equation}
where $B$ is the number of quantization bits.
In contrast to \acrshort{awgn} transformation, which adds random noise, quantization smooths out variations in the signal.
This can result in the same output property as the \acrshort{awgn} modification that makes attackers extract less sensitive information.
%

However, it is worth noting, that quantization is not a foolproof method for anonymization and has seen less research interest in contrast to noise-based anonymization.
It may be possible for an attacker to extract some speaker-specific information from the quantized \acrshort{f0} signal.
To the best of our knowledge, only the work of \cite{vq_anon} specifically used quantization methods for data anonymization, whereas much more work has been done with noise-based anonymization methods.
We consider this framework to be a good option for voice anonymization because the transformed speech using quantization-based methods will have less audible noise than speech transformed using noise-based methods.
By minimizing the perceptible noise in anonymized speech, its usefulness can be improved.

\subsection{Wrapping up \texorpdfstring{\acrshort{f0}}{F0} transformations} \locallabel{warppup_f0_trans}

\begin{figure*}[!hbt]
  \centering
  \centerline{\includegraphics[width=0.95\linewidth]{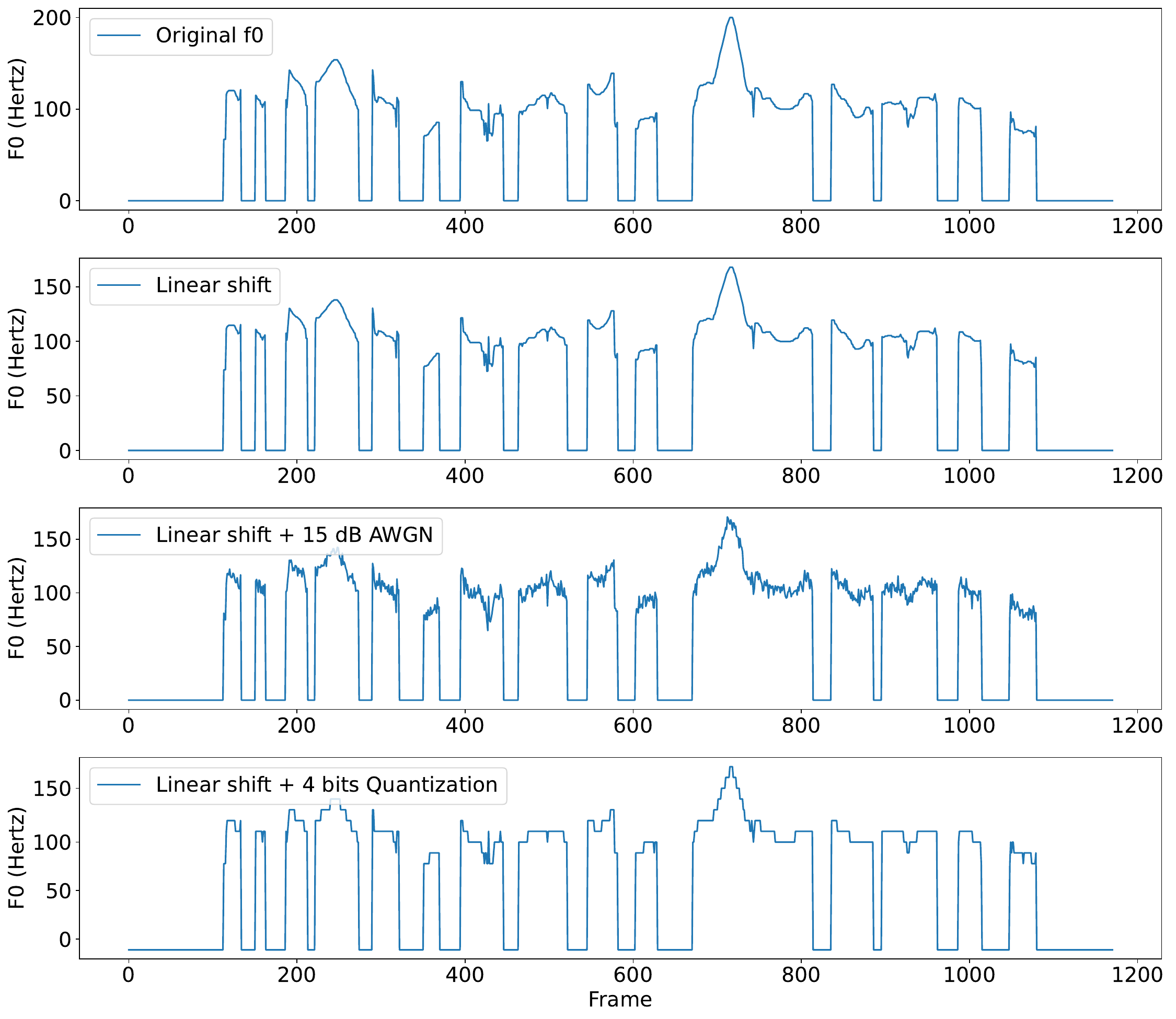}}
  \vspace{-1em}
  \caption{Example of the \acrshort{f0} transformation.
  }
  \locallabel{fig:f0_mod}
  \vspace{-0.5em}
\end{figure*}

An intuitive comparison between the types of target speaker selection strategies presented in Chapter~\ref{main:chapt4} and the \acrshort{f0} transformation presented here can be made.
The \say{random speaker} target selection strategy introduces random speaker modification for each utterance of a speaker such as it ends up being confused with other speakers.
Similarly, after shifting and scaling, the \acrshort{awgn} method,  described here has the same goal: modify the input \acrshort{f0} of a speaker such has it is more likely to be confused with other speaker's \acrshort{f0}.
In contrast, the \say{constant speaker} target selection strategy forces everyone to be modified to match a single target, aiming to reduce the number of perceived anonymized speakers to one.
Quantization works similarly, after shifting and scaling, the local change of the \acrshort{f0} trajectory that makes a \acrshort{f0} linked to an identity/group of speakers is smoothed out by the quantization process reducing the number of perceived \acrshort{f0} dependent speaker characteristics. Figure~\localref{fig:f0_mod} shows transformation examples.

%
 \locallabel{explanation:vq_noise}
  Noise-based anonymization works by adding noise in a transmission channel such that only the most relevant information stands out from the noise, whereas quantization-based anonymization works by reducing the transmission channel capacity such that the irrelevant information is not encoded.



\subsection{Experimental setup}
\vspace{-0.3em}

The baseline anonymization system is the x-vector-based provided by the \acrlong{vpc}.
With this system, two sets of experiments are performed in this section.
One set that follows all the \acrlong{vpc} 2022 requirements with the \say{farther 200 random 100} target selection strategy that allows some voice distinctiveness.
For those experiments, the gray-box {semi-informed} attacker is used to evaluate the privacy protection of the anonymization.
The other set of experiments does not comply with the voice distinctiveness requirement as it uses the \say{constant speaker} target selection strategy.
As shown in Chapter~\ref{main:chapt4}, with the \say{constant speaker} strategy, a superior white-box evaluation with better guarantees can be used to evaluate privacy protection.
Utility evaluation is done with an \acrshort{asr} model trained on anonymized speech.
The datasets used for evaluation/training/testing are the same as in both 2020 and 2022 \acrshort{vpc}.
Results are presented for the \textit{LibriSpeech test-clean} and \textit{VCTK test} datasets.
The performances are assessed in terms of \acrshort{li} and \acrshort{eer} scores for privacy and \acrshort{wer} for utility.

With the \say{farther 200 random 100} target selection strategy evaluations, the linear \acrshort{f0} transformation are evaluated against the same anonymization system that does not have it.
For the \say{constant speaker} target selection strategy evaluations, the transformations being evaluated include the linear shift, a combination of the linear shift and the \acrshort{awgn} transformation with a noise value of 15dB, and a combination of the linear shift and the quantization transformations on 4 bits.

\subsection{Experimental results}

\begin{table}[!b]
  \vspace{-1.2em}
  \caption{Privacy and utility results for the chosen target selection strategies and \acrshort{f0} transformation.    
  The interval of confidence stays within $\pm~0.40$\%~\acrshort{eer} for all experiments. Lines 2 and 4 do not involve F0 transformations.}
    \locallabel{tab:f0_modif_vpc}
  \vspace{0.2cm}
  \hspace{-1.200cm}
  \resizebox{1.14\textwidth}{!}{
    \begin{tabular}{
      l@{}@{\extracolsep{0.15mm}}
      S[table-format=1.2]@{\extracolsep{0.01mm}}
      S[table-format=2.1]@{\extracolsep{0.0in}}
      S[table-format=2.1]@{\extracolsep{3.0mm}}
      S[table-format=1.2]@{\extracolsep{0.01mm}}
      S[table-format=2.1]@{\extracolsep{0.0in}}
      S[table-format=2.1]@{\extracolsep{0.00in}}
      }
      \toprule
      \multirow{1}{0pt}{\begin{minipage}{0pt}{Dataset}\end{minipage}}
                                                                                                                                       & \multicolumn{3}{c}{  \textit{LibriSpeech test-clean}}
                                                                                                                                       & \multicolumn{3}{c}{ \textit{VCTK test}}                                                                                                                                                                                                                                                                      \\
      \midrule
      \multirow{2}{0pt}{\begin{minipage}{190pt}{\vspace{1mm}Strategy + \acrshort{f0} transformation}\end{minipage}} & \multicolumn{2}{c}{ \hspace{-2mm} Privacy}            & \multicolumn{1}{c}{ \hspace{-2.5mm} Utility}
                                                                                                                                       & \multicolumn{2}{c}{ \hspace{-2mm} Privacy cy}            & \multicolumn{1}{c}{ \hspace{-2.5mm} Utility}                                                                                                                                                                                                         \\
                                                                                                                                       & \multicolumn{1}{c}{\acrshort{li}~$\downarrow$}        & \multicolumn{1}{c}{\acrshort{eer}~$ \uparrow$} & \multicolumn{1}{c}{\acrshort{wer}~$\downarrow$} & \multicolumn{1}{c}{\acrshort{li}~$\downarrow$} & \multicolumn{1}{c}{\acrshort{eer}~$ \uparrow$} & \multicolumn{1}{c}{\acrshort{wer}~$\downarrow$} \\
      \midrule
      1. Clear speech                       & 0.93 & 4.1 & 4.1 & 0.93 & 2.7  & 12.8 \\
      \midrule
      2. Anon. \say{Farther 200 random 100}            & 0.78 & 8.3  & 4.4 & 0.73 & 10.2 & 10.7 \\
      3. Anon. \say{Farther 200 random 100} + Linear   & 0.63 & 16.5 & 4.4 & 0.66 & 13.0 & 10.4 \\
      \midrule
      4. Anon. \say{Constant speaker}                                 & 0.73 & 11.2 & 4.4 & 0.52 & 19.8 & 10.2 \\
      5. Anon. \say{Constant speaker} + Linear                     & 0.73 & 11.6 & 4.4 & 0.46 & 22.2 & 10.1 \\
      6. Anon. \say{Constant speaker} + Linear + \scshape{{awgn}}  & 0.73 & 11.2 & 4.4 & 0.47 & 21.5 & 10.2 \\
      7. Anon. \say{Constant speaker} + Linear + \scshape{{quant}} & 0.73 & 11.3 & 4.4 & 0.47 & 21.7 & 10.1 \\
      \bottomrule
    \end{tabular}
  }
\end{table}

Table~\localref{tab:f0_modif_vpc} compares the anonymization performance across \acrshort{f0} modification techniques.
The first line presents the linkability and utility of clear speech.
Clear speech encapsulates the speaker's information to a high degree as the \acrshort{li} scores are very high~$>$~0.90 for the \textit{LibriSpeech test-clean} and \textit{VCTK} datasets.
The \acrshort{wer} of this first line corresponds to our baseline utility performance, we observe that the \acrshort{wer} score for the \textit{VCTK} dataset has a higher value than on \textit{LibriSpeech}.
This utility disparity is explained by the \acrshort{asr} evaluation model which is only trained on \textit{LibriSpeech}, making it more relevant for audiobooks.


When comparing lines 2 and 3, which involve a gray-box semi-informed attacker, we observe a privacy improvement when using the linear shift \acrshort{f0} transformation for both the \textit{LibriSpeech} and \textit{VCTK} datasets as the \acrshort{li} score decrease from 0.78 to 0.63 and 0.73 to 0.66 respectively.
However, when compared to the \say{constant speaker} strategy with the white-box {informed} attacker, no privacy improvements are recorded by the linear shift \acrshort{f0} transformation on the \textit{LibriSpeech} dataset.
With those experiments, we reinforce the conclusion of Chapter~\ref{main:chapt4} by showcasing how a misunderstanding of the role of the target speaker can impact evaluation.
With the \say{farther 200 random 100} target selection strategy and the gray-box {semi-informed} attacker, one might believe that a linear shift \acrshort{f0} transformation increases the privacy performance of an anonymization system as the \acrshort{li} decreases from lines 2 and 3 of Table~\localref{tab:f0_modif_vpc}.
With proper white-box evaluation, privacy improvement is almost nonexistent (lines 4 and 5).
Due to this evaluation misunderstanding, we made inaccurate claims about the privacy improvement obtained by linear \acrshort{f0} modification in the paper: \citetitle{F0_mod_moi} \cite{F0_mod_moi}.
That is why, for the rest of this thesis, we will only use a white-box attacker with the \say{constant speaker} strategy.

Regarding the experiments with the \say{constant speaker} anonymization strategy, the results show that for all experiments, the \acrshort{wer} utility measure is the same whether the \acrshort{f0} was modified or not
As for privacy, using the \acrshort{awgn} or quantization transformations on top of the linear shift does not improve the performance compared to only using the linear shift in the \textit{VCTK} dataset as \acrshort{li} is around 0.47.
And for in the \textit{LibriSpeech} dataset, none of the transformations improves privacy.
This conclusion goes against the main hypothesis that intonation features contribute to the speaker's identity.
This disparity can be explain by the fact that the x-vector system used for finding the linkage is not heavily prosody based.
If the attacker were using a prosody based system it is more likely that we would find that the shift helps.
However, in an anonymization pipeline, privacy leakage can occur from multiple parts.
If the linguistic representation feature is the main source of speaker information leakage, modifying the \acrshort{f0} to improve \acrshort{f0}/speaker disentanglement will not result in a significative privacy performance improvement.
In the next section, we aim to measure the \acrshort{f0}/speaker disentanglement.

\subsection{\texorpdfstring{\acrshort{f0}}{F0} isolated privacy evaluation} \locallabel{isolated_privacy_eval_f0}
In this section, our focus is to measure the \acrshort{f0}/speaker disentanglement only on the \acrshort{f0} trajectory extracted from the speaker's speech, no speech transformations are applied beforehand.
By isolating the \acrshort{f0} feature alone and not using the full anonymization pipeline, we can assess and measure if any speaker information leakage occurs from the \acrshort{f0} in the pipeline.
We also put to evaluation the \acrshort{f0} transformations presented above\footnote{
The experiments presented in this section are a collaborative effort, where the experiments were done by Shalini Priya during her internship.}.

\vspace{-0.7em}
\subsubsection{Experimental setup}
\vspace{-0.5em}

The experimental setup is similar to the one in the previous experiment.
However, the \acrshort{asv} is trained to identify speakers only using the one-dimensional \acrshort{f0} features instead of the \acrshort{mfcc}.
This model is a five \acrshort{tdnn} x-vector model and is trained on the \textit{LibriSpeech train-clean-360} using the Python speaker identification sidekit tool \cite{sidekit}.
For reference, an experiment is also done with \acrshort{mfcc} as input.
When a \acrshort{f0} transformation is used, the \acrshort{asv} model is retrained to accommodate the transformations, complying with white-box attacks.
Evaluation is done on the \textit{LibriSpeech test-clean} dataset and the \acrshort{eer} metric is used to evaluate the disentanglement property of the \acrshort{f0}, a value of 50\% indicating perfect disentanglement which would provide the best downstream anonymization guarantees.

\subsubsection{Experimental results}

\vspace{-1.0em}
\begin{table}[H]
  \begin{center}
    \sisetup{
      round-mode = places, 
      round-precision = 1, 
    }
    \caption{Results in \acrshort{eer} of the privacy of the isolated \acrshort{f0} feature, with and without transformations. The confidence interval is provided for each experiment.}
    \vspace{0.2cm}
    \locallabel{tab:f0_eer}
    \begin{tabular}{lS}
      \toprule
      \multirow{2}{0pt}{\begin{minipage}{100pt}{\acrshort{f0} transformation}\end{minipage}}{}
                                                        & \multicolumn{1}{p{4cm}}{\centering \textit{LibriSpeech test-clean} \\ \acrshort{eer}~$\uparrow$ } \\
      \midrule
      {1. Clear \acrshort{mfcc}}                        & {~\num{4.3} } \textpm { \num{0.5}}                                 \\
      \midrule
      \midrule
      {2. Clear \acrshort{f0}}                          & {\num{19.6} } \textpm { \num{1.1}}                                 \\

      {3. Linear shift}                                 & {\num{19.85} } \textpm { \num{1.05}}                               \\
      {4. \scshape{{quant}$_{2bit}$}  }                 & {\num{19.2} } \textpm { \num{1.3}}                                 \\
      {5. \scshape{{awgn}$_{15dB}$}  }                  & {\num{30.898} } \textpm { \num{1.38}}                              \\
      {6. \scshape{{awgn}$_{30dB}$}  }                  & {\num{35.994} } \textpm { \num{1.456}}                             \\
      \midrule
      {7. Linear shift + \scshape{{awgn}$_{15dB}$} }    & {\num{44.227} } \textpm { \num{9.086}}                             \\
      {8. Linear shift + \scshape{{awgn}$_{30dB}$}  }   & {\num{45.984} } \textpm { \num{7.329}}                             \\
      \midrule
      {9. Linear shift + \scshape{{quant}$_{4bits}$}  } & {\num{26.7} } \textpm { \num{1.3}}                                 \\
      \bottomrule
    \end{tabular}
    \vspace{-2.0em}
  \end{center}
\end{table}

Table~\localref{tab:f0_eer} summarizes the results, and lines 1 and 2 show the \acrshort{eer} obtained with \acrshort{mfcc} and \acrshort{f0} input features respectively.
We observe that for the \textit{LibriSpeech} dataset, speakers were somewhat linkable only with their \acrshort{f0} trajectories, as the \acrshort{eer} equal 19.6\%.
As expected, the \acrshort{f0} trajectories do not enable the same degree of linkability performance as \acrshort{mfcc}.
This is expected because the \acrshort{mfcc} contains much more information.
However, an \acrshort{eer} of 19.6\% is still too high and indicates that the \acrshort{f0} is not completely disentangled from the speaker, meaning it could impact the performance of an anonymization pipeline.

Lines 3 to 6 of the table, display the \acrshort{eer} when the \acrshort{f0} is modified with either linear shift, quantization, or \acrshort{awgn} transformations.
Out of them, the \acrshort{awgn} with a noise level of 30dB shows the best disentanglement performance with a \acrshort{eer} of 36.0\%.
However, this kind of transformation might be a bit too strong and impact any potential downstream synthesis.
The \acrshort{awgn}$_{15dB}$ transformation has a lower disentanglement capability, with an \acrshort{eer} of 30.9\%, but might be more suited for downstream synthesis.
For the 4 bits quantization transformation, the \acrshort{eer} is unchanged compared to the clear \acrshort{f0}, indicating that the local (per-frame) smoothing of the \acrshort{f0} does not make the speaker less linkable in this dataset.
Finally, for the linear shift, no improvement in disentanglement is observed when all \acrshort{f0} utterances are mapped to a single \acrshort{f0} mean and variance statistics.

Lines 7 to 9 of the table, show interesting transformations combination.
It appears that combining the globally (per utterance) applied \acrshort{f0} mean and variance linear shift and locally (per frame) applied noise based \acrshort{awgn} or quantization transformations improve the disentanglement.
The best of them is linear shift + \acrshort{awgn}-based having \acrshort{eer} above 40.0\%.
While linear shift + quantization has an \acrshort{eer} of 26.7\%.

\subsection{Discussion}
In this section, which is dedicated to the \acrshort{f0} feature, we evaluated the level of disentanglement between the \acrshort{f0} and the speaker
We concluded that the \acrshort{f0} can be used to perform speaker linkability attacks, even though its performance is inferior to that of \acrshort{mfcc}.
This conclusion applies to the \textit{LibriSpeech} dataset used for evaluation and training the white-box \acrshort{asv} model, the dataset is based on audiobook reading, which may have strong biases in the way readers (and so speakers) read their chapter to match a defined prosodic style.

We presented common signal-processing-based transformations to improve the \acrshort{f0} disentanglement.
In our isolated experiment, we showed that to maximize \acrshort{f0} disentanglement, a combination of \acrshort{f0} linear shift and additive white Gaussian noise is the best.
However, applying this modification to the baseline x-vector-based anonymization system provided by the \acrshort{vpc} organizers does not increase the recorded privacy performance at all, implying that the main source of speaker leakage in this system is not the \acrshort{f0} feature.
As the main features used to perform anonymization are the target x-vector that we tackled in Chapter~\ref{main:chapt4} and the \acrshort{f0} feature studied in this section, this only leaves the linguistic representation features as the main contributor of speaker information leaker.
In the following sections, the linguistic representation features will be our subject of experiments.

Being able to evaluate the disentanglement of the features used by the speech synthesis before generating anonymized voices is a compelling option to guarantee better anonymization.
However, do all features need to be disentangled before being used by the speech synthesis to produce the most anonymized speech?
The speech synthesis itself could anonymize speech on its own, even if the input feature is not completely disentangled.
The quantized \acrshort{f0} may very well be one type of non-perfectly disentangled representation, but still improve privacy, as quantized values are easy to process.
As suggested by \cite{F0_Qian2020F0ConsistentMN,speechsplit} it appears that quantized \acrshort{f0} trajectories improve naturalness.

\vspace{\fill}
\pagebreak 

\section{Linguistic model transformation}
\vspace{-0.4em}

Until now, we have assumed that the linguistic representation features used in the baseline x-vector-based anonymization system (which is a sequence of \acrshort{asrbn}, see Section~\ref{main:chapt_2:bn}), do not convey speaker information or if so, the speech synthesis system does not rely on it to transform voices.
However, in Sections~\ref{main:sec:radar}~and~\localref{f0_modif}, we identified that out of the three features used by speech synthesis, modifying the target x-vector identity and the \acrshort{f0} does not improve privacy, indicating that the linguistic \acrshort{asrbn} representation is the primary source of speaker \acrshort{pii} leakage in this pipeline.
This could be explained by the fact that \acrshort{asrbn} is the feature with the highest dimension and is sequential.
In this section, we start by performing an isolated privacy evaluation of the \acrshort{asrbn} representation to challenge the \acrshort{asrbn}/speaker disentanglement assumption.
Then, in this section, we propose using adversarial learning to transform the \acrshort{asrbn} model to improve the disentanglement of the features it generates.

\subsection{\texorpdfstring{\acrshort{asrbn}}{ASR-BN} Isolated privacy evaluation} \locallabel{isolated_privacy_eval_asrbn}
\vspace{-0.2em}

In this section, our focus is to measure the content/speaker disentanglement only on the \acrshort{asrbn} representation, no speech transformation is applied beforehand.
The objective is to assess and measure if this representation alone leaks speaker \acrshort{pii}, if it is the case, this means the \acrshort{asrbn} could restrict the speaker concealment performance of speaker anonymization systems.
Additionally, as \acrshort{asrbn} representation also conveys the content of the message, we also evaluate utility using an \acrshort{asr} decoder to obtain the predicted sequence of words which can be compared to a reference.

\vspace{-0.2em}
\subsubsection{Experimental setup}
\vspace{-0.0em}
This experiment uses a PyTorch implementation, based on pkwrap \cite{pkwrap}, rather than the \acrshort{vpc} Kaldi \cite{KaldiPovey} implementation for the acoustic model \acrshort{asrbn} extractor.
The differences between our implementation and the one used in the \acrshort{vpc} do not affect conceptually the extraction.
Our model is trained only on \textit{LibriSpeech train-clean-100} to reduce the computation time needed for this and the following experiments.
The cost function used is the \acrshort{eelfmmi} \cite{lfmmi_flat_start_taslp}, allowing flat-start training without pre-training or prior alignment from a \acrshort{gmm} model (see Section~\ref{main:sec:lfmmi}).
According to \cite{pkwrap}, the outputs of the model are left-biphones rather than triphones and the model is composed of 15 {\acrshort{tdnnf}} layers, the \acrshort{asrbn} is extracted from the 13th.

The \acrshort{asv} model directly takes the \acrshort{asrbn} representation (of 256 dimensions) as input instead of \acrshort{mfcc}.
This model is a five \acrshort{tdnn} x-vector model and is trained on the \textit{LibriSpeech train-clean-360} using the Python sidekit \acrshort{asv} toolkit \cite{sidekit}.
For reference, an experiment is also done with \acrshort{mfcc} as input.
Evaluation is done on the \textit{LibriSpeech test-clean} dataset and the \acrshort{eer} metric is used to evaluate privacy, and the \acrshort{wer} is used to evaluate utility.

\vspace{-0.5em}
\subsubsection{Experimental result}
\vspace{-1.5em}

\begin{table}[!ht]
  \begin{center}

  \caption{Privacy and utility of the linguistic \acrshort{asrbn} representation and \acrshort{mfcc}.
    The \acrshort{asrbn} is the one being used in x-vector-based speaker anonymization systems.}
  \locallabel{tab:results}
  \vspace{0.2cm}
  \begin{tabular}{  l c c   }
    \toprule
                                    & \multicolumn{2}{c}{\textit{LibriSpeech test-clean}}                                                   \\
                                    & \multicolumn{1}{c}{\acrshort{eer}~$\uparrow$}       & \multicolumn{1}{c}{\acrshort{wer}~$\downarrow$} \\
    \midrule
    Clear \acrshort{mfcc} features  & 3.7 \footnotesize{$\pm$ 0.4}                        & 5.8 \footnotesize{$\pm$ 0.3}                    \\
    \midrule
    Clear \acrshort{asrbn} features & 6.7 \footnotesize{$\pm$ 0.8}                        & 5.8 \footnotesize{$\pm$ 0.3}                    \\
    \bottomrule
  \end{tabular}
  \end{center}
  \vspace{-1.0em}
\end{table}

Table~\localref{tab:results} shows the privacy and utility results of the experiment, we observe that the \acrshort{asrbn} feature \acrshort{eer} score is rather close to the \acrshort{mfcc} feature \acrshort{eer}, indicating that the transformation used to go from the clear raw speech signal to the \acrshort{asrbn} does not remove speaker information.
This conclusion was also observed in other studies by \cite{adiReverseGradientNot2019,mohanPrivacyPreservingAdversarialRepresentation2019_reality_adversarial} where they showed that the disentanglement property of \acrshort{asrbn} representation is limited and even nonexistent.
This indicates that, much like for the \acrshort{f0}, modifications of the extractor are needed to disentangle speaker information.
The \acrshort{wer} for utility is the same in both experiments as the \acrshort{asrbn} extractor is part of the \acrshort{asr} model.

In the next section, we introduce some transformation techniques applied to the \acrshort{asrbn} extractor and \acrshort{asrbn} feature to improve disentanglement.

\subsection{Adversarial learning model transformation} \locallabel{adv_trainin_anon}


Adversarial learning is an interesting approach for extracting disentangled representations \cite{kazemi2019style,f0_Huang2020UnsupervisedRD}.
It has also been used for anonymization purposes \cite{feutryLearningAnonymizedRepresentations2018}, where the feature to disentangle is \acrshort{pii}.
For our purpose, we apply this framework to increase the content/speaker disentanglement of the \acrshort{asrbn} feature.
Based upon the definition of adversarial learning defined in Section~\ref{main:chapt_1:adv_net}, we add to the \acrshort{asrbn} extractor an \acrshort{asv} speaker adversarial branch.
This adversarial branch adds a negative loss to the extractor during training, inducing it to remove speaker information.
As such, the loss function of the \acrshort{asr} acoustic model is expressed as:
\begin{equation}
  \min _{\theta_{bn}}\left[\min _{\theta_{\mathrm{asr}}} \mathcal{L}_{\mathrm{mmi}}\left(\theta_{bn}, \theta_{\mathrm{asr}}\right)-\alpha \min _{\theta_{\mathrm{spk}}} \mathcal{L}_{\mathrm{aam}}\left(\theta_{bn}, \theta_{\mathrm{spk}}\right)\right]
\end{equation}
where $\theta_{bn}$ denotes the weights of the \acrshort{asrbn} extractor,
$\theta_{asr}$ refers to the weights of the \acrshort{asr} decoder, 
$\theta_{spk}$ the weights of the speaker identification model, which is a five \acrshort{tdnn} sidekit x-vector speaker classifier trained with the \acrfull{aam} loss (see Section \ref{main:chapt_2:aam}),
and $\alpha$ a trade-off parameter between \acrshort{asr} and speaker loss, empirically set to 1.0.

This type of learning scheme applied to \acrshort{asr} models has been studied many times, in \cite{adiReverseGradientNot2019}, the authors slightly improved the \acrshort{asr} performance, in \cite{mohanPrivacyPreservingAdversarialRepresentation2019_reality_adversarial}, the authors showed a very small improvement in unlinkability privacy performance at the cost of a small \acrshort{asr} performance degradation.

Recently to improve adversarial training for feature disentanglement, a \say{semi-adversarial} training scheme was introduced by \cite{RyffelPartiallyEncryptedDL2019} and has shown great success when applied on a written digit dataset.
The authors succeeded in reducing the performance of an unwanted inference, i.e., the classification of fonts from a bottleneck, from a digit classifier.
As described by the \say{semi-adversarial} scheme, training
is performed in 3 steps:
\begin{enumerate}
    \vspace{-1mm}
  \item optimizes the \acrshort{asr} and speaker identification models independently until both converge.
    \vspace{-1mm}
  \item optimizes with the adversarial objective described above, inducing the \acrshort{asrbn} extractor to discard speaker information.
    \vspace{-1mm}
  \item optimizes both \acrshort{asr} decoder and speaker identification models while freezing the \acrshort{asrbn} extractor, such that both \acrshort{asr} and speaker models have the time to accommodate the adversarial training phase.
    \vspace{-1mm}
  \item perform step 2 and 3 until $\mathcal{L}_{\mathrm{aam}}$ does not evolve anymore and the $\mathcal{L}_{\mathrm{mmi}}$ is minimized.
    \vspace{-1mm}
\end{enumerate}

Our work was the first one to apply adversarial training to a traditional acoustic model instead of end-to-end (feature to text) \acrshort{asr} model, to apply the \say{semi-adversarial} training scheme for an \acrshort{asrbn} extractor, to synthesize anonymized speech given this \acrshort{asrbn} extractor.
Given the anonymized speech that this pipeline generates, we then followed with the same analysis of the privacy/utility as in the \acrshort{vpc}.
In the following sections, we present the experimental setup used. 



\subsection{Experimental setup} \locallabel{exp_imple_mine}

For evaluation, we compare the utility and privacy scores across many anonymization pipelines.
First, the baseline pipeline of the \acrshort{vpc} 2022 is compared to our implementation pipeline of the \acrshort{asrbn} extractor without adversarial training.
Then, we retrain another anonymization pipeline with the \acrshort{asrbn} extractor trained with the \say{semi-adversarial} scheme and compare our privacy/utility results.
In this section, we start by presenting the implementation details of the \acrshort{asrbn} extractor and synthesis models and later present the specificity of the adversarial learning experiment.

For the \acrshort{asrbn} extractor, the experiments use the same PyTorch \acrshort{tdnnf}-based \acrshort{asrbn} model as described in the \hyperref[main:isolated_privacy_eval_asrbn]{\say{\acrshort{asrbn} isolated privacy evaluation}} in Section~\localref{isolated_privacy_eval_asrbn}.
This is because the \acrshort{asrbn} extractor needs to be trained with multiple losses and for that purpose, it is easier to work with PyTorch than Kaldi.
Similarly, as in Section~\localref{isolated_privacy_eval_asrbn}, we only trained \acrshort{asrbn} extractors on \textit{LibriSpeech train-clean-100}, which is six times less data than what was used for training in the \acrshort{vpc}.
Due to this aspect, we expect our model to have somewhat lower performance in utility compared to the \acrshort{vpc} baseline.
{\let\thefootnote\relax\footnote{{The Python code of the experiments is available at \url{https://github.com/deep-privacy/SA-toolkit}}}}

We also reworked the speech synthesis system inspired by the HiFi-GAN-based voice conversion model presented in \cite{speech_Resynthesis}.
This model directly generates the speech signal without having to first output a Mel spectrogram (see Section~\ref{main:chapt2:speech_synt} for more details).
The training of the speech synthesis is done with the same dataset as the one defined in the \acrshort{vpc} evaluation plan, hence \textit{LibriTTS train-clean-100}.
The conclusion drawn from Chapter~\ref{main:chapt4}, about the fact that the \say{constant~speaker} target selection strategy allows better privacy evaluation, motivated us to change the speaker representation for the speech synthesis system to use a one-hot speaker embedding.
The use of one-hot embedding simplifies the anonymization pipeline, and as discussed in the conclusion of Chapter~\ref{main:chapt4}, one-hot-based \textit{any-to-many} voice conversion is usually easier to train than x-vector-based \textit{any-to-any} voice conversion.
As for \textit{any-to-one} voice conversion, this framework is usualy less robust than \textit{any-to-many} as less training data is used (the training data only comes from the \textit{one} target speaker).
During the anonymization, the same target speaker is always used. 
The target speaker selected is ID 6081, as we found through empirical analysis that the converted speech exhibited favorable utility and naturalness characteristics.
In our HiFi-GAN model, the \acrshort{f0} is mean and variance normalized, as such, it is the one-hot speaker representation that conditions the target \acrshort{f0} frequencies, the model performs the \acrshort{f0} linear shift transformation presented in Section~\localref{f0_shift_tran} on its own.
If wanted, the other \acrshort{f0} transformations presented in \localref{f0_modif} can be done on the normalized \acrshort{f0}.
Figure~\localref{image_chapt5:my_anon_model} present our modified anonymization pipeline.

\begin{figure}[htbp]
  \begin{center}
    \includegraphics[width=0.80\linewidth]{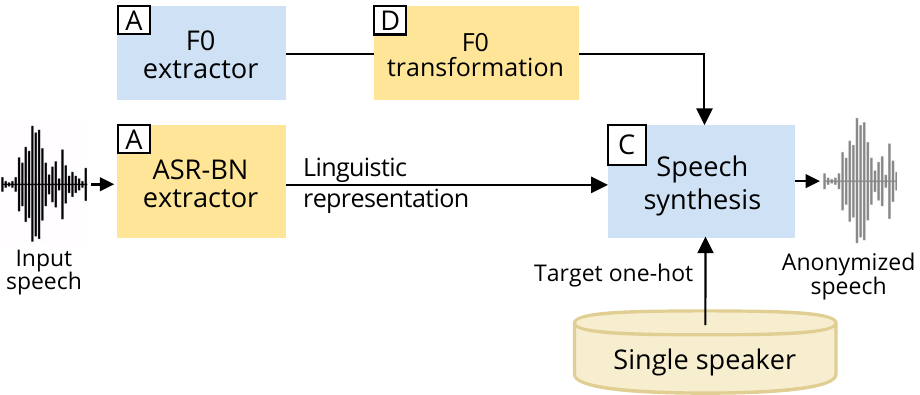}
  \end{center}
  \vspace{-5mm}
  \caption{
    Our speaker anonymization pipeline with module $D$ added and module $A$ adjusted to generate disentangled representation.
    A one-hot speaker embedding is used for speech synthesis instead of x-vector.
  }
  \locallabel{image_chapt5:my_anon_model}
\end{figure}

For evaluation, the \acrshort{vpc} toolkit is used under the white-box {informed} \acrshort{asv} attacker, and privacy results are presented in terms of \acrshort{eer} and \acrshort{li}.
As for utility, the \acrshort{asr} model is trained on anonymized speech and outputs \acrshort{wer} scores.
The datasets used for the training of the \acrshort{asv} and \acrshort{asr} evaluation models are the same as in the \acrshort{vpc} (\textit{LibriSpeech train-clean-360}).
The results are presented for the \textit{LibriSpeech test-clean} and \textit{VCKT test} datasets.

\subsubsection{Adversarial training specific experimental setup}
Our speaker model is a five-layers \acrshort{tdnn} sidekit x-vector model trained with an extension of the cross entropy cost function that encourages the network to output large angular margin between speakers \cite{aamlossArcMarginProduct}.
In our experiment, this cost function showed superior performance than the cross entropy.
We also identified that the \acrshort{tdnn} architecture was better suited than more recent ResNet architectures, this is probably due to the difference in how information is structured in bottleneck features in contrast to \acrshort{mfcc} or filterbank features.

Additionally, during our experiment, we found out that properly training the \acrshort{asv} speaker adversarial branch was challenging.
First, we identified that a larger number of speakers in the training dataset was necessary, as such, we trained the x-vector model on a combination of \textit{LibriSpeech train-clean-100} and \textit{LibriSpeech train-other-500}, for a total of 1417 speakers  (whereas the \acrshort{asr} model is trained only on \textit{LibriSpeech train-clean-100}, for a total of 251 speakers).
Second, we also identified that the way the mini-batches are created for \acrshort{asr} training does not allow proper \acrshort{asv} training.
As such, for steps 1 and 3 of the \say{semi-adversarial} training scheme, the \acrshort{asv} model is trained using large mini-batches having three seconds of audio randomly sampled per utterance, and an equal distribution of speakers in a mini-batch.
The \acrshort{asr} model is trained traditionally.
Last, during step 2, the network optimizers (Adam here \cite{kingma2014adam}) conditioning the backpropagation are separated for the \acrshort{asv} and \acrshort{asr} models to avoid the \acrshort{asv} completely collapsing during the application of the gradient reversal function.
The mini-batches are designed for \acrshort{asr} training, strongly impacting the adaptation of the \acrshort{asv} system to the \acrshort{asrbn} extractor modification during this stage.
This is supposed to be overcompensated in the \say{semi-adversarial} training scheme by the multiple repetitions of stages 2 and 3.

\subsection{Experimental results} \locallabel{sec:adv_res}

Table \localref{chapt5:bigtable_adv} presents the privacy and utility performances of the adversarial system compared to its non-adversarial counterpart and the \acrshort{vpc} 2022 baseline on \textit{LibriSpeech test-clean} and \textit{VCTK test} datasets.
The first line shows results on clear speech, where speaker verification can be addressed with very high accuracy.

The second line, for the \acrshort{vpc} 2022 baseline system provides a strong baseline, keeping the spoken content easily recognizable (absolute degradation of less than 1\% \acrshort{wer}, compared to clear speech on both datasets) while significantly increasing privacy protection.
On the \textit{LibriSpeech} dataset, privacy was increased, as the \acrshort{li} metric lowered from 0.93 down to 0.67.
On the \textit{VCTK} dataset, privacy was even more improved as the \acrshort{li} dropped from 0.93 to 0.49.
This overall trend of seeing the \textit{VCTK} dataset more unlinkable than the \textit{LibriSpeech} one can be explained by the dataset's nature and the data used to train the \acrshort{asv} attack model.
\textit{LibriSpeech} does not offer much variability within a single speaker due to the long recording sessions of audiobook chapters.
In addition, book reading speech differs from spontaneous speech, which impacts speech rate and overall intonation.
Those biases are captured by \acrshort{asv} systems \cite{rhythm_jf_odyssey,prob_x-vector}.
Additionally, the white-box attacker is trained on \textit{LibriSpeech train-clean-360}, which is of the same nature as \textit{LibriSpeech test-clean} meaning that if any of the previous biases were to be captured by the \acrshort{asv} model, they are applied to the \textit{LibriSpeech test-clean} dataset and the \textit{VCTK} one.
A similar conclusion about the \acrshort{asr} performances can be drawn, the \acrshort{wer} for \textit{LibriSpeech test-clean} is always lower than the one on \textit{VCTK}.
In the following, we primarily focus on the \textit{VCTK} results because the \textit{VCTK} dataset is less sensible to audiobook \acrshort{asv} evaluation biases.

\renewcommand*{\thefootnote}{\fnsymbol{footnote}}
\begin{table}[!htb]
  \centering

  \caption{Privacy and utility results for the adversarially trained of the \acrshort{asrbn}
   extractor\protect\footnotemark[4].
   }
  \locallabel{chapt5:bigtable_adv}
  \vspace{0.2cm}
  \begin{tabular}{
    l@{}@{\extracolsep{0.15mm}}
    S[table-format=1.2]@{\extracolsep{0.01mm}}
    S[table-format=2.1]@{\extracolsep{0.0in}}
    S[table-format=2.1]@{\extracolsep{3.0mm}}
    S[table-format=1.2]@{\extracolsep{0.01mm}}
    S[table-format=2.1]@{\extracolsep{0.0in}}
    S[table-format=2.1]@{\extracolsep{0.00in}}
    }
    \toprule
    \multirow{1}{0pt}{\begin{minipage}{60pt}{\hspace{7mm}Dataset}\end{minipage}}
   & \multicolumn{3}{c}{  \textit{LibriSpeech test-clean}}
   & \multicolumn{3}{c}{ \textit{VCTK test}} \\
    \midrule
        \multirow{1}{0pt}{\begin{minipage}{120pt}{\vspace{1mm}\vspace{0mm}\hspace{7mm}Method: \acrshort{asrbn}}\end{minipage}} & \multicolumn{2}{c}{ \hspace{-2mm} Privacy}            & \multicolumn{1}{c}{ \hspace{-2.5mm} Utility}

  & \multicolumn{2}{c}{ \hspace{-2mm} Privacy}            & \multicolumn{1}{c}{ \hspace{-2.5mm} Utility}\\
    \multirow{1}{0pt}{\begin{minipage}{120pt}{\vspace{0mm}\vspace{-1mm}\hspace{7mm}transformations}\end{minipage}} & \multicolumn{1}{c}{\acrshort{li}$~\downarrow$}                     & \multicolumn{1}{c}{\acrshort{eer}$~\uparrow$}           & \multicolumn{1}{c}{{\acrshort{wer}$~\downarrow$}}
  & \multicolumn{1}{c}{\acrshort{li}$~\downarrow$}                     & \multicolumn{1}{c}{\acrshort{eer}$~\uparrow$}           & \multicolumn{1}{c}{\acrshort{wer}$~\downarrow$}                                    \\
  \midrule
    \hspace{2mm}\textattachfileandprintout{\subfix{wavs/01-clear-p226-003-mic2.wav}}{1.} Clear speech      & 0.93 & 4.1  & 4.1 & 0.93 & 2.7  & 12.8 \\
  \midrule
      \hspace{2mm}\textattachfileandprintout{\subfix{wavs/02-vpc-baseline-p226-003-mic2.wav}}{2.} VPC 2022 baseline     & 0.67 & 13.5 & 5.1 & 0.49 & 20.6 & 13.0 \\
      \midrule
  \hspace{2mm}\textattachfileandprintout{\subfix{wavs/03-tdnnf-asrbn.wav}}{3.} \scshape{tdnnf}       & 0.81 & 8.7 & 6.9 & 0.73 & 10.8 & 19.1 \\
  \hspace{2mm}\textattachfileandprintout{\subfix{wavs/04-tdnnf-asrbn-adv.wav}}{4.} \scshape{adversarial tdnnf} \hspace{0.9cm} & 0.83 & 7.8 & 5.3 & 0.70 & 11.8 & 14.4 \\
  \bottomrule
  \end{tabular}
\end{table}
\footnotetext[4]{Audio samples can be extracted from the PDF by clicking (or double-clicking) tables rows index.}
\setcounter{footnote}{0}
\renewcommand*{\thefootnote}{\arabic{footnote}}

Experiment with our baseline \acrshort{asrbn} extractor (denoted as \say{{\scshape{tdnnf}}}, line 3) shows a very high degradation of utility compared to clear speech.
Compared to the \acrshort{vpc} baseline,  the \acrshort{wer} increases by a large margin on \textit{LibriSpeech} and \textit{VCTK} datasets.
This is because the \acrshort{asrbn} model was not trained with the non-clean speech of \textit{LibriSpeech train-other-500}.
Interestingly, with the degradation of utility, privacy is not significantly improved.

On \textit{VCTK}, the \acrshort{li} dropped from 0.93 for clear speech to 0.73 for anonymized speech, a slight improvement but far smaller than the \acrshort{vpc} baseline.
This disparity can be explained as we extracted the \acrshort{asrbn} from the 13th layer while the \acrshort{vpc} baseline extracted it from the 17th layer.
As shown in the \hyperref[main:isolated_privacy_eval_asrbn]{\say{\acrshort{asrbn} isolated privacy evaluation}}, the \acrshort{asrbn} contains a lot of speaker information which the speech synthesis model does not remove.

Before describing the result of the table, we discuss the intermediate indicators that we obtained during the training procedure of the adversarially trained \acrshort{asrbn}.
In the first step of adversarial training, the \acrshort{asrbn} extractor and the adversarial speaker model are optimized independently (by freezing a branch or the another.).
As a result, the \acrshort{asr} model (composed of the \acrshort{asrbn} extractor) outputs a \acrshort{wer} on \textit{LibriSpeech test-clean} of 5.15\%, while the speaker identification accuracy of the adversarial branch reaches 95.8\% on the \acrshort{asrbn}.
Then, in the second stage, the \acrshort{asrbn} extractor is trained to modify its weights with the negative loss of the adversarial speaker identification branch and the normal \acrshort{asr} decoder loss.
The third stage retrains both \acrshort{asr} decoder and speaker identification models on frozen \acrshort{asrbn} extractor.
In our experiment, we run the second and third stages twice, as after the second run, the speaker identification models did not converge anymore, indicating the \acrshort{asrbn} extractor representation could not be used by the speaker identification model to identify the speaker.
This could indicate that the \acrshort{asrbn} representation does not encode speaker information.
At the end of the training, the \acrshort{asr} decoder has a \acrshort{wer} of 5.38\% while the speaker recognition accuracy of the adversarial is 4.2\%.
The results of seeing the \acrshort{wer} slightly increased is similar as in \cite{mohanPrivacyPreservingAdversarialRepresentation2019_reality_adversarial}.
However, as suggested by \cite{adiReverseGradientNot2019}, using an adversarial branch trained on more data than the \acrshort{asr} decoder should provide a small \acrshort{wer} improvement as it falls into the category of semi-supervised learning where most of the labels are about the speakers rather than the text.
Overall, there is a lack of agreement in the literature for this form of training applied to this field.

Using the adversarially trained \acrshort{asrbn} extractor, we trained the same HiFi-GAN speech synthesis model and obtain the anonymization pipeline denoted as \say{{\scshape{adversarial tdnnf}}} in Table~\localref{chapt5:bigtable_adv}.
Presented in line 4 of the table, we observe that the use of adversarial learning to remove speaker information from the \acrshort{asrbn} representation is not effective to make the speech less linkable as the \acrshort{li} on \textit{LibriSpeech test-clean} and \textit{VCTK test} are the same with and without adversarial learning.
This result is the same as in \cite{mohanPrivacyPreservingAdversarialRepresentation2019_reality_adversarial}, even though we used additional training data, the more advanced \say{semi-adversarial} training scheme, an more known x-vector architecture for the adversarial model, and different optimizer and mini-batches creation to obtain the best adversarial model as possible.
One reason for this discrepancy could be inherent to adversarial training.
Before being able to properly remove information from the bottleneck, the adversarial branch must be able to characterize the information to remove.
And the model-based approach leaves room for the \acrshort{asrbn} encoder to find a type of representation that is still relevant for its primary task while adding some sort of noise that makes the particular adversarial architecture fail, similarly as in model adversarial attack \cite{Guo2019SimpleBA}.
One solution to fix this could be to use more than one adversarial network/architecture which would each capture the speaker information differently and maybe provide a better negative gradient.

While the privacy performance results are not satisfactory, the utility results are interesting.
In the table, we observe a utility improvement, the \acrshort{wer} dropped on both datasets from 6.9\% to 5.3\% for the \textit{LibriSpeech test-clean} and from 19.1\% to 14.4\% for the \textit{VCTK test} datasets when adversarial training is applied.
This conclusion is unexpected as decoding the speech directly from the adversarially trained \acrshort{asrbn} representation did not yield a \acrshort{wer} improvement (actually the \acrshort{wer} increases from 5.15\% to 5.38\%), however, when using the \acrshort{asrbn} representation to synthesize speech, the adversarially trained \acrshort{asrbn} is superior.
We hypothesize that adversarial training allows the \acrshort{asrbn} representation to be structured in a better-formatted/easy-to-process way that allows for the speech synthesis system to generate higher utility quality speech.

\subsection{Discussion}
We conducted an experiment using adversarial training to reduce speaker information encoding from the \acrshort{asrbn} extractor.
The reverse gradient of a speaker identification model was applied to the \acrshort{asrbn} extractor during training.
The results showed that this improved the utility of the \acrshort{asrbn} for speech synthesis, but not for privacy protection.
To further enhance privacy, possible extensions of this work include using a triplet loss in the adversarial model to better match unlinkability attacks and using multiple adversarial models to reduce the potential \acrshort{asrbn} specialization to one adversity.

\section{Linguistic feature transformation}
In this section, we evaluate other kinds of \acrshort{asrbn} disentanglement techniques than adversarial training.
The kind of transformation that we will be studying does not primarily rely on backpropagation to induce the network to discard speaker information from the \acrshort{asrbn}.
Instead, we will focus on transformations applied directly on the \acrshort{asrbn} features to remove speaker information,
hence the name of this section.
However, it is worth noting that for the \acrshort{asrbn} to properly capture the linguistic content information the transformations need to be applied during training to ensure the bottleneck is compatible with a given transformation.

In the following, we will start by briefly explaining an existing method based on noise perturbation used to transform the \acrshort{asrbn} feature and then propose our alternative method based on vector quantization.




\vspace{-0.8em}
\subsection{Laplace noise transformation} \locallabel{laplace}
\vspace{-0.4em}

The Laplace noise is a type of noise that can be added to a signal or a dataset to protect the privacy of the individuals associated with that data.
It is a type of \acrfull{dp} \cite{Dwork2006DifferentialP} technique that is based on the Laplace distribution \cite{Dwork2014TheAF}. 

The idea behind using Laplace noise for anonymization is to add a small amount of random noise to the data in order to make it difficult for an attacker to infer the true values of the data.
%
Usually, Laplace noise is added to data in a controlled manner, with the amount of noise determined by a parameter known as the privacy budget.
The privacy budget is a measure of the amount of noise that can be added to the data without compromising too much its utility for the intended application.

The Laplace noise to add to a sample of data is defined by:
$\text{noise} = \operatorname{Lap}(0, \frac{1}{\epsilon})$,
where $\operatorname{Lap}$ is the Laplace distribution with mean 0 and scale parameter $\frac{1}{\epsilon}$, $\epsilon$ is the privacy budget.
The value of $\epsilon$ is typically chosen based on the desired level of privacy protection, with smaller values resulting in more noise and greater privacy protection.
%

The application of Laplace noise transformation for speaker anonymization was first proposed by \citetitle{dp_vpc}, \cite{dp_vpc}, where the authors suggested adding noise to the bottleneck of two models.
The first is the bottleneck of a \acrshort{f0} auto-encoder model that transforms the \acrshort{f0} while still generating plausible trajectories.
The second is the bottleneck of the \acrshort{vpc} \acrshort{tdnnf}-based \acrshort{asrbn} extractor.
Note that in both cases, they normalize the bottleneck representation before and after applying the noise, as they found this improves training convergence.

\vspace{-0.8em}
\subsection{Vector quantization transformation} \locallabel{anon_trans_vq}
Instead of adding noise to the \acrshort{asrbn} representation, we propose constraining the layer that generates the \acrshort{asrbn} by using \acrfull{vq}.

Vector quantization approximates a continuous vector by another vector of the same dimension, but the latter belongs to a finite set of vectors, called prototype vectors, and is contained in a dictionary.
In the self-supervised framework of linguistic representation for voice conversion (see Section~\ref{main:chapt_2:vc_lin_rep} and~\ref{main:chapt3:anon_self_sup}), it has been observed that the prototype vectors learned from vector quantization primarily capture information related to the phonemes and discard some speaker information \cite{neural_disctre_vq,Unsupervised_speech_rep_vq,one-shot-vc-vector-quant}.
Similarly, our objective of applying vector quantization in the supervised linguist representation framework is to minimize the encoding of speaker information in \acrshort{asrbn}.

\vspace{-0.5em}
\subsubsection{\texorpdfstring{\acrshort{vq}}{VQ} objective}
\vspace{-0.4em}
Given the input audio sequence $s= \left(s_{1}, s_{2}, \ldots, s_{T}\right)$ of length $T$, the first {\acrshort{tdnnf}} layers produces a continuous vector $h(s) = \left(h_{1}, h_{2}, \ldots, h_{J}\right)$ of length $J$ ($J < T$ due to the subsampling performed by the network) where $h_{j} \in \mathbb{R}^{D}$ for each time step $t$, and $D$ is the size of the bottleneck representation ($D$ = 256 here).
\acrshort{vq} takes as input the sequence of continuous vectors $h(s)$ and replaces each $h_{j} \in h(s)$ by a prototype of the dictionary $E=\left\{e_{1}, e_{2}, \ldots, e_{V}\right\}$ of size $S$, each $e_{i} \in \mathbb{R}^{D}$.
\acrshort{vq} transforms $h(s)$ to $q(s) = \left(q_{1}, q_{2}, \ldots, q_{J}\right)$ with:
\vspace{-0.5em}
\begin{equation}
  \forall j \in \left\{1, 2, \ldots, J\right\}, q_j=\underset{e_i}{\arg \min }\lVert h_j-e_{i}\rVert_{2}^{2}
\vspace{-0.5em}
\end{equation}
The vector $h_{j}$ is replaced by its closest prototype vector $e_{v}$ in terms of Euclidean distance.
Since the quantization is non-differentiable (because of the $arg \min$ operation), its derivative must be approximated.
To do this, we use a \textit{straight-through estimator} \cite{strat_through_estimator} i.e.,$\frac{\partial \mathcal{L}}{\partial h(s)} \approx \frac{\partial \mathcal{L}}{\partial q(s)}$.
The prototype vectors are learned to approximate the continuous vectors which they replace by adding an auxiliary cost function:
\vspace{-0.5em}
\begin{equation}
  \mathcal{L}_{vq} = {\textstyle\sum_{j=1}^J} \lVert\operatorname{sg}\left[h_j\right]-q_j\rVert_{2}^{2}
\vspace{-0.5em}
\end{equation}
where $\mathrm{sg}[\cdot]$ denotes the stop gradient operation, blocking the update of the weights of the {\acrshort{tdnnf}} layers for this cost function (only updates the dictionary $E$).
Minimizing $\mathcal{L}_{vq}$ is a similar operation to a k-means, but applied for each mini-batch during learning, the prototypes correspond to the centroids of a k-means.

Since the volume of the continuous vector space $h(s)$ is boundless, it can grow arbitrarily if the
dictionary $E$ does not train as fast as the {\acrshort{tdnnf}}.
Adding a cost function that regularizes the {\acrshort{tdnnf}} to produce continuous vector $h(s)$ close to the prototypes of $E$ is necessary so that learning does not diverge:
\vspace{-0.5em}
\begin{equation}
  \mathcal{L}_{vq\_reg} = {\textstyle\sum_{j=1}^J} \lVert h_j-\operatorname{sg}[q_j]\rVert_{2}^{2}
\vspace{-0.5em}
\end{equation}

\subsubsection{\texorpdfstring{\acrshort{asrbn}}{ASR-BN} objective}
The cost function of the acoustic model can then be expressed as the sum of the \acrshort{mmi}, quantization and regularization functions:
\vspace{-0.5em}
\begin{equation}
  \mathcal{L}=\mathcal{L}_{mmi}+\mathcal{L}_{vq}+\beta \mathcal{L}_{vq\_reg}
\vspace{-0.5em}
\end{equation}
where $\beta$ denotes the coefficient of the regularization factor (we used $\beta = 0.25$).
In practice, we used the learning rule based on the exponential moving average (EMA) \cite{ema_vq} to update the prototypes.
EMA updates the dictionary $E$ independently of the optimizer, so learning is more robust to different optimizers and hyperparameters (e.g., learning rate, momentum).

\subsection{Wrapping up 
 \texorpdfstring{\acrshort{asrbn}}{ASR-BN} feature transformation}

Figure~\localref{fig:twod_space_anon} proposes an intuitive comparison example where the Laplace noise and vector quantization transformations are applied to a two-dimensional space.
If we consider that the gray dots encode both speaker identity and pronunciation of a particular sound, the goal of \acrshort{asrbn} transformations is to modify the speaker information from this representation to output a representation that encodes the same sound but with a different speaker identity this transformation is represented by arrows on Figure~\localref{fig:twod_space_anon}.
For the \acrshort{vq} example, the size $S$ of the dictionary equals the number of possible (reduced in the example) \acrshort{vq} prototypes.
Overall, the lower $S$ is, the more the prototypes should encode the linguistic content and not the speaker.
As \acrshort{vq} aims to select the centroid of a region, some input vectors will be more modified than others.
The output vector corresponds to the most common way to pronounce a sound.
%
In contrast, for the noise addition, all input vectors are modified with the same amount of noise.
Overall the higher the noise (lower $\epsilon$ for the Laplace noise sampling) the higher the transformation.
The output vector corresponds to how a different speaker would pronounce the sound.
The utilization of noise transformations can be associated with a \say{random speaker} target speaker selection strategy, whereas the \acrshort{vq} can be associated with a \say{constant speaker} strategy (as outlined in Chapter~\ref{main:chapt4}).
While not displayed in the figure, the shapes of the decision boundaries are affected as the models are trained with the modifications.
It is expected that for the \acrshort{vq} method, clusters of similar abstract meaning (which depends on $S$) will get closer to each other.
Whereas for the noise addition method, the decision boundary will be extended to cover a large region space to accommodate for the added noise.

\begin{figure*}[!hbtp]
  \vspace{-0.8em}
  \centering
  \begin{minipage}[b]{0.38\linewidth}
    \centering
    \centerline{\includegraphics[width=1.0\linewidth]{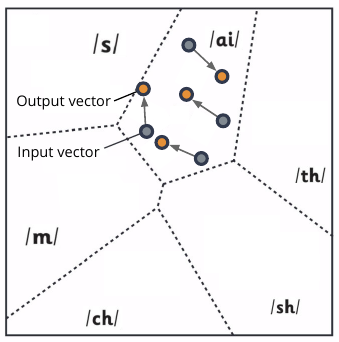}}
    \begin{minipage}[b]{\linewidth}
      \begin{center}
        Noise randomly transforms inputs hoping that the output vectors remain in the same decision regions.
      \end{center}
    \end{minipage}
  \end{minipage}
  \hspace{0.02\linewidth}
  \begin{minipage}[b]{0.38\linewidth}
    \centering
    \centerline{\includegraphics[width=1.0\linewidth]{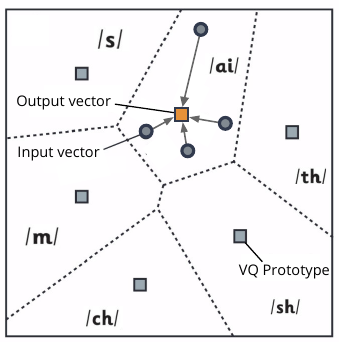}}
    \begin{minipage}[b]{\linewidth}
      \begin{center}
        \acrshort{vq} transforms inputs to discrete output vectors considered as the centroid of a quantization region.
      \end{center}
    \end{minipage}
  \end{minipage}

  \vspace{-0.2em}
  \caption{Voronoi diagrams of the noise and \acrshort{vq} based \acrshort{asrbn} transformations.
    In the figures, the annotation per region displays the phoneme obtained when decoding the \acrshort{asrbn}.
  }
  \locallabel{fig:twod_space_anon}
\vspace{-1em}
\end{figure*}

\subsection{Experimental setup}

The \acrshort{vq} and noise transformations are applied on the PyTorch \acrshort{asrbn} extractor and the HiFi-GAN speech synthesis implementation presented in the above Section~\localref{exp_imple_mine} is used.
In the experiment, we provide different sizes of dictionaries for the \acrshort{vq} transformation and different levels of noise for the Laplace transformation.
Our \acrshort{asrbn} extractors are trained only on \textit{LibriSpeech train-clean-100} to reduce the computation time.
While the HiFi-GAN is trained with \textit{LibriTTS train-clean-100}.

For evaluation, the \acrshort{vpc} toolkit is used under the white-box {informed} \acrshort{asv} attacker, and privacy results are presented in terms of \acrshort{eer} and \acrshort{li}.
As for utility, the \acrshort{asr} model is trained on anonymized speech and outputs \acrshort{wer} scores.
The datasets used for the training of the \acrshort{asv} and \acrshort{asr} evaluation models are the same as in the \acrshort{vpc} (\textit{LibriSpeech train-clean-360}).
The results are presented for the \textit{LibriSpeech test-clean} and \textit{VCKT test} datasets.

In the first experiment, we investigated the relationship between the size of the \acrshort{vq} dictionary and the privacy/utility performances obtained by a pipeline.
The results are compared to our baseline pipeline already used in Section~\localref{sec:adv_res} which does not use a vector quantization for the \acrshort{asrbn}.
Then, to compare the results with the \acrshort{vpc} 2022 baseline which trained the \acrshort{asrbn} extractor with \textit{LibriSpeech train-clean-100} and \textit{LibriSpeech train-other-500}, we retrain, with the same data, one \acrshort{asrbn} extractor constrained with a dictionary of 64 prototypes.
The speech synthesis model is also retrained.

In the second experiment, we investigated replacing the filterbank coefficients used as input features for the \acrshort{asrbn} with Wav2Vec-2.0 representation (see Section~\ref{main:chapt2:sec:wav2vec2} for more detail on this architecture).
The model topology of the \acrshort{asrbn} is adjusted, according to \cite{wav2vec2_tdnnf}, we reduced the number of the acoustic model to 9 \acrshort{tdnnf} layers.
The \acrshort{asrbn} is extracted from the 3rd layer, right before the {\acrshort{tdnnf}} downsampling layer as Wav2Vec-2.0 already downsampled the signal.
The \acrshort{asrbn} extractor is trained with \textit{LibriSpeech train-clean-100} and we fine-tune the Wav2Vec-2.0 model with a learning rate that is 20 times lower than the learning rate of the {\acrshort{tdnnf}} layers.
We used a large Wav2Vec-2.0 model pre-trained on 24.1K hours of unlabeled multilingual west Germanic speech from {V}ox{P}opuli \cite{voxpopuli}.
There is no data overlap between {V}ox{P}opuli and the data used by the VoicePrivacy evaluation plan.
We also experimented with the use of the \acrshort{f0} transformations described in Section~\localref{f0_modif}, and the one introduced by \cite{dp_vpc} authors where Laplace noise is added to the bottleneck of an auto-encoder to produce distorted but yet realistic \acrshort{f0} trajectories\footnote{We thank the authors for providing their implementation.}.

\renewcommand*{\thefootnote}{\fnsymbol{footnote}}
\footnotetext[4]{Audio samples can be extracted from the PDF by clicking (or double-clicking) tables rows index.}
\setcounter{footnote}{0}
\renewcommand*{\thefootnote}{\arabic{footnote}}

Finally, in the last experiment, we adapted the work of \cite{dp_vpc}, to transform with Laplace noise addition the \acrshort{asrbn}.
In this experiment, we used the same \acrshort{asrbn} model as previously with the Wav2Vec-2.0 representation.
We also experimented with the use of the \acrshort{f0} transformations.

\vspace{-0.4em}
\subsection{Experimental results}
\vspace{-0.4em}
In the following sections, we present the results of the experiments conducted with the vector quantization approach and outline its effectiveness by comparing it to
other methods. 

\vspace{-1em}
\subsubsection{\texorpdfstring{\acrshort{asrbn}}{ASR-BN} vector quantization} \locallabel{asrbn-vq-res}
\vspace{-0.2em}

\begin{table}[!htb]
  \vspace{-1.5em}
  \caption{Privacy and utility results for \acrlong{vq}-based anonymization. \acrshort{vq} 128 indicates the \acrshort{asrbn} extractor was constrained with a dictionary of 128 prototypes.}
  \locallabel{chapt5:bigtable_vq}
  \vspace{0.2cm}
  \centering
    \begin{tabular}{
      l@{}@{\extracolsep{0.15mm}}
      S[table-format=1.2]@{\extracolsep{0.01mm}}
      S[table-format=2.1]@{\extracolsep{0.0in}}
      S[table-format=2.1]@{\extracolsep{3.0mm}}
      S[table-format=1.2]@{\extracolsep{0.01mm}}
      S[table-format=2.1]@{\extracolsep{0.0in}}
      S[table-format=2.1]@{\extracolsep{0.00in}}
      }
      \toprule
      \multirow{1}{0pt}{\begin{minipage}{200pt}{\hspace{7mm}Dataset}\end{minipage}}
                                                                                                                                                                                                         & \multicolumn{3}{c}{  \textit{LibriSpeech test-clean}}
                                                                                                                                                                                                         & \multicolumn{3}{c}{ \textit{VCTK test}}                                                                                                                                                                                                                                                                      \\
      \midrule
      \multirow{1}{0pt}{\begin{minipage}{120pt}{\vspace{1mm}\vspace{0mm}\hspace{7mm}Method: \acrshort{asrbn}}\end{minipage}} & \multicolumn{2}{c}{ \hspace{-2mm} Privacy}            & \multicolumn{1}{c}{ \hspace{-2.5mm} Utility}
                                                                                                                                                                                                         & \multicolumn{2}{c}{ \hspace{-2mm} Privacy}            & \multicolumn{1}{c}{ \hspace{-2.5mm} Utility}                                                                                                                                                                                                         \\
                                                                                                                                                                                                      \begin{minipage}{120pt}{\vspace{1mm}\vspace{-1mm}\hspace{7mm}transformations}\end{minipage}   & \multicolumn{1}{c}{\acrshort{li}~$\downarrow$}        & \multicolumn{1}{c}{\acrshort{eer}~$ \uparrow$} & \multicolumn{1}{c}{\acrshort{wer}~$\downarrow$} & \multicolumn{1}{c}{\acrshort{li}~$\downarrow$} & \multicolumn{1}{c}{\acrshort{eer}~$ \uparrow$} & \multicolumn{1}{c}{\acrshort{wer}~$\downarrow$} \\
      \midrule
      \hspace{2mm}\textattachfileandprintout{\subfix{wavs/01-clear-p226-003-mic2.wav}}{1.} Clear speech      & 0.93 & 4.1  & 4.1 & 0.93 & 2.7  & 12.8 \\
      \midrule
      \hspace{2mm}\textattachfileandprintout{\subfix{wavs/02-vpc-baseline-p226-003-mic2.wav}}{2.} VPC 2022 baseline     & 0.67 & 13.5 & 5.1 & 0.49 & 20.6 & 13.0 \\
      \midrule
      \hspace{2mm}\textattachfileandprintout{\subfix{wavs/03-tdnnf-asrbn.wav}}{3.} \scshape{tdnnf (no vq)}  & 0.81 & 8.7  & 6.9  & 0.73 & 10.8 & 19.1 \\
      \hspace{2mm}\textattachfileandprintout{\subfix{wavs/05-tdnnf-asrbn-vq-128.wav}}{5.} \scshape{tdnnf vq} 256 & 0.62 & 16.2 & 9.9  & 0.46 & 22.9 & 24.1 \\
      \hspace{2mm}\textattachfileandprintout{\subfix{wavs/06-tdnnf-asrbn-vq-256.wav}}{6.} \scshape{tdnnf vq} 128 & 0.59 & 17.7 & 10.4 & 0.42 & 24.0 & 26.3 \\
      \hspace{2mm}\textattachfileandprintout{\subfix{wavs/07-tdnnf-asrbn-vq-64.wav}}{7.} \scshape{tdnnf vq} 64 \hspace{2.0cm}  & 0.50 & 21.1 & 12.4 & 0.29 & 30.0 & 29.1 \\
      \bottomrule
    \end{tabular}
  \vspace{-0.8em}
\end{table}

Table~\localref{chapt5:bigtable_vq} shows the privacy and utility performances of the vector quantization system compared to its non \acrshort{vq} counterpart and the \acrshort{vpc} 2022 baseline on \textit{LibriSpeech test-clean} and \textit{VCTK test} datasets.
By constraining the \acrshort{tdnnf} \acrshort{asrbn} extractor with the use of vector quantization, the performance of linkability attacks are drastically reduced the as \acrshort{li} performances for lines~5 to~7 are lower than for line 3 on all datasets.
The number $S$ of prototypes in the quantization dictionary constrains the acoustic model.
With $S$ prototype vectors, the spoken information of the speech is compressed into a discrete dictionary space of size $S$.
The smaller the dictionary, the more the network must find an efficient transformation to represent the spoken content information, leaving less room to encode the speaker's information.
We tried three dictionary sizes in our experiment: 256, 128, and 64.
The most anonymized speech was generated with $S$=64, where the \acrshort{li} dropped from 0.73 ({\scshape{no vq}}) to 0.29 ({\scshape{vq 64}}) on the \textit{VCTK} dataset.
However, this privacy improvement comes at a very high utility cost; the \acrshort{wer} raises from 19.1\% to 29.1\%.
The other dictionary sizes illustrate well the privacy utility trade-off \cite{privacy-utility-tradeoff} this model suffers when lower privacy implies better utility and vice-versa.
We hypothesize that the privacy improvement comes from the vector quantization layer, while the major utility loss comes from the small number of layers before the quantization layer and/or the amount of data used to train the model.
Constraining the network to such a few discrete vectors could be possible without significant utility loss if the network has the capabilities to transform the speech signal into a compressed high-level linguistic representation.

Table~\localref{chapt5:bigtable_vq_600} presents the experiment where more data is used to train the \acrshort{asrbn} extractor constrained with 64 prototypes.
As a result, it uses the same amount of data as in the \acrshort{vpc} evaluation.
The outcome is displayed in line 7b of the table and shows that by using more data to train the \acrshort{asrbn} extractor, higher utility is achieved, on \textit{VCTK test} the \acrshort{wer} is reduced from 29.1\% to 16.7\%; similarly, the \acrshort{wer} on \textit{LibriSpeech test-clean} almost divided by two.
This improvement in utility does not affect privacy performances, as the \acrshort{li} on \textit{VCTK test} and \textit{LibriSpeech test-clean} are not significantly different than when the network was trained on six times fewer data.
When compared with the \acrshort{vpc} 2022 baseline also trained with the same data, our model with \acrshort{vq} improves privacy performances as the \acrshort{li} is reduced from 0.49 to 0.25 on \textit{VCTK test} and 0.67 to 0.52 on \textit{LibriSpeech test-clean}.
However, this privacy improvement comes at a utility cost, as the \acrshort{wer} is increased from 13.0\% to 16.7\% on \textit{VCTK test} and 5.1\% to 6.7\% on \textit{LibriSpeech test-clean}, this small degradation surely comes from \acrshort{vq}.

\vspace{-0.8em}
\begin{table}[!htb]
  \caption{Privacy and utility results for the \acrshort{vq} 64 configurations trained with 100h (\textit{LibriSpeech train-clean-100}) and 600h (\textit{LibriSpeech train-clean-100} + \textit{LibriSpeech train-other-500}) of data.
  }
  \locallabel{chapt5:bigtable_vq_600}
  \vspace{0.2cm}
  \resizebox{1\textwidth}{!}{
  \centering
    \begin{tabular}{
      l@{}@{\extracolsep{0.15mm}}
      S[table-format=1.2]@{\extracolsep{0.01mm}}
      S[table-format=2.1]@{\extracolsep{0.0in}}
      S[table-format=2.1]@{\extracolsep{3.0mm}}
      S[table-format=1.2]@{\extracolsep{0.01mm}}
      S[table-format=2.1]@{\extracolsep{0.0in}}
      S[table-format=2.1]@{\extracolsep{0.00in}}
      }
      \toprule
      \multirow{1}{0pt}{\begin{minipage}{200pt}{\hspace{7mm}Dataset}\end{minipage}}
                                                                                                                                                                                                         & \multicolumn{3}{c}{  \textit{LibriSpeech test-clean}}
                                                                                                                                                                                                         & \multicolumn{3}{c}{ \textit{VCTK test}}                                                                                                                                                                                                                                                                      \\
      \midrule
      \multirow{1}{0pt}{\begin{minipage}{120pt}{\vspace{1mm}\vspace{0mm}\hspace{7mm}Method: \acrshort{asrbn}}\end{minipage}} & \multicolumn{2}{c}{ \hspace{-2mm} Privacy}            & \multicolumn{1}{c}{ \hspace{-2.5mm} Utility}
                                                                                                                                                                                                         & \multicolumn{2}{c}{ \hspace{-2mm} Privacy}            & \multicolumn{1}{c}{ \hspace{-2.5mm} Utility}                                                                                                                                                                                                         \\
                                                                                                                                                                                                      \begin{minipage}{120pt}{\vspace{1mm}\vspace{-1mm}\hspace{7mm}transformations}\end{minipage}   & \multicolumn{1}{c}{\acrshort{li}~$\downarrow$}        & \multicolumn{1}{c}{\acrshort{eer}~$ \uparrow$} & \multicolumn{1}{c}{\acrshort{wer}~$\downarrow$} & \multicolumn{1}{c}{\acrshort{li}~$\downarrow$} & \multicolumn{1}{c}{\acrshort{eer}~$ \uparrow$} & \multicolumn{1}{c}{\acrshort{wer}~$\downarrow$} \\
    \midrule
      \hspace{2mm}\textattachfileandprintout{\subfix{wavs/02-vpc-baseline-p226-003-mic2.wav}}{2.} VPC 2022 baseline \hspace{1mm}\textit{train-600}\hspace{2mm}     & 0.67 & 13.5 & 5.1 & 0.49 & 20.6 & 13.0 \\
      \midrule
      \hspace{2mm}\textattachfileandprintout{\subfix{wavs/07-tdnnf-asrbn-vq-64.wav}}{7.} {\scshape{tdnnf vq}} 64 \hspace{10mm}\textit{train-100}  & 0.50 & 21.1 & 12.4 & 0.29 & 30.0 & 29.1 \\
      \textattachfileandprintout{\subfix{wavs/07b-tdnnf-asrbn-vq-64.wav}}{7b.} {\scshape{tdnnf vq}} 64 \hspace{10mm}\textit{train-600}  & 0.52 & 20.0 & 6.7 & 0.25 & 32.5 & 16.7 \\
      \bottomrule
    \end{tabular}
  }
  \vspace{-0.8em}
\end{table}

\subsubsection{\texorpdfstring{\acrshort{asrbn}}{ASR-BN} Wav2Vec-2.0 vector quantization + \texorpdfstring{\acrshort{f0}}{F0} transformations} \locallabel{fully_vq}
  \vspace{-0.8em}

The experiment presented in Table~\localref{chapt5:bigtable_wav2vec_vq} evaluates the hypothesis of increasing the network depth by using a large Wav2Vec-2.0 model to replace filterbank.
Without vector quantization, our Wav2Vec-2.0 \acrshort{tdnnf} \acrshort{asrbn} extractor does not significantly improve the privacy protection (line 8); the \acrshort{li} on the \textit{VCTK} dataset reaches 0.69, far away of the 0.49 score of the \acrshort{vpc} baseline.
Interestingly, the utility improves compared to the clear speech, the \acrshort{wer} drops from 4.1\% to 3.8\% in the \textit{LibriSpeech} dataset, while in the VCTK dataset, it drops from 12.8\% to 7.8\%.
Improvement of utility is achieved because of the Wav2Vec-2.0 module; the \acrshort{asrbn} is more precise because of the network depth and amount of training data that the Wav2Vec-2.0 was trained on.
Applying voice conversion on precise \acrshort{asrbn} features enhances the speech signal allowing the \acrshort{asr} system to better recognize the linguistic content.

Applying a high vector quantization constraint on this Wav2Vec-2.0 \acrshort{tdnnf} model shows the approach's potential.
With a very small dictionary size of 48 prototypes (line 9), privacy is improved in comparison to the \acrshort{vpc} baseline; the \acrshort{li} on the \textit{VCTK} dataset reaches 0.34 while also having a good utility score of 10.0 \acrshort{wer}.

It is important to note that for utility evaluation, the comparison between the {\scshape{wav2vec2 tdnnf vq} 48} pipeline (line 9) and the \acrshort{vpc} 2022 baseline (line 2) is not fair as the \acrshort{asrbn} extractor are not trained on the same amount of data.
For the {\scshape{wav2vec2 tdnnf vq} 48} approach, the Wav2Vec-2.0 model has been pre-trained on 24.1K hours of unlabeled data, and then the {\scshape{wav2vec2 tdnnf vq} 48} extractor has been finetuned on 100h  of labeled data.
Whereas the \acrshort{vpc} 2022 baseline is trained on 600h of labeled data.
A much more appropriate comparison is between the quantized and non-quantized {\scshape{wav2vec2 tdnnf}} models, which indicates that the utility performance decreased from 7.8\% to 10.0\% of \acrshort{wer} on \textit{VCTK} data, the same can be observed on \textit{LibriSpeech} data.
The same observation can be made about the comparison of the utility performance of any anonymization system to the clear speech reference.
For clear speech, the \acrshort{asr} evaluation model is only trained on \textit{LibriSpeech train-clean-360} to generate the reference \acrshort{wer} score.
Whereas for anonymized speech, the speech is first transformed using models trained on much more than \textit{LibriSpeech train-clean-360}, then evaluated with the \acrshort{asr} evaluation model.
If the anonymization relies on \acrshort{asr} methods, and it does, the anonymization pipeline can be considered as a \acrshort{asr} enhancing preprocessor.
As such, conclusions about a utility improvement from clear speech to anonymized speech should be avoided as they are not evaluated on an even playing field.


\begin{table}[!b]
  \vspace{-1.4em}
  \caption{Privacy and utility results for \acrlong{vq}-based anonymization, using a Wav2Vec-2.0 feature extractor.
  Three kinds of \acrshort{f0} transformations were also tested, note that the linear shift is included by default in our models.}
  \locallabel{chapt5:bigtable_wav2vec_vq}
  \vspace{0.2cm}
  \hspace{-1.100cm}
  \resizebox{1.12\textwidth}{!}{
    \begin{tabular}{
      l@{}@{\extracolsep{0.15mm}}
      S[table-format=1.2]@{\extracolsep{0.01mm}}
      S[table-format=2.1]@{\extracolsep{0.0in}}
      S[table-format=2.1]@{\extracolsep{3.0mm}}
      S[table-format=1.2]@{\extracolsep{0.01mm}}
      S[table-format=2.1]@{\extracolsep{0.0in}}
      S[table-format=2.1]@{\extracolsep{0.00in}}
      }
      \toprule
      \multirow{1}{0pt}{\begin{minipage}{200pt}{\hspace{7mm}Dataset}\end{minipage}}
                                                                                                                                                                                                         & \multicolumn{3}{c}{  \textit{LibriSpeech test-clean}}
                                                                                                                                                                                                         & \multicolumn{3}{c}{ \textit{VCTK test}}                                                                                                                                                                                                                                                                      \\
      \midrule
      \multirow{2}{0pt}{\begin{minipage}{220pt}{\vspace{1mm}\vspace{1mm}\hspace{7mm}Method: \acrshort{asrbn} + \acrshort{f0} transformations}\end{minipage}} & \multicolumn{2}{c}{ \hspace{-2mm} Privacy}            & \multicolumn{1}{c}{ \hspace{-2.5mm} Utility}
                                                                                                                                                                                                         & \multicolumn{2}{c}{ \hspace{-2mm} Privacy}            & \multicolumn{1}{c}{ \hspace{-2.5mm} Utility}                                                                                                                                                                                                         \\
                                                                                                                                                                                                         & \multicolumn{1}{c}{\acrshort{li}~$\downarrow$}        & \multicolumn{1}{c}{\acrshort{eer}~$ \uparrow$} & \multicolumn{1}{c}{\acrshort{wer}~$\downarrow$} & \multicolumn{1}{c}{\acrshort{li}~$\downarrow$} & \multicolumn{1}{c}{\acrshort{eer}~$ \uparrow$} & \multicolumn{1}{c}{\acrshort{wer}~$\downarrow$} \\
      \midrule
      \hspace{2mm}\textattachfileandprintout{\subfix{wavs/01-clear-p226-003-mic2.wav}}{1.} Clear speech      & 0.93 & 4.1  & 4.1 & 0.93 & 2.7  & 12.8 \\
      \midrule
      \hspace{2mm}\textattachfileandprintout{\subfix{wavs/02-vpc-baseline-p226-003-mic2.wav}}{2.} VPC 2022 baseline              & 0.67 & 13.5 & 5.1 & 0.49 & 20.6 & 13.0 \\
      \midrule
      \hspace{2mm}\textattachfileandprintout{\subfix{wavs/08-wav2vec-asrbn-p226-003-mic2-gen.wav}}{8.} \scshape{wav2vec2 tdnnf no vq}                                     & 0.83 & 7.7  & 3.8 & 0.69 & 12.1 & 7.8  \\
      \hspace{2mm}\textattachfileandprintout{\subfix{wavs/09-wav2vec-asrbn-vq-48-p226-003-mic2-gen.wav}}{9.} \scshape{wav2vec2 tdnnf vq}  48                                    & 0.57 & 17.5 & 4.5 & 0.34 & 28.0 & 10.0 \\
      \midrule
      \textattachfileandprintout{\subfix{wavs/10-wav2vec-asrbn-vq-48-awgn-15db-p226-003-mic2-gen.wav}}{10.} \scshape{wav2vec2 tdnnf vq}  48 +  \scshape{f$_0$ {awgn}$_{15dB}$} & 0.44 & 23.4 & 4.6 & 0.12 & 40.8 & 10.3 \\

      \textattachfileandprintout{\subfix{wavs/11-wav2vec-asrbn-vq-48-dp-e1-p226-003-mic2-gen.wav}}{11.} \scshape{wav2vec2 tdnnf vq} 48 + \scshape{f$_0$ lp$_{\epsilon1}$}  & 0.46 & 22.5 & 4.6 & 0.30 & 30.2 & 9.9 \\
      \textattachfileandprintout{\subfix{wavs/12-wav2vec-asrbn-vq-48-vq-4-p226-003-mic2-gen.wav}}{12.} \scshape{wav2vec2 tdnnf vq}  48 +  \scshape{f$_0$ quant$_{4bits}$} & 0.45 & 23.0 & 4.4 & 0.14 & 39.8 & 9.9 \\
      \bottomrule
    \end{tabular}
  }
  \vspace{-1.0em}
\end{table}

The \textit{LibriSpeech} \acrshort{eer} of 17.5\% in line 9 is close to the one obtained in the \acrshort{f0} \hyperref[main:isolated_privacy_eval_f0]{\say{\acrshort{f0} isolated privacy evaluation}} of Section~\localref{isolated_privacy_eval_f0} (\acrshort{eer} of 19.9\%).
This indicates that the \acrshort{f0} trajectory might be the limiting factor of the anonymization.
As such for lines 10 to 12 we experimented with \acrshort{f0} transformations.
Line 10 presents the \acrfull{awgn} noised \acrshort{f0} transformation (see~Section~\localref{chapt5:awgn}) with a target noise power of 15dB.
Line 11 presents the Laplace noised modified bottleneck auto-encoder transformation presented in (see Section~\localref{laplace}) with an $\epsilon$ value of 1 as it appeared to be the most effective in \cite{dp_vpc}.
And finally, line 12 presents the quantization-based \acrshort{f0} transformation (see Section~\localref{chapt5:f0_vq}) with four quantization bits.

With the \acrshort{f0} \acrshort{awgn} transform, the privacy protection increase, as the \acrshort{li} on the \textit{VCTK} dataset plummeted down to 0.12 while keeping a very high utility with 10.3\% of \acrshort{wer}, similar behavior can be observed in the \textit{LibriSpeech} dataset.
The Laplace noised-based \acrshort{f0} auto-encoder transformation has similar privacy performance on the \textit{LibriSpeech} dataset as the \acrshort{awgn}.
However, it is not the case for the \textit{VCTK} dataset, where the privacy \acrshort{li} score equals 0.30 instead of being close to 0.12.

One hypothesis for this disparity is that adding noise to the bottleneck of a model to later obtain some features, is different from adding noise directly to the feature.
We hypothesize that during training, the model compensates for noise addition with a form of denoising such that for a particular noise level, the model loss still converges.
This means that it is complicated to properly define the noise level as the model convergence depends on this noise level but also on the training data.
We argue that for Laplace bottleneck noise addition transformations, it is easy to go from an appropriate transformation for a dataset (\textit{LibriSpeech} in line 11) to an inappropriate (too small) transformation for another dataset (\textit{VCTK}).
We refer to this as the sensitivity of a transformation to generalize to multiple datasets.

\renewcommand*{\thefootnote}{\fnsymbol{footnote}}
Finally, the \acrshort{f0} quantization transformation appears to be a very compelling option compared to noised-based transformations.
The privacy performances are similar to the \acrshort{awgn} \acrshort{f0} transformation while being slightly better in terms of utility.
Additionally, informal subjective listening tests report that the naturalness of the \say{{\scshape{wav2vec2 tdnnf vq}} 48 + {\scshape{f$_0$ quant$_{4bits}$}}}  pipeline, exceeded the \say{{\scshape{wav2vec2 tdnnf vq}} 48 + {\scshape{f$_0$ awgn$_{15dB}$}}} pipeline\protect\footnotemark[4].

\subsubsection{\texorpdfstring{\acrshort{asrbn}}{ASR-BN} Wav2Vec-2.0 Laplace noise + \texorpdfstring{\acrshort{f0}}{F0} transformations}

In this experiment, we attempt to reproduce the \acrshort{asrbn} bottleneck Laplace-noise transformation of \cite{dp_vpc} with our speaker anonymization pipeline presented in Figure~\localref{image_chapt5:my_anon_model}.
We use a Wav2Vec-2.0 feature extractor for a fair comparison with our best \say{{\scshape{wav2vec2 tdnnf vq}} 48 + {\scshape{f$_0$ quant$_{4bits}$}}}  pipeline referred to as \say{fully quantization-based pipeline}.
Another implementation difference to the original paper relates to how the noise is applied to the bottleneck, to match the \acrshort{f0} {\scshape{lp}} implementation, noise is sampled and applied for the whole utterance dimension, rather than every time for each frame.
Additionally, in contrast to the evaluation performed in the original paper, we use a white-box {informed} attacker and analyze our results in both \textit{LibriSpeech} and \textit{VCTK} datasets.

Table~\localref{chapt5:bigtable_wav2vec_noise} compares the privacy and utility results to our fully quantization-based pipeline.
For all Laplace noise ({\scshape{lp}}) results, we can observe that for the \textit{LibriSpeech} dataset, the pipeline has a lot of trouble preserving the utility.
Whereas for the \textit{VCTK} the strength of noise addition seemed to be appropriate as the utility is not as strongly affected.
This observation supports our previous hypothesis about the sensitivity of noise addition bottleneck transformations to generalize to multiple datasets.

We can observe that adding more noise to the bottleneck decreases utility preservation as lower $\epsilon$ values tend to have higher \acrshort{wer}.
For an $\epsilon$ of 130000 (line 13), the \acrshort{wer} is the same as our fully quantization-based pipeline with a value of 9.9\%.
Increasing the noise with an $\epsilon$ of 100000 (line 16) decreases the utility with a \acrshort{wer} value of 12.9\% in \textit{VCTK}.
We notice that compared to the \acrshort{vq} approach, \acrshort{f0} transformations tend to decrease the utility more significantly.
The {\scshape{f$0$ lp${\epsilon1}$}} transformation results in a higher \acrshort{wer} in \textit{VCTK} lines 14 to 15 and 17 to 18, indicating that it performs worse than \acrshort{awgn}.

In terms of privacy results, we also focus on the \textit{VCTK} results as the \acrshort{wer} utility performance is similar to our fully quantization-based pipeline in \textit{VCTK}.
First, we notice that \acrshort{f0} transformation is also necessary to generate the most unlinkable speech.
The best of them in privacy (and utility) seems to be the \acrshort{awgn} transformations applied with a target noise power of 5dB.
The pipeline the most comparable to our fully quantization-based pipeline is the \say{{{\scshape{wav2vec2 tdnnf lp}}$_{e130000}$ + \scshape{f$_0$ {awgn}$_{5dB}$}}} one (line 14) as the \acrshort{wer} is the same as ours.
We observe that this pipeline has a similar privacy performance with a \acrshort{li} of 0.12 (compared to 0.14).
However, we argue that this fully noised-based pipeline might produce slightly lower naturalness than the fully quantization-based pipeline, and is highly more sensible to the input dataset.
Increasing the noise with an $\epsilon$ of 100000 (line 17) when the \acrshort{f0} is transformed does not necessarily increase the privacy, as the privacy results between lines 16 and 13 are the same for the \textit{VCTK} dataset.
However, it significantly decreases utility as the \acrshort{wer} increases from 10.2\% to 13.8\%.

Regarding the results from \textit{LibriSpeech}, it is important to note that the decrease in utility performance is not always proportionate to the level of privacy enhancement.
The absolute privacy improvement between lines 12 and 13 is 0.06 of \acrshort{li}, while the absolute utility degradation is by 16.6\% of \acrshort{wer}.

\begin{table}[!bhbt]
  \vspace{-0.4em}
  \caption{Privacy and utility results for Laplace noise-based \acrshort{asrbn} transformation and the two noised based \acrshort{f0} transformations.}
  \locallabel{chapt5:bigtable_wav2vec_noise}
  \vspace{0.2cm}
  \hspace{-1.000cm}
  \resizebox{1.12\textwidth}{!}{
    \begin{tabular}{
      l@{}@{\extracolsep{0.15mm}}
      S[table-format=1.2]@{\extracolsep{0.01mm}}
      S[table-format=2.1]@{\extracolsep{0.0in}}
      S[table-format=2.1]@{\extracolsep{3.0mm}}
      S[table-format=1.2]@{\extracolsep{0.01mm}}
      S[table-format=2.1]@{\extracolsep{0.0in}}
      S[table-format=2.1]@{\extracolsep{0.00in}}
      }
      \toprule
      \multirow{1}{0pt}{\begin{minipage}{200pt}{\hspace{7mm}Dataset}\end{minipage}}
                                                                                                                                                                                                         & \multicolumn{3}{c}{  \textit{LibriSpeech test-clean}}
                                                                                                                                                                                                         & \multicolumn{3}{c}{ \textit{VCTK test}}                                                                                                                                                                                                                                                                      \\
      \midrule
      \multirow{2}{0pt}{\begin{minipage}{220pt}{\vspace{1mm}\vspace{1mm}\hspace{7mm}Method: \acrshort{asrbn} + \acrshort{f0} transformations}\end{minipage}} & \multicolumn{2}{c}{ \hspace{-2mm} Privacy}            & \multicolumn{1}{c}{ \hspace{-2.5mm} Utility}
                                                                                                                                                                                                         & \multicolumn{2}{c}{ \hspace{-2mm} Privacy}            & \multicolumn{1}{c}{ \hspace{-2.5mm} Utility}                                                                                                                                                                                                         \\
                                                                                                                                                                                                         & \multicolumn{1}{c}{\acrshort{li}~$\downarrow$}        & \multicolumn{1}{c}{\acrshort{eer}~$ \uparrow$} & \multicolumn{1}{c}{\acrshort{wer}~$\downarrow$} & \multicolumn{1}{c}{\acrshort{li}~$\downarrow$} & \multicolumn{1}{c}{\acrshort{eer}~$ \uparrow$} & \multicolumn{1}{c}{\acrshort{wer}~$\downarrow$} \\
      \midrule
      \hspace{2mm}\textattachfileandprintout{\subfix{wavs/01-clear-p226-003-mic2.wav}}{1.} Clear speech      & 0.93 & 4.1  & 4.1 & 0.93 & 2.7  & 12.8 \\
      \midrule
      \hspace{2mm}\textattachfileandprintout{\subfix{wavs/02-vpc-baseline-p226-003-mic2.wav}}{2.} VPC 2022 baseline     & 0.67 & 13.5 & 5.1 & 0.49 & 20.6 & 13.0 \\
      \midrule
      \textattachfileandprintout{\subfix{wavs/12-wav2vec-asrbn-vq-48-vq-4-p226-003-mic2-gen.wav}}{12.} \scshape{wav2vec2 tdnnf vq}  48 +  \scshape{f$_0$ quant$_{4bits}$} & 0.45 & 23.0 & 4.4 & 0.14 & 39.8 & 9.9 \\
      \midrule
      \textattachfileandprintout{\subfix{wavs/13-wav2vec-asrbn-dp-e130000-p226-003-mic2.wav}}{13.} {\scshape{wav2vec2 tdnnf lp}}$_{e130000}$                                    & 0.39 & 25.7 & 21.0 & 0.25 & 33.4 & 9.9  \\
      \textattachfileandprintout{\subfix{wavs/14-wav2vec-asrbn-dp-e130000-awgn-5db-p226-003-mic2.wav}}{14.} {\scshape{wav2vec2 tdnnf lp}}$_{e130000}$ + \scshape{f$_0$ {awgn}$_{5dB}$}   & 0.25 & 33.3 & 22.7 & 0.12 & 41.4 & 10.2 \\
      \textattachfileandprintout{\subfix{wavs/15-wav2vec-asrbn-dp-e130000-dp-e1-p226-003-mic2.wav}}{15.} {\scshape{wav2vec2 tdnnf lp}}$_{e130000}$ + \scshape{f$_0$ lp$_{\epsilon1}$} & 0.24 & 33.9 & 24.0 & 0.19 & 36.3 & 10.4 \\
      \textattachfileandprintout{\subfix{wavs/16-wav2vec-asrbn-dp-e100000-p226-003-mic2.wav}}{16.} {\scshape{wav2vec2 tdnnf lp}}$_{e100000}$                                    & 0.36 & 27.0 & 27.4 & 0.27 & 32.2 & 12.9 \\
      \textattachfileandprintout{\subfix{wavs/17-wav2vec-asrbn-dp-e100000-awgn-5db-p226-003-mic2.wav}}{17.} {\scshape{wav2vec2 tdnnf lp}}$_{e100000}$ + \scshape{f$_0$ {awgn}$_{5dB}$}   & 0.16 & 34.5 & 28.9 & 0.12 & 41.5 & 13.8 \\
      \textattachfileandprintout{\subfix{wavs/18-wav2vec-asrbn-dp-e100000-dp-e1-p226-003-mic2.wav}}{18.} {\scshape{wav2vec2 tdnnf lp}}$_{e100000}$ + \scshape{f$_0$ lp$_{\epsilon1}$} & 0.23 & 34.8 & 31.3 & 0.23 & 34.4 & 14.2 \\
      \bottomrule
    \end{tabular}
  }
  \vspace{-0.8em}
\end{table}

\subsection{Discussion}

In this section dedicated to the \acrshort{asrbn} feature, we evaluated the degree of disentanglement between the \acrshort{asrbn} and the speaker and concluded that the \acrshort{asrbn} can be used to perform speaker linkability attacker at a very high rate.
Indeed the \acrshort{asrbn} feature, commonly used by the anonymization pipeline,  leaks speaker identities as the \acrshort{eer} of the \acrshort{asrbn} in our isolated experiment was 6.7\%, very close to the \acrshort{mfcc} \acrshort{eer} score of 3.7\%.
This implies that to better guarantee proper anonymized speech generation by the speech synthesis system, the \acrshort{asrbn} should be transformed.

We presented common \acrshort{asrbn} transformation based on bottleneck Laplace noise addition used by differential privacy, and adversarial training before proposing a novel vector quantization bottleneck transformation approach.
We showed that the fully quantization-based pipeline, where the \acrshort{f0} and \acrshort{asrbn} are quantized achieves the best privacy/utility performance in the \acrshort{vpc} testing datasets.
In contrast to noise-based pipelines, the quantization-based one did not have the same sensibility issue as, good utility where recorded regardless of the dataset.

During our experiment we played with the Wav2Vec-2.0 feature extractor to replace the traditional filterbank \acrshort{asrbn} input feature.
Using the Wav2Vec-2.0 feature extractor to help process speech and output \acrshort{asrbn} representation increases the number of layers of the \acrshort{asrbn} extractor compared to previously.
Additionally, the extractor has effectively seen more training samples as Wav2Vec-2.0 requires large unlabeled data to be trained.
This new architecture allows extracting more precise \acrshort{asrbn} which is beneficial for \acrshort{wer} utility, and with such a large network, it begins to be possible to constrain the extractor to only a few discrete quantized vectors because the network has the encoding capacity to transform the speech signal into a compressed high-level representation.
In our experiment with the \acrshort{vpc} testing dataset, 48 discrete prototypes seemed to be enough with the Wav2Vec-2.0 feature extractor (the fewer the number of discrete quantized vectors, the better the privacy).
However, the dataset of the \acrshort{vpc} uses clean speech, which is favorable for this approach, under noisier environments the quantization layer has much more trouble identifying the appropriate discrete vector, leading to large utility decreases.
This could be compensated by increasing the number of discrete prototypes but at the cost of privacy degradation.
The number of discrete prototypes can be adjusted to achieve a suitable privacy/utility trade-off \cite{privacy-utility-tradeoff,understanding_tradeoff_privacy} depending on the application.
Possible extensions of this work relate to how the discrete prototypes are selected, currently, an L2 distance between the continuous and discrete vector is used but a personalized per-speaker function could be used to increase utility.
Additionally, for each frame, the number of possible discrete prototypes is currently fixed (e.g., 48), having this number dynamically increased depending on the level of uncertainty of the \acrshort{asrbn} to correctly capture the acoustic unit would be an interesting research track.

\section{Conclusion}

In this chapter, we have challenged the \acrshort{f0} and \acrshort{asrbn} attributes of voice conversion-based speaker anonymization.
We found with isolated evaluations, that they are not disentangled from speaker information limiting the performance of speaker anonymization.
As such, we presented and proposed transformations to improve disentanglement.
Our approach based on quantization has advantages compared to the more traditional one based on noise addition.
Quantization-based disentanglement allows us to achieve similar privacy performance as noised-based transformations while having superior \acrshort{wer} utility, additionally, we argue that the produced anonymized speech is also more natural which must be assessed through more perceptual tests.

One important observation made concerns the objective utility measurement.
We discourage people to claim that an anonymization system improves objective utility even though the \acrshort{wer} score might be lower.
The main reason why the \acrshort{wer} score can be lower compared to a clear speech reference comes from the disparity of training and the amount of data used to train the anonymization pipeline and the evaluation model.

\ifSubfilesClassLoaded{
  \printglossary[title=Special Terms,type=\acronymtype]
  \printbibliography
}{}

\end{document}

\clearemptydoublepage
\cleartooddpage[\thispagestyle{empty}]
\documentclass[../main.tex]{subfiles}

\ifSubfilesClassLoaded{
    \tableofcontentsfile
    \dominitoc
    \setcounter{chapter}{5} 
    \externaldocument[]{../main}
    \def\locallabelprefix{chapt_6}
}{}

\begin{document}

\selectlanguage{english}

\graphicspath{{./figures/dist}}

\chapter{New kinds of privacy and utility measurements} \locallabel{chapt6}
\vspace{-3em}
\minitoc
\section{Introduction}
This chapter unveils novel measurement metrics and techniques, offering a fresh perspective on privacy and utility assessment beyond the traditional \acrshort{eer}/\acrshort{li} and \acrshort{wer} metrics.
To begin with, we propose a new privacy measurement technique that aims to evaluate the invertibility of anonymization.
As such, we will introduce simple inversion attacks which aim to reconstruct clear x-vectors from anonymized ones.
If it is possible to invert the anonymization, even partially, it could help the attacker to better perform a linkability attack, but also it opens the room for many other undesirable use such as voice-cloning to impersonate a person's voice.
We finish this chapter by proposing a utility measurement that aims to better assess the degree of linguistic content preservation from a subjective point of view.
This measurement aims to overcome the \acrshort{wer}-based evaluation limitation, which cannot differentiate between words correctly pronounced and wrongly recognized by the \acrshort{asr} evaluation system and words that are simply not correctly pronounced.

\section{Invertibility evaluation using embedding alignment}
In Section~\ref{main:chapt3:sec:legal}, we presented the \textit{non-invertibility} and \textit{unlinkability} criteria defined by the European standard ISO/IEC 24745 \cite{iso24754} that apply to data anonymization.
Up until now, the linkability criteria is considered the main threat that speaker anonymization should defend against.
The use of the white-box \acrshort{asv} attack allowed us to well evaluate this aspect and output \acrshort{eer}/\acrshort{li} metrics used up until now.

However, when it comes to invertibility evaluation, to the best of our knowledge, no work has been done to evaluate this aspect.
As such, the subject of this section is to propose an invertibility attacker which enables an invertibility evaluation.
Conducted with Thebaud Thomas, we proposed to invert voice conversion-based anonymization using embedding alignment techniques.
Later on, other invertibility methods were extended to signal processing speaker anonymization \cite{invert_odyssey}.

\vspace{-2mm}
\subsection{Evaluation setup}
\vspace{-1mm}

We propose to invert the \acrshort{vpc} 2022 baseline and our fully quantization-based anonymization pipeline (with Wav2Vec-2.0 and a \acrshort{vq} dictionary size of 48, see Section~\ref{main:fully_vq}), both of them conditioned with the \say{constant speaker} target selection strategy.
We assume that the attacker has access to a compromised speech dataset which can be transformed with the same anonymization pipeline as the one used to anonymize the vulnerable dataset.
The attacker also trains a white-box \acrshort{asv} system that is adapted to extract x-vectors from anonymized speech.

As such, it is realistic to assume that the attacker has a dataset of clear x-vectors for which he knows the corresponding anonymized x-vectors.
Given the parallel clear/anonymized x-vectors, the attacker can approximate the anonymization function as a rotation from the clear x-vector to the anonymized one.
In this study, the anonymization function is approximated using a rotation matrix, taking the anonymized x-vector as input and producing the corresponding estimated clear x-vector.
This scenario of having clear/anonymized 1-to-1 mapping allows estimating the rotation matrix using a Procrustes Analysis~\cite{gower1975generalized}.
We refer to this scenario as the supervised scenario.

In addition to this supervised scenario, we propose a more restrictive scenario where the attacker does not know the link between the clear x-vector and the anonymized x-vector in the compromised dataset.
In this scenario, we use an unsupervised embedding alignment algorithm named Wasserstein Procrustes~\cite{grave2018unsupervised}.

With the rotation matrix estimated, the anonymized x-vector is inverted to estimate the corresponding clear x-vector. 
Then, invertibility is evaluated by measuring how well the attacker can accurately (ACC) reconstruct the clear (non-anonymized) vulnerable x-vector.
Finally, we can also evaluate linkability between clear compromised and inverted anonymized vulnerable x-vectors.
Figure \localref{img:inversion_flow} summarizes this process.

\begin{figure}[htbp]
    \begin{center}
        \includegraphics[width=1\linewidth]{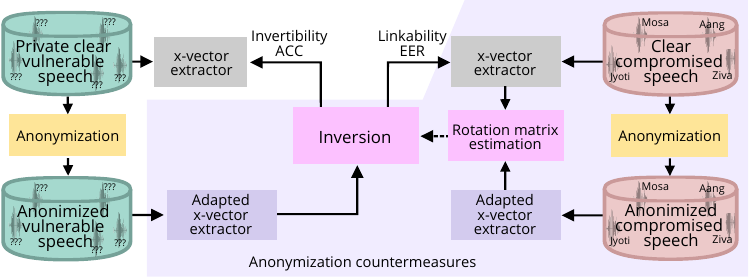}
    \end{center}
    \vspace{-6mm}
    \caption{
        Illustration of training the rotation matrix and inverting anonymized x-vectors. Invertibility and linkability measurements are then performed with clear and inverted x-vectors.
    }
    \locallabel{img:inversion_flow}
\end{figure}

\subsubsection{Metrics}

We use two metrics to evaluate our attack on the different scenarios: the \acrshort{eer} and the Top 1 speaker accuracy.
For all experiments, the \acrshort{eer} is computed by scoring the inverted x-vectors against the clear ones.
In contrast with the scoring done in previous chapters, here we use the cosine similarity instead of the \acrshort{plda}.
For this reason, lower linkability performance is expected (higher \acrshort{eer}) as the x-vector system that we are using is not optimized with angular/cosine-margin-based losses.

The Top 1 speaker accuracy is computed by comparing the inverted anonymized vulnerable utterances against their clear counterparts.
For each inverted x-vector, we looked for the nearest neighbor speaker x-vector from the clear vulnerable x-vector dataset.
The Euclidean distance was used to find the nearest neighbor because, during Procrustes analysis, the rotation matrix was estimated for that distance metric. 
The Top 1 speaker accuracy is the proportion of inverted x-vectors for which the closest clear vulnerable x-vector is from the same speaker.
A high Top 1 speaker accuracy means a high success in inverting x-vectors close to their clear counterpart and should raise concerns regarding the \textit{non-invertibility} property of the speaker anonymization system.

\vspace{-2mm}
\subsection{Suppervised and unsupervised embedding alignment}
\vspace{-2mm}

Computing the alignment of two embeddings of high dimensional real vectors is one of the fundamental problems in machine learning, with applications examples such as unsupervised word and sentence translation~\cite{rapp1995identifying,fung1995compiling,bojanowski2017unsupervised,grave2018unsupervised,biswas2020aligning}.
In this section, we introduce supervised (Procrustes) and unsupervised (Wasserstein Procrustes) embedding alignment algorithms that we will rely on to invert the anonymization.

\subsubsection{Procrustes Analysis}
\locallabel{subsec:algo_procrustes}
Let ${A}$ and ${B}$ be two sets of $N$ high dimensional real vectors of dimension $d$.
We want to find the optimal rotation ${{W}}\in \mathbb{R}^{d\times d}$ that minimizes the squared distance between both sets:
\begin{equation}
    \min_{{{W}}\in \mathbb{R}^{d\times d}} ||{A}{{W}} - {B} ||^2_2
    \locallabel{eq:proc}
\end{equation}
For correctly parallel sets ${A}$ and ${B}$ (the $n^{th}$ element of ${A}$ corresponds to the $n^{th}$ element ${B}$, $\forall n \in [\![1,N]\!]$), we can directly use Procrustes analysis~\cite{gower1975generalized} to compute an optimal ${{W}}$.
%
The operation consists of finding in the space of orthogonal matrices $\mathcal{O}_D$ the singular value decomposition:
${U} \Sigma {V^T} = {A}^T{B}$,
then later obtain ${{W}}$ with: ${{W}} = {U} {V^T}$.
This approach is well suited for supervised scenarios as it requires access to the labels of both sets to align them.
%
For two sets where the mapping is unknown, an unsupervised alignment algorithm is required. 

\subsubsection{Wasserstein Procrustes} \locallabel{subsec:algo_Wasserstein}

In the study conducted by \cite{grave2018unsupervised}, an unsupervised algorithm was proposed to align sets of language-dependent word embeddings for the purpose of unsupervised translations.
The proposed algorithm utilizes a stochastic optimization approach, alternating between minimizing the Wasserstein distance between sets and finding the optimal rotation using Procrustes analysis.
This approach enables identifying a rotation matrix that optimally lowers the distance between the two sets of embeddings, while also determining their one-to-one mapping.

\paragraph{Wasserstein distance} \cite{ruschendorf1985wasserstein} is a measure of the optimal transportation required to move one set of points to the positions of a second group.
If the $n^{th}$ element of $A$ corresponds to the $n^{th}$ element of $B$, $\forall n \in [\![1,N]\!]$, then the distance between $A$ and $B$ is: $||A-B||_2^2$.
However, if $A$ and $B$ are not ordered, it is necessary to match each point of one set with a point of the other.
To do this, we use a permutation matrix.
Let $\mathcal{P}_N$ be the set of permutation matrices that ensure a 1 to 1 mapping: $\mathcal{P}_N=\left\{P \in\{0,1\}^{N \times N}, P {1}_N={1}_N, P^T {1}_N={1}_N\right\}$.
In our problem, we are looking for a permutation matrix $P \in \mathcal{P}_N$ that will minimize the Wasserstein distance between two groups of points:
\vspace{-0.4em}
\begin{equation}
\vspace{-0.4em}
    P=\min _{P \in \mathcal{P}_N}||{A}-P {B}||_2^2
    \locallabel{eq:wassser}
\end{equation}

\paragraph{Stochastic optimization} In our problem, we aim to simultaneously solve equations \localref{eq:proc} and \localref{eq:wassser} by making use of a stochastic approach for Procrustes analysis in Wasserstein distance:
\vspace{-0.4em}
\begin{equation}
    \min _{W \in \mathcal{O}_D} \min _{P \in \mathcal{P}_N}||{A} W-P {B}||_2^2
\vspace{-0.4em}
\end{equation}

\noindent
Described in \cite{grave2018unsupervised}, the method alternate between the selection of random sub-sets of ${A}$ and ${B}$, the calculation of permutation matrix $P$ and the update of $W$.

\subsection{Experimental attack scenarios}

In this section, we first explain the scenarii and then describe the dataset accessibility hypotheses, to end with the implementation details that improve the rotation.

\vspace{-0.4em}
\subsubsection{Scenarios}
\locallabel{subsec:scenario}
\vspace{-0.4em}
In the following, we explain our motivation for experimenting with multiple attack scenarios.
First, we will present supervised and unsupervised attack conditions using Procrustes and Wasserstein Procrustes trained on compromised x-vectors.
Those two attacks are realistic, but, as they require estimating the rotation with a training dataset, the inversion attack will not be perfect. 

To evaluate the privacy protection using a more powerful attack, we use a non-realistic \textit{oracle} scenario, where the attacker has access to clear vulnerable speech to estimate the rotation.
This rotation should be close to optimum as estimating the rotation between clear vulnerable speech and anonymized vulnerable speech should give the best rotation matrix to invert the anonymized vulnerable speech.
This allows us to better challenge the \textit{non-invertibility} privacy criteria, as it directly challenges the definition requiring that \say{it should be computationally infeasible to obtain the clear data that led to any given anonymized data}.

For all the following scenarios, we approximate the anonymization function in the x-vector domain.
The rotation matrices are estimated to match each x-vector utterance of one dataset to another one.
Once the rotation matrix ${W}\in\mathbb{R}^{d\times d}$ is estimated, the  anonymized vulnerable x-vectors dataset is inverted using the transposed ${W}$:

\paragraph{Supervised scenario: Procrustes}
\locallabel{subsubsec:scenar_supervised}
This first scenario follows the dataset requirement of the \acrshort{vpc}.
To obtain the rotation matrix, a Procrustes analysis is applied to clear and anonymized compromised x-vector datasets, knowing the one-to-one correspondence between them.
Then the anonymized vulnerable x-vectors are inverted.
%
The goal of this first experiment is to estimate how well a rotation can approximate the anonymization pipeline in the x-vector domain.
During the evaluation we will evaluate:
\begin{itemize}
    \item How linkable the inverted x-vectors are to clear x-vectors? This will be compared with previous linkability evaluations under similar conditions.
    \item How many anonymized vulnerable x-vectors can be inverted well enough to recognize their clear source speaker?
\end{itemize}

\paragraph{Unsupervised scenario: Wasserstein-Procrustes}
\locallabel{subsubsec:scenar_unsupervised}
This second experiment explores the performance of an unsupervised algorithm for the invertibility attack.
In contrast to the previous one, the clear and anonymized compromised x-vector mapping is unknown.
This evaluation is the first step in knowing if unsupervised algorithms can work at all to create a clear/anonymized inversion attack.
This work could be extended into trying to match nonparallel datasets such as the anonymized vulnerable x-vector dataset and the clear or anonymized compromised x-vector datasets.




\paragraph{Oracle scenarios}
\locallabel{subsubsec:scenar_oracle}
This third and last experiment probes the optimal performances an attacker can get while approximating a speech anonymization system with rotations.
In contrast to the two previous scenarios, here the datasets used to estimate the rotations are also the test datasets (the clear and anonymized vulnerable x-vector datasets).
The two Procrustes and Wasserstein Procrustes algorithms presented above will be used here.
In those scenarios, we are more likely to evaluate the strength of the privacy protection rather than the attack/estimation of the rotation.
Here what we query is the degree of the bijection between clear and anonymized x-vectors and if the transformation between clear and anonymized space can be associated with a rotation.
If a rotation exists, it indicates an anonymization weakness, as it is possible to identify speakers in the anonymized x-vector space, and a bijective reciprocal function exists to associate each anonymized x-vector to a unique clear x-vector.

As the goal of the \say{constant speaker} target selection strategy is to generate a unique speaker, the oracle scenarios goal is to control the successfulness of the voice conversion anonymization system to generate a single and unique identity.
No bijection between one x-vector space to the other should exist once speakers are anonymized. 

\vspace{-0.4em}
\subsubsection{Dataset accessibility hypotheses}
\vspace{-0.4em}
The datasets accessibility hypotheses are summarized in Figure \localref{fig:scenars} for the different scenarios detailed above.
Purple boxes show data available to the attacker to train the rotation matrix in a given hypothesis.
Black hatched boxes show data inaccessible to the attacker in a given hypothesis.
We call supervised the scenarios where labels are available (represented as $\leftrightarrow$) to align the datasets and unsupervised the ones where the mapping is inaccessible.
The scenarios where the attacker has access to clear vulnerable speech allow testing the rotation effectiveness with close to perfect attack (named the \textit{oracle} scenario and presented below).
Regardless of the available data, the performances are evaluated using clear vulnerable speech and (inverted) anonymized vulnerable speech for invertibility assessment, and clear compromised speech and (inverted) anonymized vulnerable speech for linkability assessment.

\begin{figure}[!htb]
    \centering
    \includegraphics[width=0.77\linewidth]{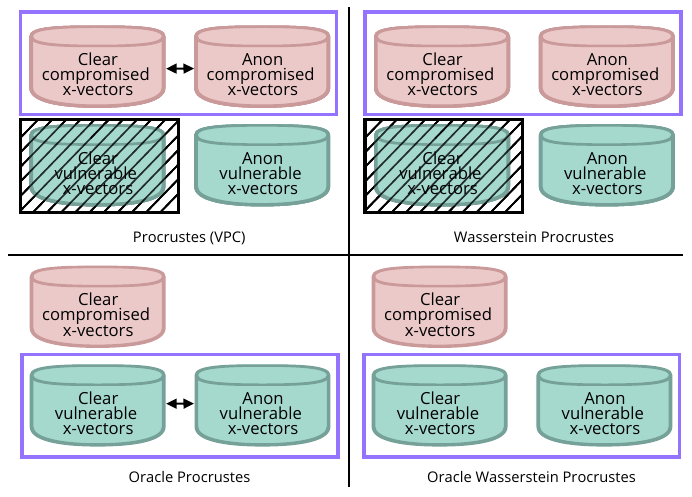}
    \caption{Schematic representation of the datasets used for different scenarios.}
    \locallabel{fig:scenars}
\vspace{-0.8em}
\end{figure}

{\let\thefootnote\relax\footnote{{The Python code of the experiments is available at \url{https://github.com/deep-privacy/x-vector-procrustes}}}}

\vspace{-0.8em}
\subsubsection{Implementation}
\vspace{-0.6em}
\locallabel{subsec:variations}
To improve the rotation performance, we extend the range of our experiments to modified x-vectors domains using principal component analysis and gender-dependent rotation estimation.
\vspace{-1.0em}
\paragraph{Principal component analysis}
\locallabel{subsec:PCA}
We apply a dimensional reduction technique to the x-vectors datasets using \acrfull{pca} \cite{pca}.
Reducing the number of dimensions reduces the candidate rotations manifold, simplifying the search for the optimal one.
The \acrshort{pca} also orders the dimensions precisely: the dimensions with the higher variance are placed first.
This means that applying \acrshort{pca} on two vector datasets acts as a pre-alignment, easing the following alignment process.
%
We reduced to 70 dimensions the originally 512 dimensions x-vectors using \acrshort{pca} the total explained variance ratio was always above 96.0\%.


\vspace{-1.0em}
\paragraph{Gender dependent training}
\locallabel{subsec:gender}
The \acrshort{vpc} evaluation allows for the attacker to have access to the gender information associated with each clear and anonymized utterance. 
We utilized this aspect to train two separate rotations, which improve the attack performance.


\vspace{-0.8em}
\subsection{Experimental results}
\vspace{-0.2em}

This section presents the experimental results for the invertibility scenarios presented above.
Table~\localref{tab:results_algn_vpc} summarizes the results for the attacks on the \acrshort{vpc} 2022 baseline, and table~\localref{tab:results_algn_fully_vq} the attacks on our fully quantization-based anonymization pipeline.

The first line of each table corresponds to the linkability assessment on clear speech.
The clear speech linkability results show a pattern similar to those presented in Table~\ref{main:table:all_eer_dsys}, although there is a noticeable increase in the \acrshort{eer} values, from 7.7\% to 10.3\% for female speakers and from 1.1\% to 2.9\% for male speakers.
This \acrshort{eer} increase is caused by the use of the cosine similarity instead of the \acrshort{plda}.

The second line of each table corresponds to the more traditional white-box linkability assessment on anonymized speech.
Similarly, as for the clear speech \acrshort{eer}, the anonymized \acrshort{eer} are degraded compared to when using the \acrshort{plda}.
For Table~\localref{tab:results_algn_vpc} on the \acrshort{vpc} baseline, the white-box \acrshort{asv} \acrshort{eer} drops of by 14.5\% of \acrshort{eer} (absolute) compare to Table~\ref{main:chapt5:bigtable_adv}.
This indicates that, especially for anonymized speech, the x-vector model that we are using indeed requires a \acrshort{plda} scoring function to better disclose the degree of privacy protection.

For lines 1 and 2, only the \acrshort{eer} is computed, because the attackers used cannot invert the anonymization.
Lines 3 to 6 explore our invertibility attacks.
We can see that for both tables, Procrustes/Wasserstein Procrustes and their oracle version reduce the \acrshort{eer} score.
This indicates that applying a rotation on the anonymized x-vector before comparing them to clear x-vectors improves the attacker compared to the (cosine similarity-based) white-box \acrshort{asv} attack, where anonymized utterances are compared with anonymized utterances.

\vspace{-0.8em}
\begin{table*}[!htb]
    \caption{Experimental results for the rotation-based invertibility attack scenarios on the \acrshort{vpc} 2022 baseline. The scoring function for \acrshort{asv} linkability assessment is cosine similarity.}
      \vspace{0.2cm}
    \centering
    \begin{tabular}{r
      l@{}@{\extracolsep{8.0mm}}
      S[table-format=2.1]@{\extracolsep{7.0mm}}
      S[table-format=2.1]@{\extracolsep{10.0mm}}
      S[table-format=2.1]@{\extracolsep{7.0mm}}
      S[table-format=2.1]@{\extracolsep{3.0mm}}
      }
    \toprule
         \multicolumn{2}{c}{} &  \multicolumn{2}{c}{\hspace{2mm}\acrshort{eer}~$\uparrow$}& \multicolumn{2}{c}{\hspace{2mm}ACC~$\downarrow$} \\
         \multicolumn{2}{c}{} &  {Female} & {Male} & {Female} & {Male} \\
    \midrule
         1 & Clear speech       & 10.3   & 2.9    &           &       \\
         2 & White-box \acrshort{asv}   &         27.9   & 27.8   &           &       \\
    \midrule
         3 & \multirow{1}{*}{Procrustes}   &         19.5   &  21.4  & 67.7   &  51.8  \\
         4 & \multirow{1}{*}{Wasserstein Procrustes} &    20.1   &  22.9   &  64.7   & 50.0  \\
    \midrule
         5 & Oracle Procrustes &  17.1 & 12.0  &  98.3   &  97.0 \\
         6 & Oracle Wasserstein Procrustes  & 17.7 & 12.8 &  99.0   &  97.2 \\
    \bottomrule
    \end{tabular}
    \locallabel{tab:results_algn_vpc}
    \vspace{-0.2em}
\end{table*}

Focusing on Table~\localref{tab:results_algn_vpc}, the Procrustes invertibility attack (line 3) manages to achieve a reconstruction accuracy of 67\% for females and 51.8\% for males speaker meaning that more than half of the time, the anonymized speaker x-vectors can be re-identified.
This raises concerns about the \textit{non-invertibility} criteria of the anonymization system.
Lines 4 explore the unsupervised scenario.
We can see that Wasserstein Procrustes gives slightly worse results than the Procrustes counterpart, as no labels are available in this scenario requiring the attacker to estimate permutation and rotation matrices.
We underline that the difference is usually around a few percent in \acrshort{eer} and inversion accuracy, so the distribution of x-vectors before and after anonymization is probably quite similar.
Similar enough to get close results to when labels are available.
Lines 5 and 6 show the results associated with the oracle approaches.
We observe that in terms of linkability, the \acrshort{eer} are getting closer to the \acrshort{plda}-based white-box \acrshort{asv}, as from line 2 to line 5, an absolute decrease of 10\% of \acrshort{eer} is observed.
Still, \acrshort{plda}-based white-box \acrshort{asv} is a better linkability evaluation, and this is because of the \acrshort{plda} calibration which relies on more data than the data used to estimate the rotation (360h vs 45min).
In terms of invertibility, the almost 100\% accuracy means that the anonymized speaker x-vectors can be re-identified using a rotation.
As the anonymization pipeline should produce a single unique voice (because of the \say{constant speaker} target selection), having a 100\% of accuracy is completely undesirable as it means that a bijection between clear and anonymized speech exists, and worse, a rotation is enough to estimate this transformation.

Finishing our invertibility attack analysis on Table~\localref{tab:results_algn_fully_vq} with our fully quantization-based anonymization pipeline.
We observe that for most linkability attacks, the \acrshort{eer} stays above 30.0\%.
The oracle Procrustes is the best with an \acrshort{eer} of 29.5\%.
However, this is still far from the \acrshort{eer} obtained with the \acrshort{plda}-based white-box \acrshort{asv} (see Table~\ref{main:chapt5:bigtable_wav2vec_vq} of Chapter~\ref{main:chapt5}), where the \acrshort{eer} was 17.5\%.
The interesting results are more about the invertibility, where for this anonymization pipeline the oracle scenarios do not achieve 100\% accuracy.
Seeing a reduction in invertibility attack confirms the privacy improvement that this pipeline proposes compared to the \acrshort{vpc} baseline presented above.
However, 85\% of accuracy is still too much.
For the non-oracle scenarios, the inverting accuracy is drastically reduced compared to the \acrshort{vpc} baseline, with more than 2x privacy improvement which is correlated with the linkability improvement.
Overall, as we are seeing \acrshort{asv} linkability attack performance decrease, invertibility attacks performance is also decreasing.

\begin{table*}[!htb]
    \caption{Experimental results for the rotation-based invertibility attack scenarios on our fully quantization-based anonymization. The scoring function for \acrshort{asv} assessment is cosine similarity.}
      \vspace{0.2cm}
    \centering
    \begin{tabular}{r
      l@{}@{\extracolsep{8.0mm}}
      S[table-format=2.1]@{\extracolsep{7.0mm}}
      S[table-format=2.1]@{\extracolsep{10.0mm}}
      S[table-format=2.1]@{\extracolsep{7.0mm}}
      S[table-format=2.1]@{\extracolsep{3.0mm}}
      }
    \toprule
         \multicolumn{2}{c}{} &  \multicolumn{2}{c}{\hspace{2mm}\acrshort{eer}~$\uparrow$}& \multicolumn{2}{c}{\hspace{2mm}ACC~$\downarrow$} \\
         \multicolumn{2}{c}{} &  F & M & F & M \\
    \midrule
         1 & Clear speech       & 10.3   & 2.9    &           &       \\
         2 & White-box \acrshort{asv}   &         37.8   & 36.4   &           &       \\
    \midrule
         3 & \multirow{1}{*}{Procrustes}   &         32.3   &  35.0  & 32.7   &  24.7  \\
         4 & \multirow{1}{*}{Wasserstein Procrustes} &    37.9   &  36.7   &  23.7   & 15.9  \\
    \midrule
         5 & Oracle Procrustes &  29.6 & 29.4  &  85.3   &  84.1 \\
         6 & Oracle Wasserstein Procrustes  & 37.8 & 33.4 &  87.8   &  82.7 \\
    \bottomrule
    \end{tabular}
    \locallabel{tab:results_algn_fully_vq}
\vspace{-0.8em}
\end{table*}

\subsection{Discussion}

In this section, we introduced a new form of attack and privacy evaluation based on the capability of an attacker to reconstruct the clear x-vector from anonymized speech.
Our proposed inversion attack is based on estimating a rotation matrix that transforms anonymized x-vectors to clear x-vectors.
Our results show that the current anonymization approaches are susceptible to this form of attack, reinforcing the fact that there is room for improvement in current speaker anonymization systems.
The oracle attack scenarios seem to be an interesting framework to evaluate the robustness of future anonymization methods against re-identification attacks as it considers the most powerful attacks.
The oracle attack scenarios even if they are unrealistic in terms of knowledge acquisition, allow us to better challenge the \textit{non-invertibility} privacy criteria, as it directly challenges the definition itself \say{it should be computationally infeasible to obtain the clear data that led to any given anonymized data}.

The simplistic way we implemented our attack by estimating rotation matrices open the possibilities for more sophisticated inversion attacks.
The following is a list of improvements that could be made.
First, in our experiments, the use of a \acrshort{plda} scoring instead of a cosine similarity scoring function could improve the linkability results we obtained.
Alternatively, a more recent x-vector model trained with more advanced architecture and cost function can also help.
Then, to improve the x-vectors transformation from anonymized to clear, the use of other methods such as normalizing flows network \cite{Kobyzev2020NormalizingFA} might also help.
Also, training the inversion attack on more data (e.g., \textit{LibriSpeech train-clean-360}) could help for the non-oracle attack scenarios.
Finally, the way we record inversion success with accuracy is limited, another metric that better reflects the actual distance from the target x-vector could be interesting.
But of course, the better would be to use the inverted x-vector for speech synthesis and compare the de-anonymized voice to the clear voice subjectively.

\vspace{\fill}
\pagebreak 

\section{Subjective mispronunciation evaluation}
In this section, we propose an early prototype to properly evaluate the main utility requirement of speaker anonymization systems: linguistic content preservation.
The way it is currently evaluated with an \acrshort{asr} system that outputs a \acrshort{wer} score on anonymized speech has one main limitation.
The \acrshort{wer} score reflects the errors of the \acrshort{asr} evaluation system to recognize correctly pronounced words, and the errors of the anonymization pipelines to correctly pronounce words.
To accommodate for this, a \acrshort{wer} reference score is computed on clear speech, and the utility performance of anonymized speech is then compared to this reference score.
Any degradation is supposed to reflect the anonymization pipeline making errors to correctly pronounce words.
The main issue with this approach comes from the fact that the most powerful anonymization pipelines rely on some sort of linguistic feature extraction before synthesizing anonymized speech.
We noticed that the linguistic feature extractors are usually better at \acrshort{asr} task than the \acrshort{asr} evaluation model itself because they are trained on more data.
For instance, in the \acrshort{vpc} plan, the linguistic feature extractor is trained with almost two times more data than the \acrshort{asr} evaluation systems.
Overall, it has been shown that the larger the corpora used to train the linguistic feature extractor, the better the anonymization system generalizes.
In Chapter~\ref{main:chapt5}, the \acrshort{wer} is usually improved after anonymization, indicating that the \acrshort{asr} evaluation models makes fewer errors in recognizing properly pronounced words.
But what about the mispronunciation errors that the anonymization pipelines make?

\subsection{Real world example}
Below, we show examples of \acrshort{wer} computation for clear speech in Figure~\localref{vb:clear}, the \acrshort{vpc} baseline anonymized speech in Figure~\localref{vb:vpc}, and our fully quantization-based anonymized speech in Figure~\localref{vb:vq}.
Here, we are interested in knowing whether the \acrshort{asr} mistakes come from the \acrshort{asr} decoding or mispronunciation errors coming from the anonymization.

\renewcommand*{\thefootnote}{\fnsymbol{footnote}}
\begin{figure}[htbp]
    \begin{center}
\begin{Verbatim}[fontsize=\footnotesize]
               ref    THE  RAINBOW   IS      A     DIVISION  OF  WHITE  LIGHT
               hyp    THE    ***    ***  RAINBOWS  DIVISION  OF  WHITE  LIGHT  [...]
               op      C      D      D       S         C      C    C      C
\end{Verbatim}
    \end{center}
    \vspace{-6mm}
    \caption{
        \textattachfileandprintoutt{\subfix{wavs/clear-p279-007-mic2.wav}}{Clear speech}\protect\footnotemark[4] \acrshort{wer} scoring with the \acrshort{asr} evaluation system.
        {\lstinline[basicstyle=\ttfamily]+ref+} indicates the ground truth, {\lstinline[basicstyle=\ttfamily]+hyp+} the output of the \acrshort{asr}, and {\lstinline[basicstyle=\ttfamily]+op+} the {\lstinline[basicstyle=\ttfamily]+S+}ubstitution, {\lstinline[basicstyle=\ttfamily]+D+}eletion, and {\lstinline[basicstyle=\ttfamily]+I+}nsertion errors or {\lstinline[basicstyle=\ttfamily]+C+}orrectly decoded words.
    }
    \locallabel{vb:clear}
    \vspace{-0.8em}
\end{figure}
\renewcommand*{\thefootnote}{\fnsymbol{footnote}}
\footnotetext[4]{Audio samples can be extracted from the PDF by clicking (or double-clicking) the text in blue.}
\setcounter{footnote}{0}
\renewcommand*{\thefootnote}{\arabic{footnote}}

In Figure~\localref{vb:clear} on clear speech, the \acrshort{asr} makes three errors (two deletions followed by a substitution) in the {\lstinline[basicstyle=\ttfamily]+RAIMBOW IS A+} segment.
Upon listening to the audio file, it appears that the reference transcript matches what has been said (even though the speech is fast-paced in this specific segment).

\begin{figure}[htbp]
    \begin{center}
\begin{Verbatim}[fontsize=\footnotesize]
                 ref    THE  RAINBOW  IS  A  DIVISION  OF  WHITE   LIGHT  INTO
                 hyp    THE  RAINBOW  IS  A  DIVISION  OF  WHITE  LICHENS  OF   [...]
                 op      C      C      C  C      C      C    C       S      S
\end{Verbatim}
    \end{center}
    \vspace{-6mm}
    \caption{
        \textattachfileandprintoutt{\subfix{wavs/vpc-anon-p279-007-mic2.wav}}{Speech anonymized with the VPC 2022 baseline} \acrshort{wer} scoring with the \acrshort{asr} evaluation system.
        {\lstinline[basicstyle=\ttfamily]+ref+} indicates the ground truth, {\lstinline[basicstyle=\ttfamily]+hyp+} the output of the \acrshort{asr}, and {\lstinline[basicstyle=\ttfamily]+op+} the {\lstinline[basicstyle=\ttfamily]+S+}ubstitution, {\lstinline[basicstyle=\ttfamily]+D+}eletion, and {\lstinline[basicstyle=\ttfamily]+I+}nsertion errors or {\lstinline[basicstyle=\ttfamily]+C+}orrectly decoded words.
    }
    \locallabel{vb:vpc}
    \vspace{-0.8em}
\end{figure}

In Figure~\localref{vb:vpc} on speech anonymized with the \acrshort{vpc} baseline, the \acrshort{asr} does not make the same mistakes.
The {\lstinline[basicstyle=\ttfamily]+RAIMBOW IS A+} segment is correctly decoded, however, an error occurs for {\lstinline[basicstyle=\ttfamily]+LIGHT INTO+} segment.
Upon listening to the audio file, this error is an \acrshort{asr} decoding error because the segment is correctly pronounced.
If we were to evaluate the utility of the anonymized speech against the clear speech using the \acrshort{wer} metric, the anonymized speech would be considered better.
However, in practice, with subjective listening tests, both anonymized and clear speech might have the same level of preservation of linguistic content.

\begin{figure}[htbp]
    \begin{center}
\begin{Verbatim}[fontsize=\footnotesize]
                 ref    THE  RAINBOW     IS    A  DIVISION  OF  WHITE  LIGHT
                 hyp    THE    RAIN   WITHOUT  A  DIVISION  OF  WHITE  LIGHT  [...]
                 op      C      S        S     C      C      C    C      C
\end{Verbatim}
    \end{center}
    \vspace{-6mm}
    \caption{
        \textattachfileandprintoutt{\subfix{wavs/vq-anon-p279-007-mic2.wav}}{Speech anonymized with our quantization-based anonymization pipeline} \acrshort{wer} scoring with the \acrshort{asr} evaluation system.
        {\lstinline[basicstyle=\ttfamily]+ref+} indicates the ground truth, {\lstinline[basicstyle=\ttfamily]+hyp+} the output of the \acrshort{asr}, and {\lstinline[basicstyle=\ttfamily]+op+} the {\lstinline[basicstyle=\ttfamily]+S+}ubstitution, {\lstinline[basicstyle=\ttfamily]+D+}eletion, and {\lstinline[basicstyle=\ttfamily]+I+}nsertion errors or {\lstinline[basicstyle=\ttfamily]+C+}orrectly decoded words.
    }
    \locallabel{vb:vq}
    \vspace{-0.8em}
\end{figure}

In Figure~\localref{vb:vq} on speech anonymized with our fully quantization-based pipeline, the \acrshort{asr} shows mistakes in the {\lstinline[basicstyle=\ttfamily]+RAINBOW IS+} segment.
Upon listening, this error comes from the anonymization pipeline, the segment is not properly pronounced.
In Chapter~\ref{main:chapt6} we concluded that this pipeline has better or similar utility as clear speech when using the \acrshort{wer} metric for utility comparison.
However, we believe that with a mispronunciation evaluation, the utility would be decreased instead of being, in the current utility evaluation, better or similar.

\vspace{-1mm}
\subsection{Mispronunciation metric proposition}
\vspace{0mm}
As we believe anonymized speech utility assessment should take into account mispronunciation errors we propose a mispronunciation metric.
To compute such a metric we would use the decoded \acrshort{asr} results and ask listeners to annotate if errors in each word come from a mispronunciation on both anonymized and clear speech.
If the error associated with a word does not come from a mispronunciation, the error ({\lstinline[basicstyle=\ttfamily]+S+}ubstitution, {\lstinline[basicstyle=\ttfamily]+D+}eletion, or {\lstinline[basicstyle=\ttfamily]+I+}nsertion) is invalidated and replaced with a {\lstinline[basicstyle=\ttfamily]+C+}orrect pronunciation label.
Then the \acrshort{wer} is calculated, with {\lstinline[basicstyle=\ttfamily]+S+}ubstitution, {\lstinline[basicstyle=\ttfamily]+D+}eletion, or {\lstinline[basicstyle=\ttfamily]+I+}nsertion errors only reflecting the actual preservation of the linguistic content as only mispronunciations errors are taken into account.
Utilizing human listeners for this form of measurement can be quite expensive. Nonetheless, the initial error detection performed by the \acrshort{asr} can serve as a useful first step in identifying errors, which can significantly decrease the total costs associated with subsequent subjective evaluations. Human listeners would know where to look for mispronunciations in a speech signal, sparing them the need to annotate everything from scratch. Furthermore, identifying audio files that are more prone to generating mispronunciations and analyzing only those files can also facilitate the evaluation process.

\subsection{Discussion}

In this section, we identified that automatic linguistic content preservation based on the \acrshort{wer} score output by an \acrshort{asr} system might lead to an imprecise evaluation.
As the current use of the \acrshort{wer} score reflects both decoding and mispronunciations errors, it is complicated to conclude if a particular anonymization system reduces decoding errors so much that the introduced mispronunciations errors are insignificant.
We argue that isolating only the mispronunciations to evaluate linguistic content preservation is the best way to evaluate utility.
With the help of the initial \acrshort{wer} score computation, we believe a subjective evaluation can isolate the actual errors we want to measure.

To conclude this section, we give an overview of the current anonymization processes, with a focus on listing them from the least to the most likely to result in mispronunciations.
This will be accompanied by our subjective comments on each pipeline.
\begin{enumerate}
        \item Signal processing anonymizations such as \cite{patino21_interspeech} are unlikely to introduce significant mispronunciation errors, as the transformation are usually non-destructive.
        \item Voice conversion anonymizations with non-disentangled features such as the baseline of the \acrshort{vpc} \cite{fangSpeakerAnonymizationUsing2019} are likely to introduce small mispronunciation errors because the synthesis system can make errors even though the \acrshort{asrbn} can theoretically encode every possible sound.
        \item Voice conversion anonymizations with disentangled features such as fully noised or fully quantized pipelines \cite{dp_vpc,pierre22_interspeech} are likely to introduce more mispronunciation errors because the \acrshort{asrbn} can no longer encode every possible sound.
        \item Voice conversion anonymizations with discrete tokens such as phonemes or words representations \cite{meyer22b_interspeech} are likely to introduce a lot of mispronunciation errors or even completely miss entire words because the linguistic feature extractor does not generate a one-on-one mapping with the real acoustic input.
\end{enumerate}
Interestingly, this ranking from the least mispronunciation errors making systems to the most is reversed when the criteria is privacy, the voice conversion anonymizations with discrete tokens being the best against linkability attacks, and signal processing anonymizations the worse.
It is also worth noting that the further we go down on the list, the more amount of training data is required for the model to generalize well.
Making them less suitable when the goal is actually to collect training data.
Overall we believe a good trade-off between utility/privacy and the amount of training data required to train the anonymization models can come out of systems in the middle of the list.

\section{Conclusion}

In this chapter, we challenged the \textit{non-invertibility} privacy criteria that speaker anonymization should comply with.
We proposed an invertibility attack based on embedding alignment techniques to invert anonymized x-vectors and reconstruct as best as possible their corresponding clear ones.
The results show that the successfulness of linkability attacks is correlated with the successfulness of our invertibility attacks.
As the main privacy evaluation of anonymization systems already relies on linkability, our invertibility evaluation does not need to be as crucial.
Even though the oracle attack is quite interesting from a formal point of view, it theoretically evaluates if the bijection (if there is one) between anonymized and clear x-vectors space corresponds to a rotation.
The best anonymization systems should not create bijections that can be estimated by rotations.

This chapter also questioned the method used to measure the preservation of linguistic content after anonymization.
We identified from a conceptual point of view that comparing \acrshort{wer} scores outputted by \acrshort{asr} evaluation models has major limitations restricting it from properly evaluating the level of linguistic content preservation.
We proposed to enhance the computation of the \acrshort{wer} to only take into account mispronunciation errors discarding inherent \acrshort{asr} decoding errors.
This allows to isolate and measure the errors made by the anonymization systems and have a utility metric that better reflects what we want to measure.
The only disadvantage of this approach is that it requires subjective human listening.

\ifSubfilesClassLoaded{
    \printglossary[title=Special Terms,type=\acronymtype]
    \printbibliography
}{}

\end{document}

\clearemptydoublepage
\backmatter

\cleartooddpage[\thispagestyle{empty}]
\documentclass[../main.tex]{subfiles}

\ifSubfilesClassLoaded{
    \tableofcontentsfile
    \dominitoc
    \setcounter{chapter}{6} 
    \def\locallabelprefix{conclusion}
    \externaldocument[]{../main}
}{}

\begin{document}

\selectlanguage{english}

\graphicspath{{./figures/dist}}

\clearemptydoublepage
\cleartooddpage[\thispagestyle{empty}]

\chapter*{Conclusion and Perspectives}
\addcontentsline{toc}{part}{Conclusion and perspectives}


\vspace{-2em}

The objective of this thesis is to propose privacy-preserving speech data collection solutions that rely on data anonymization.
The motivation behind this research is the increasing concern about privacy and security of speech data, especially with the widespread use of speech technology.
This thesis investigated the problems associated with the evaluation and design of speaker anonymization systems.
The goal is to develop speaker anonymization methods that can remove speaker identity from speech signals while preserving linguistic content and speech quality.

\vspace{-1.2em}
\section*{Summary}
\vspace{-0.8em}

In Chapter \ref{main:chapt4}, we assessed the role of the target speaker identity parameter in voice conversion-based anonymization.
For this analysis, we made the hypothesis that the voice conversion system is completely capable to remove the clear source identity and replace it with the one of a target.
Meaning, that the speaker anonymization is perfect.
Under this hypothesis, we evaluated many target speaker identity selection algorithms that are used to parameterize the voice conversion model.
We observed that most of the \acrshort{vc} parameterization used in today's speaker anonymization evaluation protocol and challenges create a link between the clear and target speakers which creates a strong bias for privacy linkability assessment.
Additionally, as an \acrshort{asv} attacker model is trained on anonymized speech to evaluate linkability, the way the attacker generates anonymized data with the target parameter matters.
From this observation, we conclude that the best target selection algorithm for privacy evaluation is the one where all source speakers aim to be converted to a single target identity or completely random identities.
With this voice conversion parametrization, there is a guarantee to generate unbiased anonymized speech where linkability assessment is correct, and where the \acrshort{asv} attacker model can be properly trained.

In the second part of this chapter, we queried if the target identity parameter affects the capability of the voice conversion model to generate anonymized voice.
In particular, we wanted to know if there is a {golden} target speaker parameter that maximizes privacy and/or utility performances.
Our conclusion showed that there is no particular target identity parameter that will improve privacy.
However, there was clear evidence that some target identity parameters were highly destructive of the utility.
We recommend manually selecting target identities that lead to good utility performance.

In Chapter \ref{main:chapt5}, we analyzed the effectiveness of a voice conversion system to anonymize speech.
The three main features used for voice conversion are the target identity studied in Chapter~\ref{main:chapt4}, the linguistic representations and \acrshort{f0} are studied in this chapter.
In order for a voice conversion system to replace a source identity with a target one, the features must be as disentangled as possible.
In this chapter, we analyzed the degree of disentanglement of the \acrshort{f0} and linguistic representations in order to assess if they contain speaker information.
It appeared that both were not freed of speaker information which might limit the performance of the anonymization.
The linguistic representation appeared to be the one containing the more speaker information.

We experimented with adversarial training to better disentangle the linguistic representation.
Here the goal is to add a negative loss when training the linguistic representation extractor to induce it to remove speaker information.
Counter-intuitively, it appears that adversarial training does not improve the privacy of anonymized speech, indicating that the linguistic representation encoded the same amount of speaker information as before.
However, we saw that the representation seemed to be more appropriate for speech synthesis as we observed an improvement in utility.

Another well-established method to extract privacy-preserving representation is to add noise to the representation.
The noise needs to be applied during training such that the model can learn to properly encode the linguistic representation in a way that only the most relevant information stands out from the noise in a transmission channel.
We reproduced this framework and showed that noise addition is very sensible to the source data and can badly impact the utility of the anonymized speech signal.

As an alternative, we proposed to use of a vector quantization technique to improve the disentanglement of the linguistic representations.
Vector quantization can improve the privacy of the representation by reducing the transmission capacity so the irrelevant information (speaker identity) cannot be encoded.
Similarly, as for noise addition, vector quantization needs to be applied during the training of the model.
The results showed that the use of vector quantization is a promising alternative to noise-based transformations, overall generating higher-quality speech, while having similar privacy properties.
The \acrshort{f0} feature can also benefit from vector quantization, indeed, our fully quantization-based anonymization pipeline showed the best performance.

In Chapter \ref{main:chapt6}, we switched from the role of playing the designer of speaker anonymization systems to the role of playing the attacker.
We introduced the first invertibility attack of a speaker anonymization system.
This invertibility attack aims to reconstruct or decrypt the anonymized speech to obtain the x-vector of the speakers to match the ones extracted on clear speech.
The method used for this attacker relies on techniques for aligning two sets of speaker embedding (a clear one, and an anonymized one).
In particular, we used Procrustes to estimate a rotation matrix that transforms an x-vector extracted from anonymized speech to the one extracted from clear speech.
Our results show that if the attacker has an unrealistic amount of information, a rotation matrix is enough to perform the inversion attack.
We must emphasize that under realistic scenarios, the degree of rotation-based inversion is much lower.
However, the only possibility that a rotation is enough to invert an anonymization system is alarming, as even if today's realistic scenarios are not successful, tomorrow's attack may get close to today's unrealistic scenario.

Lastly in this chapter, we briefly presented the current limitation of the protocol used to evaluate the preservation of the linguistic content.
We propose to perform subjective listening tests to assess if an \acrshort{asr} error comes from the \acrshort{asr} inability to decode the correctly preserved linguistic content or of the anonymization system to pronounce the linguistic content.
This would operate by utilizing human listeners to categorize whether a decoding substitution, deletion, or insertion error was produced by the \acrshort{asr} system or a mispronunciation (anonymization error).
Therefore, the \acrshort{wer} metric only informs of mispronunciation errors, providing an accurate reflection of the utility compared to clear speech, where the same evaluation procedure should be used to obtain the reference clear \acrshort{wer} score.

\makeatletter
\setlength{\bibhang}{1em}
\newlength{\bibsep}
 {\@listi \global\bibsep\itemsep \global\advance\bibsep by\parsep}
\newlist{bibsection}{itemize}{3}
\setlist[bibsection]{label=,leftmargin=\bibhang,%
        itemindent=-\bibhang,
        itemsep=\bibsep,parsep=\z@,partopsep=0pt,
        topsep=0pt}
\newlist{bibenum}{enumerate}{3}
\setlist[bibenum]{label=[\arabic*],resume,leftmargin={\bibhang+\widthof{[999]}},%
        itemindent=-\bibhang,
        itemsep=\bibsep,parsep=\z@,partopsep=0pt,
        topsep=0pt}
\let\oldendbibenum\endbibenum
\def\endbibenum{\oldendbibenum\vspace{-.6\baselineskip}}
\let\oldendbibsection\endbibsection
\def\endbibsection{\oldendbibsection\vspace{-.6\baselineskip}}
\makeatother

\vspace{-1.2em}
\section*{Perspectives and future directions}
\vspace{-0.8em}

The primary extension in the field of speaker anonymization that we believe would have the most impact would be to use other datasets.
We restricted ourselves to the ones defined in the \acrlong{vpc}, where the speech test data for which anonymization was applied is very clean (\textit{LibriSpeech test-clean}, and \textit{VCTK test}).
We believe that up until now, the usage of clean data facilitated the early development of this new field, gathering a new community.
However, as our understanding of the key challenges of speaker anonymization in terms of evaluation and design evolves, the datasets used should also evolve.
Using data that are closer to real-world application cases, (i.e., call center, voice assistant), would allow for providing a better assessment of the current effectiveness of the technology.

Another extension also relates to the datasets, to train the \acrshort{asv} model, that performs the linkability assessment.
At the moment the \textit{LibriSpeech train-clean-360} dataset is used.
However, by the nature of the dataset (audiobook), a lot of other information may be available in such a way that it helps linkability attacks.
For example, across audiobook sections or chapters, the actor might stay consistent with the way he/she narrates the story.
This raises questions about what the \acrshort{asv} attack model captures.
Are we training an automatic speaker verification system or an automatic narration-style verification system? (or even story verification?).
As anonymization techniques will get stronger and stronger, the \acrshort{asv} model is likely to focus on biased attributes to perform linkability.
We believe it is currently happening as the attacker used in the thesis is always better for \textit{LibriSpeech test-clean} dataset (i.e., 23\% of \acrshort{eer} for \textit{LibriSpeech test-clean} and 39.8\% for \textit{VCTK test} privacy results for our fully-quantization-based anonymization).
The preliminary test that we run by training the \acrshort{asv} attack model on a dataset more suited for speaker recognition (Voxceleb1) showed that the privacy protection for the \textit{VCTK test} dataset is overestimated (updated privacy score of 26.8\% of \acrshort{eer}), while the privacy protection estimation on \textit{LibriSpeech test-clean} could be biased (updated privacy score of  31.0\% of \acrshort{eer}).
Overall, knowing whether some techniques achieved better anonymization or knowing if the attack model is biased is an inherent problem in this field.

\vspace{\fill}
\pagebreak 

For privacy evaluation, this field should stay informed as much as possible with the advancement done in speaker recognition.
As of today, the \acrshort{asv} architecture/loss that we use has five years of delay.
We maintained the same \acrshort{asv} model for the purpose of comparison in this thesis, however, we believe it is necessary to adopt a more advanced \acrshort{asv} model for future advancements.
Additionally, new forms of attacks should also be designed.
Over the last few years, there has been a great number of contributions regarding developing new anonymization systems but only a very few contributions tackled the development of new attacks.
We believe that the best attacks are likely to come from the ones who understand speaker anonymization the best, that is to say, the people designing anonymization techniques.

Regarding utility evaluation, a perspective would be to use the mispronunciation \acrshort{wer} metric that we defined in Chapter~\ref{main:chapt6} to assess the preservation of the linguistic content.
We believe this form of evaluation would be fairer to compare multiple systems against each other.
Additionally, depending on the application, other utility metrics could be considered.
With Hubert Nourtel and Marie Tahon, we worked on the emotion attribute \cite{spsc,hubert_jeps} which may be important for the speech of call centers.

Another evaluation extension would be to query the ability of the speaker anonymization to work with multiple languages.
In the current state, we observe that the more private the anonymization system is, the less it works with multiple languages.
Speaker anonymization has to be designed for a specific language.
We do not necessarily consider this an issue as long as the amount of supervised data required to train a speaker anonymization system in a given language is relatively low.
Hence, investigating data-efficient model training is of great interest.

Finally, we present future directions that relate to improving the performance of the anonymization systems.

Currently, the trade-off between privacy and the utility of the systems is fixed for a given model.
Being able to dynamically adjust the level of transformation, for specific utterances or frames, is likely to improve both privacy and utility performance without requiring a complete paradigm shift.
The level of transformation applied could vary dynamically throughout the utterance, rather than being fixed, in response to the fluctuating level of confidence of \acrshort{asrbn} in correctly capturing the acoustic sound.
This approach would allow for more effective adaptation to changing conditions and increase the overall privacy/utility of the system.
To further elaborate, a dynamic adjustment of the transformation level could be achieved by utilizing a feedback loop that constantly evaluates the performance of the \acrshort{asrbn} model and adjusts the transformation accordingly.

Given a model that can dynamically adjust the transformation, being able to adapt the transformation to a speaker/group of speakers could also enable the creation of a fairer algorithm.
For example, this could be done in the vector quantization framework where in order to select the prototypes, other functions than the L2 distance could be used.

For the adversarial training approach, potential extensions of this research include incorporating a triplet loss for the adversarial model to more effectively simulate unlinkability attacks, and utilizing multiple adversarial models to counteract the tendency of the \acrshort{asrbn} models to only deceive a single adversary.

Another extension that could improve privacy, would be to train speech synthesis models to generate anonymized speech.
Currently, the speech synthesis model is parameterized and trained to reconstruct a given clear speech of a speaker from the training dataset.
This means that the anonymization transformation has to occur before intermediate representations.
Using anonymization systems that generate a small amount of mispronunciations errors to anonymize the training dataset used for speech synthesis training, and training another anonymization on that dataset could help to induce the speech synthesis model to also anonymize speech.
This step could be performed many times as long as the utility is not too impacted.

\vspace{1em}
For our final words, we want to disclose that although it is rewarding to create an anonymization system that improves privacy and or utility performances in today's evaluation protocol, this field desperately needs more understandability and explainability regarding privacy and utility evaluation.

\vspace{-1.2em}
\section*{Publication list}
\vspace{-0.7em}

\noindent
The work carried out during this thesis led to the following publications:
\vspace{0.4em}

\begin{bibenum}

    \item \cite{champion:hal-02995855}. Speaker information modification in the VoicePrivacy 2020 toolchain. In: \emph{VoicePrivacy 2020 Virtual Workshop at Odyssey 2020}.
      
        \vspace{-1em}
    \item \cite{F0_mod_moi}. A Study of F0 Modification for X-Vector Based Speech Pseudonymization Across Gender. In: \emph{The Second AAAI Workshop on Privacy-Preserving Artificial Intelligence}.
    
        \vspace{-1em}
    \item \cite{moi_specom}. Evaluating X-vector-based Speaker Anonymization under White-box Assessment. In: \emph{23rd International Conference on Speech and Computer (SPECOM)}.
      
        \vspace{-1em}
    \item \cite{invertibility_asru}. On the invertibility of a voice privacy system using embedding alignment. In: \emph{IEEE Automatic Speech Recognition and Understanding Workshop (ASRU)}.

        \vspace{-1em}
    \item \cite{mine_vq_jeps}. Privacy-Preserving Speech Representation Learning using Vector Quantization. In: \emph{Journ{\'e}es d'{\'E}tudes sur la Parole (JEP, 34e {\'e}dition)}.

        \vspace{-1em}
    \item \cite{pierre22_interspeech}. Are disentangled representations all you need to build speaker anonymization systems? In: \emph{23rd Interspeech Conference}.
        
    
    \end{bibenum}
    \vspace{0.5em}
    \vspace{1em}

\noindent
Other secondary contributions, were also done during the same period:
\vspace{0.4em}

\begin{bibenum}
\item \cite{spsc}. Evaluation of Speaker Anonymization on Emotional Speech. In: \emph{1st ISCA Symposium on Security and Privacy in Speech Communication (SPSC)}.

    \vspace{-1em}
\item \cite{hubert_jeps}. Analyse de l'anonymisation du locuteur sur de la parole {\'e}motionnelle. In: \emph{Journ{\'e}es d'{\'E}tudes sur la Parole (JEP, 34e {\'e}dition)}.

    \vspace{-1em}
\item \cite{tomashenko2020voiceprivacy_eval2022}. The VoicePrivacy 2022 challenge evaluation plan.
\end{bibenum}

\ifSubfilesClassLoaded{
	\printglossary[title=Special Terms,type=\acronymtype]
	\printbibliography
}{}

\end{document}

\setcounter{tocdepth}{-1}

\clearemptydoublepage
\cleartooddpage[\thispagestyle{empty}]
\documentclass[../main.tex]{subfiles}

\ifSubfilesClassLoaded{
    \tableofcontentsfile
    \dominitoc
    \externaldocument[]{../main}
    \def\locallabelprefix{intro_long}
}{}

\begin{document}

\selectlanguage{french}


\graphicspath{{./figures/dist}}

\chapter*{Résumé étendu en francais}
\addcontentsline{toc}{part}{Résumé étendu en francais}

\section*{1.~~~Introduction}

La communication verbale, qu'il s'agisse de parler ou d'écouter, est un moyen naturel et pratique pour nous humains d'interagir.
Au cours des conversations, les individus échangent des informations linguistiques par la parole, mais ils transmettent également des informations supplémentaires à travers des signes paralinguistiques qui permettent de reconnaître l'identité du locuteur, mais aussi ses émotions, son âge et genre, etc.
L'interaction homme-machine peut tirer parti de la richesse et de la commodité de la parole, permettant ainsi à la machine de mieux comprendre les humains et aux humains de partager facilement des informations avec les machines.
Cette approche améliore considérablement l'expérience des utilisateurs et contribue à une utilisation plus inclusive des technologies modernes pour les personnes en situation de handicap.
Cependant, la compréhension et la production de parole par les machines restent des tâches complexes qui font l'objet de recherches continues depuis de nombreuses années.
Toutefois, au cours de la dernière décennie, d'importants progrès ont été réalisés dans plusieurs tâches liées à la parole, telles que la reconnaissance automatique de la parole (transcription automatique) et la synthèse de la parole (génération de parole à partir de texte), entre autres. Ces avancées ont introduit un nouveau produit sur le marché~: les assistants vocaux.

L'objectif de nombreuses entreprises est d'établir une expérience d'interaction naturelle et pratique entre les humains et les machines, alimentée par les derniers développements technologiques en matière de traitement de la parole.
Actuellement, les consommateurs adoptent de plus en plus les différents dispositifs d'assistants vocaux.
En effet, l'étude intitulée\citetitle{smart_home_adoption_rate} révèle que 50 à 60\% de la population américaine a accès à un ou plusieurs dispositifs d'assistants vocaux.

Afin de proposer des services compétitifs, les entreprises ont recours à des techniques avancées d'apprentissage profond pour alimenter les algorithmes des assistants vocaux.
Ces algorithmes étant très gourmands en données, les performances optimales sont généralement atteintes lorsqu'une grande quantité de données est utilisée pour les entraîner.
Ceci créé une nécessité pour les entreprises de collecter, traiter et stocker les données de parole de leurs utilisateurs sur des serveurs centralisés afin d'améliorer continuellement les services proposés et de rester compétitives.
Cependant, les données de parole contenant de nombreuses informations personnelles telles que l'identité du locuteur, leur collecte soulève de sérieuses préoccupations en matière de protection des données personnelles.

Au départ, les pratiques de collecte de données des entreprises étaient inconnues ou mal comprises du public, mais cela a récemment changé.
Aux alentours de 2017, de plus en plus de personnes ont pris conscience de la situation grâce à des gros titres de presse révélant l'utilisation des données vocales collectées par les entreprises\footnote{\citetitle{verge_data_listening} (Oui, des travailleurs humains écoutent les enregistrements de Google, y compris ceux enregistrés par erreur).}. Parallèlement, l'union européenne a établi la législation sur le traitement de données la plus stricte au monde.
Cette législation, la \cite{gdpr} ou Règlement Général sur la Protection des Données (RGPD), stipule spécifiquement que les informations personnelles des citoyens européens doivent être traitées avec conformité du droit au respect de la vie privée et avec la plus grande confidentialité.

Le droit à la vie privée est un élément juridique qui vise à protéger le respect de la vie privée des individus.
C'est la capacité de contrôler qui peut accéder aux informations nous concernant, et dans quel but elles sont utilisées.
Dans le cadre du \acrshort{gdpr}, le droit à la vie privée est considéré comme un droit fondamental de l'homme, et il est essentiel pour protéger les citoyens contre l'utilisation abusive ou détournée de leurs informations personnelles.
À l'ère du numérique, le droit à la vie privée revêt une importance primordiale, car l'utilisation croissante de la technologie permet de collecter et stocker des quantités massives de données personnelles.
Alors que la technologie continue d'évoluer, notre compréhension de son influence sur le droit à la vie privée et notre engagement à la protéger doivent également évoluer.

La parole est considérée comme un type de donnée personnelle hautement sensible qui doit être protégé.
Les directives récentes émises par la Commission Nationale de l'Informatique et des Libertés (CNIL) et le Conseil Européen de la Protection des Données rappellent que les données de la parole sont intrinsèquement des données biométriques, conformément à l'article 4(14) de la RGPD\footnote{L'article 4(14) du RGPD définit les données biométriques comme "les données à caractère personnel résultant d'un traitement technique spécifique relatif aux caractéristiques physiques, physiologiques ou comportementales d'une personne physique, qui permettent ou confirment l'identification unique de cette personne physique, telles que les images faciales ou les données dactyloscopies."} \cite{white-paper-cnil,edpb-voice-assistant-guidelines}.
Étant donné que le stockage et le traitement des données biométriques sont encore plus réglementés que celui des données personnelles, il est nécessaire de développer des schémas de collecte de données respectant mieux les informations privées.

Cette thèse propose des solutions de collecte de données de parole plus respectueuse en s'appuyant sur l'anonymisation des données.
L'anonymisation des données est le processus de suppression ou d'altération des informations d'identification personnelle des données pour protéger les données personnelles des individus.
Bien que le RGPD n'impose pas l'anonymisation systématique des données personnelles, l'anonymisation constitue une solution parmi d'autres permettant de traiter les données personnelles en conformité avec les droits et la vie privée des individus.
Dans cette thèse, nous travaillons sur des méthodes d'anonymisation de locuteur qui visent à supprimer l'identité du locuteur des signaux de parole tout en préservant le contenu linguistique et la qualité de la parole.

\subsection*{1.1~~~Cadre et objectifs}

Ces dernières années, les techniques d'anonymisation de la parole ont suscité un intérêt croissant suite à la publication du \acrlong{vpc}.
La plupart des approches reposent sur des systèmes de conversion de voix pour transformer l'identité du locuteur dans le signal clair en une autre identité dans le signal anonymisé.
L'objectif est que le signal anonymisé ne puisse plus être lié à l'identité réelle du locuteur.
Pour évaluer cet aspect, des techniques de vérification automatique du locuteur (ASV -  \textit{Automatic Speaker Verification}) sont utilisées sur la parole anonymisée pour évaluer son degré de liaison, plus il est faible, mieux c'est.
La Figure~\localref{long_res:anon_plot_link} illustre cet objectif.
En ce qui concerne la préservation du contenu linguistique (utilité), elle est évaluée à l'aide de la reconnaissance automatique de la parole.

\begin{figure}[htbp]
    \begin{center}
        \includegraphics[width=0.87\linewidth]{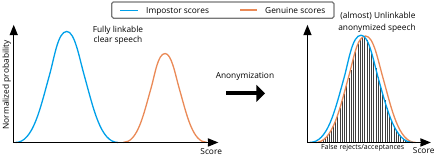}
    \end{center}
    \vspace{-4mm}
    \caption{
        Évaluation ASV de la capacité à lier un signal a un locuteur.
         Dans l'exemple, avant l'anonymisation (côté gauche), les signaux sont entièrement liables, les scores \textit{impostor} (locuteurs différents entre 2 signaux) et scores \textit{genuine} (même locuteur entre 2 signaux) sont disjoints.
         Après anonymisation (côté droit), les signaux doivent être faiblement liables, c'est-à-dire qu'il ne doit pas être possible de distinguer si deux signaux correspondent (ou pas) au même locuteur.
    }
    \locallabel{long_res:anon_plot_link}
    \vspace{-2mm}
\end{figure}

Le premier défi de l'anonymisation de la parole concerne l'évaluation de la bonne anonymisation des données personnelles (privacité).
Cette évaluation dépend du paramètre d'identité de locuteur cible de la conversion de voix.
En effet, la conversion de voix peut augmenter le dégré de liaison si elle est demandée en rendant la voix d'un locuteur très différente de celle des autres locuteurs.
Par conséquent, il est nécessaire de comprendre la conversion de voix de sorte que les voix anonymisées ne soient pas liables et que l'évaluation effectuée avec le système de vérification automatique de locuteur corresponde à l'évaluation de la privacité.

Le deuxième défi de l'anonymisation de la parole concerne l'algorithme de conversion de voix.
À l'origine, les systèmes de conversions de voix étaient conçus pour transformer le locuteur d'un signal, de sorte que les auditeurs subjectifs (humains) croient qu'un signal a été prononcé par quelqu'un d'autre.
Bien que la qualité de la transformation puisse être suffisante pour tromper les humains, ce n'est pas le cas pour les systèmes ASV.
Ainsi, des formes plus avancées de systèmes de conversion de voix doivent être utilisées pour l'anonymisation du locuteur.

Le troisième défi concerne la diversité et le nombre d'évaluations possibles pour évaluer la privacité et l'utilité.
En effet, il n'y a pas de méthode unique pour évaluer ces deux mesures.
Par exemple, pour la privacité un jeu impliquant un attaquant et un défenseur où l'attaquant vise à briser l'anonymisation et où le défenseur l'améliore doit avoir lieu afin d'adapter en continu la technologie aux menaces actuelles.
Quant à l'utilité, l'évaluation dépend toujours de l'objectif de partage de la parole.
Nous nous sommes concentrés sur la préservation du contenu linguistique, mais il existe de nombreux autres cas d'application valables, tels que la reconnaissance des émotions.

Ces trois défis ont pour objectif d'explorer l'évaluation et la conception de systèmes d'anonymisation du locuteur.
Ce résumé est organisé de la sorte pour présenter ces trois défis, les
parties~2, 3 et 4 présentent respectivement les contributions principales des chapitres~\ref{main:chapt4}, \ref{main:chapt5} et \ref{main:chapt6} la
partie 5 conclut ce résumé.

\vspace{-0.5em}
\section*{2.~~~Influence du paramétrage du locuteur cible}
\vspace{-0.5em}
Dans cette analyse, notre objectif est d'étudier l'influence du paramétrage du système de conversion de voix, avec les locuteurs cibles, sur l'évaluation de la privacité.
Pour ce faire, nous proposons de considérer que le système de conversion de voix est parfait, c'est-à-dire, qu'il est capable de remplacer totalement  l'identité du locuteur d'entrée avec l'identité choisie pour cible.
La capacité de supprimer ou remplacer les données personnelles relatives au locuteur dans un signal de parole est exactement ce qui est recherché pour l'anonymisation du locuteur.

Lors de l'évaluation de la privacité avec un système d'ASV (cf figure~\localref{long_res:anon_plot_link}), il est totalement possible que les voix anonymisées soient liables a des pseudo-identités qui correspondraient aux locuteurs cibles choisis pour chaque signal.
Cela peut poser un problème si un lien existe entre les pseudo-identités et les identités source.
Pour évaluer si un lien existe, nous proposons d'étudier les algorithmes de sélection de locuteurs cibles en vérifiant que les locuteurs cibles choisis par ceux-ci ne sont pas liables à l'identité source.

À la suite de nos expériences, nous observons que l'algorithme de sélection utilisé dans le plan d'évaluation du \acrshort{vpc} créé une telle sorte de liaison.
Ceci est mauvais, car lors de l'évaluation avec l'ASV, ce n'est plus la capacité d'évaluer le lien entre voix anonymisées et locuteur source qui est mesurée, mais le lien entre voix anonymisées et pseudo-locuteur.
De ce fait, évaluation est faussée en raison du paramétrage du système de conversion de voix.
Nous étudions d'autres alternatives d'algorithmes de sélection de locuteurs cibles qui ne créent pas de lien entre locuteur source et cible.
La conclusion est que l'algorithme qui consiste à sélectionner une seule cible vers laquelle tous les signaux doivent être convertis est plus simple et conduit à une meilleure anonymisation.

Dans ce cadre d'anonymisation/évaluation où un seul locuteur cible est utilisé pour convertir tous les signaux, nous analysons plusieurs cibles à la recherche de celle qui maximise les performances.
Nous concluons que le choix de la cible n'affecte pas la privacité mais qu'elle doit tout de même être choisie soigneusement, car un mauvais choix  peut fortement dégrader l'utilité.

\vspace{-0.5em}
\section*{3.~~Analyse des représentations utilisées pour la conversion de voix}
\vspace{-0.5em}
Au cours de cette étude notre objectif est de mesurer le degré de liaison des représentations utilisées dans le système de conversion de voix.
L'hypothèse est que si les représentations sont démêlées alors cela garanti que le système de synthèse remplace le locuteur source par par le locuteur cible.
Les représentations que nous étudions sont la fréquence fondamentale (F0) et la représentation linguistique obtenue à partir d'un modèle acoustique de reconnaissance de la parole (cf figure~\localref{long_res:image_chapt5:my_anon_model}).
Nous observons que le degré de liaison est élevé entre ces représentations et le locuteur, la représentation linguistique étant celle qui divulgue le plus d'information locuteur.

\begin{figure}[htbp]
  \begin{center}
    \includegraphics[width=0.75\linewidth]{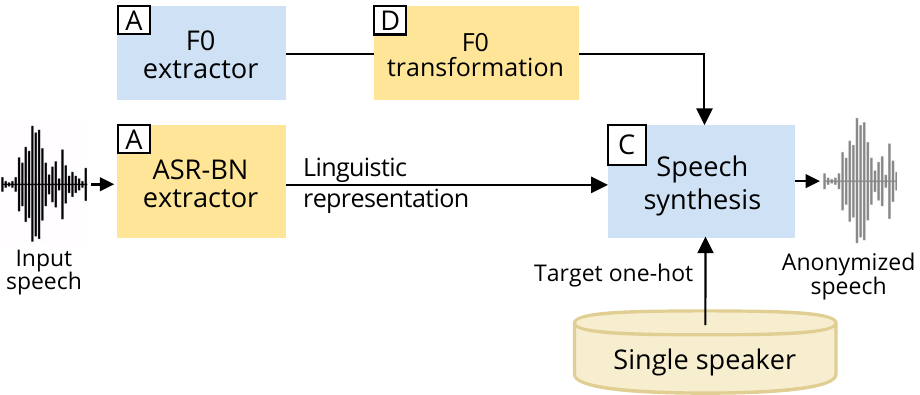}
  \end{center}
  \vspace{-5mm}
  \caption{
    Schéma représentatif d'un système de conversion de voix.
  }
  \locallabel{long_res:image_chapt5:my_anon_model}
\end{figure}

Afin d'améliorer le démêlement entre l'information locuteur et la F0 d'une part, et la représentation linguistique d'autre part, nous étudions plusieurs méthodes.
Parmi elles, l'une est basée sur l'ajout de bruit et est la méthode la plus utilisée dans le domaine de la \textit{differential privacy}. Cette méthode a été introduite pour l'anonymisation du locuteur par \cite{dp_vpc}.
L'autre méthode, que nous introduisons, est basée sur la quantification vectorielle.
L'anonymisation basée sur le bruit consiste à ajouter du bruit dans un canal de transmission de manière à ce que seules les informations les plus pertinentes ressortent du bruit, tandis que l'anonymisation basée sur la quantification consiste à réduire la capacité du canal de transmission de manière à ce que les informations non pertinentes (et personnelles) ne soient pas encodées.

Le résultat des évaluations montre que ces deux méthodes de démêlement ont des performances similaires en termes de privacité et d'utilité.
Cependant, nous observons que notre méthode basée sur la quantification est plus robuste que la méthode basée sur l'ajout de bruit, ayant de bonnes performances d'utilité à travers plusieurs jeux de tests.
Nous supposons que la quantification vectorielle généralise mieux sur les données bruitées.

\vspace{-0.5em}
\section*{4.~~Un nouveau type d'attaque et mesure d'utilité}
\vspace{-0.5em}
Dans cette partie, nous avons remis en question un autre critère de privacité auquel les systèmes d'anonymisation du locuteur doivent se conformer~: l'inversibilité.
L'inversibilité définit qu'il ne doit pas être computationnellement faisable de reconstruire la donnée source à partir de sa contrepartie anonymisée.
Afin d'évaluer ce critère, nous proposons une autre forme d'attaque basée sur des techniques d'alignement de vecteur (\textit{embedding}).
Les résultats montrent que le succès des attaques d'inversibilité est corrélé à celles des attaques de liaison.

Nous avons aussi remis en question la méthode utilisée pour mesurer la préservation du contenu linguistique après l'anonymisation.
Nous avons identifié d'un point de vue conceptuel que la comparaison des scores  (WER - \textit{Word Error Rate}) produits par les modèles d'évaluation de reconnaissance automatique de la parole présente des limitations majeures qui l'empêchent d'évaluer correctement le niveau de préservation du contenu linguistique.
Nous avons proposé d'améliorer le calcul du (WER - \textit{Word Error Rate}) pour ne prendre en compte que les erreurs de prononciation en éliminant les erreurs de décodage.
Cela permet d'isoler et de mesurer les erreurs faites par les systèmes d'anonymisation et d'obtenir une mesure d'utilité qui reflète mieux ce que nous voulons mesurer.
Le seul inconvénient de cette approche est qu'elle nécessite une écoute humaine subjective.

\vspace{-0.5em}
\section*{5.~~Conclusion}
\vspace{-0.5em}
L'objectif de cette thèse était de proposer des solutions de collecte de données vocales préservant la vie privée qui reposent sur l'anonymisation des données.
La motivation derrière cette recherche est l'inquiétude croissante quant à la collecte de données personnelles de parole, en particulier avec l'utilisation généralisée des nouvelles technologies.
Cette thèse a examiné les problèmes liés à l'évaluation et à la conception de systèmes d'anonymisation de locuteur.
L'objectif est de développer des méthodes d'anonymisation de locuteur capables de supprimer l'identité des locuteurs des signaux vocaux tout en préservant le contenu linguistique et la qualité de la parole.

Nos contributions sont diversifiées, traitant d'un point de vue conceptuel comment le paramétrage des systèmes d'anonymisation affecte l'évaluation de la privacité, comment améliorer les performances des systèmes d'anonymisation, et comment les évaluations basées à partir d'attaque ou non peuvent être étendues pour mieux refléter les performances des systèmes.
Pour conclure, nous tenons à souligner que bien qu'il soit gratifiant de créer un système d'anonymisation qui améliore la privacité et/ou l'utilité dans le protocole d'évaluation actuel, ce domaine a désespérément besoin d'une meilleure compréhension en ce qui concerne l'évaluation de la privacité et de l'utilité.

\ifSubfilesClassLoaded{
	\printglossary[title=Special Terms,type=\acronymtype]
	\printbibliography
}{}

\end{document}

\selectlanguage{english}

\clearemptydoublepage
\cleartooddpage[\thispagestyle{empty}]
\phantomsection 

\addcontentsline{toc}{part}{Bibliography} 
\printbibliography

\cleartoevenpage[\thispagestyle{empty}]
\phantomsection
\addcontentsline{toc}{part}{List of abbreviations}
\printglossary[title=List of abbreviations,type=\acronymtype]

\cleartoevenpage[\thispagestyle{empty}]
\documentclass[../main.tex]{subfiles}

\ifSubfilesClassLoaded{
    \tableofcontentsfile
    \dominitoc
    \externaldocument[]{../main}
    \def\locallabelprefix{resume}

    \graphicspath{../}
    
    \ecoledoctorale{MathSTIC}
    \etablissement{tulul}
}{
}

\begin{document}

\selectlanguage{english}

\markboth{}{}
\newgeometry{inner=30mm,outer=20mm,top=38mm,bottom=18mm}

\backcoverheader

\selectfontbackcover{ 


\phantomsection
\addcontentsline{toc}{part}{English abstract}

\titleEN{Anonymizing Speech: Evaluating and Designing Speaker Anonymization Techniques}

\keywordsEN{Speaker anonymization, Speech recognition, Speaker verification, Privacy.}

\abstractEN{
The growing use of voice user interfaces, from telephones to remote controls, automobiles, and digital assistants, has led to a surge in the collection and storage of speech data.
While data collection allows for the development of efficient tools powering most speech services, it also poses serious privacy issues for users as centralized storage makes private personal speech data vulnerable to cyber threats.
Advanced speech technologies, such as voice-cloning and personal attribute recognition, can be used to access and exploit sensitive information.
Voice-cloning technology allows an attacker to take a recording of a person's voice and use it to generate new speech that sounds like it is coming from that person.
For example, an attacker could use voice-cloning to impersonate a person's voice to gain unauthorized access to his/her financial information over the phone.
With the increasing use of voice-based digital assistants like Amazon's Alexa, Google's Assistant, and Apple's Siri, and with the increasing ease with which personal speech data can be collected and stored, the risk of malicious use of voice-cloning and speaker/gender/pathological/etc. recognition technologies have increased.
Companies and organizations need to consider these risks and implement appropriate measures to protect user data in order to prevent misuse of speech technologies and comply with legal regulations (e.g., General Data Protection Regulation (GDPR)).

To address these concerns, this thesis proposes solutions for anonymizing speech and evaluating the degree of the anonymization. In this work, anonymization refers to the process of making personal speech data unlinkable to an identity, while maintaining the usefulness (utility) of the speech signal (e.g., access to the linguistic content). The goal is to protect the privacy of individuals by removing or obscuring any Personally Identifiable Information (PPI) from the acoustic of speech. PPI includes things like a person's voice, accent, and speaking style; other personal information in the speech content like, phone number, person name, etc., is out of the scope of this thesis. 

Our research is built on top of existing anonymization methods based on voice conversion and existing evaluation protocols.
We start by identifying and explaining several challenges that evaluation protocols need to consider to evaluate the degree of privacy protection properly.
We clarify how anonymization systems need to be configured for evaluation purposes and highlight the fact that many practical deployment configurations do not permit privacy evaluation.
Furthermore, we study and examine the most common voice conversion-based anonymization system and identify its weak points, before suggesting new methods to overcome some limitations.
We isolate all components of the anonymization system to evaluate the degree of speaker PPI associated with each of them.
Then, we propose several transformation methods for each component to reduce as much as possible speaker PPI while maintaining utility.
We promote anonymization algorithms based on quantization-based transformation as an alternative to the most-used and well-known noise-based approach.
Finally, we endeavor a new attack method to invert the anonymization, creating a new threat.
In this thesis, we openly work on sharing anonymization systems and evaluation protocols to aid organizations in facilitating the preservation of privacy rights for individuals.

}

\newpage

\backcoverheader

\phantomsection
\addcontentsline{toc}{part}{French abstract}


\titleFR{Anonymisation de la parole~:~Évaluation et Conception de Techniques d'Anonymisation du Locuteur}

\keywordsFR{Anonymisation du locuteur, Reconnaissance de la parole, Vérification du locuteur, Respect vie privée.}

\abstractFR{
    L'essor de l'utilisation d'assistants vocaux, présents dans les téléphones, automobiles et autres, a augmenté la quantité de données de parole collectées et stockées.
Bien que cette collecte de données soit cruciale pour entrainer les modèles qui traitent la parole, cette collecte soulève également des préoccupations de protection de la vie privée.

Des technologies de pointe traitant la parole, telles que le clonage vocal et la reconnaissance d'attributs personnels (telles que l'identité, l'émotion, l'âge, le genre, etc.), peuvent être exploitées pour accéder et utiliser des informations personnelles.
Par exemple, un malfaiteur pourrait utiliser le clonage vocal pour se faire passer pour une autre personne afin d'obtenir un accès non autorisé à ses informations bancaires par téléphone.

Avec l'adoption croissante des assistants vocaux tels qu'Alexa, Google Assistant et Siri, et la facilité avec laquelle les données peuvent être collectées et stockées, le risque d'utilisation abusive de technologies telles que le clonage vocal et la reconnaissance d'attributs personnels augmente.
Il est donc important pour les entreprises et les organisations de prendre en compte ces risques et de mettre en place des mesures appropriées pour protéger les données des utilisateurs, en conformité avec les réglementations juridiques telles que le Règlement Général sur la Protection des Données (RGPD).

Pour répondre aux enjeux liés à la protection de la vie privée, cette thèse propose des solutions permettant d'anonymiser la parole.
L'anonymisation désigne ici le processus consistant à rendre les signaux de parole non associables à une identité spécifique, tout en préservant leur utilité, c'est-à-dire ne pas modifier le contenu linguistique du message.
L'objectif est de préserver la vie privée des individus en éliminant ou en rendant floues toutes les informations personnellement identifiables (PPI) contenues dans le signal acoustique, telles que l'accent ou le style de parole d'une personne.
Les informations linguistiques personnelles telles que numéros de téléphone ou noms de personnes ne font pas partie du champ d'étude de cette thèse.

Notre recherche s'appuie sur les méthodes d'anonymisation existantes basées sur la conversion de la voix et sur des protocoles d'évaluation existants.
Nous commençons par identifier et expliquer plusieurs défis auxquels les protocoles d'évaluation doivent faire face afin d'évaluer de manière précise le niveau de protection de la vie privée. 
Nous clarifions comment les systèmes d'anonymisation doivent être configurés pour être correctement évalués, en soulignant le fait que de nombreuses configurations ne permettent pas une évaluation adéquate de non-asociabilité d'un signal a une identité. 
Nous étudions et examinons également le système d'anonymisation basé sur la conversion de la voix le plus courant, identifions ses points faibles, et proposons de nouvelles méthodes pour en améliorer les performances.
Nous avons isolé tous les composants du système d'anonymisation afin d'évaluer le niveau de PPI encodé par chaque composant.
Ensuite, nous proposons plusieurs méthodes de transformation de ces composants dans le but de réduire autant que possible les PPI encodées, tout en maintenant l'utilité.
Nous promouvons les algorithmes d'anonymisation basés sur l'utilisation de la quantification en alternative à la méthode la plus utilisée et la plus connue basée sur le bruit.
Enfin, nous proposons une nouvelle méthode d'évaluation qui vise à inverser l'anonymisation, créant ainsi une nouvelle manière d'étudier les systèmes d'anonymisation.

}

}

\restoregeometry

\ifSubfilesClassLoaded{
}{}

\end{document}

\end{document}